\documentclass[a4paper,11pt]{article}
\pdfoutput=1 

\usepackage{jcappub} 

\usepackage[T1]{fontenc} 
\usepackage{graphicx,subcaption,multirow}
\usepackage{longtable}
\usepackage{dcolumn}
\usepackage{bm}
\usepackage[all]{xy}
\usepackage{color}
\usepackage{graphicx}
\usepackage[usenames,dvipsnames]{xcolor}
\usepackage{lscape}
\usepackage{tikz}
\usepackage{pgf}
\usepackage{booktabs}
\usepackage{arydshln}
\usepackage{placeins}

\DeclareUnicodeCharacter{2212}{\ensuremath{-}}

\definecolor{branchred}{RGB}{176,36,24}
\definecolor{branchgreen}{RGB}{74,156,52}
\definecolor{branchblue}{RGB}{48,112,173}

\newcommand{\beq}{\begin{equation}}
\newcommand{\eeq}{\end{equation}}
\newcommand{\bea}{\begin{eqnarray}}
\newcommand{\eea}{\end{eqnarray}}
\mathchardef\minus="002D

\newcommand{\MP}{M_p}

\newcommand{\exd}{{\rm d}}
\newcommand{\pref}[1]{{(\ref{#1})}}

\newcommand{\sax}{{\rm ax}}

\newcommand{\nn}{\nonumber}
\newcommand{\be}{\begin{equation}}
\newcommand{\ee}{\end{equation}}

\newcommand{\chimin}{\bar{\chi}_{\rm min}}
\newcommand{\chibf}{\bar{\chi}_{\rm bf}}

\newcommand{\mfa}{{\mathfrak a}}

\newcommand{\bfg}{{\mathbf{g}}}

\newcommand{\cH}{{\cal H}}

\newcommand{\cO}{{\cal O}}

\newcommand{\QCD}{{\scriptscriptstyle QCD}}

\newcommand{\roughly}[1]{\mathrel{\raise.3ex\hbox{$#1$\kern-0.85em
\lower1ex\hbox{$\sim$}}}}
\newcommand{\lsim}{\roughly<}

\newcommand{\ssB}{{\scriptscriptstyle B}}

\def\LCDM{$\Lambda \text{CDM}$}

\newcommand{\CLASS}{\texttt{CLASS}}

\newcommand{\FirstClass}{\texttt{21cmFirstCLASS}}

\usepackage{array}
\usepackage{tabularx}
\newcolumntype{C}[1]{>{\centering\arraybackslash}m{#1}}

\newcolumntype{Z}{>{\centering\arraybackslash}X}

\title{Enhanced Early Structure from Axiodilatons \& 21\,cm Observations}

\author[a,*]{Chan Youn Weon,}
\author[b,*]{Carsten van de Bruck,}
\author[c,d,e,*]{C.P. Burgess,}
\author[a,*]{Jeong Han Kim,}
\author[b,f,*]{and Adam Smith}
\note[*]{All authors contributed equally to this work.}

\affiliation[a]{Department of Physics, Chungbuk National University\\ Cheongju, Chungbuk 28644, Republic of Korea}

\affiliation[b]{School of Mathematical and Physical Sciences, University of Sheffield\\
Hounsfield Road, Sheffield S3 7RH, U.K.}

\affiliation[c]{Perimeter Institute for Theoretical Physics\\ 31 Caroline Street North, Waterloo ON, Canada}

\affiliation[d]{Department of Physics \& Astronomy, McMaster University\\ 1280 Main Street West, Hamilton ON, Canada}

\affiliation[e]{School of Theoretical Physics, Dublin Institute for Advanced Studies\\ 10 Burlington Rd., Dublin, Co. Dublin, Ireland}

\affiliation[f]{Department of Physics and Astronomy, University of Manchester\\ Manchester M13 9PL,
U.K.}

\emailAdd{chanyoun.weon@chungbuk.ac.kr}
\emailAdd{c.vandebruck@sheffield.ac.uk}
\emailAdd{cburgess@perimeterinstitute.ca}
\emailAdd{jeonghan.kim@cbu.ac.kr}
\emailAdd{asmith69@sheffield.ac.uk}

\begin{document}

\abstract{Axiodilatons provide a promising unified model of the Dark Sector whose properties are motivated by the use of accidental approximate symmetries to embed the light fields needed for observations into well-motivated UV physics in a technically natural way. These models turn out to help ameliorate many current puzzles in $\Lambda$CDM phenomenology ({\it e.g.} the Hubble tension and the evidence for evolving Dark Energy) `for free' despite not being designed for that purpose. We here explore in more detail another serendipitous feature of these models: their tendency to enhance early structure formation, in a way that seems suggestive given the accumulating evidence for structure arising earlier than expected in $\Lambda$CDM. Since early structure growth is the cleanest to study we explore in detail the expected 21\,cm signal produced by axiodilaton models, using the same parameter choices that improve agreement with observations elsewhere in cosmology. For fixed astrophysics, we forecast that SKA1-Low could distinguish dilaton couplings as small as a few times $10^{-2}$ from $\Lambda$CDM at $2\sigma$, mainly because of the earlier formation of structure. We provide complete derivations of the mass dependence of the relevant cosmological observables, including how dilaton-dependent baryon and electron masses affect the hyperfine transition, the Ly$\alpha$ coupling, X-ray heating and ionization during cosmic dawn and the epoch of reionization. We provide details of radiative transfer, atomic physics and the equations governing the thermal and ionization state of the intergalactic medium.
}

\maketitle

\flushbottom

\section{Introduction}
\label{sec:intro}

The only thing worse (for theorists) than a bad theory is a theory that is already invented and extremely good. Of late both particle physics and cosmology have been afflicted with very good theories indeed: the Core Theory (the Standard Model plus General Relativity) describing the known fundamental laws and the current best understanding of the universe we find around us ($\Lambda$CDM Cosmology). 

What is remarkable about these two theories is that they contradict one another: the things required for successful cosmology ({\it e.g.} Dark Matter, a technically natural small but nonzero Dark Energy and a source of primordial fluctuations) seem not to be included in our fundamental framework. But the disagreement does not yet provide enough information to suggest a definitive new fundamental approach. This tension has been the lifeblood of much of current research in cosmology: the predictions of $\Lambda$CDM are explored and tested in ever-increasing detail and many alternative models are proposed -- often motivated by the tensions {\it du jour}. When doing so one keeps in mind that such tensions are usually present at the few-sigma level in any precision comparison of data with theory involving multiple types of observations. 

The current list of cosmological tensions includes several that have proven to be quite persistent, such as
\begin{itemize}
\item{\it Hubble tension:} Early-Universe inferences of the present expansion rate from the CMB~\cite{Planck:2018vyg} within $\Lambda$CDM disagree with estimates made using late-time distance-ladder determinations \cite{Riess:2021jrx, Brout:2022vxf}, indicating that either residual systematics or new ingredients beyond the minimal model may be required (for reviews see \cite{Knox:2019rjx, Freedman:2021ahq}).  
\item{\it Time-evolving Dark Energy:} Percent-level late-time distance measurements from BAO and supernovae are beginning to directly test whether the Dark Energy density is truly constant; recent DESI BAO results, in combination with external data, appear to favor mild departures from a strictly $\Lambda$-like equation of state, with the strength of the preference depending on the adopted dataset combination and modeling \cite{DESI:2025zgx, DES:2025sig}.  
\item{\it Puzzles in structure formation:} The CMB and large-scale-structure surveys provide independent measurements of matter clustering and comparisons of measures of clustering obtained in these ways are not always completely consistent (depending on one's trust in the assessments of systematics). These continue to motivate scrutiny of both small-scale modeling and extensions of the baseline framework. Added to this is recent evidence from JWST observations indicating that structure seems to have formed earlier than expected (redshifts $z \sim 10$) in the Universe's history \cite{JWST1,Castellano:2022,Labbe:2023, casey2023cosmosweb}.
\end{itemize}

Efforts to modify cosmology using fundamental physics have generated an enormous and bespoke zoo of models (for reviews see \cite{Copeland:2006wr, Caldwell:2009ix, Silvestri:2009hh, Schoneberg:2021qvd, CosmoVerseNetwork:2025alb}). These include Early Dark Energy (EDE) models~\cite{Kamionkowski:2022pkx, Adi:2024ebl}, coupled quintessence~\cite{Amendola:1999er, Liu:2019ygl}, models with varying fundamental constants~\cite{Lopez-Honorez:2020lno}, various modified-gravity frameworks (e.g., generalized Brans--Dicke \cite{Brans:1961sx} and Horndeski-type \cite{Horndeski:2024sjk} scalar--tensor theories, $f(R)$ gravity, equation-of-state parameterizations like $w$CDM and the Chevallier--Polarski--Linder form~\cite{Sprenger:2018tdb, Wu:2021vfz} and so on). Unfortunately the entire suite of cosmological observations is unlikely ever to be granular enough to be able to decide which (if any) of these models Nature has chosen. 

But many models used in cosmology seem hard to obtain as the low-energy limit of what we know about physics at higher energies in the sense that it is not known how they can be obtained as systematic Effective Field Theory (EFT) for cosmological scales by integrating out higher-energy physics within a controlled low-energy approximation. Perhaps this is the missing clue needed to properly identify the modification of the Core Theory that can successfully describe cosmology.

The problem can be made concrete in the specific case where the new physics involves scalar fields, since for these power-counting arguments can be made very explicitly \cite{Burgess:2009ea, Adshead:2017srh}. In particular, the kinds of things (such as small scalar masses and vacuum energies) required for successful phenomenology are not generic in the very low-energy limit of more fundamental theories \cite{Gildener:1976ih}. For instance, fundamental theories can contain scalars light enough to be relevant to cosmology but when they do they typically emerge in a {\it technically natural} way, such as if they arise as Goldstone bosons or as pseudo-Goldstone bosons \cite{Weinberg:1972fn} for an approximate symmetry (see \cite{Burgess:2025vxs} for a review of the issue). 

A similarly convincing technically natural description for the small size of the Dark Energy seems much more difficult, with no consensus yet forming around any extant proposals (an observation known as the Cosmological Constant problem \cite{Weinberg:1988cp,  Burgess:2013ara}). However, the fact that this is difficult to achieve presents us with a potential opportunity: the low-energy features required by a solution to the cosmological constant problem within a sensible\footnote{We should emphasize here that consistency with UV completions as sought here is a criterion of technical naturalness (see {\it e.g.} \cite{Burgess:2013ara} for its precise description). This differs from the currently popular `swampland' program \cite{Vafa:2005ui}, which is founded on the conjecture that some otherwise reasonable low-energy theories cannot be obtained from a consistent UV completion of gravity (in practice, string theory) at the highest energies. The swampland program asserts there is an intrinsically high-energy criterion for separating bad and good theories at low energy that cannot in itself be captured purely within the low-energy limit; if true, describing the world at low energies requires intrinsically stringy information. Conceptually this differs from technical naturalness, which is a fundamentally low-energy criterion. Technical naturalness asks why integrating out the electron does not screw up the small size of the Dark Energy, and finding a way to do so requires new physics at or below the electron mass rather than at the string scale. Higher energies or UV completions only appear in the story to the extent that they can motivate the low-energy symmetry arguments that are used. } UV completion are likely to be robust and not tweakable, and so their observational signatures are unlikely to be avoidable. 

Although there is not yet a consensus around any particular theories of this type, there is a class of models that so far seem very promising \cite{Burgess:2021obw}. These suppress the Dark Energy density through a relaxation mechanism built around the low-energy interplay of approximate scale invariance and supersymmetry that is known to suppress corrections to scalar potentials and arises very commonly within higher-dimensional models \cite{Burgess:2020qsc}. In these models a pair of pseudo-Goldstone bosons (a dilaton for approximate scale invariance and an `axion' for a related internal symmetry) -- together called an axiodilaton -- provide natural candidates for the Dark Energy and Dark Matter, respectively \cite{Smith:2024ibv}.

What is most remarkable about the resulting cosmology is that it also helps understand some of the recent puzzles with observational cosmology listed earlier, despite not being designed with these anomalies in mind \cite{Smith:2025uaq}. The scale invariance associated with the dilaton makes its cosmic evolution tend to fall into a scaling `tracker' solution \cite{Ferreira:1997hj, Steinhardt:1999nw} that follows the dominant energy density at any given time. The transition from radiation to matter domination at redshifts of order 3400 therefore generically causes the dilaton to seek a new tracker solution and so transiently depart away from its earlier tracker behaviour. This quite robustly causes the Dark Energy to evolve in the later universe independent of its initial conditions. It is not uncommon for this excursion to last long enough to account for the evidence for time-dependent Dark Energy seen in galaxy surveys. Better still, the interactions between Dark Energy and Dark Matter built into the model for other purposes cause the Dark Matter to evolve in a nonstandard way and this can fake the apparently unphysical Dark Energy equation of state $w < -1$ if the data is not analysed with this in mind (see also \cite{Carroll:2004hc, Khoury:2025txd}) as is indeed done by the observational collaborations \cite{Planck:2018vyg,eBOSS:2020yzd,DES:2025upx,DESI:2025zgx,KiDS:2020ghu}.  

The model also serendipitously improves on the Hubble tension because the scale invariance implies the masses of all Standard Model particles depend on the dilaton and so evolve in time (though their ratios do not) as the dilaton undergoes its excursion \cite{Smith:2025uaq}. The transient non-tracker evolution near radiation-matter equality also makes it common for particle masses to be different at recombination than they are now and when this is so it causes the last-scattering surface to be slightly further away than it would be in $\Lambda$CDM. This brings the inferred value of $H_0$ into better agreement with measurements at smaller redshifts, along the lines that have been proposed in a more {\it ad-hoc} way in \cite{Sekiguchi:2020teg}. It is suggestive that a model designed with other criteria in mind helps with two of the current observational puzzles like this `for free'. 

\begin{figure}
    \centering
    \includegraphics[width=1\linewidth]{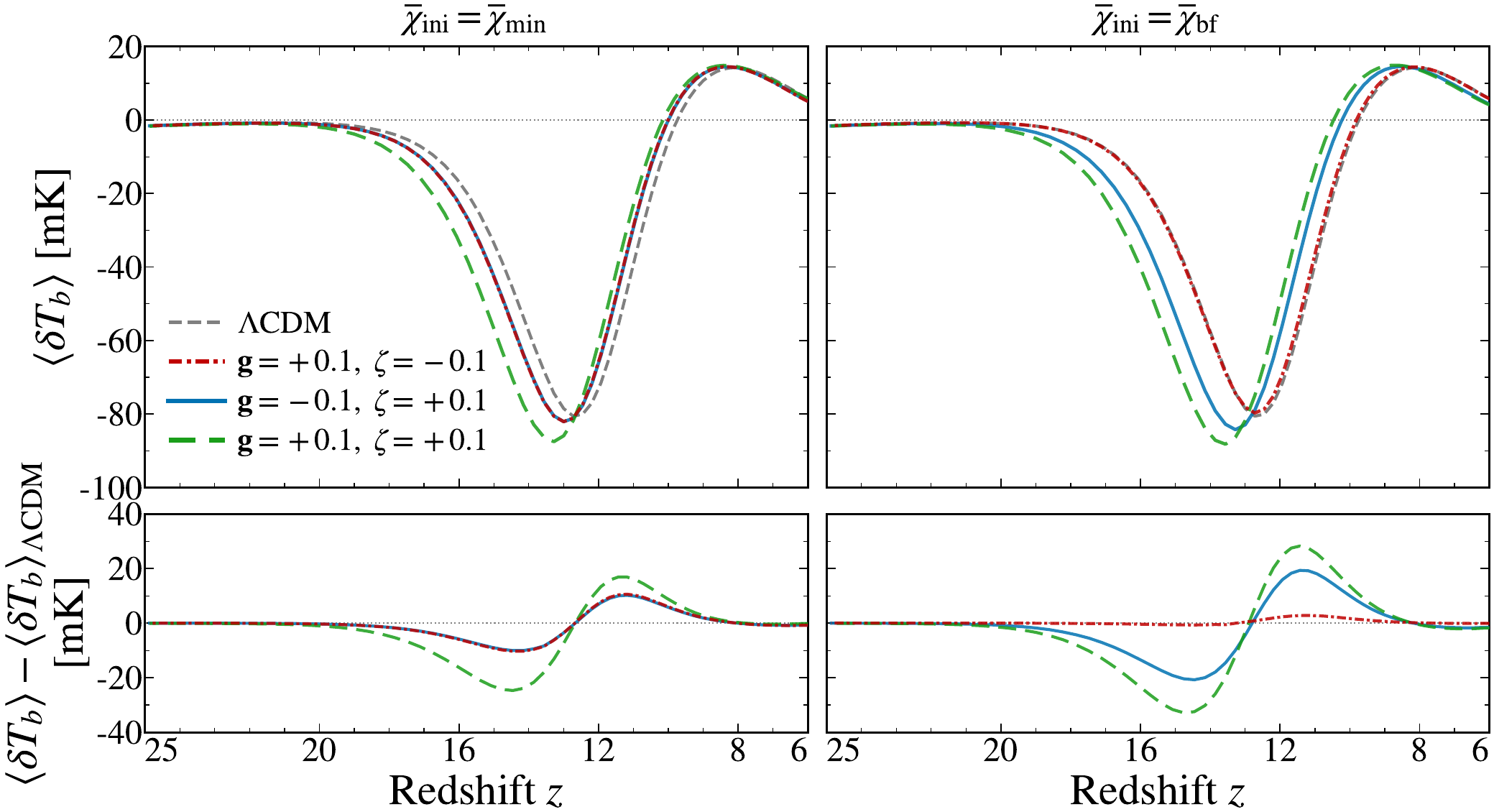}
    \caption{Volume-averaged differential 21\,cm brightness temperature $\langle\delta T_b\rangle$ (upper panels) as computed for axiodilaton and $\Lambda$CDM models as a function of redshift for $\bar\chi_{\rm ini} = \bar\chi_{\rm min} = 74.010$ chosen initially at the potential's minimum (left) or at the value $\bar\chi_{\rm ini} = \bar\chi_{\rm bf} = 74.213$ that provides the best fit to CMB, BAO, supernova and SH0ES data~\cite{Smith:2025uaq} (right). Lower panels provide the residuals between the axiodilaton and $\Lambda$CDM calculations. Colors and line styles denote the three $(\bfg,\zeta)$ and $\Lambda$CDM choices indicated in the legends. }
    \label{fig:21cmSummaryPlot}
\end{figure}

So what is new? In this paper we build on the above history and explore the implications of this same axiodilaton model for 21\,cm cosmology~\cite{Madau:1996cs, Barkana:2000fd, Furlanetto:2006jb, Morales:2009gs, Pritchard:2011xb, Furlanetto:2015apc, Mesinger:2010ne}, which uses the hyperfine transition of neutral hydrogen to map the three-dimensional distribution of matter and radiation across cosmic time. Measurements of the 21\,cm signal probe the difference between the local temperature of CMB photons and the spin temperature of neutral hydrogen gas and so can track the thermal, collisional scattering and ionization history of the intergalactic medium (and the growth of structure that can affect this history) from the dark ages ($z\sim30$--$200$) through cosmic dawn ($15\lesssim z\lesssim30$) and into the reionization regime ($6\lesssim z\lesssim12$). This is precisely when most of the dilaton's evolution would occur. 

Until recently this regime has been largely inaccessible to precision tests, but this situation is now changing and the observational prospects for 21\,cm cosmology are rapidly improving. Current radio interferometers such as the LOw-Frequency ARray (LOFAR)~\cite{LOFAR:van2013,Patil:2017zqk,Gehlot_2019} are pursuing statistical measurements over $6\lesssim z\lesssim11$, with low-frequency observations extending to $z\simeq20$--$25$. The Hydrogen Epoch of Reionization Array (HERA)~\cite{DeBoer:2016tnn,Berkhout:2024,HERA:2021bsv,HERA:2022wmy} targets a comparable window, $6\lesssim z\lesssim13$, while the wider frequency coverage of its Phase~II configuration extends its nominal redshift range toward $z\simeq27$. Future observations with the low-frequency component of the Square Kilometre Array (SKA-Low)~\cite{Mellema:2013,Cosmology-SWG:2015tjb,Koopmans:2015sua,Greig:2019tcg} are expected to probe roughly $6\lesssim z\lesssim27$. Beyond ground-based experiments, lunar farside concepts have been proposed to access lower-frequency 21\,cm signals that cannot reach the ground because of the Earth's ionosphere~\cite{Jester:2009dw}. The Lunar Crater Radio Telescope (LCRT)~\cite{Goel:2022jgw} targets the dark-ages signal over $46\lesssim z\lesssim472$, while FARSIDE~\cite{Burns:2021pkx} is designed to probe $35\lesssim z\lesssim200$.

\begin{figure}[t!]
    \centering
    \includegraphics[width=1.0\linewidth,clip]
    {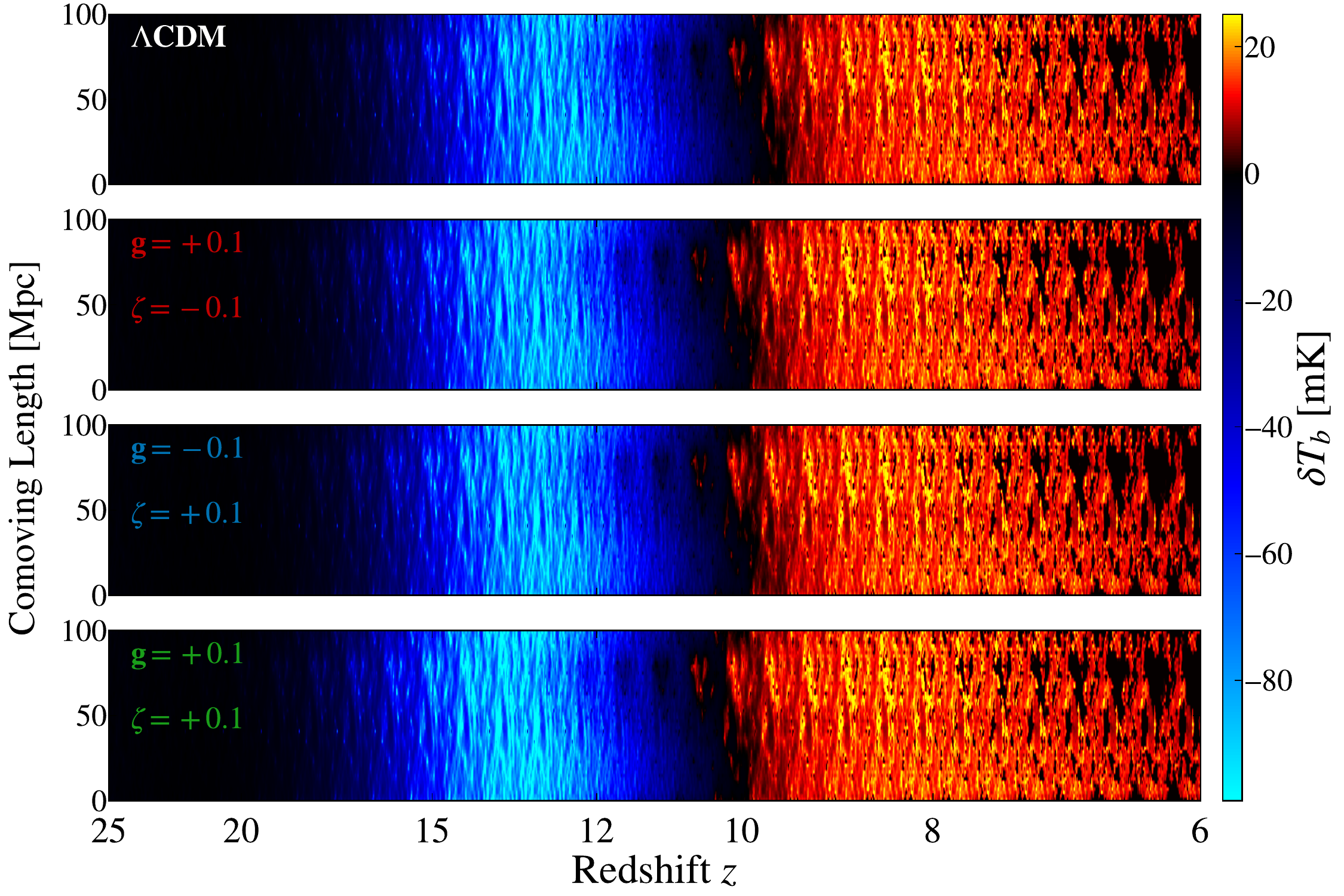}
    \caption{\small 
    Light-cone slices of the differential 21\,cm brightness
    temperature over $6\leq z\leq25$ for \LCDM\ (top) and the
    axiodilaton models with an initial background dilaton field value
    $\bar\chi_{\rm ini} = \bar\chi_{\rm min} = 74.010$ (lower panels). From the second to the
    fourth panel, the dilaton couplings to baryons and axion DM,
    denoted by $\bfg$ and $\zeta$, are
    $(\bfg,\zeta)=(+0.1,-0.1)$, $(-0.1,+0.1)$, and $(+0.1,+0.1)$,
    respectively. Each panel spans $100\,{\rm Mpc}$ along the displayed
    spatial direction, and the color scale denotes the differential
    brightness temperature relative to the CMB in units of mK.
    The axiodilaton interactions modify the timing and spatial
    morphology of the 21\,cm signal, with the most visible shift toward
    higher redshift occurring for $(\bfg,\zeta)=(+0.1,+0.1)$.
    }
    \label{fig:21cm_lightcone}
\end{figure}

Our main results are summarized in Fig.~\ref{fig:21cmSummaryPlot}, which shows the brightness temperature as a function of redshift for both axiodilaton and $\Lambda$CDM models for a selection of model parameters ($\overline{\chi}_{\rm ini}$, $\bfg$ and $\zeta$, whose definitions are explained in detail below). The figure shows how axiodilaton models differ from $\Lambda$CDM in their predictions, with the axiodilaton predicting a deeper dip that is slightly displaced to higher redshifts. A different way of presenting our results is shown in Fig.~\ref{fig:21cm_lightcone}, which in addition to the brightness temperature plots its variation in position across a comoving distance of 100 Mpc. The 21\,cm signal is inhomogeneous because Ly$\alpha$, X-ray, and ionizing UV radiation are produced by the first collapsed halos, and these form where the matter density is highest.
The number of such halos at a given redshift is fixed by how far structure has grown.
Because the halos are still rare at these epochs, a small change in that growth produces a much
larger change in their number.\footnote{The sensitivity to structure formation is not important deep in the dark ages before stars start to form, but becomes crucial for the later redshifts for which 21\,cm observations are currently possible. These are also the redshifts for which Fig.~\ref{fig:21cmSummaryPlot} shows the predictions of axiodilaton models differ the most strongly from $\Lambda$CDM. An important limitation to our preliminary analysis here is our reliance on standard $\Lambda$CDM structure formation within the nonlinear regime -- we defer a fuller re-derivation using axiodilatons to future work.} The fluctuations seen in Fig.~\ref{fig:21cm_lightcone} therefore
track the growth of structure at these redshifts. LOFAR, HERA, and SKA-Low measure these fluctuations through the 21\,cm power spectrum, which is the observable analysed in \S\ref{sec:result}.

\begin{figure}[t!]
    \centering
    \includegraphics[width=1\linewidth]{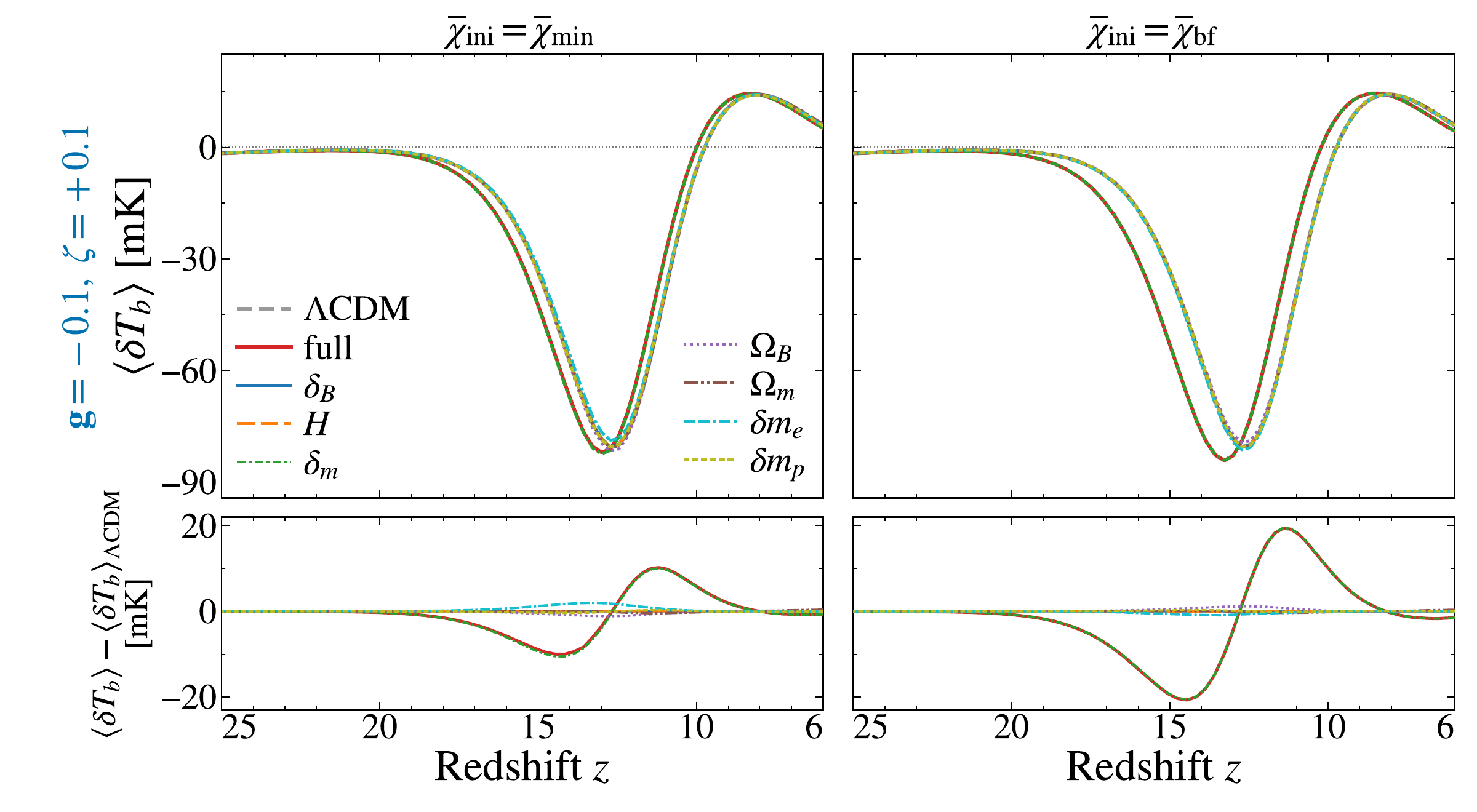}
    \caption{Volume-averaged differential 21\,cm brightness temperature $\langle\delta T_b \rangle$ for $(\bfg, \zeta)=(-0.1,+0.1)$ when the axiodilaton modifications enter the 21\,cm calculation one at a time, for $\bar\chi_{\rm ini} = \bar\chi_{\rm min} = 74.010$ chosen at the potential's minimum (left) or at the value $\bar\chi_{\rm ini} = \bar\chi_{\rm bf} = 74.213$ that provides the best fit to CMB, BAO, supernova and SH0ES data~\cite{Smith:2025uaq} (right). Current cosmological data prefer coupling combinations of opposite sign~\cite{Smith:2025uaq}, of which this is one. Colors and line styles denote the varied quantity indicated in the legend, with $\delta m_e(z)$ and $\delta m_p(z)$ given by Eq.~\eqref{eq_mass_variations}. The gray dashed curves denote \LCDM\ and the red full curves include all modifications together, with the $\delta_m$ curve following the red curve most closely.
    }
    \label{ItsTheDensity}
\end{figure}

These figures show that the signal is moved to slightly earlier redshifts relative to the $\Lambda$CDM prediction, but leave open which of the many possible effects ({\it i.e.}~varying particle masses, changed Hubble evolution, increased density perturbations, {\it etc}) is responsible for this difference. Fig.~\ref{ItsTheDensity} gives the size of the prediction when each of these contributions is turned on separately and so shows that the most important effect is the earlier growth of structure $\delta_m$ predicted by axiodilaton models. Structure growth turns out to be enhanced dominantly because of the additional attraction experienced by matter (and the axion) due to dilaton exchange, which amplifies gravitational attraction at these epochs.

We find that the same parameter choices that help ameliorate the other anomalies in $\Lambda$CDM cosmology also predict enhanced structure growth that starts slightly earlier than in vanilla $\Lambda$CDM, possibly a step towards understanding the puzzling early appearance of structure revealed by JWST measurements. This earlier growth of structure is particularly cleanly predicted at the early times relevant to 21\,cm measurements because this epoch is deep within the matter-dominated regime. Similarly clean predictions for observables like $\sigma_8$ that track structure growth at later epochs can be complicated by the timing of the onset of Dark Energy domination, which can act to suppress structure growth (along the lines discussed for example in \cite{Costa:2025kwt, Escudero:2026mgw, Weiner:2026sfm}).

Motivated by the size of these effects, we here also extend the 21\,cm and reionization pipeline to the axiodilaton scenario and compute the resulting signatures in order to forecast the sensitivity of future observations to deviations from the $\Lambda$CDM prediction. When doing so we include time-dependent baryon masses in addition to electron mass variation, since these are both predicted by the model and required to keep CMB predictions in agreement with observations. We propagate the baryon mass rescaling into the residual ionization fraction, the gas and spin temperatures, and the neutral hydrogen fraction, and we quantify the impact on the global 21\,cm signal and the power spectrum. We also generate mock observations for the SKA observatory and estimate the level at which axiodilaton cosmology can be tested. We find when doing so that SKA is likely to be sensitive to much smaller parameter ranges than those that are currently probed in the CMB and galaxy surveys.

The remainder of our presentation is organized as follows. The next section \S\ref{sec:21cm} first briefly summarizes the main features of the model (more details are given in Appendix \S\ref{sec:TheoFramework}, which for completeness reviews the definitions and cosmological evolution that appear elsewhere in the literature), zeroing in on 21\,cm physics and how this is altered by the presence of the axiodilaton. \S\ref{sec:result} summarizes the results of our analysis of the 21\,cm signal and contains our forecasts for the SKA sensitivity. Our conclusions are briefly summarized in \S\ref{sec:summary}. Appendix \ref{app:21cmMinutae} contains many detailed calculations relevant to specific changes in 21\,cm physics, while Appendix \ref{app_ska_sensitivity} collects results necessary for our forecasts for SKA.

\section{Axiodilatons and 21\,cm Physics}
\label{sec:21cm}

In this section, we describe how the axiodilaton cosmology framework is computed and how it is incorporated into the calculation of 21\,cm physics during cosmic dawn and epoch of reionization. 

\subsection{The axiodilaton Model}
\label{sec:TheADModel}

The model whose implications we explore is motivated by the Yoga models of \cite{Burgess:2021obw} (and reviewed in \cite{Burgess:2025vxs}), with a brief summary of its motivations reproduced in Appendix~\ref{sec:TheoFramework}. For the present purposes this class of models is of interest because of the properties of the Dark Sector that it suggests and the rich cosmology to which this leads. 

For the purposes of cosmology the model's core content is a pair of light pseudo-Goldstone bosons: a dilaton for an approximate scaling symmetry and an axion. In the minimal framework these couple to one another and to ordinary matter through the action
\begin{align}\label{eq:Action}
    S = \!\int \exd^4x \sqrt{-g}
    \Biggl\{
        \frac{\MP^2}{2}
        \Bigl[
            R
            &- \partial_\mu\chi\,\partial^\mu\chi
            - W^2(\chi)\,\partial_\mu\mfa\,\partial^\mu\mfa
        \Bigr]
        - V(\chi,\mfa)
    \Biggr\}
    + S_m(\psi,\tilde g)\,.
\end{align}
Here $W(\chi)$ describes the nontrivial two-derivative sigma-model interactions between the two scalars, which are important because they can compete with the two-derivative interactions of GR at low energies without undermining the semiclassical approximation (which for gravity is at its heart a derivative expansion) \cite{Burgess:2009ea, Adshead:2017srh}. We take $W$ to have the exponential form dictated by scale invariance
\begin{equation}
    W(\chi) = e^{-\zeta\chi}
\end{equation}
where $\zeta$ is a coupling parameter. 

\begin{figure}
    \centering
    \includegraphics[width=0.9\linewidth]{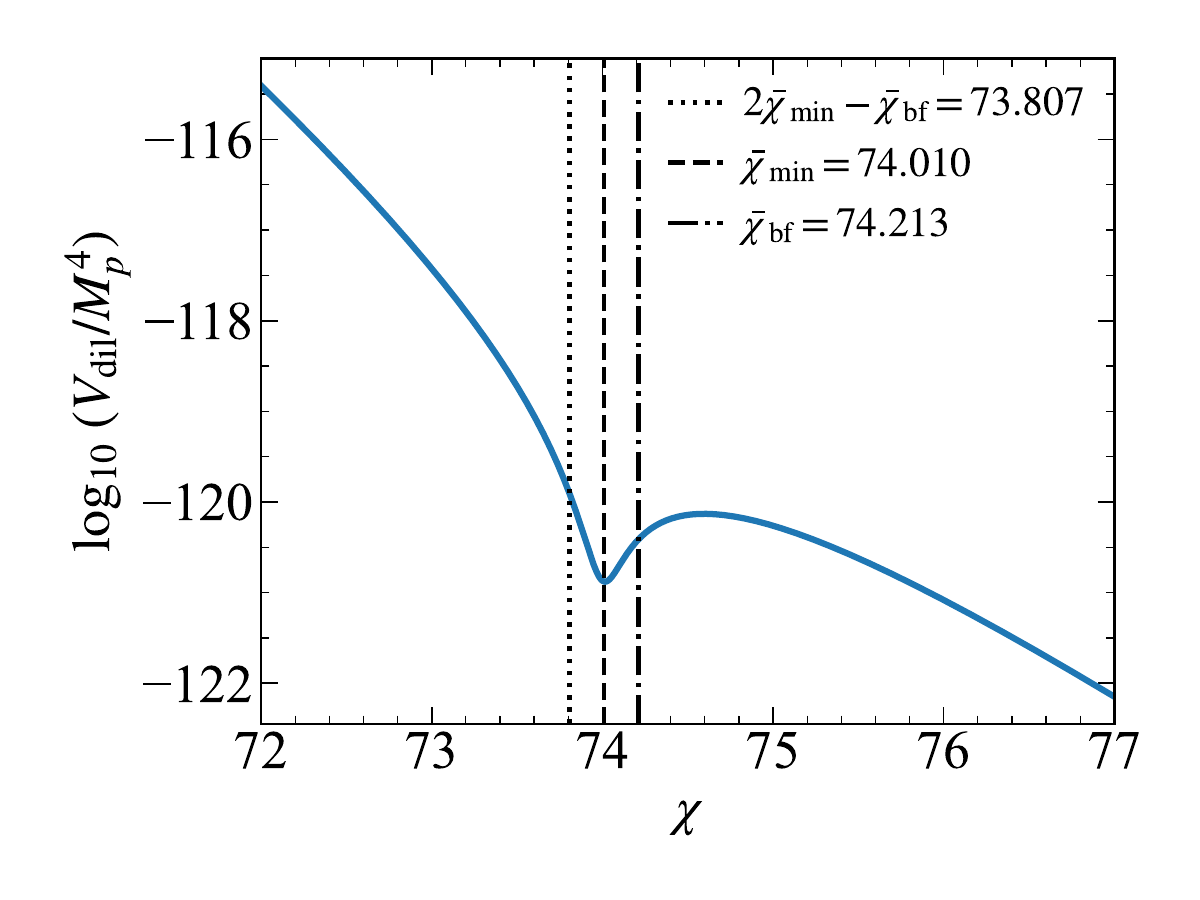} \vspace*{-0.8cm}
    \caption{A sketch of the Albrecht--Skordis potential $\log V_{\rm dil}(\chi)$ against $\chi$ for the parameters used in the analysis. The vertical lines show the position of the local minimum (dashed), the value $\bar\chi_{\rm bf}$ (dash-dotted) of the dilaton initial condition in the best-fit cosmology found in \cite{Smith:2025uaq}, and its mirror value $2\bar\chi_{\rm min}-\bar\chi_{\rm bf}$ about the minimum (dotted). (The goodness of fit does not vary strongly between having the initial condition chosen at $\chi_{\rm min}$ vs $\chi_{\rm bf}$.)}
    \label{DilatonPotential}
\end{figure}

Most of the work in \cite{Burgess:2021obw} goes into understanding the shape and size of the scalar potential, which for the present applications we take to have the form\footnote{In ultraviolet-motivated constructions the axion and dilaton actually share a joint potential for which $e^{-\lambda \chi}$ is a common factor (see \cite{vandeBruck:2026bdw} for how this impacts axion dynamics). We here for simplicity work with the simpler additive form where $m_{\mfa}$ is a given super-Hubble mass scale. The difference should not matter for our purposes provided the axion remains near its minimum and the dilaton does not roll too far.} 
\begin{align}\label{eq:joint_potential}
    V(\chi,\mfa) &\simeq V_{\rm dil}(\chi) + V_{\sax}(\mfa)  ,
    \nonumber\\
    &=V_p(\chi)\,e^{-\lambda\chi} + \tfrac12 \MP^2 m_\mfa^2 \mfa^2
\end{align}
where $V_{\rm ax}$ is only approximately quadratic near its minimum, about which the axion undergoes coherent oscillations once its effective mass satisfies $m_{\rm ax}\gg H$. In Yoga models $\lambda = 4 \zeta$ and $\zeta = \sqrt{{2}/{3}}$ but this can vary in more complicated versions so we here treat $\zeta$ as a parameter while fixing $\lambda = 4\sqrt{{2}/{3}}$.

The dilaton potential is exponential with a more slowly varying prefactor, 
\begin{equation}\label{eq:albrecht-skordis}
        V_p(\chi)=V_0\!\left[1-u_1\chi+\frac{u_2}{2}\chi^2\right]\,,
\end{equation}
(often called the Albrecht--Skordis form). The coefficients $u_1$ and $u_2$ are model-dependent and are chosen such that the dilaton potential has a local minimum. Simple choices $u_i \sim \cO(50)$ ensure that the minimum occurs for $\chi_{\rm min} \sim \cO(50)$.\footnote{In the UV motivated models large $\chi$ is responsible for both the size of the hierarchy between the electroweak and Planck scales and for the hierarchy between the electroweak scale and the size of $V_{\rm dil}$ at its minimum, but we do not use this here.} Fig.~\ref{DilatonPotential} plots the resulting shape for $V_{\rm dil}(\chi)$ using the parameter choices used in our analysis (which imply $\chi_{\rm min} = 74.010$). Because predictions for the evolution of the cosmological parameters, the CMB anisotropy power spectrum and matter power spectrum depend on the initial conditions of the dilaton and its subsequent evolution, we use three values for the initial value of the dilaton in this paper. The posterior for $\chi_{\rm ini}$ found in \cite{Smith:2025uaq} is bimodal, with two peaks of comparable height on either side of $\chi_{\rm min}$. Without the SH0ES calibration the best fit lies to the left of the minimum, and with it the best fit moves to the right, $\chi_{\rm bf}=74.213$. We therefore take one initial value on each side, placed symmetrically about the minimum, $\chi_{\rm ini}=2\chi_{\rm min}-\chi_{\rm bf}=73.807$ and $\chi_{\rm ini}=\chi_{\rm bf}=74.213$, together with $\chi_{\rm ini}=\chi_{\rm min}$. These choices differ in how robust they are. The value $2\chi_{\rm min}-\chi_{\rm bf}$ lies on the steep side of the potential, so the field rolls back to the minimum and its evolution is stable against additional effects. The value $\chi_{\rm bf}$ lies close to the local maximum of the potential to the right of the minimum, where a small perturbation can change whether the field returns to the minimum or rolls away from it. Because the cosmological evolution and the perturbations differ for these values of $\chi_{\rm ini}$, they lead to differences in the predictions for the 21\,cm physics.

Scale invariance also dictates the couplings between the dilaton and ordinary matter fields, collectively denoted here by $\psi$. These are minimally coupled to a Jordan-frame metric $\tilde g_{\mu\nu}$ that is related to the Einstein-frame metric by
\begin{equation}\label{eq:jf_metric}
    \tilde g_{\mu\nu} = A^2(\chi)\,g_{\mu\nu}, 
    \quad \hbox{where} \quad 
    A(\chi)=e^{\mathbf g\chi}.
\end{equation}
This ensures that $\chi$ couples to ordinary matter as would a Brans--Dicke scalar, with the dimensionless coupling $\mathbf g$ being related to the usual Brans--Dicke coupling.\footnote{The UV motivation predicts $\bfg < 0$ but we here take a phenomenological point of view and consider both positive and negative values. The simplest Yoga models actually predict an explicit value ($\bfg = -\sqrt{1/6}$) but this prediction is not shared by more complicated variants.} 

At face value the similarity with Brans--Dicke scalars implies that the dilaton-matter coupling is constrained by solar-system tests to satisfy $\bfg \lsim 10^{-3}$, which is smaller than the couplings we entertain for cosmology. In principle solar-system tests measure the coupling of macroscopic objects (like the Sun) to the dilaton rather than directly probing the dilaton-matter coupling $\bfg$ at the atomic level. As discussed in \cite{Brax:2023qyp}, whether these are the same depends on how matter couples to the axion because of the nonlinearity of the axiodilaton interactions. As a result we temporarily here ignore the solar-system tests when exploring the couplings favoured by cosmology, while recognizing that the apparent discrepancy must eventually be dealt with.

Crucially, the above assumptions cause the masses of all ordinary particles to vary as the dilaton varies,
\begin{equation}
\frac{m_i(x)}{m_{i,0}} = \frac{A(\chi(x))}{A(\chi_0)}
= e^{{\mathbf g}\,\Delta\chi(x)},
\label{eq:mass_ratio}
\end{equation}
where $\Delta\chi(x)\equiv \chi(x)-\chi_0$ denotes the difference between the value of $\chi(x)$ compared to a specific reference value $\chi_0\equiv \chi(x_0)$. Notice that this dependence is universal and (as described in Appendix \ref{sec:TheoFramework}) this universality is crucial for evading many otherwise very stringent phenomenological constraints.\footnote{The perceptive reader might worry that the masses of composite objects like protons and neutrons might not scale in the same way, but this is actually also easily ensured. In particular, hadron masses are proportional to the QCD scale $\Lambda_\QCD$, which itself is defined in terms of the QCD coupling constant specified at a particular energy: $g(M) = g_0$. Since $\Lambda_\QCD \propto M$ it scales as do all other masses provided $M$ does.}  The only direct probes of such a universal coupling are those that compare particle masses at different times or compare them to the Planck scale at any given time (since $M_p$ is held fixed by the conformal rescaling of
Eq.~\eqref{eq:jf_metric}). 

As we see below, it turns out that the 21\,cm signal makes such a  comparison whenever something like an atomic rate is measured against the Hubble scale $H$ (although a universal redshift-dependence of particle masses drops out of direct ratios of atomic rates at any given redshift). $A(\chi)$ also appears in gravitational clustering, since the dilaton-matter coupling it contains allows the dilaton to mediate a long-range force between non-relativistic particles. This alters the rate at which density perturbations grow, to which the 21\,cm signal responds extremely sensitively through its influence on the abundance of the first collapsed halos. 

The field equations and cosmological evolution implied by this model are summarized in Appendix \ref{sec:TheoFramework}, including the results of a comparison of its predictions to CMB and LSS data. We now turn to their implications for 21\,cm physics during cosmic dawn and the epoch of reionization.

\subsection{Semi-numerical Realization and Simulation Setup}
\label{ssec:simsetup}

Our numerical pipeline consists of two stages. We first solve the homogeneous
axiodilaton evolution and the associated linear perturbations using our
modified version of \CLASS~\cite{Blas:2011rf,Smith:2025uaq} described in Appendix~\ref{subsec:bkg_evolution}. We then pass the
relevant background and perturbation quantities to
\FirstClass~\cite{Flitter:2023mjj,Flitter:2023rzv,
Flitter:2024eay}, which performs the semi-numerical 21\,cm calculation.
In this subsection, we describe how the standard \CLASS\ outputs that
\FirstClass\ uses as cosmological inputs are replaced by the corresponding
outputs of our modified \CLASS\ calculation, and we highlight the ingredients
that differ from the $\Lambda$CDM implementation.

The background quantities supplied to initialize the 21\,cm calculation include
\begin{equation}
\left\{
H(z),\,
\bar\rho_B(z),\,
\bar\rho_{\rm ax}(z),\,
m_e(z),\,
m_p(z),\,
Y_p,\,
z_{\rm rec},\,
T_k(z_{\rm rec}),\,
\bar x_e(z_{\rm rec})
\right\}.
\label{eq_ad_background_outputs}
\end{equation}
Here $H(z)$ is the Hubble expansion rate, while $\bar\rho_B(z)$ and
$\bar\rho_{\rm ax}(z)$ are the mean baryon and axion energy densities,
respectively. These quantities follow the background evolution described in
Appendix~\ref{subsec:bkg_evolution}.
We denote the redshift-dependent electron and proton
masses by
\begin{equation}
m_e(z)=\left(1+\delta m_e(z)\right)m_{e,0},
\qquad
m_p(z)=\left(1+\delta m_p(z)\right)m_{p,0},
\label{eq_mass_variations}
\end{equation}
where $m_{e,0}$ and $m_{p,0}$ are the corresponding reference values in
$\Lambda$CDM. Applying the universal SM mass scaling in Eq.~\eqref{eq:mass_ratio} to the
electron and proton gives
\begin{equation}
\delta m_e (z)
=
\delta m_p (z)
=
e^{\bfg\Delta\bar\chi(z)} - 1.
\label{eq_common_mass_variation}
\end{equation}

The primordial helium mass fraction is denoted by
$Y_p=4\bar n_{\rm He}/(\bar n_H+4\bar n_{\rm He})$, where $\bar n_H$ and
$\bar n_{\rm He}$ are the mean number densities of hydrogen and helium
nuclei, respectively. Assuming
$\bar\rho_H\simeq m_p(z)\bar n_H$ and
$\bar\rho_{\rm He}\simeq4m_p(z)\bar n_{\rm He}$, the mean baryon energy
density is
\begin{equation}
    \bar\rho_B(z)
    =
    \bar\rho_H(z)+\bar\rho_{\rm He}(z)
    \simeq
    m_p(z)
    \left[
    \bar n_H(z)+4\bar n_{\rm He}(z)
    \right].
\end{equation}
It then follows that the mean  densities of hydrogen and helium nuclei can be
written as
\begin{equation}
    \bar n_H(z)
    =
    \frac{(1-Y_p)\bar\rho_B(z)}{m_p(z)},
    \qquad
    \bar n_{\rm He}(z)
    =
    \frac{Y_p\bar\rho_B(z)}{4m_p(z)}.
    \label{eq_mean_nH_nHe_rhoB}
\end{equation}
Substituting the background solution
$\bar\rho_{\ssB}(z)=\bar\rho_{\ssB,0}a^{-3}
e^{\bfg\Delta\bar\chi(z)}$ from
Eq.~\eqref{eq:baryon_background_density_solution} and the proton mass
scaling $m_p(z)=m_{p,0}e^{\bfg\Delta\bar\chi(z)}$ from
Eq.~\eqref{eq:mass_ratio} into the expressions above, the common factor
$e^{\bfg\Delta\bar\chi(z)}$ cancels, giving
\begin{equation}
    \bar n_H(z)
    =
    (1-Y_p)\bar n_{\ssB,0}(1+z)^{3},
    \qquad
    \bar n_{\rm He}(z)
    =
    \frac{1}{4} Y_p\bar n_{\ssB,0} (1+z)^{3}.
    \label{eq_mean_nH_nHe}
\end{equation}
The mean hydrogen and helium nuclei densities therefore retain their
standard $(1+z)^{3}$ scaling and coincide with their reference
$\Lambda$CDM values for the same present-day baryon density.

The redshift $z_{\rm rec}$ denotes the recombination redshift and is also used
as the initial redshift for the semi-numerical 21\,cm evolution. The
quantities $T_k(z_{\rm rec})$ and $\bar x_e(z_{\rm rec})$ are the homogeneous
gas kinetic temperature and the mean free electron fraction at this initial
redshift. The free electron fraction is normalized per hydrogen nucleus as
\begin{equation}
    \bar x_e(z_{\rm rec})
    \equiv
    \frac{\bar n_e(z_{\rm rec})}{\bar n_H(z_{\rm rec})}\;.
    \label{eq_xe_initial}
\end{equation}
In \CLASS, the homogeneous recombination and thermal histories are computed
by the thermodynamics module using
\texttt{HyRec}~\cite{Ali_Ha_moud_2011,Lee_2020}. We modify this calculation to account
for the axiodilaton expansion history and the redshift dependence of the
SM particle masses, while adopting a fixed primordial helium mass fraction
$Y_p\simeq0.245$ for all models.\footnote{In principle, axiodilaton cosmology could also alter the primordial value of $Y_p$, though whether this occurs largely depends on the value of particle masses at BBN compared to now (and so is dependent on initial conditions). Ref.~\cite{Smith:2024ibv} show that cosmologies exist for which masses at BBN agree with present values despite them taking an excursion at recombination (due to the transient associated with radiation-matter equality). In this work we assume initial conditions at BBN allow us to adopt the standard value $Y_p=0.245$ as a common primordial input for
all models.}
For each simulated evolution, $z_{\rm rec}$ is defined as the redshift at which
$\bar x_e(z_{\rm rec})=0.1$. The resulting values of
$T_k(z_{\rm rec})$ and $\bar x_e(z_{\rm rec})$ are used as the initial
thermodynamic conditions for the subsequent evolution of the gas kinetic
temperature and the free electron fraction, described in
Appendices~\ref{subsec:Tk} and \ref{subsec:xe_evolution}, respectively.

The modified linear outputs include
\begin{equation}
\left\{
D_B(k,z),\,
D_{\rm ax}(k,z),\,
\theta_B(\mathbf{k},z),\,
\theta_{\rm ax}(\mathbf{k},z),\,
P_m(k,z),\,
\sigma(M_R,z)
\right\}.
\label{eq:ad_perturb_outputs}
\end{equation}
Here $D_B(k,z)$ and $D_{\rm ax}(k,z)$ are the baryon and axion DM growth functions. 
The quantities $\theta_B(\mathbf{k},z)$ and $\theta_{\rm ax}(\mathbf{k},z)$ denote the velocity divergences of baryons and axion DM, respectively, defined in Eq.~\eqref{eq:fluid_pert_def}. 

From the matter power spectrum $P_m(k,z)$, we compute the variance of the smoothed linear
density field,
\begin{equation}
    \sigma^2(M_R,z)
    =
    \int d\ln k\,
    \frac{k^3}{2\pi^2}
    P_m(k,z)
    \left|W(kR)\right|^2 ,
    \label{eq_sigma_R}
\end{equation}
where $R$ is the smoothing scale associated with the mass relation
$M_R=4\pi R^3\bar\rho_m(0)/3$, and $\bar\rho_m(0)$ is the mean matter
density at $z=0$. The real-space top-hat window function and its Fourier
transform are given by
\begin{equation}
W_R(r)
=
\frac{3}{4\pi R^3}
\Theta(R-r),
\qquad
W(kR)
=
\frac{3}{(kR)^3}
\left[
    \sin(kR)-kR\cos(kR)
\right].
\label{eq_tophat_window}
\end{equation}
The top-hat window function suppresses modes with $k\gtrsim 1/R$, so a
larger smoothing radius removes more small-scale power from the density
field and generally gives a smaller $\sigma(M_R,z)$. Conversely, a smaller
$R$ retains fluctuations on smaller scales and gives a larger
$\sigma(M_R,z)$. In the axiodilaton cosmology, the modified evolution of
the baryon and axion perturbations changes the shape and amplitude of the
linear matter power spectrum.

For each parameter choice, the precomputed background and
perturbation quantities are tabulated over the redshift range relevant for the
21\,cm calculation and supplied to \FirstClass. In the second stage,
\FirstClass\ generates three-dimensional realizations of the baryon and axion
DM density fields in a periodic comoving box of volume $V=L^3$. Our fiducial
simulations use
\begin{equation}
    L=200\,{\rm Mpc},\qquad
    N_{\rm hi}=300^3,\qquad
    N_{\rm lo}=100^3 .
    \label{eq:box_setup}
\end{equation}
Here $N_{\rm hi}$ and $N_{\rm lo}$ denote the numbers of cells in the
high-resolution and low-resolution grids, respectively. The high-resolution
grid is used to construct the perturbed density and velocity fields, while the
low-resolution grid is used for the computationally expensive source, thermal,
radiative, ionization, and 21\,cm calculations. The fundamental mode of the box
is
\begin{equation}
    k_{\rm f}=\frac{2\pi}{L}
    \simeq 0.031\,{\rm Mpc}^{-1}.
\end{equation}
The high-resolution grid has a cell size
$\Delta x_{\rm hi}=L/300\simeq0.67\,{\rm Mpc}$, while the low-resolution grid
has $\Delta x_{\rm lo}=L/100=2\,{\rm Mpc}$ and a Nyquist wavenumber
\begin{equation}
    k_{\rm Ny,lo}
    =
    \frac{\pi}{\Delta x_{\rm lo}}
    \simeq
    1.57\,{\rm Mpc}^{-1}.
\end{equation}

At each redshift, the linear baryon and axion DM density fields are generated
from the common primordial seed using the scale-dependent growth functions
$D_B(k,z)$ and $D_{\rm ax}(k,z)$. Mild nonlinear evolution is then included
through second-order Lagrangian perturbation theory. In this approach, fluid
elements are moved from their initial Lagrangian coordinates $\mathbf q$ to
Eulerian coordinates $\mathbf x$ according to
\begin{equation}
    \mathbf x_i(\mathbf q,z)
    =
    \mathbf q
    +
    \mathbf s_i(\mathbf q,z),
    \qquad
    i\in\{B,{\rm ax}\},
    \label{eq:lpt_mapping}
\end{equation}
where $\mathbf s_i$ is the displacement field. The first-order displacement is
computed in Fourier space as
\begin{equation}
    \mathbf s_i^{(1)}(\mathbf k,z)
    =
    \left[
        D_i(k,z)-D_i(k,z_{\rm rec})
    \right]
    \frac{i\mathbf k}{k^2}
    \delta_m(\mathbf k,0),
    \qquad
    i\in\{B,{\rm ax}\}.
    \label{eq:lpt_first_order}
\end{equation}
Here $z_{\rm rec}$ is the initial redshift at which the Lagrangian grid is defined and
the displacement field is set to zero. In our implementation,
$z_{\rm rec}$ is determined from the recombination
history of each axiodilaton model. The second-order displacement is included
using the 2LPT implementation~\cite{Bouchet:1994xp} inherited from
\FirstClass.\footnote{A fully consistent second-order treatment in the
axiodilaton cosmology would require rederiving the coupled 2LPT
equations. We retain the standard 2LPT correction implemented in
\FirstClass. The axiodilaton dependence is therefore included in the
first-order displacement, with corrections to the second-order kernel
neglected.} The resulting three-dimensional density fields are denoted by
\begin{equation}
\left\{
\delta_B^{\rm NL}(\mathbf r,z),
\delta_{\rm ax}^{\rm NL}(\mathbf r,z)
\right\}.
\label{eq_nonlinear_density_fields}
\end{equation}
These fields are referred to as nonlinear density fields because they include
second-order mode coupling and the nonlinear Lagrangian-to-Eulerian density
mapping. This terminology does not imply a fully nonlinear $N$-body treatment.

For computational tractability, the astrophysical parameters are fixed
throughout this work to the fiducial EOS2021~\cite{Munoz:2021psm} values
adopted in \FirstClass. We collect them as
\begin{eqnarray}
    \Theta_{\rm astro}
    &=&
    \left\{
    \log_{10}f_{\ast}^{\rm II},\,
    \alpha_{\ast}^{\rm II},\,
    \log_{10}f_{\rm esc}^{\rm II},\,
    \alpha_{\rm esc}^{\rm II},\,
    \log_{10}\left(\frac{L_X^{\rm II}}{{\rm SFR}}\right),\,
    \log_{10}\left(\frac{M_{\rm turn}}{M_\odot}\right),\,
    N_{\gamma},\,
    t_\ast
    \right\}
    \nonumber\\
    &=&
    \left\{
    -1.25,\,
    0.5,\,
    -1.35,\,
    -0.3,\,
    40.5,\,
    8.7,\,
    5000,\,
    0.5
    \right\}.
    \label{eq:astro_params_schematic}
\end{eqnarray}
Here, $f_{\ast}^{\rm II}$ is the Pop II star-formation efficiency, and
$\alpha_{\ast}^{\rm II}$ controls its dependence on halo mass. The parameter
$f_{\rm esc}^{\rm II}$ is the fraction of ionizing photons that escape from
their host galaxies into the IGM, and $\alpha_{\rm esc}^{\rm II}$ controls
its dependence on halo mass. The ratio $L_X^{\rm II}/{\rm SFR}$ is the
soft-band X-ray luminosity per unit star-formation rate (SFR). The parameter
$M_{\rm turn}$ is the characteristic halo mass below which star formation is
exponentially suppressed. The quantity $N_\gamma$ is the number of ionizing
photons produced per stellar baryon, and $t_\ast$ specifies the
star-formation timescale as a fraction of the Hubble time. The parameters
$f_{\ast}^{\rm II}$, $\alpha_{\ast}^{\rm II}$,
$f_{\rm esc}^{\rm II}$, $\alpha_{\rm esc}^{\rm II}$, $N_\gamma$, and
$t_\ast$ are dimensionless. The logarithm of
$L_X^{\rm II}/{\rm SFR}$ is evaluated with the ratio expressed in units of
${\rm erg}\,{\rm s}^{-1}(M_\odot\,{\rm yr}^{-1})^{-1}$.

A joint MCMC analysis varying both the astrophysical and axiodilaton
parameters will be presented in follow-up work. Such an analysis will quantify their degeneracies and determine the extent
to which variations in the astrophysical parameters can mimic the changes in the 21\,cm signal induced by the axiodilaton cosmology.

\subsection{21\,cm Brightness Temperature}
\label{sec:21cm_brightness}

The 21\,cm signal originates from the hyperfine splitting of the ground state of neutral hydrogen in the intergalactic medium (IGM). The hydrogen atom has two hyperfine levels, a triplet state in which the electron and proton spins are aligned and a singlet state with anti-aligned spins. 
The transition from the triplet to the singlet state emits a photon with rest-frame wavelength $\lambda_{21,0}\equiv\lambda_{21} (m_{e,0},m_{p,0}) \simeq 21.1~\mathrm{cm}$. In standard cosmology (where particle masses do not vary) this corresponds to a frequency $\nu_{21,0} \equiv \nu_{21} (m_{e,0},m_{p,0})\simeq 1420.4~\mathrm{MHz}$. 

In general, the photon frequency observed at the Earth, $\nu_{\mathrm{obs}}$, is related to the rest-frame 21\,cm frequency
$\nu_{21}(z)=\nu_{21}\!\bigl(m_e(z),m_p(z)\bigr)$ of radiation emitted
from the IGM at redshift $z$ through
\begin{equation}
    \nu_{\mathrm{obs}} = \frac{\nu_{21}(z)}{1+z}.
\end{equation}
In the $\Lambda$CDM case $\nu_{21}(z) = \nu_{21,0}$, so one has $z = \nu_{21,0}/\nu_{\mathrm{obs}} - 1$. 
If the electron and proton masses vary with redshift, the rest-frame
21\,cm frequency becomes redshift dependent. Since it is a hyperfine splitting it is proportional to $m_e^2/m_p$ and so scales with redshift like 
\begin{equation}
    \nu_{21}(z)
    =
    \nu_{21,0}
    \left[
        \frac{m_e(z)}{m_{e,0}}
    \right]^2
    \left[
        \frac{m_p(z)}{m_{p,0}}
    \right]^{-1}
    =
    \nu_{21,0}e^{\bfg\Delta\bar\chi(z)}
    \label{eq:nu21_mass_scaling}
\end{equation}
(see Appendix~\ref{app:nu21} for a detailed derivation). As expected, $\nu_{\mathrm{obs}}$ scales with $z$ as do other atomic scales.

With the hyperfine frequency specified, we next define the spin temperature $T_s(\mathbf{r},z)$ at the position $\mathbf{r}=\chi(z) \hat{\mathbf{n}}$, where $\hat{\mathbf{n}}$ denotes the line of sight and $\chi(z)$ is the comoving radial distance to redshift $z$. It is defined via the relative populations of the hyperfine levels in the IGM as
\begin{equation}
    \frac{n_1(\mathbf{r},z)}{n_0(\mathbf{r},z)} \equiv \frac{g_1}{g_0}\exp\!\left(-\frac{T_\star(z)}{T_s(\mathbf{r},z)}\right),
\label{eq:Tsdef}
\end{equation}
where $n_1(\mathbf{r},z)$ and $n_0(\mathbf{r},z)$ denote the local number densities of hydrogen atoms in the triplet and singlet states, and $g_1 = 3$ and $g_0 = 1$ are their statistical degeneracies, respectively. Here $T_\star(z) \equiv h\nu_{21}(z)/k_B$ is the 21\,cm transition temperature. In the $\Lambda$CDM case, one has $T_{\star,0}\equiv h\nu_{21,0}/k_B\simeq 68.2 ~\mathrm{mK}$. Since $T_\star(z)=h\nu_{21}(z)/k_B$, it inherits the same mass
dependence as the hyperfine frequency,
\begin{equation}
    T_\star(z)
    =
    T_{\star,0}
    \left[
        \frac{m_e(z)}{m_{e,0}}
    \right]^2
    \left[
        \frac{m_p(z)}{m_{p,0}}
    \right]^{-1}
    =
    T_{\star,0}e^{\bfg\Delta\bar\chi(z)}.
    \label{eq:Tstar}
\end{equation}

With these definitions, we now write the 21\,cm brightness temperature using radiative transfer.
Treating the CMB as a backlight along the line of sight $\hat{\mathbf{n}}$, the radiative transfer of the 21\,cm line through a patch of the IGM with optical depth $\tau_{21}(\mathbf{r},z)$ and spin temperature $T_s(\mathbf r,z)$ yields the local-frame brightness temperature at redshift $z$
\begin{equation}
T_{b,\mathrm{loc}}(\mathbf{r},z)
=
\bar T_\gamma(z)\,e^{-\tau_{21}(\mathbf{r},z)}
+
T_s(\mathbf{r},z)\,\bigl(1-e^{-\tau_{21}(\mathbf{r},z)}\bigr),
\label{eq:deltaTbLoc}
\end{equation}
where $\bar T_\gamma (z)=T_0 (1+z)$ is the background CMB temperature with $T_0 \simeq2.726~{\rm K}$.
The observed brightness temperature at the Earth is redshifted relative to the local one according to $T_{b}(\mathbf{r},z) = T_{b,\mathrm{loc}}(\mathbf{r},z)/(1+z)$. 
Therefore, the observed differential brightness temperature relative to the
CMB is
\begin{align}
    \delta T_{b}(\mathbf r,z)
    &\equiv
    \frac{
       \left( T_{b,{\rm loc}}(\mathbf r,z)-\bar T_\gamma(z) \right)
    }{
        1+z
    }
    \nonumber\\
    &=
    \frac{
       \left( T_s(\mathbf r,z)-\bar T_\gamma(z) \right)
    }{
        1+z
    }
    \left(
        1-e^{-\tau_{21}(\mathbf r,z)}
    \right),
    \label{eq:deltaTbObs}
\end{align}
where the optical depth is the dimensionless measure of 21\,cm absorption
along the line of sight,
\begin{equation}
    \tau_{21}(\mathbf r,z)
    =
    \frac{3}{16}
    \frac{\hbar c^3 A_{10}(z)}
    {k_B T_s(\mathbf r,z)\nu_{21}^2(z)}
    \frac{
        n_{\rm HI}(\mathbf r,z)
    }{
        (1+z)\,dv_\parallel/dr_\parallel(\mathbf r,z)
    }.
    \label{eq_tau21_general}
\end{equation}
Here $A_{10}(z)$ is the Einstein coefficient for spontaneous decay of the
hyperfine triplet state. Its present-day value is
$A_{10,0}\simeq2.85\times10^{-15}~{\rm s}^{-1}$. As derived in
Appendix~\ref{sec:A10}, its redshift dependence is
\begin{equation}
    A_{10}(z)
    =
    A_{10,0}
    \left[
        \frac{m_e(z)}{m_{e,0}}
    \right]^4
    \left[
        \frac{m_p(z)}{m_{p,0}}
    \right]^{-3}
    =
    A_{10,0}e^{\bfg\Delta\bar\chi(z)}.
    \label{eq:A10_mass_scaling}
\end{equation}

We use the comoving line-of-sight coordinate
$r_\parallel$\footnote{The subscripts $\parallel$ and $\perp$ denote
components parallel and perpendicular to $\hat{\mathbf n}$, respectively.},
so that the velocity gradient contains both the Hubble flow and the baryon
peculiar velocity contribution~\cite{Furlanetto:2006jb}, 
\begin{equation}
\frac{dv_\parallel}{dr_\parallel}(\mathbf r,z)
=
\frac{H(z)}{1+z}
+
\frac{\partial v_{B,\parallel}}
{\partial r_\parallel}(\mathbf r,z),
\label{eq:los_velocity_gradient}
\end{equation}
where $v_{B,\parallel}\equiv\mathbf v_B\cdot\hat{\mathbf n}$ is the
line-of-sight component of the baryon peculiar velocity.\footnote{The
peculiar velocity changes the 21\,cm signal in two ways. It enters
$\tau_{21}$ through Eq.~\eqref{eq:los_velocity_gradient}, and it shifts the
observed frequency of each cell. Together the two effects are the
redshift-space distortion of the signal. \FirstClass\
includes the second effect by moving each cell along $\hat{\mathbf n}$ to
the position set by its observed frequency and summing the cells that
arrive at the same position~\cite{Mao2012}.} In the
axiodilaton cosmology, both terms in
Eq.~\eqref{eq:los_velocity_gradient} are modified. The Hubble rate follows
the altered background evolution in Fig.~\ref{fig:bkg}, while the baryon peculiar velocity is
determined by the modified growth of baryon perturbations in Eq.~\eqref{eq:axion_baryon_growth_quasistatic}.

In Eq.~\eqref{eq_tau21_general}, the local neutral hydrogen density is
\begin{equation}
n_{\rm HI}(\mathbf r,z)
=
x_{\rm HI}(\mathbf r,z)n_H(\mathbf r,z)
=
x_{\rm HI}(\mathbf r,z)\bar n_H(z)
\left(
1+\delta_B^{\rm NL}(\mathbf r,z)
\right),
\label{eq_nHI_local}
\end{equation}
where $x_{\rm HI}(\mathbf r,z)$ is the local neutral hydrogen fraction. The
second equality follows from the local hydrogen nuclei density,
\begin{equation}
n_H(\mathbf r,z)
=
\bar n_H(z)
\left(
1+\delta_B^{\rm NL}(\mathbf r,z)
\right).
\label{eq_local_nH}
\end{equation}
Here $\bar n_H(z)$ is the background mean hydrogen nuclei density defined in
Eq.~\eqref{eq_mean_nH_nHe}, and $\delta_B^{\rm NL}(\mathbf r,z)$ is the
nonlinear baryon overdensity defined in
Eq.~\eqref{eq_nonlinear_density_fields}, constructed from the linear baryon
overdensity $\delta_B$ whose evolution is shown in
Fig.~\ref{fig:delta_evolution}.

\begin{figure}[t!]
\centering
\includegraphics[width=1\linewidth]
{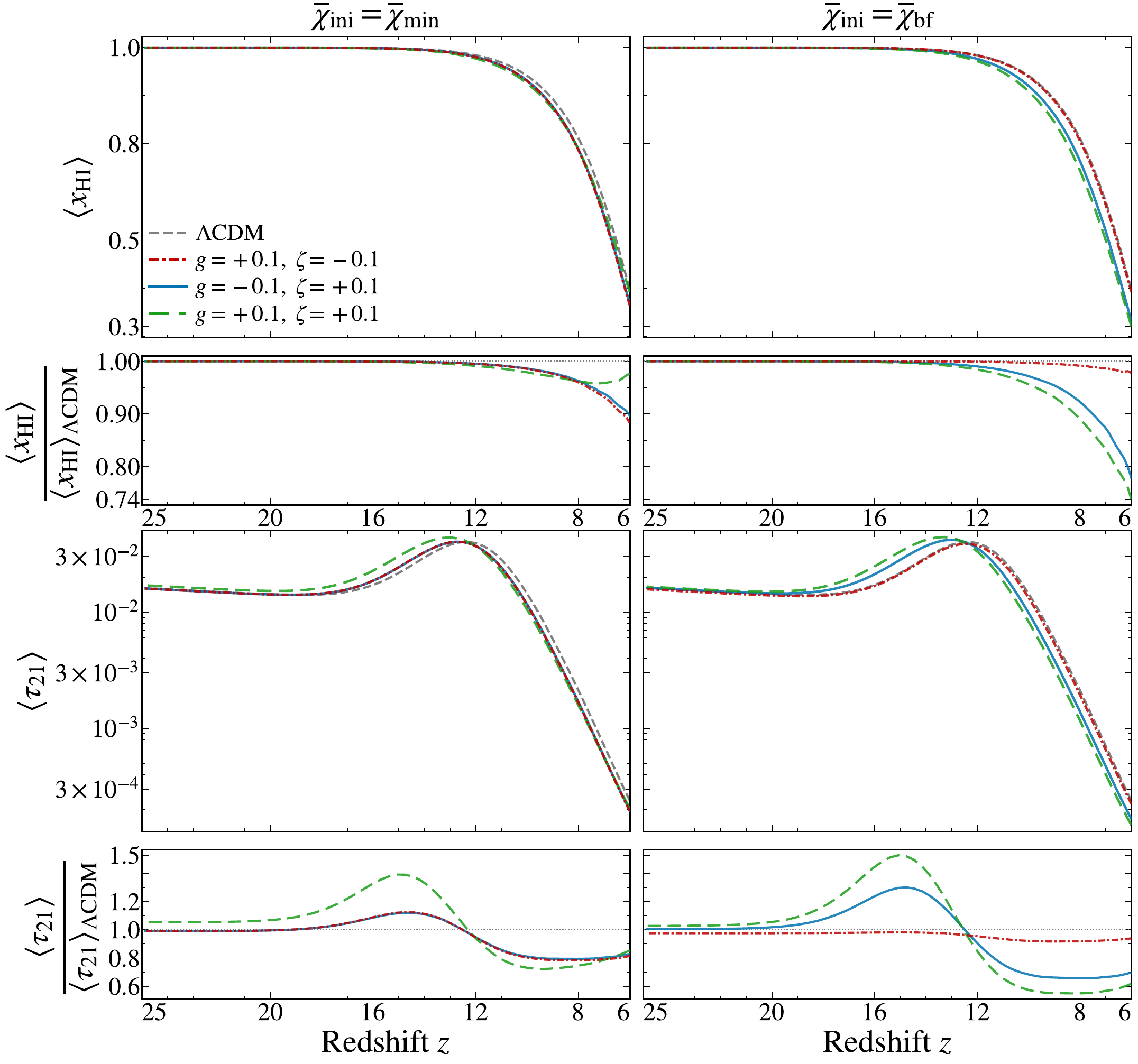}\vspace*{-0.2cm}
\caption{
Redshift evolution of the volume-averaged neutral hydrogen fraction
$\langle x_{\rm HI}(\mathbf r,z)\rangle$ (top) and the volume-averaged 21\,cm optical depth
$\langle\tau_{21}(\mathbf r,z)\rangle$ (bottom) for \LCDM\ (dashed
gray) and representative axiodilaton models. For each quantity, the lower subpanel shows the ratio of the axiodilaton predictions to
the \LCDM\ results. The left and right columns correspond to the initial
dilaton field values $\bar\chi_{\rm ini} = \bar\chi_{\rm min} = 74.010$ and $\bar\chi_{\rm ini} = \bar\chi_{\rm bf}=74.213$,
respectively. The red dash-dotted, blue solid, and green dashed
curves represent $(\bfg,\zeta)=(+0.1,-0.1)$,
$(-0.1,+0.1)$, and $(+0.1,+0.1)$, respectively.
}
\label{fig:x_HI}
\end{figure}

\begin{figure}[!htbp]
    \centering
    \includegraphics[width=1\linewidth]{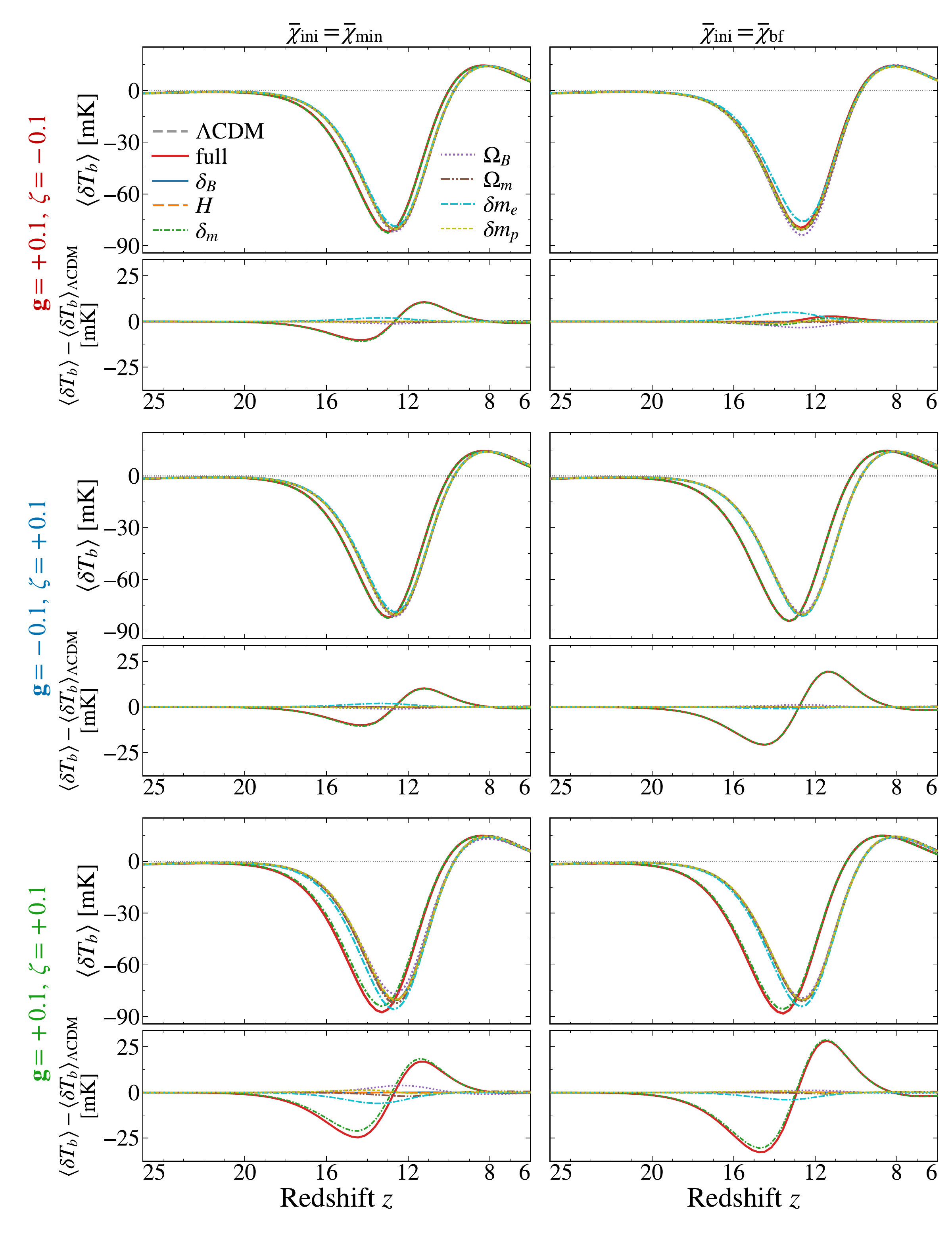} \vspace*{-1.0cm}
    \caption{Volume-averaged differential 21\,cm brightness temperature
    $\langle\delta T_b\rangle$ as a function of redshift when the
    axiodilaton modifications enter the 21\,cm calculation one at a
    time. From top to bottom, the rows show $(\bfg,\zeta)=(+0.1,-0.1)$,
    $(-0.1,+0.1)$, and $(+0.1,+0.1)$, for $\bar\chi_{\rm ini} = \bar\chi_{\rm min} = 74.010$ (left) and $\bar\chi_{\rm ini} = \bar\chi_{\rm bf}=74.213$ (right). Colors and line styles denote the varied
    quantity indicated in the legend. The labeled quantities are the
    baryon and total matter density perturbations $\delta_{\ssB}$ and
    $\delta_m$, the Hubble rate $H(z)$, the fractional energy densities
    $\Omega_{\ssB}(z)$ and $\Omega_{\rm m}(z)$, and the mass variations
    $\delta m_e(z)$ and $\delta m_p(z)$ of Eq.~\eqref{eq_mass_variations}.
    The red full curves include all of these modifications together.}
    \label{fig:varyparams}
\end{figure}
Before astrophysical ionization becomes important, the neutral fraction
$x_{\rm HI}(\mathbf r,z)$ in Eq.~\eqref{eq_nHI_local} remains close to
unity. Once ionizing sources form, it is determined by the local ionized
hydrogen fraction $x_{\rm HII}(\mathbf r,z)$ through
\begin{equation}
x_{\rm HI}(\mathbf r,z)
=
1-x_{\rm HII}(\mathbf r,z),
\label{eq_xHI_from_xHII}
\end{equation}
where $x_{\rm HII}(\mathbf r,z)$ includes contributions from UV and X-ray
ionization,
\begin{equation}
x_{\rm HII}(\mathbf r,z)
=
x_{\rm HII}^{\rm UV}(\mathbf r,z)
+
x_{\rm HII}^{\rm X}(\mathbf r,z).
\label{eq_xHII_total}
\end{equation}
The UV contribution  $x_{\rm HII}^{\rm UV}(\mathbf r,z)\simeq x_e^{\rm UV}(\mathbf r,z)$ is described in
Appendix~\ref{app:UV_ionization}, and the X-ray contribution
$x_{\rm HII}^{\rm X}(\mathbf r,z)\simeq x_e^{\rm X}(\mathbf r,z)$ is
discussed in Appendix~\ref{subsec:xe_evolution}. 
We distinguish these contributions because UV and X-ray photons produce
qualitatively different ionization fields. UV photons are absorbed close
to their sources and form highly ionized regions, while X-ray photons have
much higher energies and smaller photoionization cross sections, allowing
them to propagate over larger distances and generate a partially
ionized component in the IGM.
In the axiodilaton
cosmology, these ionized fractions are modified through the altered
formation and distribution of astrophysical sources and the mass-dependent
atomic processes.
Therefore, the local free-electron fraction
per hydrogen nucleus is
\begin{equation}
x_e(\mathbf r,z)
\simeq
x_{\rm HII}(\mathbf r,z).
\label{eq_xe_total}
\end{equation}
The associated local number densities are, respectively, 
\begin{equation}
n_{\rm HII}(\mathbf r,z)
=
x_{\rm HII}(\mathbf r,z)n_H(\mathbf r,z), \qquad
n_e(\mathbf r,z)
=
x_e(\mathbf r,z)n_H(\mathbf r,z).
\label{eq_ne_nHII}
\end{equation}

We show in Fig.~\ref{fig:x_HI} (top) the volume-averaged neutral
hydrogen fraction $\langle x_{\rm HI}(\mathbf r,z)\rangle$.
At a given redshift, the axiodilaton models generally have a smaller neutral fraction than \LCDM. The relative ordering among the models broadly follows the evolution of the baryon and total matter overdensities and the linear matter power spectrum shown in
Figs.~\ref{fig:delta_evolution} and
\ref{fig:matter_power_spectrum}. This correspondence reflects the
earlier formation and greater abundance of source-hosting halos in
models with enhanced perturbation growth.
For $\bar\chi_{\rm ini}=\chimin$, deviations from \LCDM\ become visible at
$z\simeq15$. The green branch with
$(\bfg,\zeta)=(+0.1,+0.1)$ initially shows the largest reduction, but
its ratio turns upward below $z\simeq7$ and approaches approximately
$0.96$ at $z\simeq6$. The red and blue branches continue to decrease
toward lower redshift, reaching approximately $0.90$. For $\bar\chi_{\rm ini}=\chibf$, the separation is much
larger. The neutral-fraction ratios of the green and blue branches
decrease to approximately $0.73$ and $0.78$ at $z\simeq6$, while the
red branch remains within a few percent of the \LCDM\ result over most
of the redshift range shown.

A quantitative evaluation of the 21\,cm optical depth in
Eq.~\eqref{eq_tau21_general} requires the spin temperature
$T_s(\mathbf r,z)$, whose calculation is described in
Sec.~\ref{sec_21cm_spin}. Here, we first summarize the resulting
evolution of the optical depth.
We show in the bottom panels of Fig.~\ref{fig:x_HI} the
volume-averaged optical depth $\langle\tau_{21}(\mathbf r,z)\rangle$
and its ratio to the \LCDM\ prediction. Over the redshift range of
primary interest, $z<25$, the optical depth remains of order
$\mathcal{O}(10^{-2})$ or smaller and exhibits a broad maximum around
$z\sim12$--$14$. As the Ly$\alpha$ background develops, the
Wouthuysen--Field coupling drives the spin temperature toward the
colder gas kinetic temperature, increasing the optical depth through
its inverse dependence on $T_s$. Toward lower redshift, X-ray heating
raises $T_s$, while reionization reduces the neutral hydrogen density.
These effects cause the optical depth to decrease rapidly.
As in the case of the neutral hydrogen fraction, the relative ordering among the models broadly follows the evolution
of the baryon and total matter overdensities.

As shown above, the 21\,cm optical depth remains small over the redshift
range of primary interest in this work. In the optically thin limit,
$\tau_{21}(\mathbf r,z)\ll1$, Eq.~\eqref{eq:deltaTbObs} reduces to
\begin{align}
\delta T_{b}(\mathbf r,z)
&\simeq
\frac{
\left(T_s(\mathbf r,z)-\bar T_\gamma(z) \right)
}{
1+z
}
\tau_{21}(\mathbf r,z).
\label{eq_deltaTbObs_thin}
\end{align}
This form is useful for interpreting the lower-redshift behavior of the
21\,cm signal. However, all numerical results presented below are obtained using
the full expression in Eq.~\eqref{eq:deltaTbObs}.
The remaining quantity required to evaluate the differential brightness
temperature is the spin temperature
$T_s(\mathbf r,z)$, whose evolution is discussed in
Sec.~\ref{sec_21cm_spin}.

Fig.~\ref{fig:varyparams} shows which axiodilaton modification produces these differences. The matter density contrast $\delta_m$ changes the number of star-forming halos (Appendix~\ref{app:f_coll}). The baryon density contrast $\delta_\ssB$ changes the local gas density (Sec.~\ref{sec:21cm_brightness}). The Hubble rate $H(z)$ changes the 21\,cm optical depth and the rate of thermal and ionization evolution (Sec.~\ref{sec:21cm_brightness} and Appendix~\ref{subsec:Tk}). The fractional energy densities $\Omega_\ssB(z)$ and $\Omega_{\rm m}(z)$ change the 21\,cm optical depth through the mean baryon density and the expansion rate, and therefore the amplitude of $\delta T_{b}$ (Sec.~\ref{sec:21cm_brightness}). The mass variations $\delta m_e(z)$ and $\delta m_p(z)$ change the hyperfine transition, the Wouthuysen--Field coupling, and the recombination and heating rates (Appendices~\ref{app:nu21}--\ref{subsec:xalpha}). For all benchmark points, the $\delta_m$ curve lies closest to the full curve and moves the absorption trough to higher redshift.
The shift occurs because star-forming halos form earlier.
The modified growth of the matter density contrast therefore produces the largest part of the change in $\langle\delta T_{b}\rangle$.
In addition, both $\delta_m$ and $\delta m_e(z)$ deepen the absorption trough.

\subsection{Spin Temperature}
\label{sec_21cm_spin}

We now summarize how the spin temperature $T_s(\mathbf r,z)$, defined in
Eq.~\eqref{eq:Tsdef} through the hyperfine level populations, is expressed
in terms of the gas thermal state, the ionization state, and the Ly$\alpha$
radiation field. 

The spin temperature is set by the competition among three processes.
CMB photons tend to drive $T_s$ toward the CMB temperature
$\bar T_\gamma(z)$. Collisions between hydrogen atoms, electrons, and
protons tend to drive $T_s$ toward the gas kinetic temperature
$T_k(\mathbf r,z)$. Ly$\alpha$ scattering mixes the hyperfine states
through the Wouthuysen--Field effect and drives $T_s$ toward the color
temperature $T_\alpha(\mathbf r,z)$ of the Ly$\alpha$ radiation field.
In the limit $T_s\gg T_\star(z)$, the Boltzmann equations for the
hyperfine level populations lead to the quasi-equilibrium solution
\begin{equation}
T_s^{-1}(\mathbf r,z)
=
\frac{
x_{\rm CMB}(\mathbf r,z)\bar T_\gamma^{-1}(z)
+
x_{\rm coll}(\mathbf r,z)T_k^{-1}(\mathbf r,z)
+
x_\alpha(\mathbf r,z)T_\alpha^{-1}(\mathbf r,z)
}{
x_{\rm CMB}(\mathbf r,z)
+
x_{\rm coll}(\mathbf r,z)
+
x_\alpha(\mathbf r,z)
}.
\label{eq:Ts}
\end{equation}
The dimensionless CMB coupling coefficient is defined as
\begin{equation}
x_{\rm CMB}(\mathbf r,z)
\equiv
\frac{1-e^{-\tau_{21}(\mathbf r,z)}}
{\tau_{21}(\mathbf r,z)}.
\label{eq:x_CMB}
\end{equation}
This coefficient accounts for the radiative transfer of CMB photons at
the 21\,cm frequency through the IGM. 
As shown in Fig.~\ref{fig:x_HI}, the 21\,cm optical depth is typically
much smaller than unity over the redshift range considered here.
Consequently, $x_{\rm CMB}(\mathbf r,z)\simeq1$.

The collisional coupling coefficient is
\begin{eqnarray}\label{eq:x_coll}
x_{\rm coll}(\mathbf r,z)
=
\frac{T_\star(z)}
{A_{10}(z)\bar T_\gamma(z)}
\big[
\kappa_{10}^{\rm HH}\!\left(T_k(\mathbf r,z)\right)
n_{\rm HI}(\mathbf r,z)
&+&
\kappa_{10}^{\rm eH}\!\left(T_k(\mathbf r,z)\right)
n_e(\mathbf r,z) \\
&+&
\kappa_{10}^{\rm pH}\!\left(T_k(\mathbf r,z)\right)
n_p(\mathbf r,z)
\big].\nonumber
\end{eqnarray}
The coefficients $\kappa_{10}^{i{\rm H}}$ denote the collisional
de-excitation rates for scattering between hydrogen atoms and the species
$i\in\{{\rm H},e,p\}$, where $p$ denotes a free proton and
$n_p(\mathbf r,z)\simeq n_{\rm HII}(\mathbf r,z)$. These coefficients
have units of ${\rm cm}^{3}\,{\rm s}^{-1}$ and are functions of the
local gas kinetic temperature $T_k(\mathbf{r},z)$ (see Appendix~\ref{subsec:xray} for details). Their calculation requires solving the
corresponding quantum mechanical two-body scattering problems \cite{Zygelman_2005,Furlanetto:2006su,Furlanetto:2007te}.
At very low temperatures, $T_k\lesssim3\,{\rm K}$,
$\kappa_{10}^{\rm pH}$ is the largest of the three coefficients, whereas
$\kappa_{10}^{\rm eH}$ becomes the largest at higher temperatures.
The coefficient $\kappa_{10}^{\rm HH}$ is strongly suppressed at low
temperatures and decreases rapidly as $T_k\rightarrow1\,{\rm K}$.
We adopt the tabulated values of $\kappa_{10}^{i{\rm H}}$ over
$T_k\in[1,10^4]\,{\rm K}$ from
Refs.~\cite{Zygelman_2005,Furlanetto:2006su,Furlanetto:2007te}, as
implemented in \FirstClass.
Although the coefficients typically lie in the range
$\kappa_{10}^{i{\rm H}}\sim10^{-13}$--$10^{-8}\,
{\rm cm}^{3}\,{\rm s}^{-1}$, their contributions to
$x_{\rm coll}$ are determined by the products
$n_i(\mathbf r,z)\kappa_{10}^{i{\rm H}}$. The much larger abundance of
neutral hydrogen generally makes ${\rm HH}$ collisions the main
collisional contribution in the weakly ionized IGM. The low particle
densities in the diffuse IGM nevertheless suppress all three products,
so collisional coupling is generally subdominant at lower redshifts $z\lesssim20$. It becomes important mainly at substantially
higher redshift or in highly overdense regions such as minihalos \cite{Furlanetto:2006jb, Shapiro2006}.

Here we retain the explicit redshift dependence
of $T_\star(z)$ and $A_{10}(z)$ in Eq.~\eqref{eq:x_coll}. 
The remaining axiodilaton dependence of $x_{\rm coll}$ enters through
$T_k(\mathbf r,z)$ and the local densities $n_{\rm HI}(\mathbf r,z)$,
$n_e(\mathbf r,z)$, and $n_p(\mathbf r,z)$. 
The scattering coefficients may also carry an implicit dependence on
$m_e(z)$ and $m_p(z)$ because these masses enter the corresponding
two-body scattering problems. We neglect this additional dependence and
use the standard tabulated coefficients throughout our calculation%
\footnote{Previous studies have examined the dependence of collisional
de-excitation rates and the resulting 21\,cm signal on variations in the
fine structure constant
\cite{Khatri:2007yv,Flambaum:2010um,Khatri:2010ns}. These analyses do not
determine the response of $\kappa_{10}^{i{\rm H}}$ to the correlated
variations of $m_e(z)$ and $m_p(z)$ with fixed fine structure constant
considered here. Such a treatment requires a dedicated two-body
scattering calculation and is not included in this work.}. Since $x_{\rm coll}\ll1$ over the redshift range relevant to our analysis, we expect any resulting correction to have a negligible effect on the spin-temperature
evolution.

\begin{figure}[htbp!]
    \centering
    \includegraphics[width=0.95\linewidth]
    {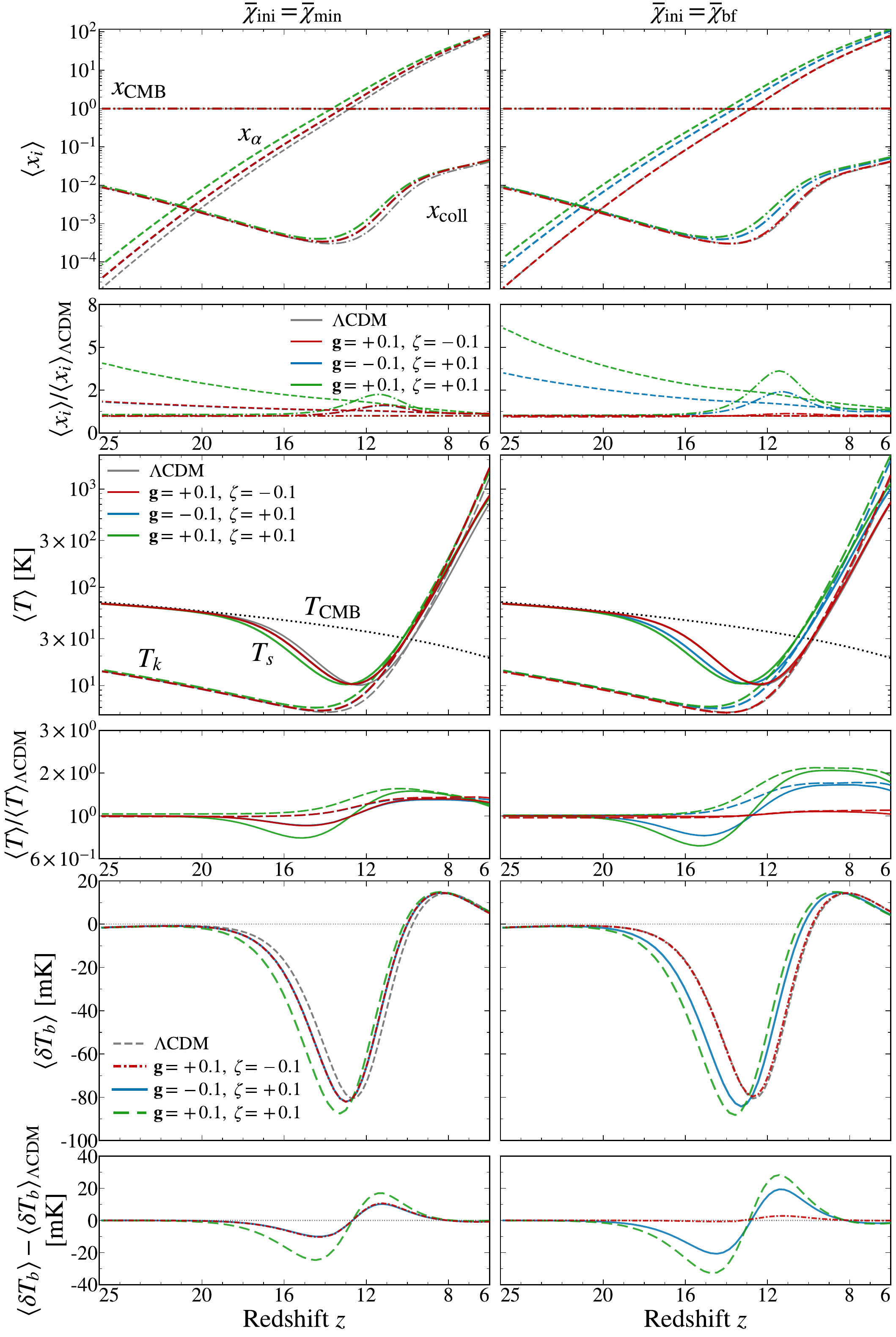}\vspace*{-0.2cm}
    \caption{
    Redshift evolution of the volume-averaged 21\,cm coupling coefficients
$\langle x_i\rangle$ (top), spin and gas kinetic temperatures
$\langle T_s\rangle$ and $\langle T_k\rangle$ (middle), and differential
brightness temperature $\langle\delta T_b\rangle$ (bottom) for
$\bar\chi_{\rm ini} = \bar\chi_{\rm min} = 74.010$ (left) and $\bar\chi_{\rm ini} = \bar\chi_{\rm bf}=74.213$ (right). See the text for a
detailed discussion.
    }
    \label{fig:x_Ts_Tb}
\end{figure}

The detailed expression of the Wouthuysen--Field
coupling coefficient $x_\alpha$ entering Eq.~\eqref{eq:Ts} is presented in
Appendix~\ref{subsec:xalpha}. The evolution of the gas kinetic
temperature and the Ly$\alpha$ radiation temperature
are discussed in Appendices~\ref{subsec:Tk} and \ref{sec:Salpha}, respectively.
Here, we first summarize the
calculated coupling coefficients and their combined consequences for the
spin temperature and the differential brightness temperature.

Fig.~\ref{fig:x_Ts_Tb} (top) shows the volume-averaged coupling coefficients. 
The CMB coupling coefficient remains very close to unity throughout the
redshift range shown, $\langle x_{\rm CMB}\rangle\simeq1$, as expected from
the small 21\,cm optical depth.
The collisional coefficient $\langle x_{\rm coll}\rangle$ remains well
below unity. It initially decreases as the gas density falls, reaches a
minimum around $z\simeq14$--$15$, and rises modestly toward lower redshift
as radiative heating and ionization alter the scattering rates and the
abundances of electrons and protons. Its magnitude nevertheless remains
too small for collisions to have an appreciable effect on the spin
temperature in this redshift range.

The principal transition is governed by the Wouthuysen--Field coefficient
$x_\alpha$. At high redshifts, $z\gtrsim20$,
$\langle x_\alpha\rangle\ll1$, and the CMB coupling keeps
$\langle T_s\rangle$ close to $\bar T_\gamma$, as shown in the middle panel
of Fig.~\ref{fig:x_Ts_Tb}. As the first radiative
sources generate a Ly$\alpha$ background, $\langle x_\alpha\rangle$
increases and exceeds unity around $z\simeq13$--$15$. Repeated Ly$\alpha$
scattering then drives $\langle T_s\rangle$ toward the Ly$\alpha$ color
temperature, which closely approaches the gas kinetic temperature
$\langle T_k\rangle$. Since the gas is colder than the CMB during this
period, $\langle T_s\rangle$ falls below $\bar T_\gamma$, and the
21\,cm absorption signal deepens. As shown in
Fig.~\ref{fig:x_Ts_Tb}, both $\langle T_s\rangle$ and
$\langle\delta T_{b}\rangle$ reach their minima around
$z\simeq13$--$14$.

Toward lower redshift, X-ray heating raises the gas kinetic temperature,
and the strong Wouthuysen--Field coupling causes
$\langle T_s\rangle$ to follow this rise. As shown by the temperature
curves in the middle panel of Fig.~\ref{fig:x_Ts_Tb}, $\langle T_s\rangle$ turns upward
after reaching its minimum and crosses $\bar T_\gamma$ around
$z\simeq10$--$11$. The $\langle\delta T_{b}\rangle$ curves (bottom panel)
therefore cross zero, marking the transition of the 21\,cm signal from
absorption to emission. The emission signal reaches its maximum around
$z\simeq8$--$9$. Toward still lower redshift, the curves turn downward
as reionization gradually reduces the neutral hydrogen abundance.

The differences in the timing of radiative source formation among the models can be traced dominantly to the model-dependent evolution of the baryon and total matter perturbations, which is discussed in Appendix~\ref{subsec:perturbation_evolution} and shown in Fig.~\ref{fig:varyparams}. When the axiodilaton modifications enter the 21\,cm calculation one at a time, the shifts in $\langle\delta T_b\rangle$ are reproduced mainly by $\delta_m$.
Stronger structure growth increases the
abundance of star-forming halos at a given redshift, causing the radiation backgrounds to develop earlier. This
connection is most clearly reflected in the onset of the
Wouthuysen--Field coupling in Fig.~\ref{fig:x_Ts_Tb}. For $\bar\chi_{\rm ini}=\chimin$, the green
branch exhibits the earliest and strongest increase in
$\langle x_\alpha\rangle$, while the red and blue branches remain closer
to the \LCDM\ prediction. For $\bar\chi_{\rm ini}=\chibf$, the green
branch again shows the largest departure, followed by the blue branch,
whereas the red branch closely tracks \LCDM.

The model-dependent depths of the
$\langle\delta T_{b}\rangle$ absorption feature shown in the bottom
panels of Fig.~\ref{fig:x_Ts_Tb} arise from the competing effects of
Ly$\alpha$ coupling and X-ray heating.
While the gas remains colder than the CMB, an earlier and stronger Ly$\alpha$
background rapidly drives $\langle T_s\rangle$ closer to
$\langle T_k\rangle$, thereby increasing the magnitude of the absorption
signal. Earlier source formation also advances X-ray heating, although its
effect develops more gradually because it requires the cumulative deposition
of X-ray energy into the IGM. In the models considered here, the enhanced
Ly$\alpha$ coupling therefore deepens the absorption signal before X-ray
heating raises $\langle T_k\rangle$ and $\langle T_s\rangle$.
Toward lower redshift, the differences in
$\langle\delta T_{b}\rangle$ among the models become smaller.
Since $\tau_{21}\propto T_s^{-1}$, the optically thin expression in
Eq.~\eqref{eq_deltaTbObs_thin} depends on the temperature factor
$1-\bar T_\gamma/T_s$, which approaches unity once
$T_s\gg\bar T_\gamma$. The 21\,cm signal then becomes less sensitive to
model-dependent variations in $T_s$ and is governed mainly by the remaining
neutral hydrogen and its spatial distribution.

\begin{figure}[t!]
    \centering
    \includegraphics[width=1.0\linewidth]{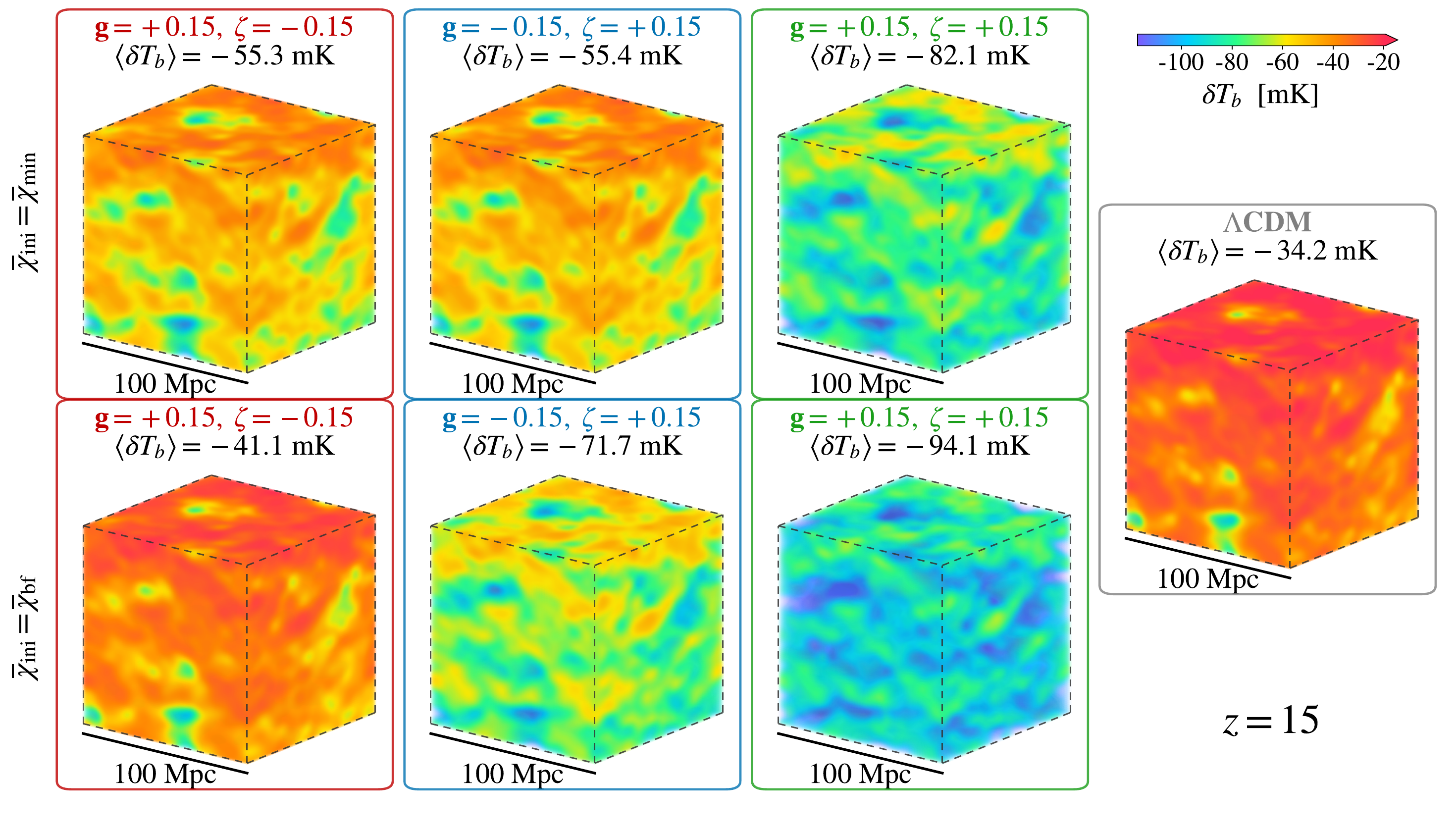}
    \caption{Three-dimensional renderings of the 21\,cm brightness-temperature field $\delta T_b$ at $z=15$ for the axiodilaton models and \LCDM. Each cube shows a $(100\,{\rm Mpc})^3$ comoving volume generated with \FirstClass. From left to right, the first three columns correspond to
    $(\bfg,\zeta)=(+0.15,-0.15)$, $(-0.15,+0.15)$, and
    $(+0.15,+0.15)$, respectively. The upper and lower rows show
    $\bar\chi_{\rm ini} = \bar\chi_{\rm min} = 74.010$ (top) and $\bar\chi_{\rm ini} = \bar\chi_{\rm bf}=74.213$ (bottom), respectively, while the
    rightmost panel shows the \LCDM\ reference. The volume-averaged
    brightness temperature $\langle\delta T_b\rangle$ is indicated above
    each cube.}
    \label{fig:dTb_3D_z15}
\end{figure}

Fig.~\ref{fig:dTb_3D_z15} complements that result by displaying the
spatial distribution of $\delta T_b$ at $z=15$. 
The \LCDM\ model has the largest, or least negative, mean brightness
temperature, $\langle\delta T_b\rangle=-34.2\,{\rm mK}$, whereas the
green branch, $(\bfg,\zeta)=(+0.15,+0.15)$, produces the most negative
signal for both values of $\bar\chi_{\rm ini}$. The red and blue branches lie between these cases, with their relative separation depending on $\bar\chi_{\rm ini}$.

\section{Results}
\label{sec:result}

In this section, we examine how axiodilaton cosmology modifies the
21\,cm signal during cosmic dawn and reionization relative to \LCDM.
Our main focus is the spatial fluctuations of the 21\,cm brightness
temperature, characterized by the 21\,cm power spectrum. We assess the deviations of the 
model from \LCDM\ using the projected
sensitivity of SKA1-Low. 
Here, we fix the astrophysical parameters to
the fiducial EOS2021 values~\cite{Munoz:2021psm}, vary only the axiodilaton
couplings $\bfg$ and $\zeta$, and determine the allowed parameter space using a
simple chi-squared analysis.

\subsection{The 21\,cm Power Spectrum}
\label{sec:21cm_power}

We define
the fluctuation field as
\begin{equation}
    \Delta T_{21}(\mathbf r,z)
    \equiv
    \delta T_{b}(\mathbf r,z)
    -
    \left\langle
    \delta T_{b}(\mathbf r,z)
    \right\rangle .
    \label{eq_DeltaT21_field}
\end{equation}
The corresponding 21\,cm power spectrum is defined through
\begin{equation}
    \left\langle
    \Delta T_{21}(\mathbf k,z)
    \Delta T_{21}^{*}(\mathbf k',z)
    \right\rangle
    =
    (2\pi)^3
    \delta_D^{(3)}(\mathbf k-\mathbf k')
    P_{21}(k,z),
    \label{eq_P21_def}
\end{equation}
and we present our results in terms of the dimensionless power spectrum
\begin{equation}
    \Delta_{21}^2(k,z)
    \equiv
    \frac{k^3}{2\pi^2}P_{21}(k,z),
    \label{eq_Delta21_def}
\end{equation}
which has units of ${\rm mK}^2$.

\begin{figure}[t!]
    \centering
    \includegraphics[width=1.0\linewidth]
    {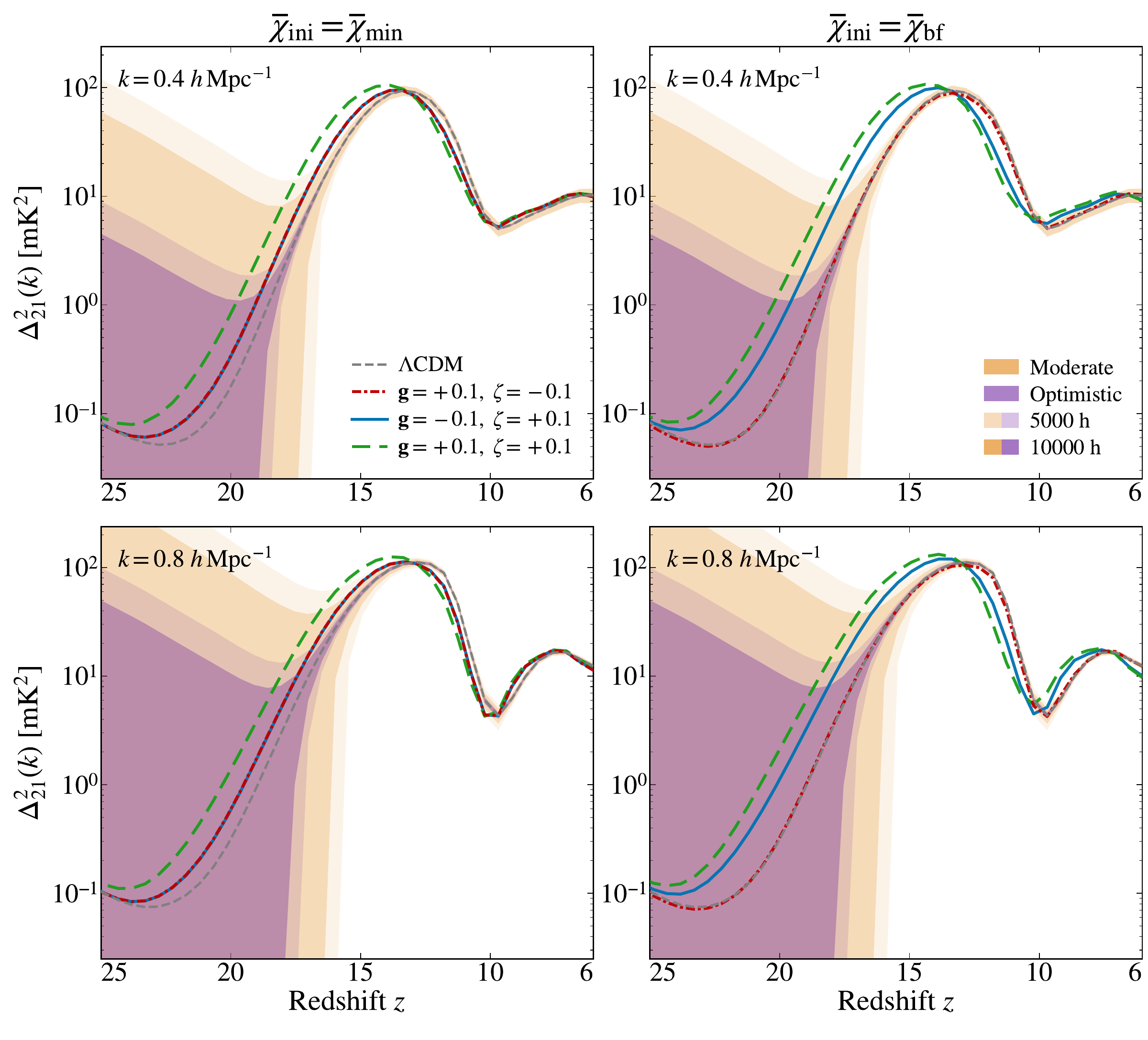}
    \caption{
    Redshift evolution of the dimensionless 21\,cm power spectrum
    $\Delta_{21}^2(k,z)$ at
    $k=0.4\,h\,{\rm Mpc}^{-1}$ (upper row) and
    $k=0.8\,h\,{\rm Mpc}^{-1}$ (lower row).
    The left and right columns correspond to
    $\bar\chi_{\rm ini} = \bar\chi_{\rm min} = 74.010$ and $\bar\chi_{\rm ini} = \bar\chi_{\rm bf}=74.213$, respectively.
    The gray dashed curves represent \LCDM. The red dash-dotted, blue solid,
    and green dashed curves correspond to
    $(\bfg,\zeta)=(+0.1,-0.1)$, $(-0.1,+0.1)$, and $(+0.1,+0.1)$,
    respectively. The shaded regions centered on the \LCDM\ prediction
    show the projected total uncertainties for SKA1-Low. Orange and purple
    correspond to the moderate and optimistic sky-temperature models,
    respectively. For each sky-temperature model, the lighter and darker
    regions correspond to total observing times of $5000\,{\rm hr}$ and
    $10000\,{\rm hr}$, respectively.
    }
    \label{fig:P21_chi}
\end{figure}

\begin{figure}[htbp!]
    \centering
    \includegraphics[width=0.8\linewidth]{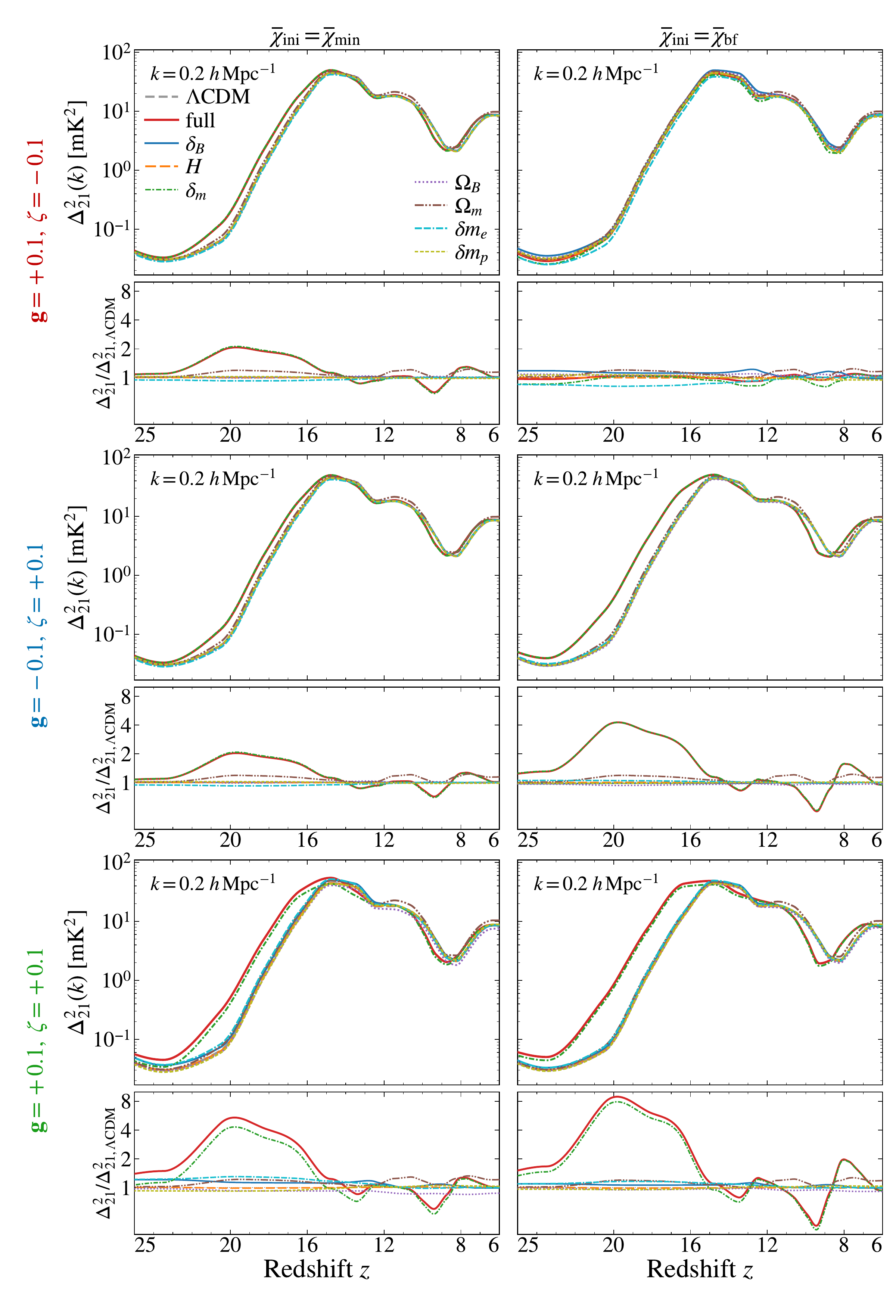}
    \caption{Dimensionless 21\,cm power spectrum $\Delta_{21}^2(k,z)$ at
    $k=0.2\,h\,{\rm Mpc}^{-1}$ as a function of redshift when the
    axiodilaton modifications enter the 21\,cm calculation one at a
    time. From top to bottom, the rows show $(\bfg,\zeta)=(+0.1,-0.1)$,
    $(-0.1,+0.1)$, and $(+0.1,+0.1)$, for $\bar\chi_{\rm ini} = \bar\chi_{\rm min} = 74.010$ (left) and $\bar\chi_{\rm ini} = \bar\chi_{\rm bf}=74.213$ (right). Colors and line styles denote the varied
    quantity indicated in the legend. The labeled quantities are the
    baryon and total matter density perturbations $\delta_{\ssB}$ and
    $\delta_m$, the Hubble rate $H(z)$, the fractional energy densities
    $\Omega_{\ssB}(z)$ and $\Omega_{\rm m}(z)$, and the mass variations
    $\delta m_e(z)$ and $\delta m_p(z)$ of Eq.~\eqref{eq_mass_variations}.
    The red full curves include all of these modifications together.}
    \label{fig:ablation_P21}
\end{figure}

Fig.~\ref{fig:P21_chi} shows the redshift evolution of
$\Delta_{21}^2(k,z)$ at two representative wavenumbers,
$k=0.4\,h\,{\rm Mpc}^{-1}$ and $0.8\,h\,{\rm Mpc}^{-1}$. The overall features are similar at both scales. At $z\gtrsim22$, radiative sources are rare and the 21\,cm
fluctuations remain small. As the first galaxies generate spatially
inhomogeneous Ly$\alpha$ and X-ray backgrounds, variations in
Wouthuysen--Field coupling and gas heating cause the power spectrum to rise
rapidly. The power reaches its main maximum around $z\simeq13$--$15$, when
the radiation backgrounds and the gas temperature remain highly
inhomogeneous. The subsequent spread of X-ray heating reduces the
temperature contrast across the IGM, producing the minimum near
$z\simeq9$--$10$. At lower redshifts, ionized regions grow and become more widespread,
generating a second rise through spatial fluctuations.

For $\bar\chi_{\rm ini}=\chimin$, shown in the left column of
Fig.~\ref{fig:P21_chi}, the red and blue curves, corresponding to
$(\bfg,\zeta)=(+0.1,-0.1)$ and $(-0.1,+0.1)$, respectively, are nearly
degenerate and remain close to the \LCDM\ prediction. Their power begins to increase slightly earlier than in \LCDM, while the locations and amplitudes of the later features differ only modestly. The green curve, corresponding to $(\bfg,\zeta)=(+0.1,+0.1)$, displays a clearer departure at high redshift. Its rise begins earlier, and both the main maximum and the subsequent minimum are displaced toward higher redshift. Since this initial condition places the dilaton at the minimum of its potential, the ordering closely reflects the modified growth of matter perturbations. In particular, the positive product $\bfg\zeta$ for $(\bfg,\zeta)=(+0.1,+0.1)$ enhances the effective interaction between baryon and axion perturbations, leading to earlier halo and source formation (see {\it e.g.}~Eq.~\pref{eq:effective_gravitational_couplings}). 
For the red and blue models, both of which have $\bfg\zeta<0$, this effective coupling is weakened, resulting in slower growth of matter overdensities and correspondingly smaller 21\,cm power at high redshift. 
All three curves approach one another as
reionization progresses.

For $\bar\chi_{\rm ini}=\chibf$, shown in the right column of
Fig.~\ref{fig:P21_chi}, the red curve,
$(\bfg,\zeta)=(+0.1,-0.1)$, remains close to the \LCDM\ prediction
throughout the redshift range. The blue and green curves,
corresponding to $(\bfg,\zeta)=(-0.1,+0.1)$ and $(+0.1,+0.1)$,
respectively, begin to rise and reach their main features at higher
redshifts. The main maxima and subsequent heating minima of the blue
and green curves are shifted toward higher redshifts relative to
\LCDM, typically by $\Delta z\simeq2$--$3$. These shifts lead to
substantial differences in $\Delta_{21}^2$ at a fixed redshift. The
relative ordering among the models broadly follows the evolution of
the baryon and total matter overdensities and the linear matter power
spectrum. Among the three axiodilaton models,
the green model evolves earliest, followed by the blue model, while
the red model remains closest to \LCDM. Fig.~\ref{fig:ablation_P21} shows that the matter density contrast determines how early each model departs from \LCDM\
for both initial values.
When the axiodilaton modifications enter one at a time, the $\delta_m$ curve lies closest to the full curve and moves the rise and the main maximum to higher redshift, as for $\langle\delta T_{b}\rangle$ in Fig.~\ref{fig:varyparams}. Since $\Delta_{21}^2$ measures spatial fluctuations, $\delta_m$ affects $\Delta_{21}^2$ through both the number and the locations of star-forming halos, which form in overdense regions (Appendix~\ref{app:conditional_halo_mass_function}).

To compare these deviations with the expected observational sensitivity, we
assign an effective uncertainty to each $(k,z)$ bin,
\begin{equation}
    \sigma_{\rm obs}^2(k,z)
    =
    \sigma_{\rm SKA}^2(k,z)
    +
    \sigma_{\rm theo}^2(k,z),
    \qquad
    \sigma_{\rm theo}(k,z)
    =
    f_{\rm theo}\,
    \Delta^2_{21,\Lambda{\rm CDM}}(k,z).
    \label{eq_noise_obs}
\end{equation}
Here $\sigma_{\rm SKA}(k,z)$ is the SKA1-Low uncertainty on
$\Delta_{21}^2(k,z)$, computed with \textsc{21cmSense}~\cite{Murray:2024}
using the AA4 layout~\cite{SKAO:2023}. The uncertainty $\sigma_{\rm SKA}$ includes both the thermal noise and the sample variance of the $\Lambda$CDM 21\,cm power spectrum, as described in Appendix~\ref{app_ska_sensitivity}. The baseline mapping, observing
configuration, foreground assumptions, and system-temperature model are also
described in Appendix~\ref{app_ska_sensitivity}. We adopt $f_{\rm theo}=\{0, 0.05, 0.1\}$ to represent theoretical uncertainties in the semi-numerical predictions of the 21\,cm
signal~\cite{Zahn:2011,Schneider:2023ciq,Park:2025phj}.

The projected uncertainty grows rapidly toward high redshift. As shown
in Eq.~\eqref{eq_21cmsense_noise_cell}, the thermal-noise power depends
on the sky temperature. Since
$T_{\rm sky}\propto\nu_{\rm obs}^{-2.55}$, its contribution becomes
increasingly important at high redshift. Here, the moderate and
optimistic forecasts refer to the sky-temperature models given in
Eqs.~\eqref{eq_Tsky_moderate} and \eqref{eq_Tsky_optimistic},
respectively. The moderate model gives a higher sky temperature and
hence a larger thermal uncertainty. 
Each sky-temperature model is
evaluated for both $5000\,{\rm hr}$ and $10000\,{\rm hr}$ of observing
time. A longer observing time narrows the uncertainty bands.
For reference, the benchmark observing strategy adopted in the
\textsc{21cmSense} SKA-AA4 forecast tutorial assumes
\begin{equation}
    t_{\rm obs}
    =
    180\times6\,{\rm hr}
    =
    1080\,{\rm hr}.
\end{equation}
Here, $180$ days approximates the portion of a year during which the
chosen sky field is observable at night, while $6\,{\rm hr}$ denotes
the effective daily observing time for that field. Scaling from this estimate,
$5000\,{\rm hr}$ and $10000\,{\rm hr}$ correspond to approximately
$4.6$ and $9.3$ observing years.
Finally, the uncertainty bands also broaden with
increasing wavenumber. As shown in Eq.~\eqref{eq_DeltaN_cell_def}, the
thermal uncertainty of an individual cell scales as $k^3$.

\subsection{SKA1-Low Sensitivity to Axiodilaton Models}
\label{sec:Sensitivity}

We quantify the expected distinguishability of the axiodilaton model
from \LCDM\ using the 21\,cm power spectrum presented above. We take
the \LCDM\ power spectrum as a reference data set.
For an axiodilaton model specified by the coupling-parameter vector
$\boldsymbol{\theta}\equiv(\bfg,\zeta)$, with
$\bar\chi_{\rm ini}$ held fixed, we define
\begin{equation}
    \chi^2(\boldsymbol{\theta})
    =
    \sum_{p=1}^{6}
    \sum_{q=1}^{10}
    \frac{
    \left[
    \Delta^2_{21,{\rm AD}}
    (k_p,z_q;\boldsymbol{\theta})
    -
    \Delta^2_{21,\Lambda{\rm CDM}}(k_p,z_q)
    \right]^2
    }
    {\sigma_{\rm obs}^2(k_p,z_q)}.
    \label{eq_chi2_AD}
\end{equation}
Here $\sigma_{\rm obs}(k_p,z_q)$ is the total uncertainty defined in
Eq.~\eqref{eq_noise_obs}. We neglect correlations
between different wavenumber and redshift bins.
We use logarithmically spaced wavenumber bins centered at
\begin{equation}
    k_p
    =
    (0.192,\ 0.288,\ 0.432,\ 0.648,\ 0.972,\ 1.457)
    \,h\,{\rm Mpc}^{-1},
\end{equation}
where the dimensionless Hubble parameter $h=0.6766$ is listed in
Table~\ref{tab:inputs}. The centers of the adjacent bins satisfy
$k_{p+1}/k_p=1.5$. The resulting widths of the wavenumber-bin are larger than
the line-of-sight wavenumber spacing $\Delta k_\parallel$ given in
Eq.~\eqref{eq_kparallel_spacing}.
We use redshift bins centered at
\begin{equation}
    z_q
    =
    (19.292,\ 17.210,\ 15.516,\ 14.111,\ 12.926,\ 11.913,\
    11.037,\ 10.273,\ 9.600,\ 9.003),
    \label{eq_chi2_redshift_bins}
\end{equation}
which corresponds to observed 21\,cm frequencies
$\nu_{\rm obs}=\nu_{21,0}/(1+z_q)$ spaced uniformly in steps of
$8\,{\rm MHz}$, from $70$ to $142\,{\rm MHz}$. The redshift-bin widths are
comparable to those given in Eq.~\eqref{eq:zbin}.

The data vector therefore contains
$N_{\rm bin}=6\times10=60$ power-spectrum measurements. The SKA1-Low noise enters the analysis only through the uncertainties $\sigma_{\rm obs}(k_p,z_q)$ in Eq.~\eqref{eq_chi2_AD}. The mock data themselves are set equal to the \LCDM\ prediction, without adding a random fluctuation drawn from this noise. The minimum of Eq.~\eqref{eq_chi2_AD} over $\boldsymbol{\theta}$ is
therefore attained in the \LCDM\ limit, where it is close to zero,
and we define the region consistent with \LCDM\ by
\begin{equation}
    \Delta\chi^2(\boldsymbol{\theta})
    \equiv
    \chi^2(\boldsymbol{\theta})
    -\min_{\boldsymbol{\theta}'}\chi^2(\boldsymbol{\theta}')
    \le
    \Delta\chi^2_{\rm crit},
    \label{eq_delta_chi2}
\end{equation}

We determine the critical value using the survival probability of the
$\chi^2$ distribution,
\begin{equation}
    \int_{\chi^2_{\rm crit}}^\infty
    f_{\nu_{\rm dof}}(x)\,dx
    =
    p,
    \qquad
    f_{\nu_{\rm dof}}(x)
    =
    \frac{
    x^{\nu_{\rm dof}/2-1}e^{-x/2}
    }{
    2^{\nu_{\rm dof}/2}
    \Gamma(\nu_{\rm dof}/2)
    }.
    \label{eq_chi2_critical}
\end{equation}
For $\nu_{\rm dof}=2$, the probability density reduces to $f_2(x)=e^{-x/2}/2$, and Eq.~\eqref{eq_chi2_critical} gives $\Delta\chi^2_{\rm crit} = -2{\rm ln}p$. We take $p=0.0455$, which corresponds to the 2$\sigma$ significance level of a Gaussian distribution, so that
\begin{equation}
    \Delta\chi^2_{\rm crit}
    \simeq
    6.18.
    \label{eq_chi2_critical_value}
\end{equation}

This threshold is the same as that used to define the $2\sigma$
contours in the \texttt{21cmfish} Fisher-matrix forecasts of the
21\,cm power spectrum~\cite{Mason:2022obt}.
The $\nu_{\rm dof}=2$ distribution assumed for $\Delta\chi^2$ holds only if the power spectrum responds approximately linearly to $(\bfg,\zeta)$, which fails near the origin. We therefore perform a coverage test with noisy mock spectra at points on the contours, following Ref.~\cite{Feldman:1997qc}. The contours contain the true point in $96\%$ of mocks for $\bar\chi_{\rm ini}=\chimin$ and at least $93\%$ for $\bar\chi_{\rm ini}\neq\chimin$, compared with the expected $95.45\%$, so the threshold is reliable. A full treatment of non-Gaussian features is left to a future MCMC analysis that also varies the astrophysical parameters.
\begin{figure}[t!]
    \centering
    \includegraphics[width=1.0\linewidth]
    {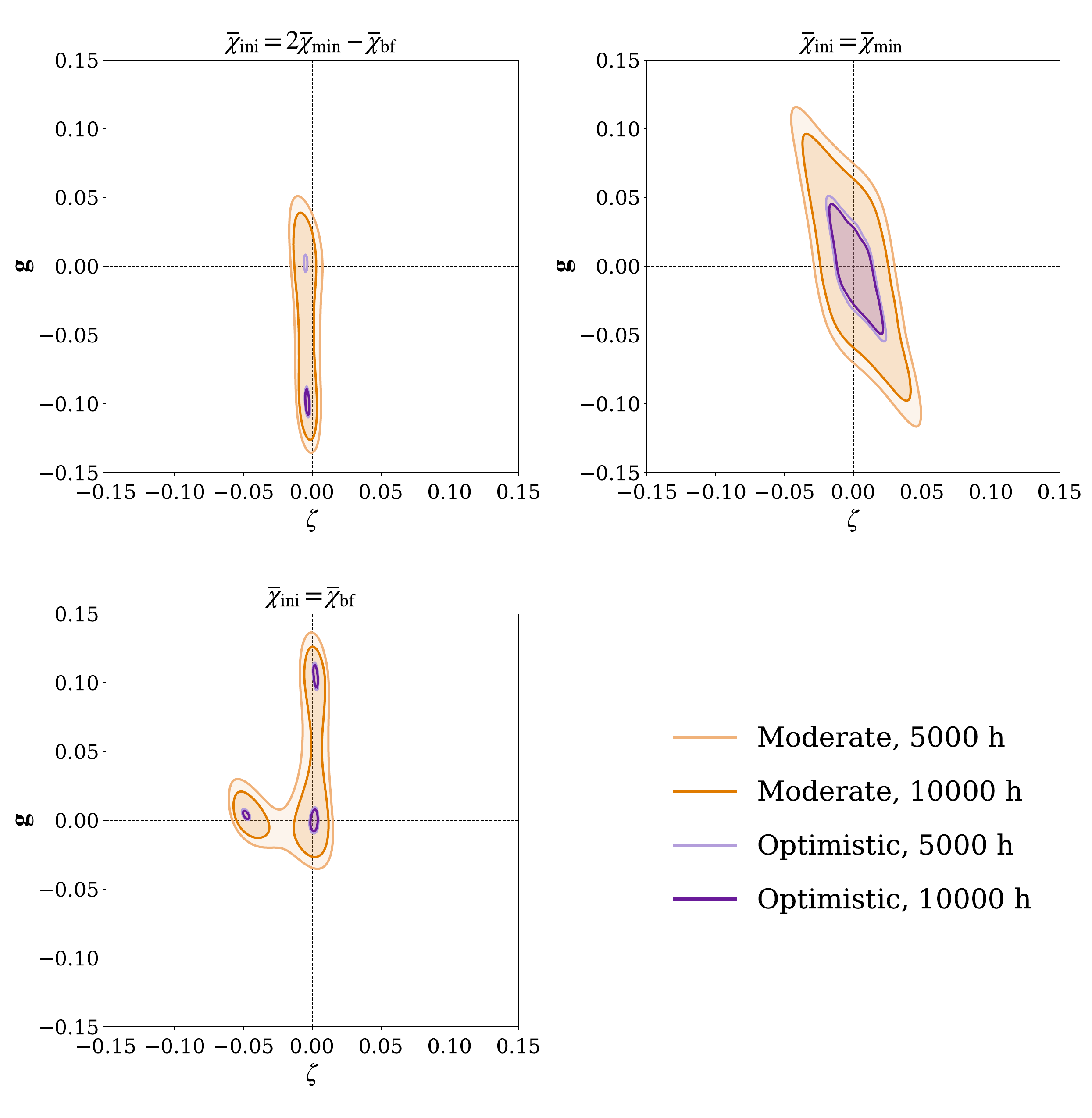}
    \caption{
    Projected SKA1-Low sensitivity to the axiodilaton
    parameters in the $(\bfg,\zeta)$ plane for
    $\bar\chi_{\rm ini} = 2\bar\chi_{\rm min}-\bar\chi_{\rm bf} = 73.807$ (top left), $\bar\chi_{\rm ini} = \bar\chi_{\rm min} = 74.010$ (top right), and $\bar\chi_{\rm ini} = \bar\chi_{\rm bf}=74.213$ (bottom left).
    Parameter values outside the contours produce a statistically distinguishable
    21\,cm power spectrum from \LCDM\ at the expected $2\sigma$ level.
    The light and dark orange contours correspond
    to the moderate sky-temperature model with observing times of
    $5000\,{\rm hr}$ and $10000\,{\rm hr}$, respectively. The light and
    dark purple contours show the corresponding results for the
    optimistic sky-temperature model. The theoretical
    uncertainty is set to $f_{\rm theo}=0$ to show the idealized
    instrumental sensitivity.
    }
    \label{fig:chi2_contour}
\end{figure}

Fig.~\ref{fig:chi2_contour} shows the regions of the
$(\bfg,\zeta)$ parameter space that remain consistent with \LCDM\ at
the expected $2\sigma$ level, so that models outside the contours differ significantly from the \LCDM\ prediction.
For $\bar\chi_{\rm ini}=\chimin$, the contours are anti-correlated, extending along the direction in which $\bfg$ and $\zeta$ have opposite signs. This is because $\bfg$ and $\zeta$ enter the dilaton equation of motion, Eq.~\eqref{eq:dilaton_background}, through the combined source term $-\bfg\bar\rho_{\ssB}-\zeta\bar\rho_{\sax}$, so the effect of one coupling can be compensated by an oppositely signed value of the other, leaving the 21\,cm signal close to the \LCDM\ prediction. When $\bfg$ and $\zeta$ have the same sign, their effects add, leading to larger changes in both the background evolution and the matter perturbations.
For this initial value the contours are also approximately symmetric about the origin. Because the dilaton starts at the minimum, changing the signs of both couplings, $(\bfg,\zeta)\to(-\bfg,-\zeta)$, reverses the sign of the source term, so the dilaton is displaced by the same amount in the opposite direction. Since the particle masses depend on $\bfg\Delta\bar\chi$, Eq.~\eqref{eq:mass_ratio}, and the potential is close to symmetric near its minimum, the resulting changes in the SM masses and in the Hubble rate $H$ at high redshift are nearly the same for $(\bfg,\zeta)$ and $(-\bfg,-\zeta)$, as shown in Fig.~\ref{fig:bkg}. The two choices also produce nearly the same change in the matter power spectrum shown in Fig.~\ref{fig:matter_power_spectrum}, which governs the formation of the structures responsible for the 21\,cm signal.

Taking the broader $5000\,{\rm hr}$ contours as a conservative
observational scenario, the moderate forecast can distinguish all
models from \LCDM\ outside the region
\begin{equation}
    -0.12\lesssim\bfg\lesssim0.12,
    \qquad
    -0.05\lesssim\zeta\lesssim0.05.
\end{equation}
For the optimistic forecast with $10000\,{\rm hr}$, the corresponding
region contracts to
\begin{equation}
    -0.05\lesssim\bfg\lesssim0.05,
    \qquad
    -0.02\lesssim\zeta\lesssim0.02.
\end{equation}
For $\bar\chi_{\rm ini}=\chimin$, however, it is difficult to distinguish $(\bfg,\zeta)$ from $(-\bfg,-\zeta)$, which limits the determination of the individual signs of the two couplings. This degeneracy is broken when the dilaton starts away from the minimum.

For $\bar\chi_{\rm ini}=\chibf$, 
the region with $\bfg<0$ and $\zeta>0$ produces a clear
departure from \LCDM, whereas a much larger portion of the region with
$\bfg>0$ and $\zeta<0$ remains statistically similar to \LCDM. 
This behavior agrees with the power-spectrum results presented in
Fig.~\ref{fig:matter_power_spectrum}: the red branch,
$(\bfg,\zeta)=(+0.1,-0.1)$, remains closer to \LCDM, while the blue
branch, $(-0.1,+0.1)$, shows a substantially larger deviation because
its higher axion density shifts matter--radiation equality to an earlier
time and allows more matter to cluster efficiently before recombination.
Further details are discussed in
Appendix~\ref{subsec:perturbation_evolution}.

For the moderate forecast with $5000\,{\rm hr}$, the models outside the region
\begin{equation}
-0.04\lesssim\bfg\lesssim0.14,
\qquad
-0.06\lesssim\zeta\lesssim0.02,
\end{equation}
can be distinguished from \LCDM.
This asymmetry suggests that $(\bfg,\zeta)$ may be distinguishable from
$(-\bfg,-\zeta)$, providing some sensitivity to the individual signs
of the two couplings.
The optimistic forecast with $10000\,{\rm hr}$ reduces the
region to
\begin{equation}
-0.01\lesssim\bfg\lesssim0.11,
\qquad
-0.05\lesssim\zeta\lesssim0.01\,.
\end{equation}
Consequently, a large portion of the parameter space is expected to produce a statistically significant difference from \LCDM.

For $\bar\chi_{\rm ini}=2\chimin-\chibf=73.807$, the dilaton starts to the left of the potential minimum at the same distance from $\chimin$ as $\chibf$. The particle masses depend on $\bfg\Delta\bar\chi(z)$, Eq.~\eqref{eq:mass_ratio}, and the axion energy density depends on $\zeta\Delta\bar\chi(z)$, Eq.~\eqref{eq:axion_background_density_solution}. In the perturbation equations, Eq.~\eqref{eq:axion_baryon_growth_quasistatic}, the friction terms depend on $\bfg\bar\chi'$ and $\zeta\bar\chi'$, and the effective couplings in Eq.~\eqref{eq:effective_gravitational_couplings} depend on $\zeta^2$, $\bfg\zeta$, and $\bfg^2$. All of these terms are unchanged under $(\bfg,\zeta,\Delta\bar\chi)\to(-\bfg,-\zeta,-\Delta\bar\chi)$. If the potential were symmetric about $\chimin$, $(\bfg,\zeta)$ with $\bar\chi_{\rm ini}=\chibf$ and $(-\bfg,-\zeta)$ with $\bar\chi_{\rm ini}=2\chimin-\chibf$ would therefore have the same background and perturbation evolution. The contours for $\bar\chi_{\rm ini}=73.807$ therefore have approximately the shape of the $\chibf$ contours reflected through the origin of the $(\bfg,\zeta)$ plane. For this initial value the contours lie close to $\zeta\simeq0$ and extend much further towards $\bfg<0$ than towards $\bfg>0$. The two sets of contours are not exactly symmetric because the potential in Eqs.~\eqref{eq:joint_potential} and \eqref{eq:albrecht-skordis} is not symmetric about $\chimin$ (see {\it e.g.}~Fig.~\ref{DilatonPotential}).

\begin{figure}[t!]
    \centering
    \includegraphics[width=1.0\linewidth]
    {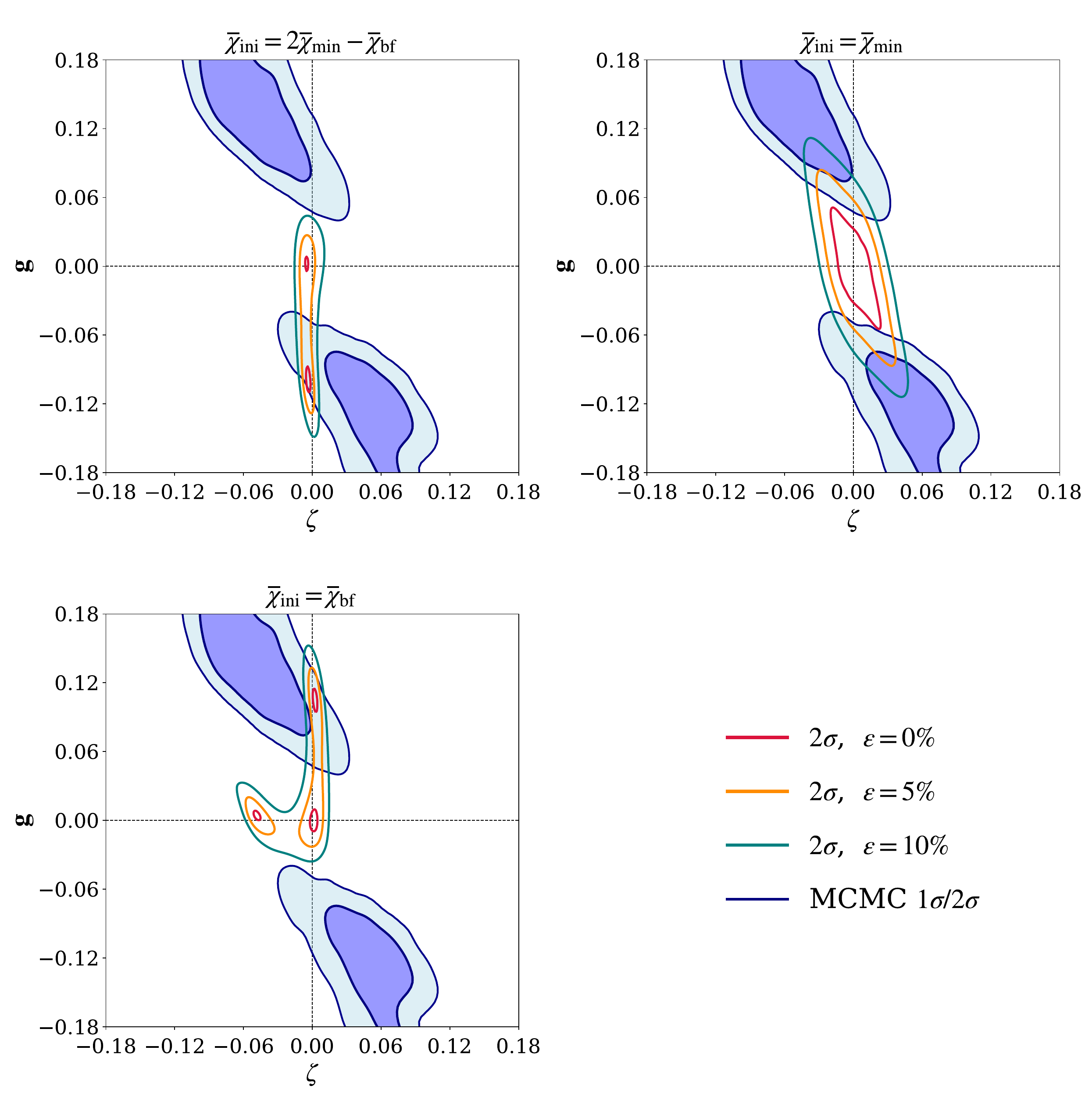}
    \caption{
    Same as Fig.~\ref{fig:chi2_contour}, but for the optimistic forecast
    with $5000\,{\rm hr}$ and theoretical uncertainties of
    $f_{\rm theo}=0$, $0.05$, and $0.1$ (labeled $\varepsilon$ in the legend), shown by the red, orange, and
    green contours, respectively. 
    Models outside the contours can be distinguished from
    \LCDM\ at the $2\sigma$ level.
    Each panel corresponds to
    $\bar\chi_{\rm ini} = 2\bar\chi_{\rm min}-\bar\chi_{\rm bf} = 73.807$ (top left), $\bar\chi_{\rm ini} = \bar\chi_{\rm min} = 74.010$ (top right), and $\bar\chi_{\rm ini} = \bar\chi_{\rm bf}=74.213$ (bottom left), respectively. The blue shaded regions, which are identical in every panel, show the current cosmological constraints obtained from the
    SH0ES-calibrated MCMC analysis of Ref.~\cite{Smith:2025uaq} after
    marginalizing over $\bar\chi_{\rm ini}$. The darker and lighter blue regions represent the $1\sigma$ and $2\sigma$ credible regions, respectively.
    }
    \label{fig:chi2_theoretical_error}
\end{figure}

Fig.~\ref{fig:chi2_theoretical_error} compares the projected
SKA1-Low sensitivity for different levels of theoretical uncertainty.
As $f_{\rm theo}$ increases to $10\%$, the contours expand, increasing
the fraction of parameter space that remains unconstrained by the
21\,cm signal. If future SKA1-Low measurements agree with the \LCDM\
mock data assumed here, models outside the corresponding contours could
be excluded at the expected $2\sigma$ level.
Since the astrophysical parameters are held fixed, these contours represent the most optimistic sensitivity, and varying the astrophysics jointly with $(\bfg,\zeta)$ would be expected to enlarge them.

The blue regions address a different statistical question. They show
the parameter values favored by currently available cosmological data
in the SH0ES-calibrated MCMC analysis of
Ref.~\cite{Smith:2025uaq}. This analysis combines Planck~2018,
ACT~DR6 lensing, DESI~DR2, and Pantheon+ data and marginalizes over
$\bar\chi_{\rm ini}$. Because the two analyses use different data sets
and statistical assumptions, their contour sizes cannot be compared
directly. Nevertheless, the projected 21\,cm sensitivity provides
orthogonal information. In particular, for fixed $\bar\chi_{\rm ini}=\chibf$, we could distinguish between the two opposite-sign combinations of $\bfg$ and
$\zeta$.

\section{Summary and Outlook}
\label{sec:summary}

In this work we explored the implications of a light axiodilaton scalar for 21\,cm physics. Our work is prompted by the Yoga framework of Ref.~\cite{Burgess:2021obw} and by the tentative observational successes of the resulting cosmologies \cite{Smith:2024ibv, Smith:2025uaq}.

Specifically, in the axiodilaton model studied here, the light dilaton is a natural candidate for Dark Energy, whereas the axion partner plays the role of Dark Matter. 21\,cm observables provide opportunities to test this model beyond CMB anisotropies, because its success relies on the couplings of the dilaton to matter, and cosmic dawn and the epoch of reionization are precisely when most of the dilaton's evolution occurs. The advent of high-precision measurements of this epoch therefore promises strong constraints or even potential discovery. We find that the dilaton couplings modify the standard $\Lambda$CDM predictions in three ways, through a modified recombination history arising from the dilaton-dependent electron and baryon masses, a modified evolution of the Hubble scale $H(z)$, and new scalar-mediated attractive forces between matter fields.

We computed the background and linear perturbations of the axiodilaton cosmology with our modified version of \CLASS\ and used them as input to a modified version of the semi-numerical code \FirstClass, in which the dilaton-dependent masses are carried through the recombination history, the gas and spin temperatures, and the neutral hydrogen fraction. This allowed a fairly detailed simulation of the hydrogen physics during cosmic dawn and the epoch of reionization, to compare the predictions of the axiodilaton and $\Lambda$CDM models. Calculating the volume-averaged 21\,cm brightness temperature (Fig.~\ref{fig:21cmSummaryPlot}) and the 21\,cm power spectra (Fig.~\ref{fig:P21_chi}), we find that the predictions of the axiodilaton model are similar in shape to the predictions in the $\Lambda$CDM model but displaced to higher redshifts, mainly due to the fact that structures form earlier in the axiodilaton model.

When comparing the axiodilaton model to $\Lambda$CDM, we used three benchmark pairs of couplings, $(\bfg,\zeta)=(+0.1,-0.1)$, $(-0.1,+0.1)$ and $(+0.1,+0.1)$, of the size preferred by the analysis of CMB and LSS observables in \cite{Smith:2025uaq}. We evaluated each for three initial values of the dilaton, one at the minimum of its potential, $\bar\chi_{\rm ini}=\chi_{\rm min} = 74.010$, and one on each side of it, $\bar\chi_{\rm ini}=73.807$ and $74.213$, where that analysis finds two comparably preferred solutions. That work found that the axiodilaton model provided a better fit to CMB, LSS and supernovae data than the $\Lambda$CDM model, because the evolution of particle masses helps to partially alleviate the Hubble tension. Scanning the full $(\bfg,\zeta)$ plane for $\bar\chi_{\rm ini}=\chimin$, we found that SKA1-Low could distinguish from $\Lambda$CDM at the $2\sigma$ level all models outside the region $-0.12\lesssim\bfg\lesssim0.12$, $-0.05\lesssim\zeta\lesssim0.05$ for the moderate forecast with $5000\,{\rm hr}$, and that this region contracts to $-0.05\lesssim\bfg\lesssim0.05$, $-0.02\lesssim\zeta\lesssim0.02$ for the optimistic forecast with $10000\,{\rm hr}$ (Fig.~\ref{fig:chi2_contour}). As Fig.~\ref{fig:chi2_theoretical_error} shows, all three benchmark pairs lie outside the contours for all three initial values even with a 10\% theoretical uncertainty on the predicted power spectrum, so couplings of the size preferred by current data would be distinguishable from $\Lambda$CDM, provided the astrophysics is known.

To determine which feature of the axiodilaton model is most important in differentiating it from $\Lambda$CDM, we isolated the individual contributions of various physical mechanisms (such as the varying $H(z)$, evolving particle masses, and so on). The results of this comparison, shown in Fig.~\ref{fig:varyparams}, demonstrate that the dominant effect comes from the earlier formation of structure in the axiodilaton framework. This earlier formation of structure can be seen in Fig.~\ref{fig:delta_evolution} and is driven by an additional attractive force between matter fields mediated by the dilaton. This prominent role of early structure formation is particularly intriguing in light of recent JWST observations, which hint at an abundance of high-redshift structure exceeding standard \(\Lambda \)CDM expectations \cite{JWST1,Castellano:2022,Labbe:2023, casey2023cosmosweb}. These findings open several avenues for future inquiry, notably whether axiodilaton dynamics contribute to other emerging cosmological anomalies—such as the apparent preference for negative neutrino squared-masses or discrepancies in late-time structure growth—that currently reside just below the threshold of statistical significance.

Our calculation has a number of weaknesses on which it would be good to improve. First, the astrophysical parameters in this work are held fixed at the fiducial EOS2021 values~\cite{Munoz:2021psm}, so the SKA1-Low contours represent the most optimistic sensitivity, and a joint analysis varying the astrophysical and axiodilaton parameters together, which we leave to future work, is needed to determine how much of this sensitivity survives their degeneracies. In addition, the critical overdensity for spherical collapse is kept at its $\Lambda$CDM value, $\delta_c=1.686$, in both the collapsed fraction and the halo mass function, even though the additional attractive force mediated by the dilaton could modify it. Understanding how this critical overdensity is modified in the presence of light coupled scalars is therefore another avenue that warrants further investigation.

We also make several approximations whose effect on our results is small. We use the X-ray energy-deposition fractions $f_{\rm heat}$, $f_{{\rm ion},j}$, and $f_{\rm Ly\alpha}$, the Ly$\alpha$ heating tables, and the collisional de-excitation rates $\kappa_{10}^{i{\rm H}}$ computed at the present-day masses. We omit the term from the varying proton mass in the $T_k$ evolution, Eq.~\eqref{eq_Tk_evolution}, and the mass dependence of $z_n^{\rm max}$ in Eq.~\eqref{eq:zmax_axiodilaton}. The electron-mass dependence of the deposition fractions mostly cancels, Appendix~\ref{app:fheat_scaling}. Ly$\alpha$ heating is several orders of magnitude smaller than X-ray heating, Fig.~\ref{fig:heat_terms}, and $x_{\rm coll}\ll1$ over the redshifts of Fig.~\ref{fig:x_Ts_Tb}. The proton-mass term is small, and the masses change little between the emission and absorption redshifts of a Lyman-series photon. The rates $\kappa_{10}^{i{\rm H}}$ become important in the dark ages, when collisions alone couple $T_s$ to $T_k$~\cite{Furlanetto:2006jb}. Predictions for FARSIDE~\cite{Burns:2021pkx}, which targets $35\lesssim z\lesssim200$, therefore require $\kappa_{10}^{i{\rm H}}$ computed with the varying masses, and we leave this to future work. We have not estimated the effect of the other approximations. All models use the same primordial helium fraction $Y_p=0.245$, and the second-order displacement uses the standard 2LPT kernel. The stellar spectral template and the sub-grid density distribution and self-shielding fit keep their fiducial forms. The conversion between observed frequency and redshift uses the constant $\nu_{21,0}$.

A final weakness is the discrepancy between the size of the coupling $\bfg$ entertained in cosmology relative to the constraints coming from solar system tests. We find the cosmology of these models to be of sufficient interest to warrant a more systematic study of how it can accommodate these potential problems.

\acknowledgments
JHK thanks Hyunbae Park and Kyungjin Ahn for helpful discussions. AS and CvdB thank Eleonora Di Valentino for the insightful comments and suggestions.
We thank APCTP, Pohang, Korea, for its hospitality during the TRP program APCTP-2026-T02, from which this work benefited.
JHK and CYW are supported in part by the National Research Foundation of Korea (NRF) Grant No. RS-2026-25484206, the BK-21 FOUR program through NRF, and the Institute of Information \& Communications Technology Planning \& Evaluation (IITP)-Information Technology Research Center (ITRC) Grant No. IITP-2025-RS-2024-00437284. CvdB is partly supported by the Lancaster–Sheffield Consortium for Fundamental Physics under STFC grant: ST/X000621/1. AS is supported by the W.D.Collins Scholarship. CB's research was partially supported by funds from the Natural Sciences and Engineering Research Council (NSERC) of Canada. Research at the Perimeter Institute is supported in part by the Government of Canada through NSERC and by the Province of Ontario through MRI.

\appendix

\addtocontents{toc}{\setcounter{tocdepth}{2}}

\section{Theoretical Framework}
\label{sec:TheoFramework}

As mentioned in the introduction (and reviewed in \cite{Burgess:2025vxs}), the class of models we study is chosen because it appears to help ameliorate the cosmological constant problem. It does so by incorporating the accidental approximate symmetries that are known to arise very generically in higher-energy UV completions of GR \cite{Burgess:2020qsc}. They also play an important role in earlier related approaches to the cosmological constant problem based on supersymmetric large extra dimensions (SLED) \cite{Aghababaie:2003wz}, which also suggest the importance of relaxation in suppressing the energy densities seen in cosmology \cite{Burgess:2006ds, Tolley:2007et, vandeBruck:2026bdw}. These models generically predict a very rich Dark Sector that is approximately supersymmetric but coupled gravitationally to a non-supersymmetric particle physics sector \cite{Burgess:2004yq, Burgess:2021juk, Smith:2024ayu}. 

This section describes in more detail the parts of the model relevant for the cosmological applications used here, including a review of the relevant equations. This section is mostly review material that is included for the sake of completeness (for further details of the model and its dynamics see Refs.~\cite{Smith:2025uaq,Smith:2024ibv}). Summaries of the model's cosmological evolution are deferred to the next subsection, including the motivation for testing it with 21\,cm observations.

\subsection{The axiodilaton Model}
\label{subsec:axio_dilaton_model}
\label{sec:Models}

The core ingredient of the model is a very light spinless dilaton, $\chi$, which is the pseudo-Goldstone boson for an approximate scaling symmetry whose interplay with supersymmetry is part of what allows progress on the cosmological constant problem. Because the Dark Sector is gravitationally coupled and approximately supersymmetric (unlike the particle sector) this dilaton gets paired with another low-energy Goldstone scalar $\mfa$, an axion. These two fields have nontrivial two-derivative sigma-model interactions that can compete with GR at low energies without undermining the semiclassical approximation (unlike what typically happens in single-scalar models) \cite{Burgess:2009ea, Adshead:2017srh}.

In the minimal picture these two fields are respectively the Dark Energy and Dark Matter \cite{Smith:2024ibv}. Their action in the Einstein frame is given by
\begin{align}\label{eq:ActionApp}
    S = \!\int \exd^4x \sqrt{-g}
    \Biggl\{
        \frac{\MP^2}{2}
        \Bigl[
            R
            &- \partial_\mu\chi\,\partial^\mu\chi
            - W^2(\chi)\,\partial_\mu\mfa\,\partial^\mu\mfa
        \Bigr]
        - V(\chi,\mfa)
    \Biggr\}
    + S_m(\psi,\tilde g)\,,
\end{align}
where $W(\chi)$ is a dilaton-dependent kinetic prefactor that encodes the curvature of the two-field target-space metric and $S_m$ is the action describing the Standard Model degrees of freedom. We work with dimensionless fields normalised in reduced Planck units.
For the models considered here, scale invariance requires $W$ to take the exponential form 
\begin{equation}
    W(\chi) = e^{-\zeta\chi}
\end{equation}
where $\zeta$ controls the strength of derivative coupling between the axion and dilaton. There is no loss of generality in omitting a constant prefactor for the exponential in $W(\chi)$ because of the freedom to shift $\chi$ so that\footnote{Another convenient choice instead chooses $\chi_{\rm now} = 0$ at the current epoch, but we do not use those conventions here.}  $W(0) = 1$.  

To a good approximation matter fields $\psi$ are minimally coupled to the Jordan-frame metric $\tilde g_{\mu\nu}$, which is related to the Einstein-frame metric by
\begin{equation}\label{eq:jf_metricApp}
    \tilde g_{\mu\nu} = A^2(\chi)\,g_{\mu\nu}, 
    \qquad 
    A(\chi)=e^{\mathbf g\chi},
\end{equation}
where scale invariance also ensures an exponential form.\footnote{There is again no loss of generality in assuming $A(0) = 1$ because this can be ensured by performing a constant rescaling of the metric.} This ensures that $\chi$ couples to ordinary matter as would a Brans--Dicke scalar, with the dimensionless coupling $\mathbf g$ being related to the usual Brans--Dicke coupling.
The Jordan frame is the metric seen by these matter species and defines physical rulers and clocks built from them, while the Einstein frame is convenient for writing the gravitational dynamics in canonical form.

The above assumptions produce a universal coupling to all matter species, and cause all non-relativistic SM mass scales to evolve with the dilaton field as
\begin{equation}
\frac{m_i(z)}{m_{i,0}}
=
\frac{A(\bar\chi(z))}{A(\bar\chi_0)}
=
e^{{\mathbf g}\,\Delta\bar\chi(z)},
\label{eq:mass_ratioApp}
\end{equation}
where $\Delta\bar\chi(z)\equiv \bar\chi(z)-\bar\chi_0$ denotes the difference between the value of $\bar\chi(z)$ at redshift $z$ and its present-day value $\bar\chi_0\equiv \bar\chi(z=0)$.
Here $i$ labels a non-relativistic SM species, and $m_{i,0}\equiv m_i(z=0)$ is fixed to the corresponding present-day value used in the reference \LCDM\ cosmology.

Having particle masses be time-dependent at late times due to their coupling with a quintessence scalar seems like it should be hard to reconcile with observations in the present Universe, however, because the rescaling is universal it drops out of most late-time spectroscopic and equivalence principle observables that depend only on dimensionless ratios (such as $\mu=m_e/m_p$ and $Q=m_p/\Lambda_{\rm QCD}$). These are not directly sensitive to $A(\chi)$ in the minimal universal-coupling limit.
The only direct probes of such a universal coupling are those that compare particle masses at different times or compare them to the Planck scale at any given time (since $M_p$ is held fixed by the conformal rescaling of
Eq.~\eqref{eq:jf_metric}). 

As discussed in the main text we assume the potential for the two scalars is separable,
\begin{eqnarray}\label{eq:joint_potentialApp}
    V(\chi,\mfa) &\simeq& V_{\rm dil}(\chi) + V_{\rm ax}(\mfa)\nn\\
                  &=& V_p(\chi)\,e^{-\lambda\chi} + \tfrac12 \MP^2 m_\mfa^2 \mfa^2 ,
\end{eqnarray}
where $V_{\rm ax}$ is taken to be approximately quadratic near its minimum, about which the axion undergoes coherent oscillations and behaves as a pressureless fluid component once its effective mass $m_{\rm ax}\gg \cH$. For the dilaton potential we adopt the Albrecht--Skordis form found in \cite{Burgess:2021obw}, 
\begin{equation}\label{eq:albrecht-skordisApp}
        V_p(\chi)=V_0\!\left[1-u_1\chi+\frac{u_2}{2}\chi^2\right]\,,
\end{equation}
where the coefficients $u_1$ and $u_2$ are chosen as described in the main text and the normalization $V_0$ is fixed by the shooting procedure described in Appendix~\ref{subsec:bkg_evolution}. Fig.~\ref{DilatonPotential} plots the resulting shape for $V_{\rm dil}(\chi)$ using the parameter choices made here (which imply $\chi_{\rm min} = 74.010$). 

Varying the action in Eq.~\eqref{eq:Action} with respect to the Einstein-frame metric gives the Einstein equation
\begin{align}
G_{\mu\nu}
+
\frac{V(\chi,\mfa)}{\MP^2}\,g_{\mu\nu}
&-\left(
    \partial_\mu\chi\,\partial_\nu\chi
    -\frac{1}{2}g_{\mu\nu}\,
     \partial_\sigma\chi\,\partial^\sigma\chi
 \right)
\nonumber\\
&
-W^2(\chi)\left(
    \partial_\mu \mfa\,\partial_\nu \mfa
    -\frac{1}{2}g_{\mu\nu}\,
     \partial_\sigma \mfa\,\partial^\sigma \mfa
 \right)
=
\frac{1}{\MP^2}\,T_{\mu\nu}\,,
\label{eq:einstein_equation}
\end{align}
where $G_{\mu\nu}$ is the metric's Einstein tensor, $T_{\mu\nu}$ is the total Einstein-frame energy--momentum
tensor of all SM matter and radiation species obtained
from the matter action $S_{\rm m}$.

Variation with respect to the axion and dilaton fields gives
\begin{equation}
\Box \mfa
+2\frac{W_{,\chi}}{W}\,
  \partial_\mu\chi\,\partial^\mu \mfa
-\frac{V_{,\mfa}}{\MP^2 W^2}
=0\,,
\label{eq:axion_covariant}
\end{equation}
and
\begin{equation}
\Box\chi
-W W_{,\chi}\,
 \partial_\mu \mfa\,\partial^\mu \mfa
-\frac{V_{,\chi}}{\MP^2}
=
-\frac{\bfg}{\MP^2}\,T\,,
\label{eq:dilaton_covariant}
\end{equation}
respectively, where $T\equiv g^{\mu\nu}T_{\mu\nu}$ denotes the Einstein-frame trace of the matter stress tensor.

Notice that the two-derivative sigma-model interaction encapsulated by the function $W(\chi)$ implies a dilaton-dependence to many properties in the axion sector. The canonically normalised axion fluctuation is proportional to $W(\chi)\mfa$, for $\chi$ some convenient value for the dilaton, making the physical axion mass depend on $\chi$ as
\begin{equation}
m_{\rm ax}(z)
=
\frac{m_{\mfa}}{W(\bar\chi(z))} .
\label{eq:axion_effective_mass}
\end{equation}
For $W(\chi)=e^{-\zeta\chi}$, this can be written in
terms of the present-day value as
\begin{equation}
\frac{m_{\rm ax}(z)}{m_{{\rm ax},0}}
=
\frac{W(\bar\chi_0)}{W(\bar\chi(z))}
=
e^{\zeta\left(\bar\chi(z)-\bar\chi_0\right)}
\equiv
e^{\zeta\,\Delta\bar\chi(z)},
\label{eq:axion_mass_ratio}
\end{equation}
with $m_{{\rm ax},0}\equiv m_{\rm ax}(z=0)$.  

When the axion oscillates rapidly compared to the Hubble time, its averaged equation of state approaches that of cold matter. As a result, the averaged axion energy density can be written in the convenient form
\begin{equation}
  \label{eq:axion_background_density}
  \bar\rho_{\sax}(z) \;=\; \frac{C\,m_{\rm ax}(z)}{a^3}\,,
\end{equation}
where $C$ is fixed  by the initial conditions and $a$ is the scale factor. In particular, the dilaton dependence of $W(\chi)$ can cause the axion mass to evolve with time,  $m_{\rm ax}(z)$, and this can cause the evolution of the axion energy density to deviate from the standard $a^{-3}$ scaling (and this is important when using data to infer the Dark Energy equation of state). 

Throughout this paper we focus on axion masses large enough that the axion is in the oscillatory regime throughout the cosmology after Big Bang nucleosynthesis (BBN). This requires $m_{\rm ax}\gtrsim \cH$ at the end of BBN, which corresponds to $m_{\mfa}\gtrsim 10^{-15}\,{\rm eV}$ for the benchmark choices considered here.
In this regime the detailed 21\,cm predictions become essentially insensitive to the precise value of $m_{\mfa}$, with the dominant effects arising from the coupling $\mathbf g$ and the $\chi$-dependence of $W$.

\subsection{Axiodilaton Cosmology}
\label{sec:ADCosmo}

In this section we review the main features of axiodilaton cosmology. This largely repeats results from earlier work (see \cite{Smith:2024ayu,Smith:2024ibv,Smith:2025uaq}) but we include it here for completeness as the results are relevant for 21\,cm physics.

\subsubsection{Background Evolution}
\label{subsec:bkg_evolution}

We start by summarizing the system of
background evolution equations used in this work. A complete derivation and detailed discussion of these equations are provided in Ref.~\cite{Smith:2025uaq}. Here, we present only the main quantities relevant to the 21\,cm calculation, focusing on the redshift range $8\lesssim z\lesssim20$. 
We work in conformal Newtonian gauge and write scalar
metric perturbations as
\begin{equation}
    {\rm d}s^2
    =
    a^2(\eta)
    \left[
    -\left(1+2\Phi(\eta,\mathbf{x})\right){\rm d}\eta^2
    +
    \left(1-2\Psi(\eta,\mathbf{x})\right)
    \delta_{ij}\,{\rm d}x^i{\rm d}x^j
    \right],
    \label{eq:newtonian_gauge_metric}
\end{equation}
where $\eta$ is conformal time, primes denote
${\rm d}/{\rm d}\eta$, and $\cH\equiv a'/a$ is the conformal Hubble
rate.
We decompose the axiodilaton fields as
\begin{equation}
    \chi(\eta,\mathbf{x})
    =
    \bar\chi(\eta)+\delta\chi(\eta,\mathbf{x}),
    \qquad
    \mfa(\eta,\mathbf{x})
    =
    \bar\mfa(\eta)+\delta\mfa(\eta,\mathbf{x}) .
    \label{eq:scalar_background_perturbation_split}
\end{equation}
For the fluid components, we define
\begin{equation}
    \rho_I(\eta,\mathbf{x})
    =
    \bar\rho_I(\eta)\left[1+\delta_I(\eta,\mathbf{x})\right],
    \qquad
    \theta_I(\eta,\mathbf{x})
    \equiv
    \nabla\cdot{\bf v}_I(\eta,\mathbf{x}) .
    \label{eq:fluid_pert_def}
\end{equation}
Here $I$ labels the relevant effective fluid species, including the
coarse-grained axion fluid and the SM fluids.  Barred quantities
denote homogeneous background values, $\delta_I$ is the density contrast,
and $\theta_I$ is the velocity divergence.  In the previous subsection, the
bars on homogeneous background quantities were omitted for notational
simplicity.

Evaluating the homogeneous $(00)$ component of
Eq.~\eqref{eq:einstein_equation} and averaging over the rapid axion
oscillations gives the Friedmann equation
\begin{equation}
  \label{eq:friedmann_background}
  \cH^2
  =
  \frac{1}{3\MP^2}
  \left[
    \frac{\MP^2}{2}\,\bar\chi'^2
    +a^2 V_{\rm dil}(\bar\chi)
    +a^2\bar\rho
    +a^2\bar\rho_{\sax}
  \right],
\end{equation}
where $\bar\rho=\bar\rho_r+\bar\rho_{\ssB}$ denotes the combined
energy density of radiation (photons and neutrinos) and baryons. 
The average axion energy density $\bar{\rho}_{\rm ax}$ defined in
Eq.~\eqref{eq:axion_background_density} satisfies
\begin{equation}
    \bar{\rho}_{\rm ax}'
    +
    3\mathcal{H}\bar{\rho}_{\rm ax}
    =
    -\zeta\bar{\chi}'\bar{\rho}_{\rm ax}.
    \label{eq:axion_background_continuity}
\end{equation}
This equation has the solution
\begin{equation}
    \bar{\rho}_{\rm ax}(z)
    =
    \bar{\rho}_{{\rm ax},0}\,a^{-3}
    e^{\zeta\Delta\bar{\chi}(z)},
    \label{eq:axion_background_density_solution}
\end{equation}
where $\bar{\rho}_{{\rm ax},0}$
is the present-day background axion energy density,
$\Delta\bar{\chi}(z)\equiv\bar{\chi}(z)-\bar{\chi}_0$,
and $\bar{\chi}_0\equiv\bar{\chi}(z=0)$ is the present-day
dilaton field value.
The exponential factor describes the departure from the
standard $a^{-3}$ scaling caused by the axion--dilaton kinetic coupling
$W^2(\chi)$ in Eq.~\eqref{eq:Action}.

The evolution of the remaining background fluids follows from their
stress--energy conservation equations. For relativistic species, the
trace of the energy--momentum tensor vanishes, leading to
\begin{equation}\label{eq:radiation_background_density}
    \bar\rho_{\gamma}' + 4\cH\,\bar\rho_{\gamma}=0\,,
    \qquad
    \bar\rho_{\nu}' + 4\cH\,\bar\rho_{\nu}=0\,.
\end{equation}
Here we neglect a possible dilaton coupling to semi-relativistic
neutrinos. 

The non-relativistic baryon density obeys
\begin{equation}
    \bar\rho_{\ssB}' + 3\cH\,\bar\rho_{\ssB}
    =
    \bfg\,\bar\chi'\,\bar\rho_{\ssB}\,.
    \label{eq:baryon_background_density}
\end{equation}
This equation has the solution
\begin{equation}
    \bar\rho_{\ssB}(z)
    =
    \bar\rho_{\ssB,0}\,a^{-3}
    e^{\bfg\Delta\bar\chi(z)}\,,
    \label{eq:baryon_background_density_solution}
\end{equation}
where $\bar{\rho}_{\ssB,0}$
is the present-day background baryon energy density, and
the additional exponential factor arises
from the evolution of the baryon masses induced by the conformal
coupling $A(\chi)=e^{\bfg\chi}$ in Eq.~\eqref{eq:jf_metric}.

The energy exchange with the axion and baryon sectors is balanced by
the homogeneous dilaton field, whose equation of motion is
\begin{equation}
  \label{eq:dilaton_background}
  \bar\chi'' + 2\cH\,\bar\chi'
  + \frac{a^2}{\MP^2}\,V_{,\chi}
  =
  \frac{a^2}{\MP^2}
  \left(
      -\bfg\,\bar\rho_{\ssB}
      - \zeta \,\bar\rho_{\sax}
  \right).
\end{equation}
The term $2\cH\bar\chi'$ describes Hubble damping, and
$V_{,\chi}$ determines the force associated with the dilaton
potential. The terms on the right-hand side are sourced by the baryon
and axion energy densities through their respective couplings to the
dilaton. The combined source is
$-\bfg\bar\rho_{\ssB}-\zeta\bar\rho_{\sax}$ and its magnitude and sign
control the matter-induced evolution of the dilaton. In particular,
opposite signs of $\bfg$ and $\zeta$ allow the baryonic and axionic
contributions to partially cancel.
For the numerical analysis, we fix the initial dilaton velocity to
$\bar\chi'_{\rm ini}=0$ and consider three benchmark initial conditions,
$\bar\chi_{\rm ini}=2\chimin-\chibf=73.807$,
$\bar\chi_{\rm ini}=\bar\chi_{\rm min}=74.010$ and
$\bar\chi_{\rm ini}=\chibf=74.213$.  The value $\chibf$ is the best-fit
initial dilaton value obtained with a free initial condition from the
CMB-A, DESI, Pantheon$+$, and SH0ES data combination in
Ref.~\cite{Smith:2025uaq}. Without the SH0ES calibration the best fit lies to the left of the minimum, and $2\chimin-\chibf$ is the mirror of $\chibf$ about $\bar\chi_{\rm min}$.

We solve the cosmological evolution using our modified version of
\CLASS~(v3.2)~\cite{Blas:2011rf}, developed in
Ref.~\cite{Smith:2025uaq}, which implements the axiodilaton
cosmology.
We present the resulting background solutions below. The
perturbation variables and their evolution equations are introduced in
Appendix~\ref{subsec:perturbation_evolution}.

\begin{table}[t]
\centering
\small
\renewcommand{\arraystretch}{1.25}
\setlength{\dashlinedash}{3pt}\setlength{\dashlinegap}{2pt}\setlength{\arrayrulewidth}{0.4pt}
\newcommand{\tblab}[1]{\multirow{2}{*}{\textsc{#1}}}
\begin{tabular*}{\linewidth}{@{}l@{\hspace{1.5em}}@{\extracolsep{\fill}}ccccccccc@{}}
\toprule
\tblab{Cosmology} & $\Omega_{\rm m,0}$ & $\Omega_{\ssB,0}$ & $h$ & $10^{9}A_s$ & $n_s$ & $k_\ast\;[{\rm Mpc}^{-1}]$ & $m_\nu\;[{\rm eV}]$ & $N_{\rm eff}$ & $Y_p$\\
\cdashline{2-10}
& 0.3111 & 0.04897 & 0.6766 & 2.105 & 0.9665 & 0.05 & 0.06 & 3.044 & 0.245\\
\end{tabular*}
\begin{tabular*}{\linewidth}{@{}l@{\hspace{1.5em}}@{\extracolsep{\fill}}ccccc@{}}
\midrule
\tblab{axiodilaton} & $\lambda$ & $2\bar\chi_{\rm min}-\chibf$ & $\bar\chi_{\rm min}$ & $\chibf$ & $\bar\chi'_{\rm ini}$\\
\cdashline{2-6}
& $4\sqrt{2/3}$ & 73.807 & 74.010 & 74.213 & 0\\
\end{tabular*}
\begin{tabular*}{\linewidth}{@{}l@{\hspace{1.5em}}@{\extracolsep{\fill}}ccccccc@{}}
\midrule
\tblab{Simulation} & $L\;[{\rm Mpc}]$ & $N_{\rm hi}$ & $N_{\rm lo}$ & $\Delta x_{\rm hi}\;[{\rm Mpc}]$ & $\Delta x_{\rm lo}\;[{\rm Mpc}]$ & $k_{\rm f}\;[{\rm Mpc}^{-1}]$ & $k_{\rm Ny,lo}\;[{\rm Mpc}^{-1}]$\\
\cdashline{2-8}
& 200 & $300^{3}$ & $100^{3}$ & 0.67 & 2 & 0.031 & 1.57\\
\bottomrule
\end{tabular*}
\caption{Fixed inputs used throughout this work: the standard cosmological
parameters, set to the Planck 2018 best-fit values~\cite{Planck:2018vyg};
the axiodilaton inputs, with $\bar\chi_{\rm min}$ the local minimum of the
dilaton potential and $\chibf$ the best-fit initial value of Ref.~\cite{Smith:2025uaq} with the SH0ES calibration; and the box size, grid resolutions, and derived
cell sizes, fundamental mode, and low-resolution Nyquist wavenumber of the
semi-numerical simulation (Sec.~\ref{ssec:simsetup}).}
\label{tab:inputs}
\end{table}

The axiodilaton calculation adopts five standard cosmological quantities,
$\Omega_{\rm m,0}$, $\Omega_{\ssB,0}$, $h$, $A_s$, and $n_s$, that
specify the present matter content, the Hubble constant, and the primordial
curvature perturbations. We fix them to the Planck 2018 best-fit
values~\cite{Planck:2018vyg} listed in Table~\ref{tab:inputs},
which also gives the simulation settings of Sec.~\ref{ssec:simsetup}.\footnote{In standard \CLASS\ calculations, the optical depth 
$\tau_{\rm reio}$ is used to account for the effects of reionization on the
CMB anisotropy spectra. Since this study does not calculate the CMB
anisotropy spectra, $\tau_{\rm reio}$ is not included among the input
parameters.}
These quantities enter our modified \CLASS\ calculation in the same
form as in the baseline $\Lambda$CDM cosmology. The reduced Hubble
parameter fixes
$H_0=100h\,{\rm km\,s^{-1}\,Mpc^{-1}}$ and hence the present critical
density,
$\bar\rho_{\rm crit,0}=3H_0^2/(8\pi G_N)$. The parameters
$\Omega_{\rm m,0}$ and $\Omega_{\ssB,0}$ are imposed as target
present-day density parameters, corresponding to
$\bar\rho_{\rm m,0}=\Omega_{\rm m,0}\bar\rho_{\rm crit,0}$ and
$\bar\rho_{\ssB,0}=\Omega_{\ssB,0}\bar\rho_{\rm crit,0}$.
We retain a standard neutrino sector with one massive eigenstate of
mass $m_\nu=0.06\,{\rm eV}$ and $N_{\rm eff}=3.044$. Since
$\Omega_{\rm m,0}$ includes the massive-neutrino contribution, the axion component is identified with
\begin{equation}
    \Omega_{\rm ax,0}
    =
    \Omega_{\rm m,0}
    -
    \Omega_{\ssB,0}
    -
    \Omega_{\nu,0},
\label{eq:axion_cdm_abundance}
\end{equation}
where $\bar\rho_{\rm ax,0}=\Omega_{\rm ax,0}\bar\rho_{\rm crit,0}$.  
The initial baryon and axion densities are adjusted simultaneously through a shooting procedure to reproduce the target values
$\Omega_{\ssB,0}$ and $\Omega_{\rm ax,0}$ at $z=0$.
A third shooting condition adjusts the normalization $V_0$ of the
dilaton potential to reproduce the target present-day DE
abundance,
$\Omega_{\chi,0}
=1-\Omega_{\rm m,0}-\Omega_{\rm r,0}$,
thereby ensuring $\Omega_{\rm tot,0}=1$.
The parameters $A_s$ and $n_s$ specify the primordial curvature
power spectrum at the pivot scale $k_\ast=0.05\,{\rm Mpc}^{-1}$. The same primordial
spectrum is used for all axiodilaton branches and for the $\Lambda$CDM
reference. 
The recombination history
and $z_{\rm rec}$ are calculated independently by the modified
\texttt{HyRec}~\cite{Ali_Ha_moud_2011,Lee_2020} module using the redshift-dependent electron mass.

The axiodilaton inputs varied in the present analysis are
\begin{equation}
    \Theta_{\rm AD}
    =
    \left\{
    \bfg,\,
    \zeta,\,
    \bar\chi_{\rm ini}
    \right\}.
\label{eq:YOGA_parameters}
\end{equation}
We use the following three coupling branches
as representative examples:
\begin{equation}
    (\bfg,\zeta)
    =
    (+0.1,-0.1),\qquad
    (-0.1,+0.1),\qquad
    (+0.1,+0.1).
\label{eq:YOGA_benchmark_branches}
\end{equation}
These benchmark branches are used to illustrate the characteristic
behavior of the model in the main text. In the subsequent $\chi^2$ analysis, the two couplings are varied independently over
$\bfg\in[-0.5,0.5]$ and $\zeta\in[-0.5,0.5]$.
The $\Lambda$CDM reference is
calculated using the same standard cosmological and neutrino inputs.

\begin{figure}[htbp!]
\centering
    \includegraphics[width=0.9\linewidth]{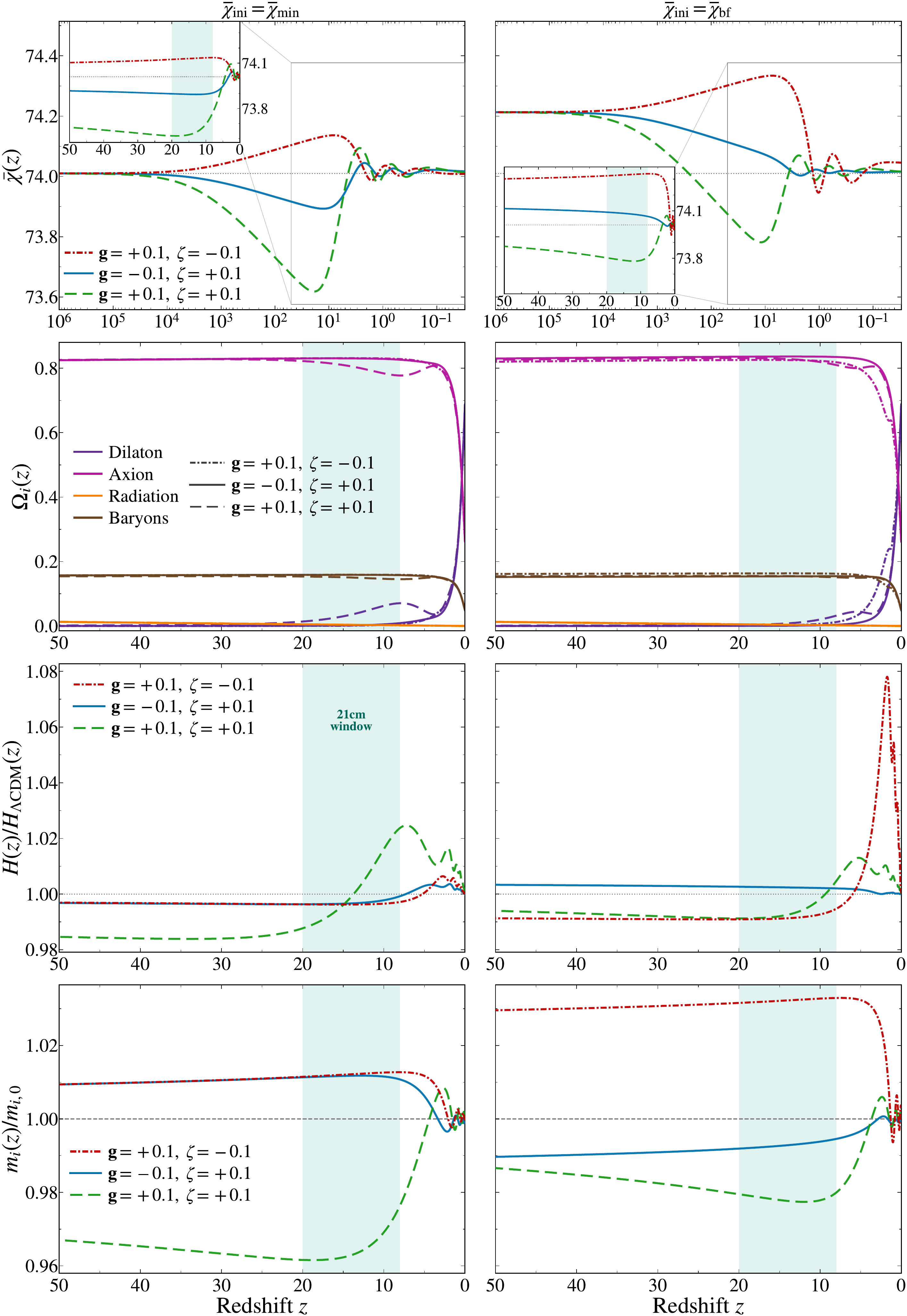}
\caption{
Background evolution of the relevant cosmological quantities for the initial
dilaton field values $\bar\chi_{\rm ini} = \bar\chi_{\rm min} = 74.010$ (left) and $\bar\chi_{\rm ini} = \bar\chi_{\rm bf}=74.213$ (right). From top to bottom, the rows show $\bar\chi(z)$,
the fractional energy densities $\Omega_i(z)$, the physical Hubble-rate
ratio $H/H_{\Lambda{\rm CDM}}$, and the SM particle mass ratios $m_i(z)/m_{i,0}$.
Colors and line styles denote the three $(\bfg,\zeta)$ choices indicated
in the legends. The detailed evolution is discussed in the text. The light-blue shaded region marks the redshift range $8\leq z\leq20$ targeted in our 21\,cm analysis.
}
\label{fig:bkg}
\end{figure}
Table~\ref{tab:background_evolution_summary} summarizes the resulting high-redshift behavior of each branch.
The top row of Fig.~\ref{fig:bkg} shows the redshift evolution of the
background dilaton field. For
$\bar{\chi}_{\rm ini}=\chimin$ (left panel), the dilaton starts at the local minimum
of its potential, so $V_{,\chi}$ is initially negligible and the early
evolution of $\bar{\chi}$ is governed mainly by the competition between
the baryonic and axionic source terms in
Eq.~\eqref{eq:dilaton_background}. Opposite-sign choices of $\bfg$ and
$\zeta$ therefore partially cancel the net force acting on the
dilaton. This explains why the red $(\bfg,\zeta)=(+0.1,-0.1)$ and blue 
$(\bfg,\zeta)=(-0.1,+0.1)$ branches remain close to the initial field value
$\bar{\chi}_{\rm ini}=\chimin$. The cancellation is only approximate
because $\bar{\rho}_{\rm ax}$ is larger than $\bar{\rho}_{\ssB}$, so
the axionic source term determines the direction of the residual
force. Consequently, the negative value of $\zeta$ in the red
$(\bfg,\zeta)=(+0.1,-0.1)$ branch produces a small net force toward
larger $\bar{\chi}$, whereas the blue $(\bfg,\zeta)=(-0.1,+0.1)$ branch evolves slightly
toward smaller $\bar{\chi}$. For the green $(\bfg,\zeta)=(+0.1,+0.1)$ branch, the baryonic
and axionic source terms act with the same sign. Since
$V_{,\chi}\simeq0$ initially, the right-hand side of
Eq.~\eqref{eq:dilaton_background} is negative, driving $\bar{\chi}$
toward smaller values, to the left of the local minimum.

For $\bar{\chi}_{\rm ini}=\bar{\chi}_{\rm bf}$ (right panel), the field is initialized to the right
of the local minimum, where $V_{,\chi}>0$. The potential gradient
therefore drives $\bar{\chi}$ toward smaller values. Compared with the
$\bar{\chi}_{\rm ini}=\chimin$ case, the field begins farther from the
local minimum and undergoes a larger excursion as it evolves toward
$\bar{\chi}_{\rm min}$. The three branches otherwise exhibit broadly
similar qualitative behavior, with their relative evolution governed by
the same competition between the baryonic and axionic source terms.

\begin{table}[t!]
\centering
\renewcommand{\arraystretch}{1.25}
\resizebox{\textwidth}{!}{
\begin{tabular}{c c c c c c c c}
\hline
$\bar\chi_{\rm ini}$
&
Plot Style
&
Branch
&
$\Delta\bar\chi$
&
$\bar\rho_{\rm ax}$
&
$\bar\rho_{\ssB}$
&
$H/H_{\Lambda{\rm CDM}}$
&
$m_i(z)/m_{i,0}$
\\
\hline
\multirow{3}{*}{$74.010$}
&
\tikz[baseline=-0.5ex]
\draw[
branchred,
line width=1.2pt,
dash pattern=on 4pt off 2pt on 1.2pt off 2pt
] (0,0)--(1.0,0);
&
$(\bfg,\zeta)=(+0.1,-0.1)$
&
$>0$
&
Reduced
&
Enhanced
&
$<1$
&
$>1$
\\
&
\tikz[baseline=-0.5ex]
\draw[
branchblue,
solid,
line width=1.2pt
] (0,0)--(1.0,0);
&
$(\bfg,\zeta)=(-0.1,+0.1)$
&
$<0$
&
Reduced
&
Enhanced
&
$<1$
&
$>1$
\\
&
\tikz[baseline=-0.5ex]
\draw[
branchgreen,
line width=1.2pt,
dash pattern=on 10pt off 5pt
] (0,0)--(1.0,0);
&
$(\bfg,\zeta)=(+0.1,+0.1)$
&
$<0$
&
Reduced
&
Reduced
&
$<1$
&
$<1$
\\
\hline
\multirow{3}{*}{$74.213$}
&
\tikz[baseline=-0.5ex]
\draw[
branchred,
line width=1.2pt,
dash pattern=on 4pt off 2pt on 1.2pt off 2pt
] (0,0)--(1.0,0);
&
$(\bfg,\zeta)=(+0.1,-0.1)$
&
$>0$
&
Reduced
&
Enhanced
&
$<1$
&
$>1$
\\
&
\tikz[baseline=-0.5ex]
\draw[
branchblue,
solid,
line width=1.2pt
] (0,0)--(1.0,0);
&
$(\bfg,\zeta)=(-0.1,+0.1)$
&
$>0$
&
Enhanced
&
Reduced
&
$>1$
&
$<1$
\\
&
\tikz[baseline=-0.5ex]
\draw[
branchgreen,
line width=1.2pt,
dash pattern=on 10pt off 5pt
] (0,0)--(1.0,0);
&
$(\bfg,\zeta)=(+0.1,+0.1)$
&
$<0$
&
Reduced
&
Reduced
&
$<1$
&
$<1$
\\
\hline
\end{tabular}
}
\caption{Summary of the background evolution of the three coupling branches at
high redshift, $z\gtrsim10$. Here,
$\Delta\bar\chi(z)\equiv\bar\chi(z)-\bar\chi_0$, where
$\bar\chi_0$ is the present-day field value. Enhanced and reduced denote
changes relative to the corresponding $\Lambda$CDM quantities at the
same redshift. The entries describe the broad high-redshift behavior and
do not reflect the sharp changes and subsequent damped oscillations that
occur at lower redshifts.}
\label{tab:background_evolution_summary}
\end{table}

The second row of Fig.~\ref{fig:bkg} shows the fractional energy
densities of the dilaton, axion, radiation, and baryon components. At redshifts of $\mathcal{O}(10)$, the energy density is dominated by the axion and baryon
components, while the dilaton contribution is negligible. According to
Eqs.~\eqref{eq:axion_background_density_solution} and
\eqref{eq:baryon_background_density_solution}, the axion and baryon
densities scale as $a^{-3}e^{\zeta\Delta\bar\chi}$ and
$a^{-3}e^{\bfg\Delta\bar\chi}$, respectively. The signs of
$\zeta\Delta\bar\chi$ and $\bfg\Delta\bar\chi$ therefore determine
whether the corresponding densities are enhanced or reduced.
For $\bar\chi_{\rm ini}=\chimin$ (left panel), the dash-dotted
$(\bfg,\zeta)=(+0.1,-0.1)$ and solid $(\bfg,\zeta)=(-0.1,+0.1)$
branches remain nearly indistinguishable. Although the signs of
$\Delta\bar\chi$ are opposite, both branches satisfy
$\bfg\Delta\bar\chi>0$ and $\zeta\Delta\bar\chi<0$. They therefore
have an enhanced baryon density and a reduced axion density. The dashed $(\bfg,\zeta)=(+0.1,+0.1)$ branch exhibits a transient
redistribution of the fractional energy densities around $z\sim10$. In this
branch, $\Delta\bar\chi<0$ gives both
$\bfg\Delta\bar\chi<0$ and $\zeta\Delta\bar\chi<0$, reducing the baryon
and axion densities and increasing the relative contribution of the dilaton.

For $\bar\chi_{\rm ini}=74.213$ (right panel), at redshifts of
$\mathcal{O}(10)$, the dash-dotted branch,
$(\bfg,\zeta)=(+0.1,-0.1)$, has
$\Delta\bar\chi>0$, giving
$\bfg\Delta\bar\chi>0$ and
$\zeta\Delta\bar\chi<0$. This results in an enhanced baryon density and a
reduced axion density.
Since the axion constitutes most of the matter
density, the total matter contribution is reduced. 
In the solid $(\bfg,\zeta)=(-0.1,+0.1)$ branch,
$\Delta\bar\chi>0$ gives
$\bfg\Delta\bar\chi<0$ and
$\zeta\Delta\bar\chi>0$. The resulting enhancement of the dominant axion
density maintains a larger matter contribution and delays the relative
growth of the dilaton component.
In the dashed $(\bfg,\zeta)=(+0.1,+0.1)$ branch,
$\Delta\bar\chi<0$ gives
$\bfg\Delta\bar\chi<0$ and
$\zeta\Delta\bar\chi<0$, reducing both the baryon and axion densities.
Since the axion constitutes most of the matter density, this branch also
has a reduced total matter contribution.
Toward lower redshift, $\Delta\bar\chi$ approaches zero. Consequently, all
three $(\bfg,\zeta)$ choices in both panels approach the same present-day
fractional energy densities.

The third row of Fig.~\ref{fig:bkg} shows the physical Hubble rate $H$
relative to the $\Lambda$CDM prediction. The conformal Hubble rate
$\mathcal{H}$ appearing in Eq.~\eqref{eq:friedmann_background} is related
to $H$ by $\mathcal{H}=H/(1+z)$. The broad deviation at high
redshift is mainly determined by the modified axion and baryon densities
in Eqs.~\eqref{eq:axion_background_density_solution} and
\eqref{eq:baryon_background_density_solution}. In this regime, the dilaton
accounts for only a small fraction of the total energy density, and its
contribution to the Hubble rate is therefore negligible. Since the axion
constitutes most of the matter density, the sign and magnitude of
$\zeta\Delta\bar{\chi}$ largely determine whether the expansion rate is
enhanced or reduced.
For $\bar\chi_{\rm ini}=\chimin$ (left panel), at redshifts of $\mathcal{O}(10)$, both the red
$(\bfg,\zeta)=(+0.1,-0.1)$ and blue
$(\bfg,\zeta)=(-0.1,+0.1)$ branches satisfy
$\bfg\Delta\bar\chi>0$ and $\zeta\Delta\bar\chi<0$. Consequently, the
baryon density is enhanced, while the axion density is reduced in both
branches.
Since the axion dominates the total matter density, the reduction in
the axion density outweighs the increase in the baryon density. The red
and blue branches consequently yield similar Hubble rates, both
slightly below the $\Lambda$CDM prediction.
In the green $(\bfg,\zeta)=(+0.1,+0.1)$ branch,
$\Delta\bar\chi<0$, so both $\bfg\Delta\bar\chi$ and
$\zeta\Delta\bar\chi$ are negative, reducing the baryon and axion
densities and producing the largest suppression of the Hubble rate at
high redshift. At lower redshift, the fractional contribution of the dilaton to the total energy density increases. Its rapid evolution toward the local minimum
enhances its kinetic energy, causing the sharp rise in the Hubble rate.

For $\bar\chi_{\rm ini}=\chibf$ (right panel), at redshifts of $\mathcal{O}(10)$, the red
$(\bfg,\zeta)=(+0.1,-0.1)$ branch has
$\Delta\bar\chi>0$. Hence,
$\bfg\Delta\bar\chi>0$ and $\zeta\Delta\bar\chi<0$, leading to an
enhanced baryon density and a reduced axion density.
Since the axion constitutes most of the matter density,
the reduction in the axion density dominates, resulting in a Hubble rate
below the $\Lambda$CDM prediction. 
In the green $(\bfg,\zeta)=(+0.1,+0.1)$ branch,
$\Delta\bar\chi<0$, giving
$\bfg\Delta\bar\chi<0$ and $\zeta\Delta\bar\chi<0$. This reduces both
the axion and baryon densities, resulting in a Hubble rate below the
$\Lambda$CDM prediction.
The blue $(\bfg,\zeta)=(-0.1,+0.1)$ branch has
$\Delta\bar\chi>0$, giving $\zeta\Delta\bar\chi>0$ and
$\bfg\Delta\bar\chi<0$. The resulting increase in the axion density
outweighs the reduction in the baryon density, raising the Hubble rate
above the $\Lambda$CDM prediction. As the field approaches its
present-day value, $\Delta\bar\chi$ approaches zero and the axion and
baryon density rescalings approach unity. 
Meanwhile, the transient features at low redshift arise as the dilaton
contribution to the Friedmann equation becomes increasingly important.
In the red $(\bfg,\zeta)=(+0.1,-0.1)$ and green
$(\bfg,\zeta)=(+0.1,+0.1)$ branches, the field moves rapidly toward the
local minimum over a narrow redshift interval, causing simultaneous changes in
the dilaton kinetic and potential energies. Their combined evolution produces the sharp features in
$H/H_{\Lambda{\rm CDM}}$.

The last row of Fig.~\ref{fig:bkg} shows the redshift evolution of the
SM particle masses relative to their present-day values. As shown in
Eq.~\eqref{eq:mass_ratio}, whether the masses are enhanced or reduced is
determined by the sign of $\bfg\,\Delta\bar{\chi}(z)$.
For $\bar\chi_{\rm ini}=\chimin$ (left panel), at redshifts of $\mathcal{O}(10)$, the red $(\bfg,\zeta)=(+0.1,-0.1)$ branch has
$\Delta\bar\chi>0$ and hence $\bfg\Delta\bar\chi>0$, giving
$m_i(z)/m_{i,0}>1$. In the blue
$(\bfg,\zeta)=(-0.1,+0.1)$ branch, both
$\Delta\bar\chi$ and $\bfg$ are negative, so their product is again positive
and $m_i(z)/m_{i,0}>1$. In both branches,
$\lvert\Delta\bar\chi\rvert$ remains small, and the mass ratio therefore
lies only slightly above unity. In the green
$(\bfg,\zeta)=(+0.1,+0.1)$ branch, the larger excursion toward smaller field
values gives $\Delta\bar\chi<0$ and $\bfg\Delta\bar\chi<0$, resulting in
$m_i(z)/m_{i,0}<1$ and a more pronounced deviation from unity.
At lower redshift, the dilaton field evolves toward its present-day value, reducing
$\lvert \bfg\Delta\bar\chi\rvert$. Consequently, the mass ratios in all three
branches approach unity.

For $\bar\chi_{\rm ini}=\chibf$ (right panel), at redshifts of $\mathcal{O}(10)$, the red
$(\bfg,\zeta)=(+0.1,-0.1)$ branch has $\Delta\bar\chi>0$ and hence
$\bfg\Delta\bar\chi>0$, giving $m_i(z)/m_{i,0}>1$. In the green
$(\bfg,\zeta)=(+0.1,+0.1)$ branch, $\Delta\bar\chi<0$ and
$\bfg\Delta\bar\chi<0$, resulting in $m_i(z)/m_{i,0}<1$. In the blue
$(\bfg,\zeta)=(-0.1,+0.1)$ branch, $\Delta\bar\chi>0$ while $\bfg<0$, so
$\bfg\Delta\bar\chi<0$ and the mass ratio is also below unity. At lower
redshift, $\bar\chi$ approaches its present-day value in all three branches,
causing $\lvert \bfg\Delta\bar\chi\rvert$ to decrease and the corresponding
mass ratios to converge to unity.

\subsubsection{Perturbation Evolution}
\label{subsec:perturbation_evolution}

The complete linear perturbation system and its derivation have been
presented in Ref.~\cite{Smith:2025uaq}. Here we give a concise summary of
the perturbation variables that enter the 21\,cm calculation, with
particular emphasis on the baryon density contrast, the total matter
density contrast, and the linear matter power spectrum. For clarity, we
organize the discussion around the quasi-static forms of the axion and
baryon growth equations, which make the effects of the dilaton couplings
transparent. All numerical results shown
below are obtained by evolving the complete perturbation system.

We consider the subhorizon quasi-static regime, defined by
\begin{equation}
    |X'|\lesssim\mathcal{H}|X|,
    \qquad
    |X''|\lesssim\mathcal{H}^2|X|,
    \qquad
    k^2\gg\mathcal{H}^2,
\end{equation}
for $X\in\{\delta\chi,\Phi,\Psi\}$. Under these conditions, the time
derivatives of the dilaton and metric perturbations are subdominant.
In this regime, the coupled perturbation system reduces to the following approximate growth equations for the axion and baryon
fluids.
\begin{equation}
\begin{aligned}
\delta_{\rm ax}''
+
\left(
\mathcal{H}
-
\zeta\bar{\chi}'
\right)
\delta_{\rm ax}'
+
\frac{k^4}{4a^2m_{\rm ax}^2}\,
\delta_{\rm ax}
&\simeq
4\pi a^2
\left[
G_{\mfa\mfa}(k,z)\,
\bar{\rho}_{\rm ax}\delta_{\rm ax}
+
G_{\mfa B}(k,z)\,
\bar{\rho}_{\ssB}\delta_{\ssB}
\right],
\\[0.5em]
\delta_{\ssB}''
+
\left(
\mathcal{H}
+
\bfg\bar{\chi}'
\right)
\delta_{\ssB}'
&\simeq
4\pi a^2
\left[
G_{\mfa B}(k,z)\,
\bar{\rho}_{\rm ax}\delta_{\rm ax}
+
G_{BB}(k,z)\,
\bar{\rho}_{\ssB}\delta_{\ssB}
\right].
\end{aligned}
\label{eq:axion_baryon_growth_quasistatic}
\end{equation}
In deriving the baryon equation, we have neglected the
baryon pressure and Thomson drag terms, which are subdominant after
recombination on the scales considered here.
The second term on the left-hand side of each equation shows that the
background dilaton evolution changes the axion and baryon friction
coefficients by $-\zeta\bar{\chi}'$ and $\bfg\bar{\chi}'$, respectively.
The terms on the right-hand side of these equations show that dilaton
exchange also modifies the gravitational response of the two fluids.
It is convenient to define these effective gravitational couplings as
\begin{gather}
G_{\mfa\mfa}(k,z)
\equiv
G_N\left[1+2\zeta^2\frac{k^2}{a^2k_\chi^2}\right],
\qquad
G_{\mfa B}(k,z)
\equiv
G_N\left[1+2\zeta\bfg\frac{k^2}{a^2k_\chi^2}\right],
\nonumber\\
G_{BB}(k,z)
\equiv
G_N\left[1+2\bfg^2\frac{k^2}{a^2k_\chi^2}\right].
\label{eq:effective_gravitational_couplings}
\end{gather}
Here,
\begin{equation}
k_\chi^2
\equiv
\frac{k^2}{a^2}
+
m_{\chi,\mathrm{eff}}^2,
\qquad
m_{\chi,\mathrm{eff}}^2
=
\frac{1}{\MP^2}
\left(
V_{,\bar{\chi}\bar{\chi}}
-
\zeta^2\bar{\rho}_{\rm ax}
\right).
\label{eq:dilaton_interaction_scale}
\end{equation}
The scale dependence of the dilaton-mediated interaction follows from
\begin{equation}
    \frac{k^2}{a^2k_\chi^2}
    =
    \frac{k^2/a^2}
    {k^2/a^2+m_{\chi,\mathrm{eff}}^2}.
\end{equation}
For $m_{\chi,\mathrm{eff}}^2>0$ and higher $k$-modes $k/a\gg m_{\chi,\mathrm{eff}}$, the physical wavelength of the mode is shorter than the dilaton Compton wavelength, and the factor $k^2/(a^2k_\chi^2)$ approaches unity. The dilaton-mediated force then reaches its unsuppressed limit.
For lower $k$-modes $k/a\ll m_{\chi,\mathrm{eff}}$, the physical wavelength exceeds the Compton wavelength, and
$k^2/(a^2k_\chi^2)\ll1$, corresponding to the Yukawa
suppression of the force. The effective dilaton mass therefore determines
the transition scale $k/a\simeq  m_{\chi,\mathrm{eff}}$. 

Most of the scales and redshifts considered in this study lie in the
unsuppressed regime, where the effective axion and baryon self-couplings approach
$G_{\mfa\mfa}\simeq G_N(1+2\zeta^2)$ and
$G_{BB}\simeq G_N(1+2\bfg^2)$, respectively.
The mixed
axion--baryon interaction, $G_{\mfa B}(k,z)$, is enhanced for
$\zeta \bfg>0$ and weakened for $\zeta \bfg<0$. 
This mixed coupling affects the baryon perturbation particularly strongly
because it enters the baryon equation through
$G_{\mfa B}(k,z)\bar{\rho}_{\rm ax}\delta_{\rm ax}$, while the corresponding
term in the axion equation is
$G_{\mfa B}(k,z)\bar{\rho}_{\ssB}\delta_{\ssB}$, with
$\bar{\rho}_{\rm ax}>\bar{\rho}_{\ssB}$.

The axion gradient term $k^4/(4a^2m_{\rm ax}^2)$ increases rapidly
toward larger $k$ and acts as an effective pressure against gravitational
collapse. For the axion mass adopted here, this term becomes important only
at wavenumbers well above the range considered in this study and therefore
has a negligible effect on the perturbations and matter power spectra
presented below.

\begin{figure}[t!]
\centering
\includegraphics[width=1.0\textwidth]{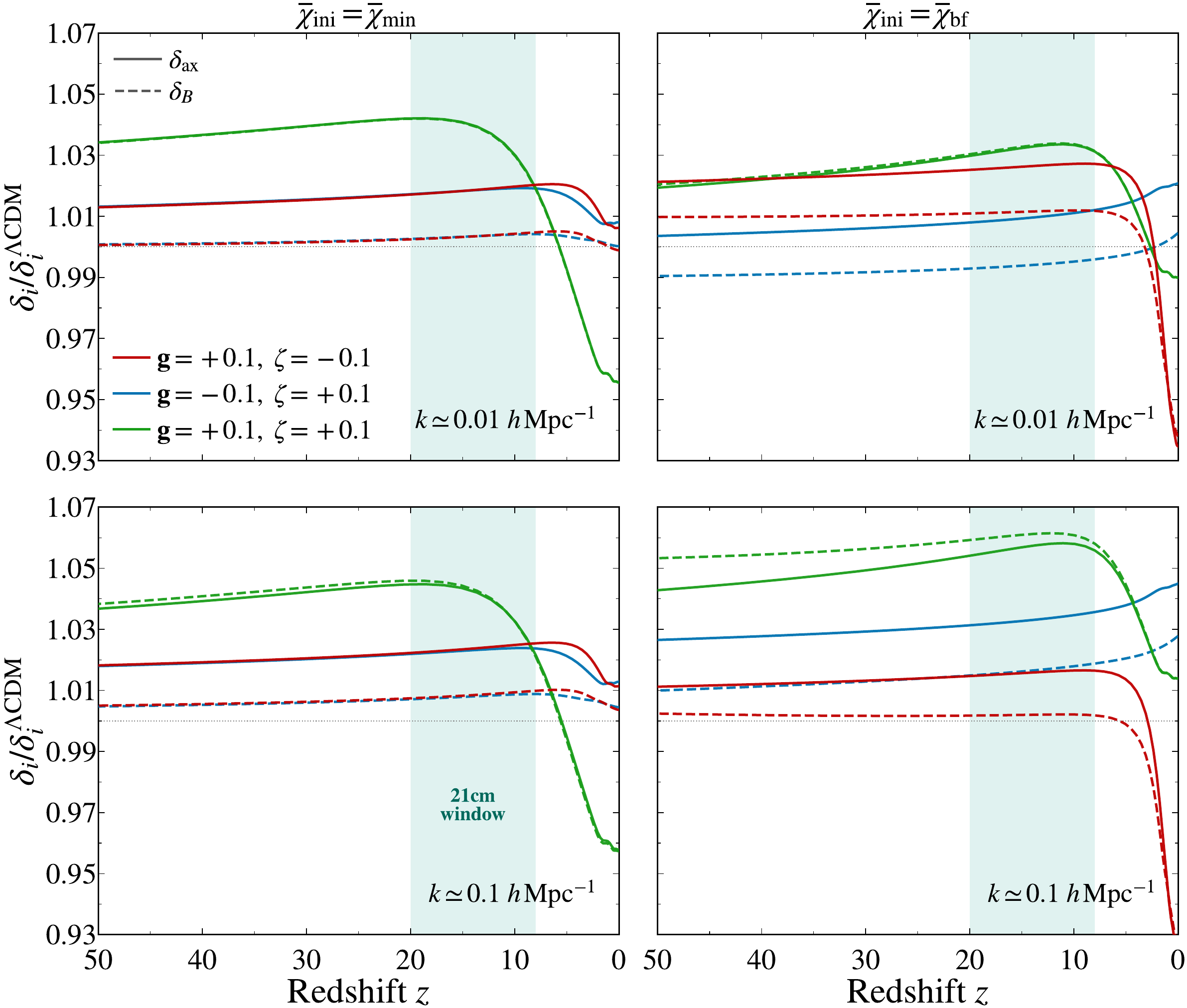}
\caption{Redshift evolution of the axion density contrast
$\delta_{\rm ax}$ (solid lines) and baryon density contrast
$\delta_{\ssB}$ (dashed lines), normalized to their corresponding
$\Lambda$CDM values. The upper and lower panels show
$k=0.01$ and $0.1\,h\,{\rm Mpc}^{-1}$, respectively, while the left
and right panels correspond to $\bar\chi_{\rm ini} = \bar\chi_{\rm min} = 74.010$ and $\bar\chi_{\rm ini} = \bar\chi_{\rm bf}=74.213$, respectively. The red, blue, and green curves represent
$(\bfg,\zeta)=(+0.1,-0.1)$, $(-0.1,+0.1)$, and
$(+0.1,+0.1)$, respectively. The light-blue shaded region marks the redshift range $8\leq z\leq20$ targeted in our 21\,cm analysis.}
\label{fig:delta_evolution}
\end{figure}

Fig.~\ref{fig:delta_evolution} shows the redshift evolution of the axion
and baryon density contrasts relative to their corresponding
$\Lambda$CDM values. The upper and lower rows correspond to
$k=0.01$ and $0.1\,h\,{\rm Mpc}^{-1}$, respectively, while the left
and right columns show the results for
$\bar\chi_{\rm ini}=\chimin$ and $\chibf$.
For $\bar\chi_{\rm ini}=\chimin$, the red
$(\bfg,\zeta)=(+0.1,-0.1)$ and blue
$(\bfg,\zeta)=(-0.1,+0.1)$ branches produce nearly identical axion
and baryon density contrasts. Their background evolution is also nearly identical, as shown in
Fig.~\ref{fig:bkg}, and they have the same coupling
combination, $\bfg\zeta=-0.01$, which weakens the
gravitational coupling $G_{\mfa B} < G_N$. This results in the overall reduction in baryon and axion density contrasts. Since
the axion density $\bar \rho_{\rm ax}$ is much larger than the baryon density $\bar \rho_{\ssB}$, the baryon density contrast is more strongly suppressed than the axion density contrast. In the green
$(\bfg,\zeta)=(+0.1,+0.1)$ branch, the positive mixed coupling
correction $G_{\mfa B}>G_N$ enhances the mutual gravitational interaction between the
two matter components. Its smaller Hubble rate also reduces expansion
damping, yielding the largest axion and baryon density contrasts.

For $\bar\chi_{\rm ini}=\chibf$, the ratio $k_\chi/(k/a)$ remains
close to unity over the scales shown for both the red
$(\bfg,\zeta)=(+0.1,-0.1)$ and blue
$(\bfg,\zeta)=(-0.1,+0.1)$ branches. Their effective
scalar-mediated couplings consequently remain close to their
unsuppressed values. The separation between these branches is
primarily governed by their different background densities and
Hubble rates.

At $k=0.1\,h\,{\rm Mpc}^{-1}$, the larger axion density in the blue $(\bfg,\zeta)=(-0.1,+0.1)$
branch moves matter--radiation equality to a higher redshift. This
mode enters the horizon during radiation domination, and the earlier
equality in the blue $(\bfg,\zeta)=(-0.1,+0.1)$ branch shortens the period over which its growth
is suppressed by the radiation background. The larger axion density $\bar\rho_{\rm ax}$
also increases the fraction of matter that can cluster efficiently.
These effects give larger axion and baryon density contrasts 
in the blue $(\bfg,\zeta)=(-0.1,+0.1)$ branch than in the red $(\bfg,\zeta)=(+0.1,-0.1)$ branch at high $k$.

At $k=0.01\,h\,{\rm Mpc}^{-1}$, the mode enters the horizon after
matter--radiation equality, so the shift in the equality redshift has
only a small effect on its perturbation transfer function. The lower
Hubble rate in the red $(\bfg,\zeta)=(+0.1,-0.1)$ branch reduces expansion damping and leads to
larger density contrasts than in the blue $(\bfg,\zeta)=(-0.1,+0.1)$ branch. 
The green $(\bfg,\zeta)=(+0.1,+0.1)$ branch exhibits enhanced axion
and baryon density contrasts at both scales, owing primarily to the
positive mixed coupling $G_{\mfa B}>G_N$.


\begin{figure}[t!]
    \centering
    \includegraphics[width=\textwidth]{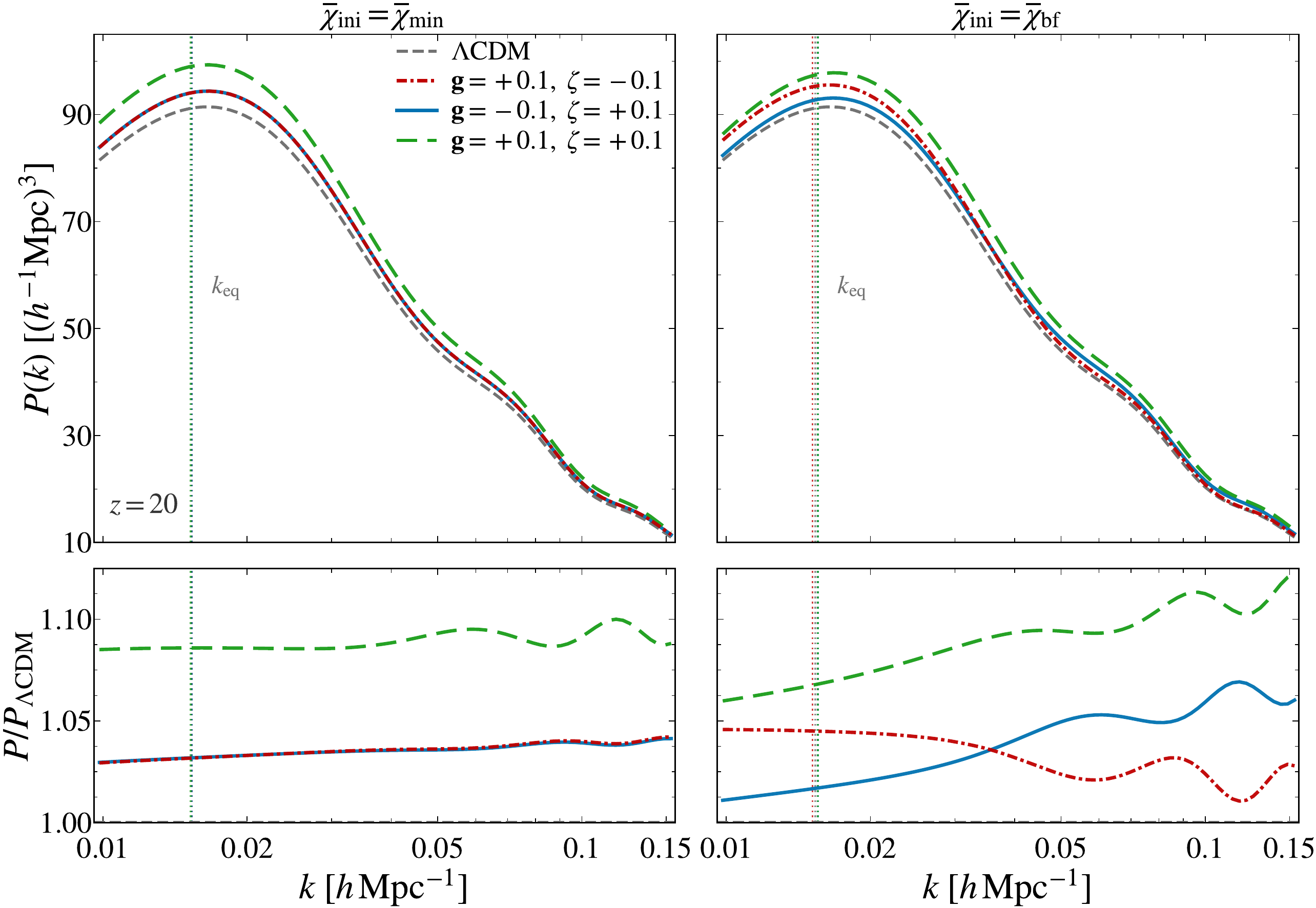}
    \caption{Linear total matter power spectra in the
    axiodilaton cosmology at $z=20$. The upper panels show
    $P_m(k)$, and the lower panels show its ratio to the corresponding
    $\Lambda$CDM prediction. The left and right panels correspond to
    $\bar\chi_{\rm ini} = \bar\chi_{\rm min} = 74.010$ and $\bar\chi_{\rm ini} = \bar\chi_{\rm bf}=74.213$, respectively. The red dash-dotted, blue solid, and green dashed curves represent
    $(\bfg,\zeta)=(+0.1,-0.1)$, $(-0.1,+0.1)$, and
    $(+0.1,+0.1)$, respectively. The gray dashed curves show the
    $\Lambda$CDM prediction, and the colored vertical dotted lines mark
    the matter--radiation equality scale $k_{\rm eq}$ for the
    corresponding branches.}
    \label{fig:matter_power_spectrum}
\end{figure}

In the following we focus on adiabatic initial conditions for the cosmological perturbations. Fig.~\ref{fig:matter_power_spectrum} shows the linear total matter
power spectrum and its ratio to the corresponding $\Lambda$CDM prediction
at $z=20$. Its overall behavior follows the evolution of the density
contrasts discussed in Fig.~\ref{fig:delta_evolution}. For
$\bar\chi_{\rm ini}=\chimin$, the red
$(\bfg,\zeta)=(+0.1,-0.1)$ and blue
$(\bfg,\zeta)=(-0.1,+0.1)$ branches remain nearly degenerate, while the
green $(\bfg,\zeta)=(+0.1,+0.1)$ branch shows the largest enhancement
owing primarily to the positive mixed coupling correction
$G_{\mfa B}>G_N$.

For $\bar\chi_{\rm ini}=\chibf$, the ordering of the red
$(\bfg,\zeta)=(+0.1,-0.1)$ and blue
$(\bfg,\zeta)=(-0.1,+0.1)$ branches changes with scale. The red
$(\bfg,\zeta)=(+0.1,-0.1)$ branch has more power at low $k$, where the
lower Hubble rate reduces expansion damping. 
Toward larger $k$, the blue
$(\bfg,\zeta)=(-0.1,+0.1)$ branch shows more power because its higher
axion density shifts matter--radiation equality to an earlier time and
allows more matter to cluster efficiently before recombination.
The green $(\bfg,\zeta)=(+0.1,+0.1)$ branch remains enhanced across the
full range of scales, primarily due to $G_{\mfa B}>G_N$. The oscillatory
residuals at larger $k$ reflect shifts in the baryon acoustic features
relative to $\Lambda$CDM.

\begin{figure}
    \centering
    \includegraphics[width=1\linewidth]{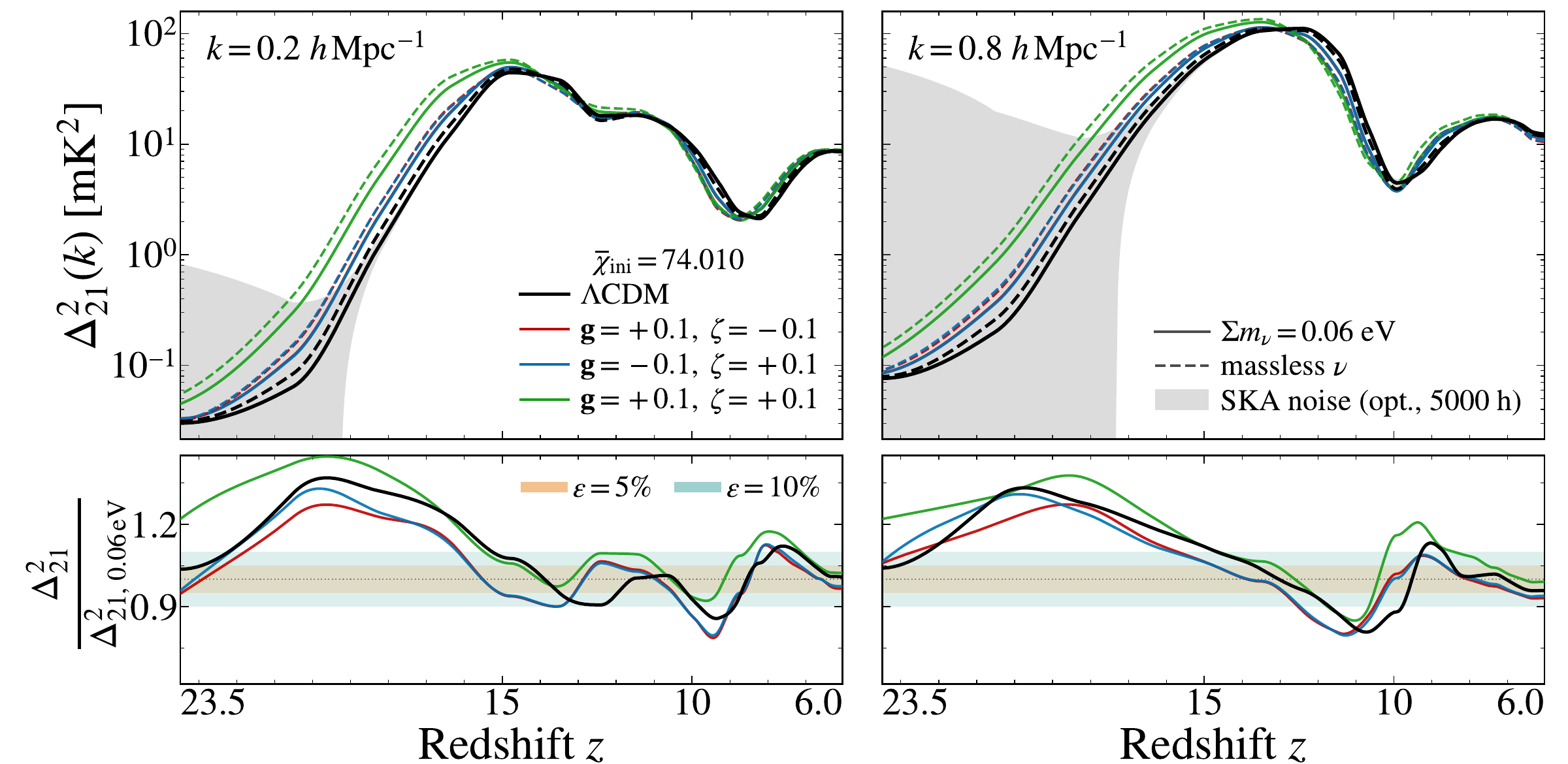}
    \caption{Effect of the neutrino mass on the dimensionless 21\,cm power spectrum
    for $\bar\chi_{\rm ini}=\bar\chi_{\rm min}=74.010$, at $k=0.2\,h\,{\rm Mpc}^{-1}$ (left) and
    $k=0.8\,h\,{\rm Mpc}^{-1}$ (right). The upper panels show $\Delta_{21}^2(k,z)$ for \LCDM\ (black)
    and for $(\bfg,\zeta)=(+0.1,-0.1)$ (red), $(-0.1,+0.1)$ (blue), and $(+0.1,+0.1)$ (green),
    computed with one massive neutrino of $\Sigma m_\nu=0.06\,{\rm eV}$ (solid) and with massless
    neutrinos (dashed). The gray shaded regions show the projected SKA1-Low noise for the
    optimistic sky-temperature model with $5000\,{\rm hr}$ of observing time. The lower panels show
    the ratio of the massless-neutrino result to the $\Sigma m_\nu=0.06\,{\rm eV}$ result for each
    model. The shaded bands mark theoretical uncertainties of $5\%$ and $10\%$, labeled
    $\varepsilon$ in the legend and denoted $f_{\rm theo}$ in Eq.~\eqref{eq_noise_obs}.}
    \label{fig:neutrino_mass}
\end{figure}

Fig.~\ref{fig:neutrino_mass} shows how the neutrino mass affects the 21\,cm power spectrum for
$\bar\chi_{\rm ini}=\chimin$. Replacing the massive neutrino with $\Sigma m_\nu=0.06\,{\rm eV}$ by massless
neutrinos raises $\Delta_{21}^2$ by up to about $30\%$ at $z\simeq20$, where the SKA1-Low noise is large,
and changes it by up to about $15\%$ for $8\lesssim z\lesssim15$. The size of this change is similar for
\LCDM\ and for the three axiodilaton branches.

\subsection{Current Cosmological Constraints}
\label{subsec:Cosmoconstraints}

The implications of the minimal axiodilaton model of the Dark Sector were worked out for the CMB and early structure formation in Ref.~\cite{Smith:2024ibv}, with the results compared with observations in \cite{Smith:2025uaq}. We briefly summarize these results here as a benchmark for the projected 21\,cm constraints, covering the preferred matter couplings, the dilaton's initial condition, and the extent to which current data already constrain the growth of structure.

The main takeaway is that the axiodilaton model fits the CMB and late-time structure better than $\Lambda$CDM: combining Planck 2018 \cite{Planck:2018nkj,Planck:2019nip}, ACT lensing \cite{ACT:2023kun,ACT:2023dou}, SPT+lensing \cite{SPT-3G:2025bzu,SPT-3G:2025zuh}, Pantheon+ \cite{Brout:2022vxf}, and DESI DR2 \cite{DESI:2025zgx,DESI:2025zpo} gives $\Delta\chi^2 \approx -7$, rising to $-19$ with SH0ES-calibrated supernovae. This mainly reflects the model letting particle masses differ at recombination from today, exposing the $H_0$--$m_e$ degeneracy so that a larger $m_e$ at recombination permits a larger $H_0$ today, reducing the Hubble tension to $3\sigma$.

\begin{figure}
    \centering

    \begin{subfigure}{0.49\textwidth}
        \centering
        \includegraphics[width=\linewidth]{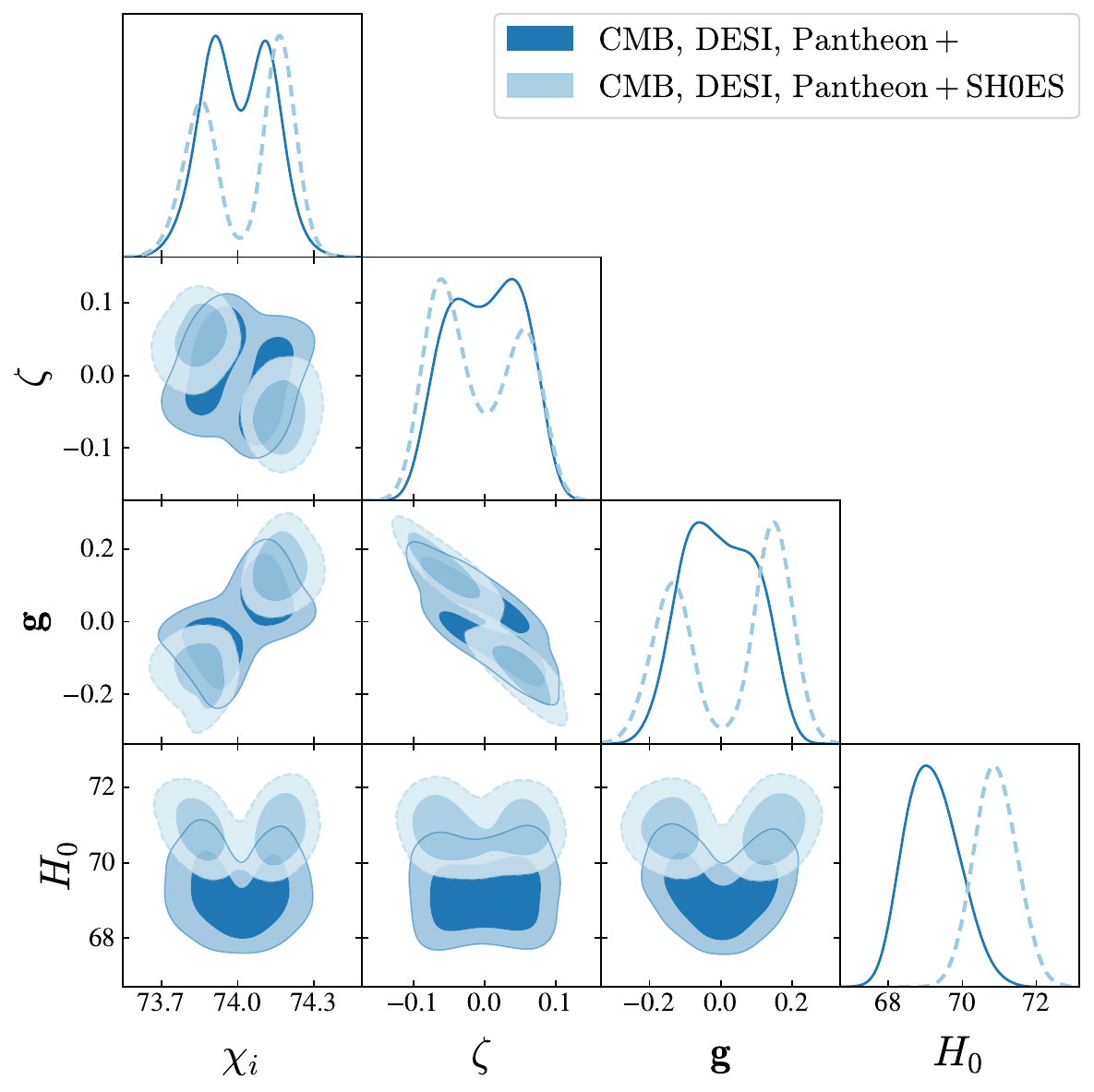}
    \end{subfigure}
    \hfill
    \begin{subfigure}{0.49\textwidth}
        \centering
        \includegraphics[width=\linewidth]{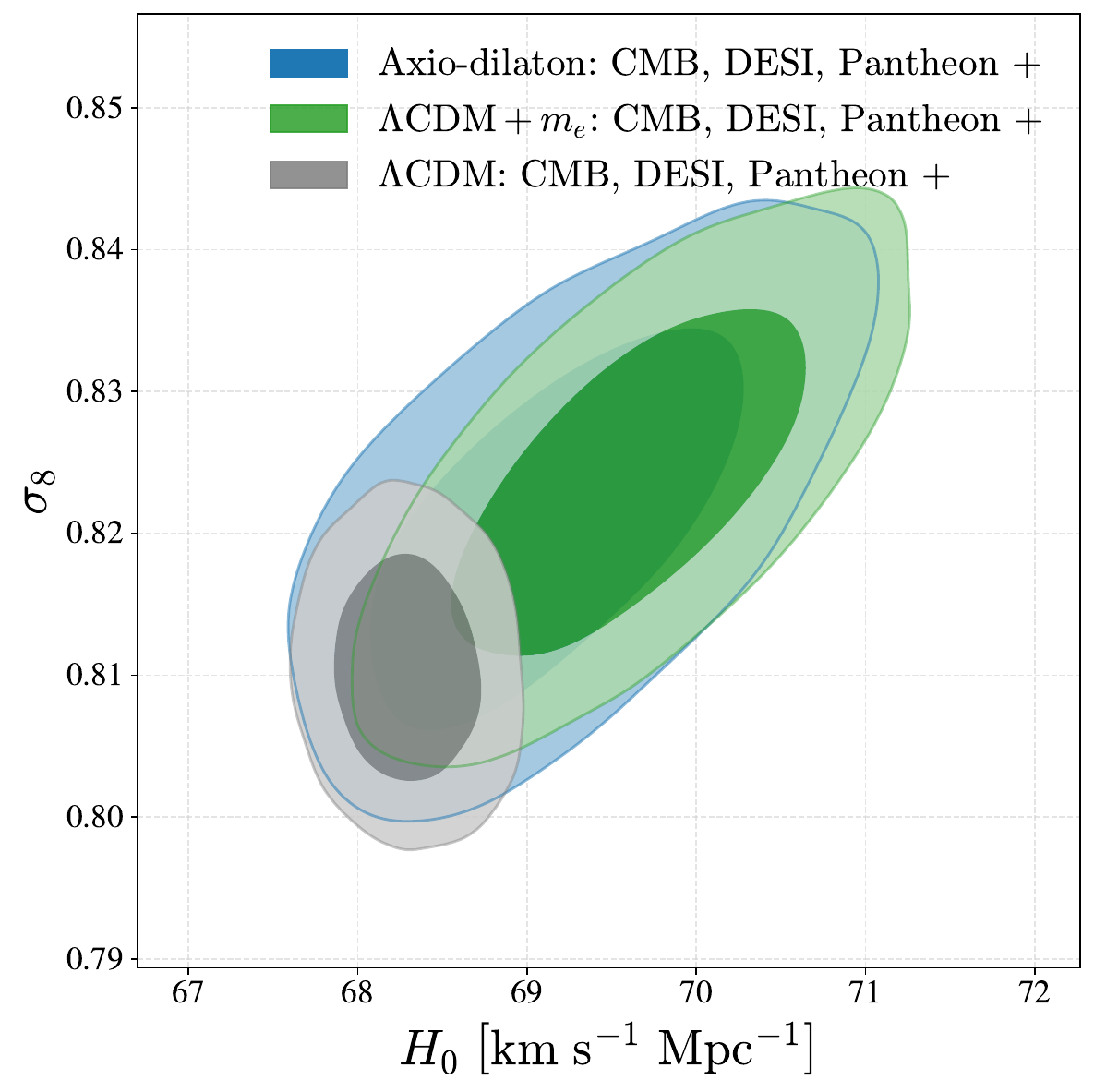}
    \end{subfigure}

    \caption{
    Triangle plot showing constraints on the minimal axiodilaton model couplings, dilaton displacement and Hubble constant, with and without the SH0ES calibration, alongside constraints in the $H_0$--$\sigma_8$ plane for the axiodilaton model, \LCDM\ with a varying electron mass ($\Lambda{\rm CDM}+m_e$), and \LCDM. The constraints use the Planck 2018, ACT DR6 Lensing, DESI DR2, and Pantheon+ supernovae datasets, with the triangle plot taken from \cite{Smith:2025uaq}. Imposing the SH0ES calibration amplifies the preference for $\mathcal{O}(0.1)$ couplings and initial dilaton displacements.
    }
    \label{fig:cosmo_constraints}
\end{figure}

Fig.~\ref{fig:cosmo_constraints} shows that the fits prefer a nonzero coupling $\bfg$ of size $\mathcal{O}(0.1)$ (best fit $\bfg = 0.14$, $1\sigma$ range $0.00 \pm 0.10$ \cite{Smith:2025uaq}). Since $\bfg$ controls the dilaton dependence of Standard Model masses, larger values allow a larger difference between $m_e$ at recombination and today, providing the freedom needed for the improved CMB fit described above. At the same time, $\bfg$ and $\zeta$ enter the dilaton's equation of motion through the combined source term additively, so a large coupling to baryonic matter needed to move Standard Model masses can be compensated by a large coupling of opposite sign to the axion, leaving little net effect on the dilaton when it is dynamically rolling at late times, and keeping the background close to $\Lambda$CDM. This compensation produces the butterfly-shaped, anti-correlated contours in the $\bfg$--$\zeta$ panel, consisting of two separate peaks connected through a shallow valley near the origin. As mentioned in the main text, at face value couplings of this size exceed solar-system bounds so using both at the same time assumes the particle couplings seen in cosmology differ from the ones relevant for macroscopic objects (like the Sun) whose couplings are probed in the solar system (perhaps along the lines of the screening mechanism proposed in \cite{Brax:2023qyp}).

Since the difference in particle masses cares about the combination $\mathbf{g}(\bar\chi-\bar\chi_0)$, the dilaton's initial condition is similarly constrained, with two comparably high peaks on either side of $\bar\chi_{\rm ini} \simeq 74$ and a deeper minimum between them, mirroring the bimodality already seen in $\bfg$ and $\zeta$. For the CMB-B DESI Pantheon+ combination, the best fit is $\bar\chi_{\rm ini}=73.844$ (to the left of the minimum). With the SH0ES calibration included, the best fit shifts to $\bar\chi_{\rm ini}=\chibf$ (to the right of the minimum). In both cases the posterior remains strongly bimodal, with a second comparably preferred solution on the opposite side of the central minimum.

The same coupling that shifts recombination-era masses also affects late-time growth, directly constrained by DESI DR2 and Pantheon+ in Fig.~\ref{fig:cosmo_constraints}. During matter domination, the preferred cosmologies can enhance $\delta_m$ growth, driving earlier structure formation and possibly the early formation of galaxies. As the dilaton approaches its late-time minimum it undergoes shallow oscillations, driven by the competing matter couplings, that generate transient enhancements in the Dark Energy density and hence in the Hubble rate. The resulting increase in expansion damping suppresses structure growth at late times, counteracting the earlier enhancement accumulated during matter domination. This matches the right panel of Fig.~\ref{fig:cosmo_constraints}, where the allowed $\sigma_8$ values remain broadly overlapping with the $\Lambda$CDM constraint despite the enhanced growth at earlier times, motivating a direct search for this earlier enhancement through the 21\,cm signal.

\section{\texorpdfstring{21\,cm Minutiae}{21cm Minutiae}}
\label{app:21cmMinutae}
\label{app:21cm}

This appendix shows how each ingredient of the 21\,cm signal depends on the electron and
proton masses, which vary with redshift in the axiodilaton cosmology. For every quantity we
derive its scaling with $m_e(z)$ and $m_p(z)$ and state what is implemented numerically. We start with the hyperfine transition in Appendix~\ref{app:Tb_minimal},
then treat the three ingredients of the spin temperature, namely the gas kinetic temperature
in Appendix~\ref{subsec:Tk}, the free-electron fraction from X-ray and UV ionization
in Appendices~\ref{subsec:xe_evolution} and~\ref{app:UV_ionization}, and the Ly$\alpha$
background with its coupling and heating
in Appendices~\ref{app:lya_scattering}--\ref{app:lya_coupling_heating}.
The 21\,cm physics in this appendix is well established. The spin-temperature physics and the formalism for the global signal and its fluctuations were developed in Refs.~\cite{Field:1958rri,Madau:1996cs,Zaldarriaga:2003du,Furlanetto:2004nh}; for derivations and details we refer to the reviews~\cite{Barkana:2000fd,Furlanetto:2006jb,Morales:2009gs,Pritchard:2011xb,Loeb:2013,Furlanetto:2015apc,Barkana:2016nat,Furlanetto:2019dhm}. The thermal and ionization history of the intergalactic medium follows Refs.~\cite{Furlanetto:2006global,Pritchard:2006sq,Mesinger:2010ne,Park:2018ljd}. All quantities are computed with \FirstClass~\cite{Flitter:2023mjj,Flitter:2023rzv,Flitter:2024eay}, which is built on \textsc{CLASS} and \textsc{21cmFAST}~\cite{Mesinger:2010ne,Murray:2020trn}. Where we depart from the fiducial \FirstClass\ treatment, we say so.

\subsection{Hyperfine Transition of Neutral Hydrogen}
\label{app:Tb_minimal}
\label{app:nu21}
\label{sec:A10}

This appendix summarizes the mass dependence of the hyperfine transition frequency $\nu_{21}(z)$ and the Einstein coefficient $A_{10}(z)$, quoted in Eqs.~\eqref{eq:nu21_mass_scaling} and \eqref{eq:A10_mass_scaling} of the main text. The hyperfine splitting of the hydrogen ground state arises from the Fermi contact interaction between the magnetic moments of the electron and the proton. For the $1s$ state, the splitting between the triplet and singlet energy levels is proportional to the electron probability density at the position of the proton. A detailed derivation is given in Ref.~\cite{griffiths_introduction_2018}, with the result
\begin{equation}
\Delta E_{\rm hf}
=
\frac{8\pi}{3}\,g_e\,g_I\,\mu_B(z)\,\mu_N(z)\,|\psi_{1s}(r_{ep}=0,z)|^2,
\label{eq:energy_splitting}
\end{equation}
where $g_e\simeq2$ and $g_I\simeq5.586$ are the electron and proton $g$-factors, and $\mu_B(z)=e\hbar/(2m_e(z)\,c)$ and $\mu_N(z)=e\hbar/(2m_p(z)\,c)$ are the Bohr and nuclear magnetons. The factor $|\psi_{1s}(r_{ep}{=}0,z)|^2=1/(\pi a_0^3(z))=m_e^3(z)\,e^6/(\pi\hbar^6)$ is the electron probability density evaluated at electron--proton separation $r_{ep}$, with $a_0(z)=\hbar^2/(m_e(z)\,e^2)$ the Bohr radius. Substituting these expressions and using $e^2=\alpha\hbar c$, where $\alpha$ is the fine-structure constant, gives the transition frequency
\begin{equation}
\nu_{21}(z)
=
\frac{\Delta E_{\rm hf}(z)}{h}
=
\frac{g_e\,g_I}{3\pi}\,
\frac{\alpha^4 c^2}{\hbar}\,
m_e^2(z)\,m_p^{-1}(z).
\label{eq:nu21}
\end{equation}
The $g$-factors are dimensionless and unaffected by the mass rescaling, so normalizing to the measured present-day values $\nu_{21,0}\simeq 1420.4 \rm MHz$ and applying Eq.~\eqref{eq:mass_ratio} to both masses gives
\begin{equation}
\nu_{21}(z)
=
\nu_{21,0}
\left[\frac{m_e(z)}{m_{e,0}}\right]^2
\left[\frac{m_p(z)}{m_{p,0}}\right]^{-1}
=
\nu_{21,0}\,e^{\bfg\Delta\bar\chi(z)}.
\label{eq:nu21_z_app}
\end{equation}
A larger electron mass increases $\nu_{21}(z)$, while a larger proton mass lowers it through the nuclear magneton. The 21\,cm transition temperature $T_\star(z)=h\nu_{21}(z)/k_B$ inherits the same scaling, which is Eq.~\eqref{eq:Tstar} of the main text.
The spontaneous decay of the triplet state proceeds through a magnetic dipole transition, in which the electron spin flips while the spatial wavefunction remains in the $1s$ orbital. The evaluation of the magnetic dipole matrix element, detailed in Refs.~\cite{Field:1958rri, Wild1952}, is then
\begin{equation}
A_{10}(z)
=
\frac{g_e^2\,\omega_{21}^3(z)\,\mu_B^2(z)}{3\hbar c^3}
=
\frac{2}{81}\,
g_e^5\,g_I^3\,
\frac{\alpha^{13}\,c^2}{\hbar}\,
m_e^4(z)\,m_p^{-3}(z),
\label{eq:A10_mass}
\end{equation}
where $\omega_{21}(z)=2\pi\nu_{21}(z)$ is the angular frequency of the transition, and the second line follows from substituting Eq.~\eqref{eq:nu21} and using $e^2=\alpha\hbar c$ once more. Normalizing to the present-day value $A_{10,0}\simeq2.85\times10^{-15}\,{\rm s^{-1}}$ and applying Eq.~\eqref{eq:mass_ratio} to both masses gives
\begin{equation}
A_{10}(z)
=
A_{10,0}
\left[\frac{m_e(z)}{m_{e,0}}\right]^4
\left[\frac{m_p(z)}{m_{p,0}}\right]^{-3}
=
A_{10,0}\,e^{\bfg\Delta\bar\chi(z)},
\label{eq:A10_z_app}
\end{equation}
showing that a larger electron mass increases $A_{10}(z)$ and shortens the lifetime of the triplet state, while a larger proton mass lowers it through the nuclear magneton.
\subsection{Kinetic Temperature Evolution}
\label{subsec:Tk}

The gas kinetic temperature of the IGM,
$T_k(\mathbf r,z)$, enters the spin temperature in
Eq.~\eqref{eq:Ts} through collisional coupling and through the
Ly$\alpha$ color temperature, which approaches $T_k$ under repeated
Ly$\alpha$ scattering. Its evolution follows from the first law of
thermodynamics and can be written on redshift slices as~\cite{Mesinger:2010ne,Pritchard:2011xb}\footnote{The gas kinetic temperature measures the mean kinetic energy per particle, $\tfrac{3}{2}k_BT_k(\mathbf r,z)=\langle\tfrac12 m_p(z)v^2\rangle$, where $v$ is the thermal velocity of a particle. The varying proton mass $m_p(z)$ therefore adds a term to Eq.~\eqref{eq_Tk_evolution}. We have checked that the effect of this term is small, and we neglect it in this work.}
\begin{align}
\frac{dT_k(\mathbf r,z)}{dz}
&=
\frac{2T_k(\mathbf r,z)}{1+z}
+
\frac{2T_k(\mathbf r,z)}
{3\left[1+\delta_\ssB^{\rm NL}(\mathbf r,z)\right]}
\frac{d\delta_\ssB^{\rm NL}(\mathbf r,z)}{dz}
-
\frac{T_k(\mathbf r,z)}
{1+f_{\rm He}+x_e^{\rm X}(\mathbf r,z)}
\frac{d x_e^{\rm X}(\mathbf r,z)}{dz}
\nonumber\\
&+
\frac{2}{3k_B\left[1+f_{\rm He}+x_e^{\rm X}(\mathbf r,z)\right]}
\frac{dt}{dz}
\left[
\epsilon_{\rm rad}(\mathbf r,t)
+
\epsilon_{\rm comp}(\mathbf r,t)
+
\epsilon_X(\mathbf r,t)
+
\epsilon_{\mathrm{Ly}\alpha}(\mathbf r,t)
\right],
\label{eq_Tk_evolution}
\end{align}
where $dt/dz=-[(1+z)H(z)]^{-1}$ and
\begin{equation}
f_{\rm He}
\equiv
\frac{\bar n_{\rm He}}{\bar n_H}
=
\frac{Y_p}{4(1-Y_p)}.
\label{eq_fHe}
\end{equation}
The quantity $x_e^{\rm X}(\mathbf r,z)$ denotes the local free-electron fraction per
hydrogen nucleus, including residual and X-ray-induced ionization. Its
evolution is discussed in Appendix~\ref{subsec:xe_evolution}.\footnote{Throughout this paper, $x_e$ and $f_{\rm He}$ are normalized per hydrogen nucleus, following the convention of \CLASS\ and \texttt{HyRec}~\cite{Ali_Ha_moud_2011,Lee_2020}. \FirstClass\ computes with $x_e$ normalized per baryon, $x_e=n_e/(n_H+n_{\rm He})$.}

The first term in Eq.~\eqref{eq_Tk_evolution} describes adiabatic
cooling due to cosmic expansion. The second term accounts for the
thermal response to local baryon compression or rarefaction. The third
term describes the change in the heat capacity of the gas as the number
of free electrons evolves.
The last line contains four radiative
contributions. 
The term $\epsilon_{\rm rad}(\mathbf{r},t)$ denotes the radiative heating rate of the gas associated with the CMB-driven 21\,cm hyperfine transition.
The term
$\epsilon_{\rm comp}$ accounts for Compton energy exchange through
Thomson scattering, while $\epsilon_X$ and
$\epsilon_{\mathrm{Ly}\alpha}$ describe heating by X-ray
photoelectrons and energy exchange through repeated Ly$\alpha$
scattering, respectively. 
For each channel
$\lambda\in\{{\rm rad},{\rm comp},X,\mathrm{Ly}\alpha\}$, we denote by
$\epsilon_\lambda(\mathbf r,t)$ the net energy injection rate per
hydrogen nucleus per unit time. Its redshift-space counterpart is
defined by
\begin{equation}
\tilde\epsilon_\lambda(\mathbf r,z)\,dz
\equiv
-
\epsilon_\lambda(\mathbf r,t)\,dt .
\label{eq_eps_redshift_def}
\end{equation}
Thus, $\epsilon_\lambda$ has units of energy per unit time per hydrogen
nucleus, while $\tilde\epsilon_\lambda$ has units of energy per hydrogen
nucleus.

The CMB-driven radiative heating rate associated with the 21\,cm
hyperfine transition is 
\begin{equation}
\epsilon_{\rm rad}(\mathbf r,t)
=
\frac{3}{4}\,
\frac{\bar T_\gamma(t)}{T_\star(t)}\,
A_{10}(t)\,
\frac{h^2}{m_H(t)\,\lambda_{21}^2(t)}
\left[
1+\frac{2T_k(\mathbf r,t)}{T_\star(t)}
\right],
\label{eq_eps_rad_time}
\end{equation}
where $m_H(t)\simeq m_p(t)$ is the hydrogen-atom mass and $\lambda_{21}(t)=c/\nu_{21}(t)$ is the rest-frame 21\,cm wavelength. Eq.~\eqref{eq_eps_rad_time} is the 21\,cm term of the heating rate in Ref.~\cite{Meiksin:2021cuh}. A CMB photon scattered in the 21\,cm line gives the atom a recoil. The recoil energy heats the gas. The factor $3A_{10}(t)\bar T_\gamma(t)/[4T_\star(t)]$ is the number of 21\,cm scatterings per hydrogen atom per unit time. This uses the Rayleigh--Jeans occupation number $\bar T_\gamma(t)/T_\star(t)$ of the CMB at $\nu_{21}(t)$. The factor $h^2/[m_H(t)\lambda_{21}^2(t)]=[h\nu_{21}(t)]^2/[m_H(t)c^2]$ is twice the recoil energy of one scattering. Both terms in the bracket are positive, so $\epsilon_{\rm rad}$ never cools the gas. Ref.~\cite{Venumadhav:2018uwn} computed the heating rate from the energy stored in the hyperfine level populations. That energy does not reach the gas once the populations are in steady state~\cite{Meiksin:2021cuh}. In the axiodilaton cosmology, the explicit mass dependence of the prefactor is
\begin{equation}
\frac{\bar T_\gamma(z)\,A_{10}(z)\,h^2}{T_\star(z)\,m_H(z)\,\lambda_{21}^2(z)}
=
\frac{\bar T_\gamma(z)\,A_{10,0}\,h^2}{T_{\star,0}\,m_{p,0}\,\lambda_{21,0}^2}
\left[
\frac{m_e(z)}{m_{e,0}}
\right]^{6}
\left[
\frac{m_p(z)}{m_{p,0}}
\right]^{-5}
=
\frac{\bar T_\gamma(z)\,A_{10,0}\,h^2}{T_{\star,0}\,m_{p,0}\,\lambda_{21,0}^2}\,
e^{\bfg\Delta\bar\chi(z)},
\label{eq_Tstar_A10_mass_scaling}
\end{equation}
where $T_{\star,0}$, $A_{10,0}$, and $\lambda_{21,0}$ denote the present-day values. Additional model dependence enters through the evolved $T_k$.

The Compton contribution from Thomson scattering between free
electrons and CMB photons is
\begin{equation}
\epsilon_{\rm comp}(\mathbf r,t)
=
\frac{3}{2}k_B x_e^{\rm X}(\mathbf r,t)
\frac{
\bar T_\gamma(t)-T_k(\mathbf r,t)
}{
t_\gamma(t)
},
\qquad
t_\gamma(t)
=
\frac{3m_e(t)c}
{8\sigma_T(t)a_R\bar T_\gamma^4(t)} .
\label{eq_eps_comp_time}
\end{equation}
The sign of $\epsilon_{\rm comp}$ is determined by
$\bar T_\gamma-T_k$. Compton scattering heats the gas when the CMB is
hotter and cools it when the gas is hotter.

The Thomson cross section is
\begin{equation}
\sigma_T(z)
=
\frac{8\pi}{3}r_e^2(z),
\qquad
r_e(z)
=
\frac{\alpha\hbar}{m_e(z)c}.
\label{eq_sigmaT_mass}
\end{equation}
In the axiodilaton cosmology it therefore
scales as
\begin{equation}
\sigma_T(z)
=
\sigma_{T,0}
\left[
\frac{m_e(z)}{m_{e,0}}
\right]^{-2}
=
\sigma_{T,0}
\exp\left[-2\bfg\Delta\bar\chi(z)\right],
\label{eq_sigmaT_mass_scaling}
\end{equation}
where $\sigma_{T,0}$ denotes the present-day value of the Thomson cross section.
The radiation constant
\begin{equation}
a_R
=
\frac{\pi^2k_B^4}{15\hbar^3c^3}
\end{equation}
has no dependence on $m_e(z)$ or $m_p(z)$. It follows that
\begin{equation}
t_\gamma(z)
=
t_{\gamma,0}(z)
\left[
\frac{m_e(z)}{m_{e,0}}
\right]^3
=
t_{\gamma,0}(z)
\exp\left[3\bfg\Delta\bar\chi(z)\right],
\label{eq_tgamma_mass_scaling}
\end{equation}
where
\begin{equation}
t_{\gamma,0}(z)
\equiv
\frac{3m_{e,0}c}
{8\sigma_{T,0}a_R\bar T_\gamma^4(z)}
\end{equation}
denotes the reference Compton timescale evaluated using the present-day
electron mass and Thomson cross section.
A smaller electron mass shortens the Compton timescale and strengthens
the thermal coupling between the gas and the CMB, while a larger
electron mass weakens this coupling. The Compton contribution is also
modified through the mass-dependent residual ionization history
$x_e^{\rm X}$, the expansion rate $H(z)$, and the resulting evolution
of $T_k$.

The term $\epsilon_{\mathrm{Ly}\alpha}$ describes the net energy
exchange between Ly$\alpha$ photons and the gas through repeated resonant
scattering. Its detailed calculation and the corresponding modifications
in the axiodilaton cosmology are presented in
Appendix~\ref{subsubsec:lya_heating}. The X-ray heating term $\tilde\epsilon_X$
becomes important after the first astrophysical sources form, raising the
gas temperature and driving its subsequent thermal evolution. This term
is described in Appendix~\ref{subsec:xray}.

\begin{figure}[t]
    \centering
    \includegraphics[width=1.0\linewidth]{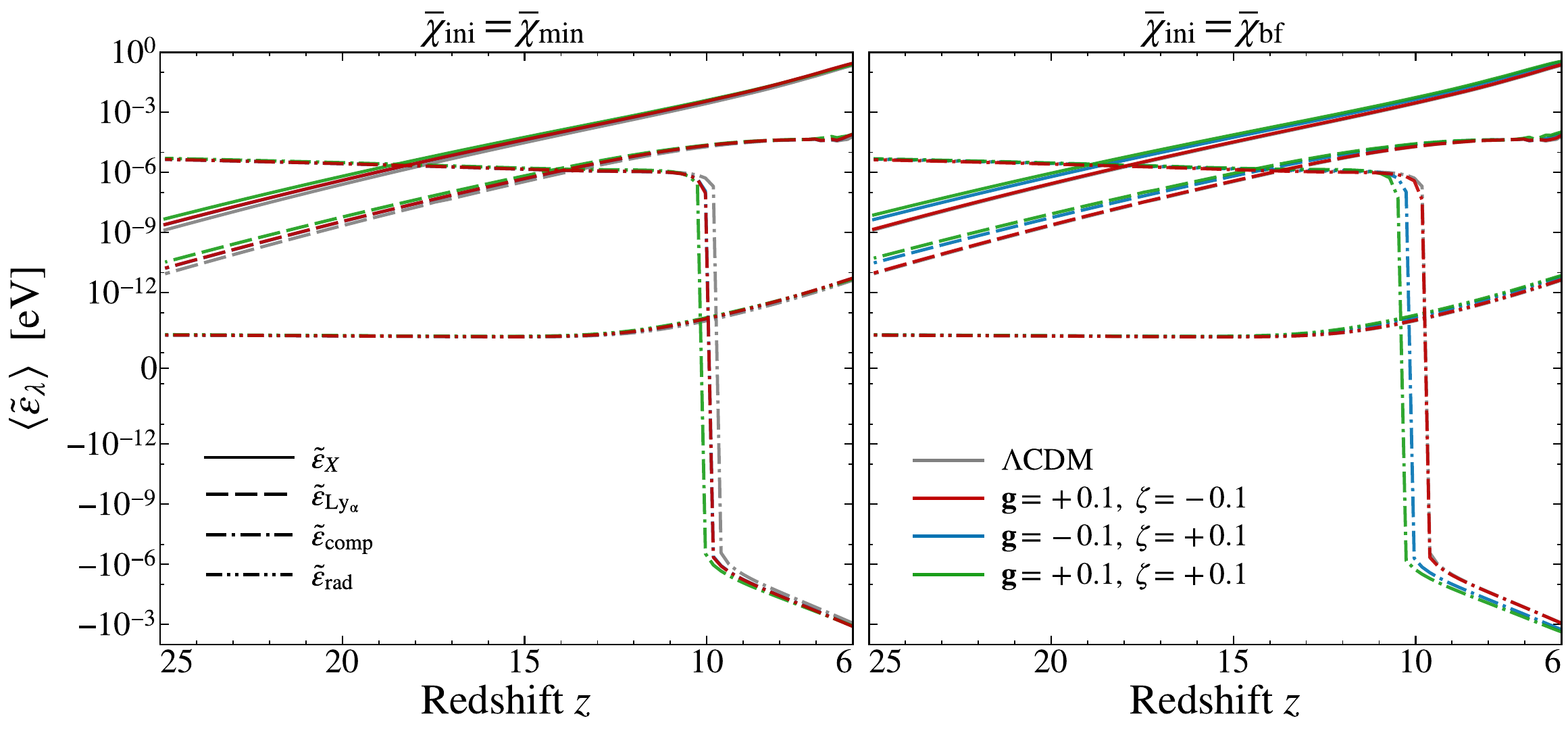}
    \caption{Volume-averaged energy-exchange terms $\langle\widetilde{\epsilon}_\lambda\rangle$, in units of energy per hydrogen nucleus, contributing to the gas kinetic-temperature evolution in Eq.~\eqref{eq_Tk_evolution}. The left and right panels correspond to $\bar\chi_{\rm ini} = \bar\chi_{\rm min} = 74.010$ (left) and $\bar\chi_{\rm ini} = \bar\chi_{\rm bf}=74.213$ (right), respectively. The solid, dashed, dash-dotted, and dash-dot-dotted curves show the X-ray heating $\widetilde{\epsilon}_{X}$, Ly$\alpha$ heating $\widetilde{\epsilon}_{\mathrm{Ly}\alpha}$, Compton energy exchange $\widetilde{\epsilon}_{\rm comp}$, and CMB-driven radiative heating $\widetilde{\epsilon}_{\rm rad}$, respectively. Gray denotes \LCDM, while red, blue, and green correspond to $(\bfg,\zeta)=(+0.1,-0.1)$, $(-0.1,+0.1)$, and $(+0.1,+0.1)$. Positive and negative values represent net heating and cooling, respectively. X-ray heating becomes the dominant contribution after the formation of the first sources, while the Compton term changes sign when the gas becomes hotter than the CMB.}
    \label{fig:heat_terms}
\end{figure}

For convenience, we summarize the volume-averaged evolution of
the four energy-exchange terms in Fig.~\ref{fig:heat_terms}. 
At $z\gtrsim20$, Compton scattering provides the largest positive
contribution because the gas remains colder than the CMB. As the first
astrophysical sources develop, $\tilde\epsilon_{X}$ increases rapidly,
overtakes the other terms around $z\sim18$, and subsequently dominates
the thermal evolution. The Ly$\alpha$ contribution $\tilde\epsilon_{\mathrm{Ly}\alpha}$ also grows with the
radiation background but remains several orders of magnitude smaller
than the X-ray term, while $\tilde\epsilon_{\rm rad}$ is negligible throughout
the redshift range shown. Once X-ray heating raises the gas temperature
above the CMB temperature near $z\sim10$, the Compton term $\tilde\epsilon_{\rm comp}$ changes sign
and begins to cool the gas. 
For
$\bar\chi_{\rm ini}=\chimin$, the green branch exhibits the earliest
enhancement of the source-dependent heating terms, while the remaining
branches stay close to \LCDM. For $\bar\chi_{\rm ini}=\chibf$, the green
branch again develops the earliest and strongest heating, followed by
the blue branch, whereas the red branch remains closest to \LCDM.

The middle panels of Fig.~\ref{fig:x_Ts_Tb} show the volume-averaged
gas kinetic temperature $\langle T_k(\mathbf r,z)\rangle$. Before the
onset of significant X-ray heating, all axiodilaton predictions remain
close to the \LCDM\ evolution, with the gas cooling predominantly through
adiabatic expansion. 
The direct modifications to Compton coupling, as shown in
Fig.~\ref{fig:heat_terms}, and to the other thermal rates induced by the
varying SM masses therefore have only a small effect on the mean gas
temperature over this redshift range. The thermal histories begin to separate as the X-ray background develops.
For $\bar\chi_{\rm ini}=\chimin$, the green branch exhibits the earliest
temperature increase, while the red and blue branches remain closer to
the \LCDM\ evolution. For $\bar\chi_{\rm ini}=\chibf$, the temperature
rises earliest in the green branch, followed by the blue branch, whereas
the red branch closely tracks \LCDM. This ordering reflects the
model-dependent growth of structure, which changes the abundance of
X-ray-emitting sources and the timing of X-ray heating.

\subsubsection{X-ray Heating}
\label{subsec:xray}

The X-ray heating term $\widetilde\epsilon_X$ in
Eq.~\eqref{eq_Tk_evolution} describes the energy deposition per
hydrogen nucleus into the IGM by X-ray photons produced by early radiative
sources. 
We denote by $z$ the redshift at which the X-ray photons are absorbed and
the gas is heated, and by $z'$ the redshift at which the photons are
emitted, with $z'\geq z$. Following the shell-based formalism implemented
in \FirstClass, the redshift-space X-ray heating term at the target cell
$(\mathbf r,z)$ is obtained by integrating over the photon frequency and
the source redshift,
\begin{align}
\widetilde\epsilon_X(\mathbf r,z)
={}&
\int_{\nu_X}^{\infty}d\nu
\sum_{i\in\{{\rm HI},{\rm HeI},{\rm HeII}\}}
\left[h\nu-E_i^{\rm th}(z)\right]
f_{\rm heat}
\left(
h\nu-E_i^{\rm th}(z),
x_e^{\rm X}(\mathbf r,z)
\right)
\nonumber\\
&\times
x_i^{\rm X}(\mathbf r,z)
\sigma_i(\nu,z)
\widetilde{\mathcal J}_X(\mathbf r,\nu,z).
\label{eq_eps_X_shell}
\end{align}
Here, $E_i^{\rm th}(z)$ is the ionization threshold energy of species $i$,
and $h\nu-E_i^{\rm th}(z)$ is the kinetic energy of the primary
photoelectron produced by photoionization. The dimensionless function
$f_{\rm heat}$ gives the fraction of this energy that is deposited as heat,
evaluated at the local ionized fraction
$x_e^{\rm X}(\mathbf r,z)$. The quantity $\sigma_i(\nu,z)$ is the
photoionization cross section of species $i$ for a photon with frequency
$\nu$ at the absorption redshift.
The quantity $x_i^{\rm X}(\mathbf r,z)$ denotes the local abundance of the
absorbing species $i$ per hydrogen nucleus,
\begin{align}
x_{\rm HI}^{\rm X}(\mathbf r,z)
&=
1-x_e^{\rm X}(\mathbf r,z),
\nonumber\\
x_{\rm HeI}^{\rm X}(\mathbf r,z)
&=
f_{\rm He}
\left[1-x_e^{\rm X}(\mathbf r,z)\right],
\nonumber\\
x_{\rm HeII}^{\rm X}(\mathbf r,z)
&=
f_{\rm He}x_e^{\rm X}(\mathbf r,z),
\label{eq_xabs_approx}
\end{align}
where $f_{\rm He}$ is the mean helium
abundance relative to hydrogen, Eq.~\eqref{eq_fHe}.

The varying SM masses directly modify the ionization thresholds and
photoionization cross sections appearing in
Eq.~\eqref{eq_eps_X_shell}. To make this dependence explicit, consider the
ground-state photoionization cross section of a hydrogenic ion with nuclear
charge $Z$~\cite{Mo2010,Osterbrock1989},
\begin{equation}
\sigma_{\rm pi}(Z,\nu,z)
=
\frac{\xi(z)}{Z^2}
\left[
\frac{\nu^{\rm th}(Z,z)}{\nu}
\right]^4
\frac{
\exp\left[
4-\dfrac{4\tan^{-1}\varepsilon}{\varepsilon}
\right]
}{
1-\exp\left(-\dfrac{2\pi}{\varepsilon}\right)
},
\qquad
\varepsilon
=
\sqrt{
\frac{\nu}{\nu^{\rm th}(Z,z)}-1
},
\label{eq_sigma_hydrogenic}
\end{equation}
where $\nu\geq\nu^{\rm th}(Z,z)$. The threshold frequency is defined by
\begin{equation}
E^{\rm th}(Z,z)
=
\frac{1}{2}\mu_Z(z)c^2\alpha^2 Z^2
\equiv
h\nu^{\rm th}(Z,z),
\qquad
\mu_Z(z)
=
\frac{m_e(z)m_N(z)}
{m_e(z)+m_N(z)} \simeq m_e(z),
\label{eq_Eth_def}
\end{equation}
where $m_N(z)$ is the nuclear
mass. The normalization of the cross section is
\begin{equation}
\xi(z)
=
\frac{2^9\pi}{3{\rm e}^4}\,
\alpha\pi a_0^2(z),
\label{eq_xi_def}
\end{equation}
where ${\rm e}$ denotes Euler's number and $a_0(z)$ is the Bohr radius defined in Appendix~\ref{app:Tb_minimal}.

In the axiodilaton cosmology, all non-relativistic SM masses undergo the
same rescaling given in Eq.~\eqref{eq:mass_ratio}. The reduced masses satisfy
$\mu_Z(z)/\mu_{Z,0}={m_e(z)}/{m_{e,0}} = e^{{\mathbf g}\,\Delta\bar\chi(z)}$. The relevant atomic quantities therefore obey
\begin{equation}
\frac{E_i^{\rm th}(z)}{E_{i,0}^{\rm th}}
=
\frac{\nu_i^{\rm th}(z)}{\nu_{i,0}^{\rm th}}
=
e^{{\mathbf g}\,\Delta\bar\chi(z)},
\qquad
\frac{a_0(z)}{a_{0,0}}
=
e^{{-\mathbf g}\,\Delta\bar\chi(z)},
\qquad
\frac{\xi(z)}{\xi_0}
=
e^{{-2\mathbf g}\,\Delta\bar\chi(z)}.
\label{eq_X_atomic_scalings}
\end{equation}
We apply the atomic scalings derived in
Eq.~\eqref{eq_X_atomic_scalings} to the species-dependent
photoionization cross sections implemented in \FirstClass. The cross
sections in the axiodilaton cosmology are evaluated as
\begin{equation}
\sigma_i(\nu,z)
=
e^{-2{\mathbf g}\,\Delta\bar\chi(z)}\,
\sigma_i^{\rm fid}
\!\left(
\frac{\nu}{e^{{\mathbf g}\,\Delta\bar\chi(z)}}
\right),
\qquad
i\in\{{\rm HI},{\rm HeI},{\rm HeII}\},
\label{eq:sigma_rescale_nu}
\end{equation}
where $\sigma_i^{\rm fid}(\nu)$ denotes the corresponding cross section
evaluated using the present-day SM masses.

The dependence of the heating fraction
$f_{\rm heat}\!\left(h\nu-E_i^{\rm th}(z),\,x_e^{\rm X}(\mathbf r,z)\right)$
on the varying electron mass is presented in
Appendix~\ref{app:fheat_scaling}. The estimate shows that the explicit
$m_e$ dependence largely cancels among the dominant energy-loss channels,
up to logarithmic corrections and effects near atomic thresholds.
Accordingly, we use the tabulated heating fractions implemented in
\FirstClass, following Refs.~\cite{Lopez-Honorez:2020lno, Furlanetto:2009uf}, while evaluating
the primary photoelectron energy using the mass-dependent threshold
$E_i^{\rm th}(z)$.

In Eq.~\eqref{eq_eps_X_shell},
$\widetilde{\mathcal J}_X$ is the redshift-space spectral X-ray photon field,
\begin{equation}
\widetilde{\mathcal J}_X(\mathbf r,\nu,z)
\equiv
\int_z^\infty dz'\,
\frac{1}{4\pi r_p^2(z,z')}
\frac{
d\widetilde\phi_X(\mathbf r,\nu,z,z')
}{dz'}.
\label{eq_JX_def}
\end{equation}
It has units of ${\rm cm}^{-2}\,{\rm Hz}^{-1}$ and represents the X-ray
photon fluence arriving at the target cell from all source shells. The
proper radial separation between the absorption redshift $z$ and the
emission redshift $z'$ is
\begin{equation}
r_p(z,z')
=
\frac{c}{1+z}
\int_z^{z'}
\frac{d\bar z}{H(\bar z)}.
\label{eq_rp_def}
\end{equation}
The modified expansion history in the axiodilaton cosmology therefore
changes both the propagation distance and the geometric dilution of the
X-ray photon field.

The redshift-space X-ray photon number arriving from sources in the shell
$[z',z'+dz']$ is modeled as
\begin{align}
\frac{
d\widetilde\phi_X(\mathbf r,\nu,z,z')
}{dz'}
={}
\left[-\frac{dt}{dz}\right]_z
\frac{d\dot N_X(\mathbf r,z,z')}{dz'}
\alpha_X\nu_X^{-1}
\left(\frac{\nu}{\nu_X}\right)^{-\alpha_X-1}
\left(\frac{1+z'}{1+z}\right)^{-\alpha_X-1}
e^{-\tau_X(\nu,z,z')},
\label{eq_dphidxdz_shell}
\end{align}
where the factor
$\alpha_X\nu_X^{-1}(\nu/\nu_X)^{-\alpha_X-1}$ is the normalized photon
number spectrum of the X-ray sources, with $\alpha_X=1.0$ and
$h\nu_X=500\,{\rm eV}$ adopted in our analysis. The factor involving
$(1+z')/(1+z)$ accounts for the cosmological redshifting of the photon
frequency between emission and absorption, while $e^{-\tau_X}$ accounts
for absorption during propagation through the IGM. \FirstClass\ replaces $e^{-\tau_X}$ by a step function, as in Ref.~\cite{Mesinger:2010ne}. Photons with $\tau_X(\nu,z,z')>1$ are fully absorbed, otherwise photons are not absorbed. 

The X-ray optical depth accumulated between $z'$ and $z$ is
\begin{equation}
\tau_X(\nu,z,z')
=
\int_z^{z'}d\tilde z\,
\frac{c\,\bar n_H(\tilde z)}
{(1+\tilde z)H(\tilde z)}
\sigma_{\rm eff}(\tilde\nu,\tilde z),
\label{eq_tauX_meanIGM}
\end{equation}
where $\tilde\nu=\nu(1+\tilde z)/(1+z)$ is the photon frequency at the
intermediate redshift $\tilde z$.
The effective photoionization cross section per hydrogen nucleus is
\begin{equation}
\sigma_{\rm eff}(\tilde\nu,\tilde z)
=
\sum_{i\in\{{\rm HI},{\rm HeI},{\rm HeII}\}}
\left\langle
x_i^{\rm X}(\mathbf r,\tilde z)
\right\rangle
\sigma_i(\tilde\nu,\tilde z),
\label{eq_sigma_eff}
\end{equation}
where the angle brackets denote a volume average.
The axiodilaton dependence of the optical depth arises from the modified
expansion rate $H(\tilde z)$, the volume-averaged abundances of HI, HeI,
and HeII, and the mass-dependent photoionization cross sections given in
Eq.~\eqref{eq:sigma_rescale_nu}.

The total X-ray photon emission rate from sources in the shell is written
as
\begin{align}
\frac{d\dot N_X(\mathbf r,z,z')}{dz'}
={}
f_*^{\rm II}\zeta_X
\Bigg\{
\Omega_{\ssB,0}\bar\rho_{\rm crit,0}
\left[
1+\delta_{\ssB,R_{\rm sh}}^{\rm NL}(\mathbf r,z')
\right]
\frac{dV(z,z')}{dz'}
\Bigg\}
\dfrac{H(z')}{t_\ast}\,f_{\rm coll,*}^{\rm eff}(\mathbf r,z',R_{\rm sh})
\label{eq_dNxdz_source}
\end{align}
Here, $f_*^{\rm II}$ is the Pop II star-formation efficiency. The X-ray
photon yield per unit stellar mass is defined as
\begin{equation}
\zeta_X
\equiv
\frac{L_X^{\rm II}/{\rm SFR}}{h\nu_X\,\ln\left(2\,{\rm keV}/h\nu_X\right)},
\label{eq:zeta_X}
\end{equation}
where $L_X^{\rm II}/{\rm SFR}$ is the X-ray luminosity per unit
star-formation rate in the band $h\nu_X\le h\nu\le2\,{\rm keV}$~\cite{Park:2018ljd}. With $\alpha_X=1$, the logarithm makes the X-ray energy emitted in this band per unit stellar mass equal to $L_X^{\rm II}/{\rm SFR}$.
We adopt the fixed values of $f_*^{\rm II}$ and
$L_X^{\rm II}/{\rm SFR}$ specified in
Eq.~\eqref{eq:astro_params_schematic}.
The quantity $f_{\rm coll,*}^{\rm eff}$ is the effective collapsed
fraction defined in Eq.~\eqref{eq_fcoll_star}. The product $f_*^{\rm II}f_{\rm coll,*}^{\rm eff}H(z')/t_\ast$ is the fraction of baryons converted into stars per unit time~\cite{Park:2018ljd}. Here $t_\ast$ is the star-formation timescale as a fraction of the Hubble time, Eq.~\eqref{eq:astro_params_schematic}. The quantity
$\delta_{\ssB,R_{\rm sh}}^{\rm NL}$ is the nonlinear baryon overdensity
smoothed on the source-shell scale.
The comoving source-shell radius and differential volume element are given by
\begin{equation}
R_{\rm sh}(z,z')
=
c\int_z^{z'}\frac{d\bar z}{H(\bar z)},
\qquad
\frac{dV(z,z')}{dz'}
=
4\pi R_{\rm sh}^2(z,z')\frac{c}{H(z')}.
\label{eq:source_shell_geometry}
\end{equation}
Therefore, the combination
$\Omega_{\ssB,0}\bar\rho_{\rm crit,0}
[1+\delta_{\ssB,R_{\rm sh}}^{\rm NL}]\,dV/dz'$
gives the baryonic mass assigned to the source shell.

As summarized in Fig.~\ref{fig:heat_terms}, the volume-averaged X-ray
heating term $\widetilde\epsilon_X$ increases rapidly after the first
radiative sources form and subsequently dominates the gas thermal evolution.
Its dependence on the axiodilaton cosmology is discussed in the main text
accompanying Fig.~\ref{fig:heat_terms}.

\subsubsection{Parametric Estimate of the Heating Fraction $f_{\rm heat}$}
\label{app:fheat_scaling}

In this subsection, we give a parametric estimate of the heating fraction
$f_{\rm heat}(E,x_e)$ that appears in the X-ray heating rate in Eq.~\eqref{eq_eps_X_shell},
with particular emphasis on its dependence on the varying electron mass in axiodilaton cosmology. 
The heating fraction is defined as the fraction of the initial kinetic energy of a primary electron that is ultimately deposited as heat. 
In general, for a primary electron of kinetic energy $E$ in a medium with $x_e$, one may write the heating fraction as the ratio of the electron--electron energy loss rate to the total energy loss rate~\cite{Furlanetto:2009uf}
\begin{equation}
f_{\rm heat}(E,x_e)\ \sim\ 
\frac{\left|dE/dt\right|_{ee}}
{\left|dE/dt\right|_{ee}+\left|dE/dt\right|_{\rm ion}+\left|dE/dt\right|_{\rm exc}}\ ,
\label{eq:fheat_rate_def}
\end{equation}
where $|dE/dt|_{ee}$ denotes the Coulomb stopping power due to scattering off ambient electrons (which thermalizes as heat), while $|dE/dt|_{\rm ion}$ and $|dE/dt|_{\rm exc}$ are the energy loss rates associated with collisional ionization and excitation of H/He respectively.
For electron--electron interactions, the non-relativistic Coulomb stopping power can be written in the form of
\begin{equation}
\left|\frac{dE}{dt}\right|_{ee}
\ =\ 
\frac{4 \pi n_e\,e^4}{m_e\,v_e}\, \ln\!\left[
\frac{2E}{\hbar}
\left(\frac{4\pi n_e e^2}{m_e}\right)^{-1/2}
\right] ,
\label{eq:de_dt_ee_start}
\end{equation}
where $n_e$ is the ambient electron density and $v_e$ is the speed of the fast electron. Using $v_e=\sqrt{2E/m_e}$ and expanding the logarithmic factor, we find
\begin{equation}
\left|\frac{dE}{dt}\right|_{ee}
\ \propto\ 
m_e^{-1/2}\,E^{-1/2}\ \Big[ \ln E
+\frac{1}{2}\ln m_e
+\cdots \Big] \;,
\label{eq:de_dt_ee_scaling}
\end{equation}
where the ellipsis denotes terms that are independent of $E$ and $m_e$.

The energy loss rate associated with collisional ionization can be estimated as
\begin{equation}
\left|\frac{dE}{dt}\right|_{\rm ion}
\ =\ 
n_{\rm target}\,v_e\,\sigma_{\rm ion}(E)\,
\langle \Delta E_{\rm ion}\rangle\ ,
\label{eq:de_dt_ion_general}
\end{equation}
where $n_{\rm target}$ is the target density (e.g.\ $n_{\rm HI}$, $n_{\rm HeI}$, $n_{\rm HeII}$),
and $\langle \Delta E_{\rm ion}\rangle$ is the mean energy removed from the primary electron per ionization event.
For a parametric estimate, we take $\langle \Delta E_{\rm ion}\rangle \sim E_i^{\rm th}$, which captures the mandatory energy cost of removing a bound electron and is most appropriate when $E$ is not far above threshold. Since $E_i^{\rm th}\propto m_e$, this implies $\langle \Delta E_{\rm ion}\rangle \propto m_e$.
At sufficiently high energy, the ionization cross section can be approximated by a Bethe-like form
\begin{equation}
\sigma_{\rm ion}(E)
\ \sim\ 
\pi a_0^2\,
\frac{E_i^{\rm th}}{E}\,
\Big[
A\,\ln\!\Big(\frac{E}{E_i^{\rm th}}\Big)+B
\Big]\ ,
\label{eq:sigma_ion_bethe_form}
\end{equation}
where $A, B$ are dimensionless coefficients.
Using $a_0^2 \propto m_e^{-2}$ and $E_i^{\rm th} \propto m_e$, we obtain the parametric scaling
\begin{equation}
\left|\frac{dE}{dt}\right|_{\rm ion}
\ \propto\ 
m_e^{-1/2}\,E^{-1/2}\,
\Big[
\ln E - \ln m_e+\cdots
\Big].
\label{eq:de_dt_ion_scaling}
\end{equation}

A similar scaling holds for collisional excitation.
The energy loss rate associated with collisional excitation can be estimated as
\begin{equation}
\left|\frac{dE}{dt}\right|_{\rm exc}\ \sim\ n_{\rm target}\,v_e\,\sigma_{\rm exc}(E)\,\langle\Delta E_{\rm exc}\rangle\;.
\end{equation}
The collisional excitation cross sections of H/He are modeled in Ref.~\cite{Stone2002}. Parametrically, in the Born/Bethe regime, one expects $\sigma_{\rm exc}(E)\propto a_0^2\,(E_i^{\rm exc}/E)$, where $E_i^{\rm exc}\propto m_e$ is the excitation energy of species $i$, up to slowly varying logarithmic factors and transition-dependent dimensionless coefficients.
Since $\langle\Delta E_{\rm exc}\rangle\propto m_e$ and $a_0^2\propto m_e^{-2}$, combining these scalings yields
\begin{equation}
\left|\frac{dE}{dt}\right|_{\rm exc}
\ \propto\
m_e^{-1/2}\,E^{-1/2}\,
\Big[\ln E-\ln m_e+\cdots\Big]\;.
\label{eq:de_dt_exc_scaling}
\end{equation}
Therefore, $\left|dE/dt\right|_{ee}$, $\left|dE/dt\right|_{\rm ion}$, and $\left|dE/dt\right|_{\rm exc}$ share the same leading prefactor $m_e^{-1/2}E^{-1/2}$ (up to logarithmic corrections), so the explicit $m_e$-dependence largely cancels in the ratio defining $f_{\rm heat}(E,x_e)$, except near thresholds where kinematic opening/closing of channels can induce a stronger sensitivity. The remaining logarithmic dependence on $m_e$ is relatively weak. 
We note that a quantitatively accurate determination of $f_{\rm heat}(E,x_e)$ requires a numerical evaluation of the full energy deposition cascade, rather than a purely parametric estimate. 
A dedicated analysis of this kind is beyond the scope of this work.
Accordingly, in our implementation we use the tabulated heating fractions provided in  \FirstClass, following Refs.~\cite{Furlanetto:2009uf,Lopez-Honorez:2020lno}.

\subsubsection{Effective Collapsed Fraction}
\label{app:f_coll}

The effective collapsed fraction represents the fraction of matter that has collapsed into halos capable of hosting radiative sources within a region of comoving radius $R$ centered at $\mathbf r$,
\begin{equation}
\begin{aligned}
f_{\rm coll}^{\rm eff}(\mathbf r,z,R)
&\equiv
\int_{M_{\min}}^{M_R} dM_h\,
\frac{M_h}{\bar\rho_m(0)}
\frac{dn_{\rm cond}}{dM_h}
\left(\mathbf r,z,M_h,R\right)
\left(
\frac{M_h}{10^{10}M_\odot}
\right)^{\alpha_*^{\rm II}+\alpha_{\rm esc}^{\rm II}}
\exp\left(-\frac{M_{\rm turn}}{M_h}\right).
\end{aligned}
\label{eq_fcoll_eff_definition}
\end{equation}
Here, $dn_{\rm cond}/dM_h$ is the conditional halo mass function described in Appendix~\ref{app:conditional_halo_mass_function}.
The parameters $\alpha_*^{\rm II}$ and $\alpha_{\rm esc}^{\rm II}$
specify the halo-mass dependence of the Pop II star-formation efficiency
and ionizing escape fraction, respectively. The turnover mass
$M_{\rm turn}$ controls the exponential suppression of star formation in
low-mass halos, and the lower integration limit is set to
$M_{\min}=M_{\rm turn}/50$. 
We adopt these astrophysical parameter values specified in
Eq.~\eqref{eq:astro_params_schematic}.
We also define the effective collapsed fraction with the
escape-fraction weighting removed as
\begin{equation}
f_{\rm coll,*}^{\rm eff}(\mathbf r,z,R)
\equiv
\left.
f_{\rm coll}^{\rm eff}(\mathbf r,z,R)
\right|_{\alpha_{\rm esc}^{\rm II}=0}.
\label{eq_fcoll_star}
\end{equation}

\begin{figure}[t!]
    \centering
    \includegraphics[width=0.9\linewidth]
    {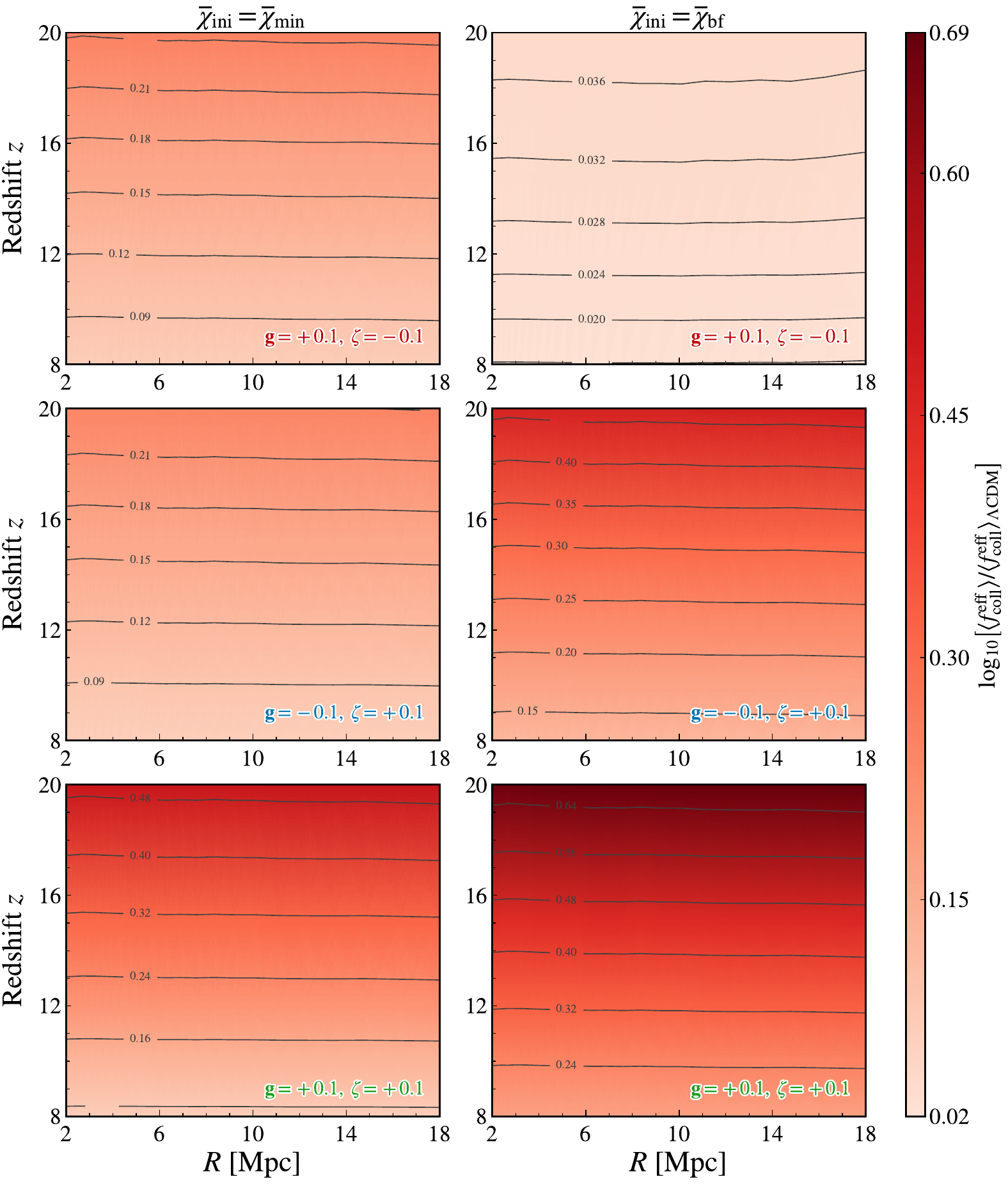}
    \caption{Ratio of the volume-averaged effective collapsed fraction $\langle f_{\rm coll}^{\rm eff}\rangle$ in the axiodilaton cosmology to its \LCDM\ counterpart as a function of redshift $z$ and comoving radius $R$. The left and right columns correspond to $\bar\chi_{\rm ini} = \bar\chi_{\rm min} = 74.010$ (left) and $\bar\chi_{\rm ini} = \bar\chi_{\rm bf}=74.213$ (right), respectively. The top, middle, and bottom rows show $(\bfg,\zeta)=(+0.1,-0.1)$, $(-0.1,+0.1)$, and $(+0.1,+0.1)$, respectively. The effective collapsed fraction is enhanced in all benchmarks, with a stronger enhancement at higher redshift and a comparatively weak dependence on $R$.}

    \label{fig:fcoll_eff}
\end{figure}

Fig.~\ref{fig:fcoll_eff} shows that the axiodilaton models increase the effective collapsed fraction throughout the plotted range, where $\langle f_{\rm coll}^{\rm eff}\rangle$ denotes the volume average over the simulation box. The enhancement becomes larger toward higher redshift. At these redshifts, source-hosting halos occupy the rare-peak tail of the halo mass function, whose abundance depends exponentially on the matter variance. A modest enhancement of the matter perturbations can therefore produce a considerably larger fractional increase in the abundance of these halos. The nearly horizontal contours show that the ratio varies much more strongly
with redshift than with the comoving length scale $R$.

For $\bar\chi_{\rm ini}=\chimin$, the two benchmarks with
$\bfg\zeta<0$ give nearly identical ratios, and the benchmark with
$\bfg\zeta>0$ produces a larger enhancement. For
$\bar\chi_{\rm ini}=\chibf$, the $(\bfg,\zeta)=(+0.1,-0.1)$ benchmark
remains close to \LCDM, followed by progressively larger enhancements
for $(-0.1,+0.1)$ and $(+0.1,+0.1)$. This ordering follows the
model-dependent matter growth shown in
Fig.~\ref{fig:delta_evolution}.

\subsubsection{Conditional Halo Mass Function}
\label{app:conditional_halo_mass_function}

In this section, we summarize the conditional halo mass function
$dn_{\rm cond}/dM_h(\mathbf r,z,M_h,R)$. It describes the comoving
number density of halos in the mass interval $[M_h,M_h+dM_h]$ within a
spherical region of radius $R$ centered at $\mathbf r$.

We begin with the extended Press--Schechter conditional halo mass
function~\cite{Lacey:1994su,Somerville:1997df}
\begin{equation}
\frac{dn_{\rm cond}^{\rm EPS}}{dM_h}
\left(\mathbf r,z,M_h,R\right)
=
\frac{\bar\rho_m(0)}{M_h}
f\left(\mathbf r,z,M_h,R\right)
\left|
\frac{d\sigma^2(M_h,z)}{dM_h}
\right|,
\label{eq:eps_conditional_hmf}
\end{equation}
where $f(\mathbf r,z,M_h,R)$ is the conditional first-crossing
distribution per unit variance,
\begin{equation}
f\left(\mathbf r,z,M_h,R\right)
=
\frac{1}{\sqrt{2\pi}}
\frac{\delta_c-\delta_{m,R}(\mathbf r,z)}
{\left[\sigma^2(M_h,z)-\sigma^2(M_R,z)\right]^{3/2}}
\exp\left[
-\frac{
\left[\delta_c-\delta_{m,R}(\mathbf r,z)\right]^2
}{
2\left[\sigma^2(M_h,z)-\sigma^2(M_R,z)\right]
}
\right].
\label{eq:conditional_function}
\end{equation}
The total matter mass associated with the smoothing region is
$M_R=(4\pi/3)R^3\bar\rho_m(0)$. Likewise, a halo of mass $M_h$ is
associated with the comoving radius $R_h$ through
$M_h=(4\pi/3)R_h^3\bar\rho_m(0)$. We restrict the halo masses considered to $M_h\leq M_R$, or equivalently
$R_h\leq R$, which ensures that
$\sigma^2(M_h,z)-\sigma^2(M_R,z)\geq0$. The variance
$\sigma^2(M,z)$ is defined in Eq.~\eqref{eq_sigma_R}. The smoothed linear
matter overdensity $\delta_{m,R}(\mathbf r,z)$ is defined by
\begin{equation}
\delta_{m,R}(\mathbf r,z)
=
\int d^3\mathbf y\,
W_R(|\mathbf r-\mathbf y|)
\delta_m(\mathbf y,z),
\label{eq_delta_smoothed_top_hat}
\end{equation}
where $W_R$ is the top-hat window function defined in
Eq.~\eqref{eq_tophat_window}.
The critical
linear overdensity for spherical collapse is fixed at
$\delta_c=1.686$.

Since the Sheth--Tormen mass function provides a better fit to
numerical \LCDM\ simulations, we rescale the extended Press--Schechter
conditional mass function as~\cite{Park:2018ljd,Mesinger:2010ne}
\begin{equation}
\frac{dn_{\rm cond}}{dM_h}
\left(\mathbf r,z,M_h,R\right)
=
\frac{
\displaystyle \frac{dn^{\rm ST}}{dM_h}(M_h,z)
}{
\displaystyle \frac{dn^{\rm PS}}{dM_h}(M_h,z)
}
\frac{dn_{\rm cond}^{\rm EPS}}{dM_h}
\left(\mathbf r,z,M_h,R\right).
\label{eq:renormalized_conditional_hmf}
\end{equation}
The Press--Schechter and Sheth--Tormen halo mass functions are given
by~\cite{Sheth:1999su,Jenkins:2000bv}
\begin{equation}
\frac{dn^{\rm PS}}{dM_h}(M_h,z)
=
\frac{\bar\rho_m(0)}{M_h}
\frac{1}{\sqrt{2\pi}}
\frac{\delta_c}{\sigma^3(M_h,z)}
\exp\left[
-\frac{\delta_c^2}{2\sigma^2(M_h,z)}
\right]
\left|
\frac{d\sigma^2(M_h,z)}{dM_h}
\right|,
\label{eq:PS_halo_mass_function}
\end{equation}
and
\begin{equation}
\begin{aligned}
\frac{dn^{\rm ST}}{dM_h}(M_h,z)
={}&
A_{\rm ST}
\frac{\bar\rho_m(0)}{M_h}
\sqrt{\frac{a_{\rm ST}}{2\pi}}
\frac{\delta_c}{\sigma^3(M_h,z)}
\left[
1+
\left(
\frac{a_{\rm ST}\delta_c^2}{\sigma^2(M_h,z)}
\right)^{-p_{\rm ST}}
\right]
\\
&\times
\exp\left[
-\frac{a_{\rm ST}\delta_c^2}{2\sigma^2(M_h,z)}
\right]
\left|
\frac{d\sigma^2(M_h,z)}{dM_h}
\right|,
\end{aligned}
\label{eq:ST_halo_mass_function}
\end{equation}
respectively. We adopt the Sheth--Tormen fitting parameters
$A_{\rm ST}=0.353$, $a_{\rm ST}=0.73$, and $p_{\rm ST}=0.175$.

The halo mass function in
Eq.~\eqref{eq:renormalized_conditional_hmf}, adopted in \FirstClass,
is calibrated against \LCDM\ simulations. In our calculation, the
smoothed matter overdensity and matter variance are evaluated separately
for each axiodilaton model, while the standard \LCDM\ forms of the
extended Press--Schechter and Sheth--Tormen mass functions are retained.
We therefore regard this treatment as an approximation. Dedicated
axiodilaton simulations will be required to calibrate the halo mass
function for this cosmology.

\subsection{Evolution of the X-ray Free-Electron Fraction}
\label{subsec:xe_evolution}

We now describe the evolution of the X-ray contribution to the
free-electron fraction, denoted by $x_e^{\rm X}(\mathbf r,z)$, which was first introduced in Eq.~\eqref{eq_xHII_total}. The same quantity also enters the evolution of the gas kinetic temperature in
Eq.~\eqref{eq_Tk_evolution}. The UV-ionized component is treated separately, with $x_{\rm HII}^{\rm UV}(\mathbf r,z)\simeq x_e^{\rm UV}(\mathbf r,z)$, as specified in Appendix~\ref{app:UV_ionization}.

The initial condition is spatially homogeneous and is determined from the mean residual free-electron fraction $\bar x_e(z)$ introduced in
Eq.~\eqref{eq_xe_initial}. It is given by
\begin{equation}
x_e^{\rm X}(\mathbf r,z_{\rm rec})
=
\bar x_e(z_{\rm rec})
=
0.1 \;.
\label{eq_xeX_initial}
\end{equation}
Here, $z_{\rm rec}$ denotes the recombination redshift and is used as the initial redshift for the semi-numerical 21\,cm evolution.
The local evolution is written as
\begin{equation}
\frac{d x_e^{\rm X}(\mathbf r,z)}{dz}
=
\widetilde{\Lambda}_{\rm rec,X}(\mathbf r,z)
-
\widetilde{\Lambda}_{\rm ion,X}(\mathbf r,z).
\label{eq_xe_evolution}
\end{equation}
The recombination sink term is written as
\begin{equation}
\widetilde{\Lambda}_{\rm rec,X}(\mathbf r,z)
=
\left[-\frac{dt}{dz}\right]_z
\alpha_{B}\!\left[T_k(\mathbf r,z),m_e(z)\right]\,
C(z)\,
n_H(\mathbf r,z)\,
\left[x_e^{\rm X}(\mathbf r,z)\right]^2,
\label{eq_xe_local_rec}
\end{equation}
where $\alpha_B$ is the case-B
recombination coefficient with units of ${\rm cm^3\,s^{-1}}$, and $C(z)$
is the dimensionless Peebles coefficient \cite{Peebles:1968ja}, the probability that an atom
in the $n=2$ level reaches the ground state before a CMB photon reionizes the atom.
The Peebles coefficient is evaluated at each redshift step by \FirstClass, and approaches
unity once the CMB photons are too cold to reionize the $n=2$ level, so that $C(z)\simeq1$
over the range $z\lesssim500$ relevant to the 21\,cm signal.


The leading dependence of $\alpha_B$ on the electron mass is derived in
Appendix~\ref{sec:recombination}, and we apply it using the redshift-dependent electron mass $m_e(z)$.

The X-ray ionization contribution in Eq.~\eqref{eq_xe_evolution} is
\begin{equation}
\begin{aligned}
\widetilde{\Lambda}_{\rm ion,X}(\mathbf r,z)
={}
\int_{\nu_X}^{\infty}d\nu\,
\widetilde{\mathcal J}_X(\mathbf r,\nu,z)
\sum_{i\in\{{\rm HI},{\rm HeI},{\rm HeII}\}}
x_i^{\rm X}(\mathbf r,z)\,
\sigma_i(\nu,z)\,
F_i(\nu,\mathbf r,z).
\end{aligned}
\label{eq_Lambda_ion_X_local}
\end{equation}
Here, $\widetilde{\mathcal J}_X(\mathbf r,\nu,z)$ is the X-ray photon
field in units of ${\rm cm}^{-2}\,{\rm Hz}^{-1}$, defined in
Eq.~\eqref{eq_JX_def}. The local abundance
$x_i^{\rm X}(\mathbf r,z)$ of absorber species $i$ and its
photoionization cross section $\sigma_i(\nu,z)$ are given in
Eqs.~\eqref{eq_xabs_approx} and \eqref{eq:sigma_rescale_nu},
respectively. The lower integration limit $\nu_X$ is the low-energy
cutoff of the X-ray source spectrum, $h\nu_X=500\,{\rm eV}$.

The dimensionless ionization yield $F_i$ gives the total number of
ionizations produced following the absorption of an X-ray photon by
species $i$ and is written as
\begin{equation}
\begin{aligned}
F_i(\nu,\mathbf r,z)
\equiv
1+
\sum_{j\in\{{\rm HI},{\rm HeI},{\rm HeII}\}}
f_{{\rm ion},j}
\left(
h\nu-E_i^{\rm th}(z),
x_e^{\rm X}(\mathbf r,z)
\right)
\frac{h\nu-E_i^{\rm th}(z)}
{E_j^{\rm th}(z)} .
\end{aligned}
\label{eq_Fi_definition}
\end{equation}
The first term represents the primary ionization of species $i$. The
second term accounts for secondary ionizations produced by the
photoelectron with kinetic energy $h\nu-E_i^{\rm th}(z)$. The function
$f_{{\rm ion},j}(E,x_e)$ specifies the fraction of the photoelectron
energy deposited into the ionization of species $j$. 

\begin{figure}[t!]
    \centering
    \includegraphics[width=0.9\linewidth]{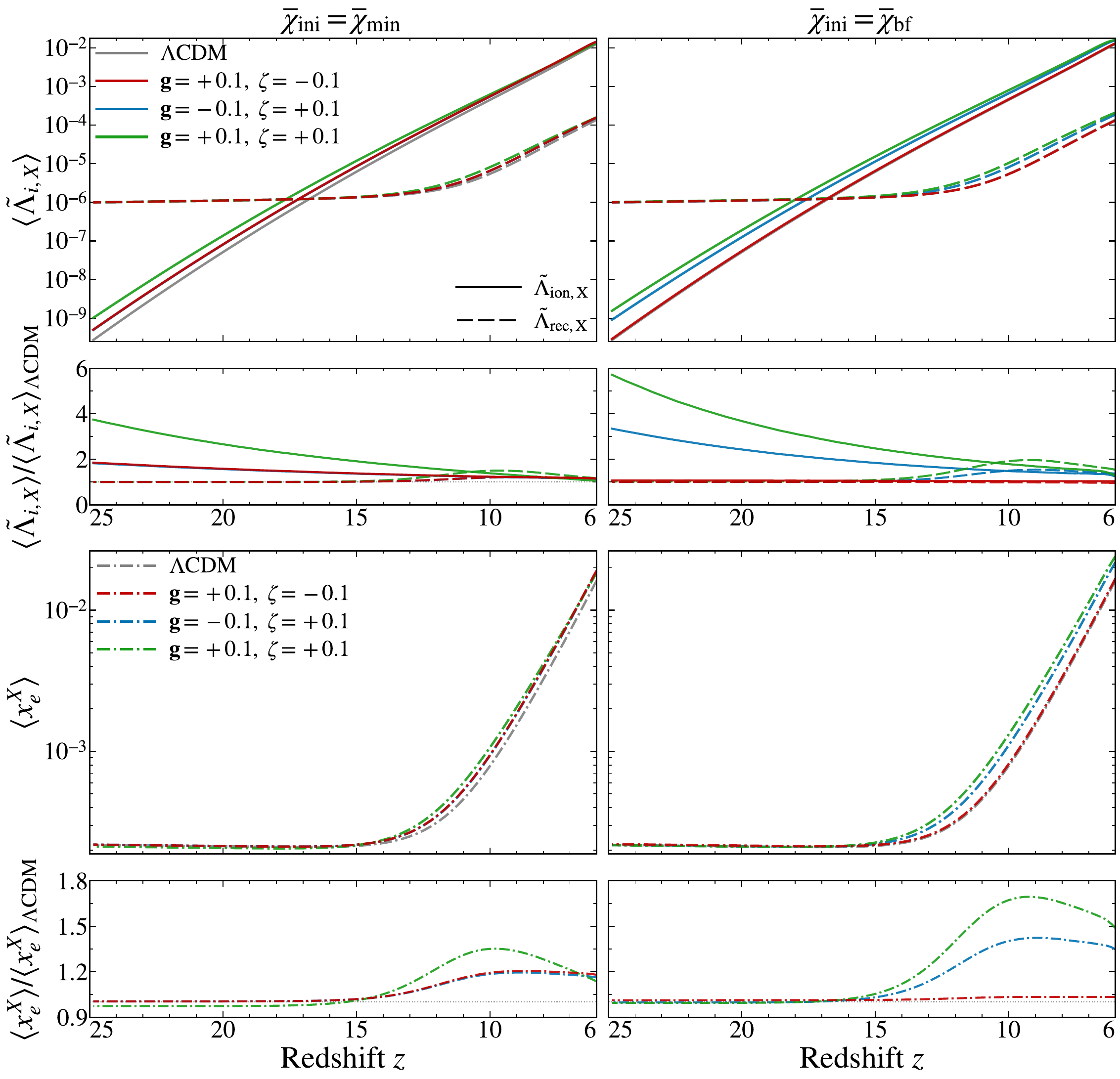}
    \caption{
    Evolution of the volume-averaged X-ray ionization and recombination
    terms and the resulting X-ray free-electron fraction for
    $\bar\chi_{\rm ini} = \bar\chi_{\rm min} = 74.010$ (left) and $\bar\chi_{\rm ini} = \bar\chi_{\rm bf}=74.213$ (right). The upper panel shows
    $\langle\widetilde{\Lambda}_{\rm ion,X}\rangle$ and
    $\langle\widetilde{\Lambda}_{\rm rec,X}\rangle$ as solid and dashed
    curves, respectively, together with their ratios to the corresponding
    \LCDM\ values. The lower panel shows
    $\langle x_e^{\rm X}\rangle$ and its ratio to the \LCDM\ result.
    The gray curves represent \LCDM, while the red, blue, and green curves
    denote $(\bfg,\zeta)=(+0.1,-0.1)$, $(-0.1,+0.1)$, and
    $(+0.1,+0.1)$, respectively.
    }
    \label{fig:xray_ion_rec_xe}
\end{figure}

Eq.~\eqref{eq_Lambda_ion_X_local} describes the ionization produced by
the same X-ray background that gives rise to the heating rate in
Eq.~\eqref{eq_eps_X_shell}. The two expressions differ only in their
energy-deposition kernels.
For heating, the factor multiplying the photon field is the
deposited thermal energy
$\left[h\nu-E_i^{\rm th}(z)\right]f_{\rm heat}$, while for ionization it
is the ionization yield $F_i$. The same X-ray background therefore heats
and ionizes the neutral IGM simultaneously. In our implementation, we follow the same strategy as in Appendix~\ref{subsec:xray} and keep the deposition
fractions $f_{\rm heat}$ and $f_{{\rm ion},j}$ equal to their fiducial \FirstClass\ tabulations.

Fig.~\ref{fig:xray_ion_rec_xe} shows how the competition between
X-ray ionization and recombination determines the evolution of the
volume-averaged X-ray free-electron fraction. At high redshift,
$\langle\widetilde{\Lambda}_{\rm rec,X}\rangle$ exceeds
$\langle\widetilde{\Lambda}_{\rm ion,X}\rangle$. The abundance of
X-ray-emitting sources is still very small during this period, leaving
the X-ray background and the associated primary and secondary
ionization rates strongly suppressed. Recombination can nevertheless
proceed through the residual free electrons, and its rate is enhanced by the larger physical hydrogen
density at high redshift. Consequently,
$\langle x_e^{\rm X}\rangle$ initially decreases slightly as the
Universe evolves toward lower redshift.

As structure formation progresses, the rapidly increasing abundance of
source-hosting halos raises the X-ray emissivity and causes
$\langle\widetilde{\Lambda}_{\rm ion,X}\rangle$ to grow by several
orders of magnitude. The ionization term overtakes the recombination
term at $z\simeq17$. The recombination term also increases toward lower redshift because its
rate scales approximately as $\left(x_e^{\rm X}\right)^2$. Its growth is
relatively slow because the physical hydrogen density decreases as the
Universe expands and, once X-ray heating becomes effective, the rising gas
temperature reduces $\alpha_B$. The ionization term therefore grows much more rapidly, producing
the sharp rise of $\langle x_e^{\rm X}\rangle$ at lower redshift.

The ordering among the axiodilaton benchmarks follows the amplitude of
their matter overdensities. The green model,
$(\bfg,\zeta)=(+0.1,+0.1)$, has the largest overdensities and forms
X-ray-emitting sources most rapidly, giving the largest ionization rate
and free-electron fraction. The other benchmarks follow the
corresponding ordering of their density fluctuations.

\subsubsection{Recombination Coefficient $\alpha_B$}
\label{sec:recombination}

The hydrogen case-B coefficient accounts for radiative recombinations
into excited states, excluding direct captures to the ground state.
Detailed calculations of the hydrogenic case-B coefficient are given in
Refs.~\cite{Seaton:1959,Burgess:1960,Nahar:2021}. These studies imply the approximate leading low-temperature scaling
\begin{equation}
\alpha_B(T_k,m_e)
\propto
m_e^{-3/2}T_k^{-1/2}.
\label{eq:alphaB_me_scaling}
\end{equation}
Detailed numerical calculations of the hydrogen case-B recombination
coefficient are presented in Ref.~\cite{Hummer:1994MNRAS}. For our
numerical calculation, we adopt the standard fitting function given in
Ref.~\cite{Pequignot:1991}, as implemented in
Ref.~\cite{Seager:1999km},
\begin{equation}
\alpha_B^{(0)}(T_k)
=
\frac{a\,\left(T_k/10^4\,{\rm K}\right)^{b}}
{1+c\,\left(T_k/10^4\,{\rm K}\right)^{d}},
\label{eq:alphaB_fit}
\end{equation}
where
$a=4.309\times10^{-13}\,{\rm cm^3\,s^{-1}}$,
$b=-0.6166$, $c=0.6703$, and $d=0.5300$.
Here, $\alpha_B^{(0)}$ denotes the coefficient evaluated at the
present-day electron mass $m_{e,0}$. In the low-temperature limit
$T_k\ll10^4\,{\rm K}$, the denominator approaches unity and the fit scales as
$\alpha_B^{(0)}\propto T_k^{-0.6166}$. This behavior is close to the
$T_k^{-1/2}$ dependence in Eq.~\eqref{eq:alphaB_me_scaling}.
Since Eq.~\eqref{eq:alphaB_fit} contains no variation of the electron
mass, we extend it using the leading mass dependence in
Eq.~\eqref{eq:alphaB_me_scaling}
\begin{equation}
\alpha_B(T_k,m_e)
\simeq
F\,\alpha_B^{(0)}(T_k)
\left(\frac{m_e(z)}{m_{e,0}}\right)^{-3/2}.
\label{eq:alphaB_fit_me}
\end{equation}
This prescription incorporates the leading explicit dependence on
$m_e$ and should be regarded as an approximation to a full
multilevel calculation with varying atomic parameters. The factor $F$
is an empirical correction that allows an effective three-level
treatment to match multilevel recombination calculations
\cite{Seager:1999km}. Following the implementation in \FirstClass, we
adopt $F=1.125$.

\begin{figure}[t!]
\centering
\includegraphics[width=0.9\linewidth]{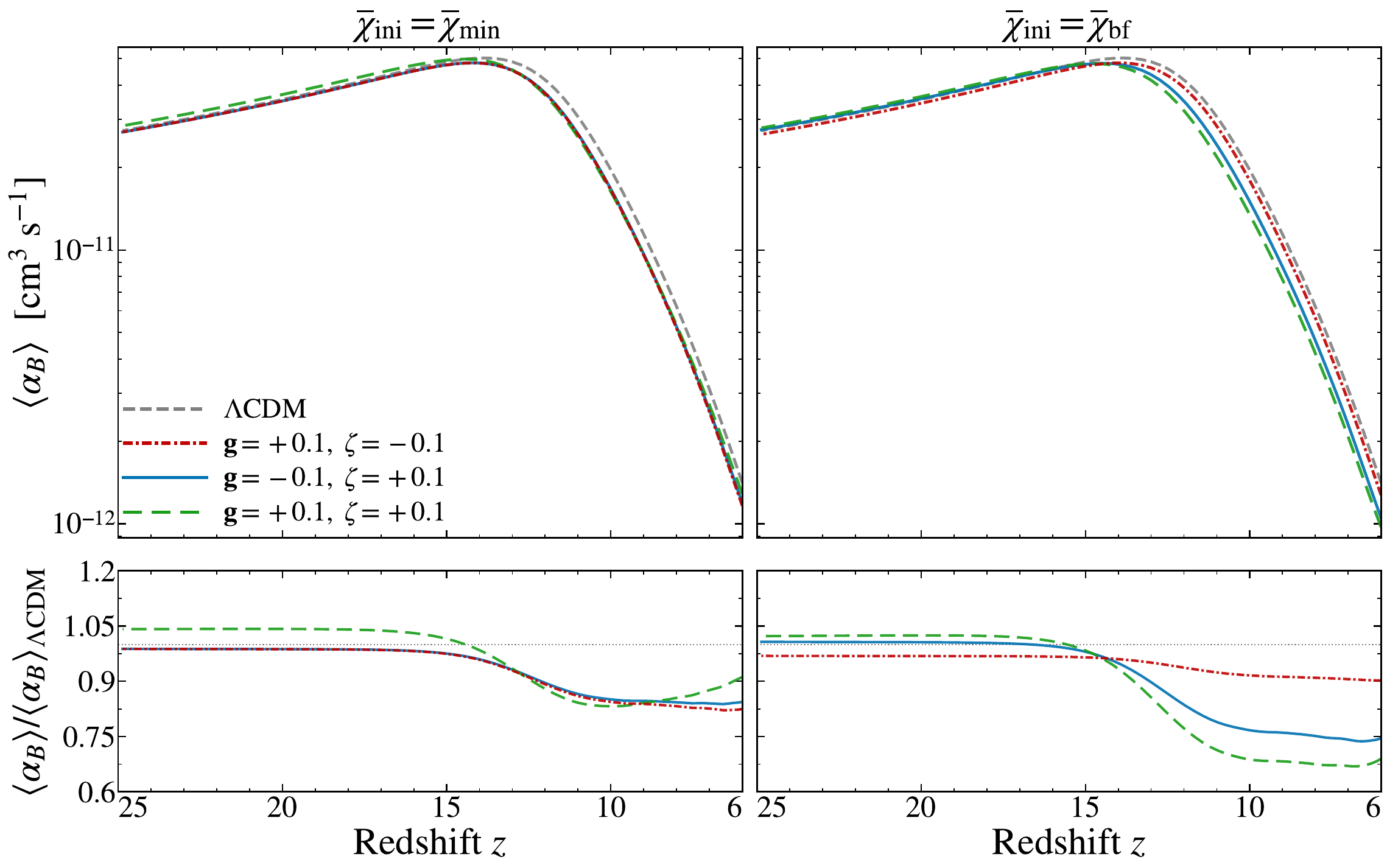}
\caption{
Evolution of the volume-averaged case-B recombination coefficient
$\langle\alpha_B\rangle$ (upper panels) and the corresponding ratios
relative to the \LCDM\ value (lower panels). Results for
$\bar\chi_{\rm ini} = \bar\chi_{\rm min} = 74.010$ (left) and $\bar\chi_{\rm ini} = \bar\chi_{\rm bf}=74.213$ (right) are shown in the left and right
columns, respectively. The gray dashed curves represent \LCDM.
The red dash-dotted, blue solid, and green dashed curves denote
$(\bfg,\zeta)=(+0.1,-0.1)$, $(-0.1,+0.1)$, and $(+0.1,+0.1)$,
respectively.
}
\label{fig:alphaB_evolution}
\end{figure}

As shown in Fig.~\ref{fig:alphaB_evolution},
$\langle\alpha_B\rangle$ initially increases toward lower redshift as
the gas cools. It reaches a maximum near the onset of X-ray heating and
subsequently decreases as the gas temperature rises. The differences
among the models are determined by the explicit $m_e^{-3/2}$ dependence
associated with the electron-mass evolution in Fig.~\ref{fig:bkg} and
the temperature dependence associated with the thermal histories in
Fig.~\ref{fig:x_Ts_Tb}.

For $\bar\chi_{\rm ini}=\chimin$, the green case lies above \LCDM\ at
high redshift because its electron mass is smaller, while the red and
blue cases remain close to \LCDM. Before the formation of the first
sources, the gas temperatures of all three models closely follow the
\LCDM\ evolution, and their effect on the relative values of
$\langle\alpha_B\rangle$ is therefore small. Once X-ray heating becomes
effective at $z\lesssim13$, the gas temperature rises above the \LCDM\
value in all three models. The inverse temperature dependence of
$\alpha_B$ then reduces $\langle\alpha_B\rangle$ relative to its
high-redshift behavior.

For $\bar\chi_{\rm ini}=\chibf$, the electron-mass effect produces a
clearer separation at high redshift. The green and blue cases lie above
\LCDM\ because they have smaller electron masses, while the red case
lies below \LCDM\ because its electron mass is larger. At
$z\lesssim13$, X-ray heating substantially raises the gas temperature
in the green and blue cases, leading to a pronounced decrease in their ratios. The
temperature of the red case remains close to the \LCDM\ value near the
onset of heating and becomes slightly higher at lower redshift, producing
a correspondingly smaller decrease in the ratio.

\subsection{UV Ionization}
\label{app:UV_ionization}

The UV contribution $x_{\rm HII}^{\rm UV}(\mathbf r,z)$ introduced in
Eq.~\eqref{eq_xHII_total} is determined using an excursion-set
photon-counting criterion. For a spherical region of comoving radius $R$
centered at $\mathbf r$, the number of ionizing photons produced per
hydrogen nucleus is written as
\begin{equation}
n_{\rm ion}(\mathbf r,z,R)
=
\zeta_{\rm ion}
f_{\rm coll}^{\rm eff}(\mathbf r,z,R),
\label{eq_nion}
\end{equation}
where
\begin{equation}
\zeta_{\rm ion}
\equiv
f_*^{\rm II}f_{\rm esc}^{\rm II}N_{\gamma}.
\label{eq_zeta_ion}
\end{equation}
Here, $f_*^{\rm II}$ and $f_{\rm esc}^{\rm II}$ are the normalization
parameters for the star-formation and escape fractions, respectively,
and $N_\gamma$ is the number of hydrogen-ionizing photons produced per
stellar baryon. These quantities are included among the astrophysical
parameters defined in Eq.~\eqref{eq:astro_params_schematic}. The
dimensionless effective collapsed fraction
$f_{\rm coll}^{\rm eff}(\mathbf r,z,R)$ is defined in
Eq.~\eqref{eq_fcoll_eff_definition}.

The number of photons in Eq.~\eqref{eq_nion} is compared with the photon
budget required to ionize the hydrogen within the region. This budget
accounts for the fraction of hydrogen already ionized by X-rays and the
additional photons consumed by recombinations. An ionizing UV photon that
reaches an ionization front converts neutral hydrogen into ionized
hydrogen and contributes to the growth of the HII region. A photon
absorbed by a hydrogen atom that has recombined inside an existing HII
region restores its ionized state without advancing the ionization front.
Recombinations therefore increase the number of photons required to
maintain and expand the ionized region. This contribution is particularly
important in small-scale dense structures~\cite{Park:2016jjg} and is evaluated using the
inhomogeneous-recombination prescription described in
Appendix~\ref{app:inhomogeneous_recombination}.

The required number of UV photons per hydrogen nucleus is expressed as
\begin{equation}
n_{\rm ion}^{\rm req}(\mathbf r,z,R)
=
\left[
1-x_{e,R}^{\rm X}(\mathbf r,z)
\right]
\left[
1+n_{{\rm rec},R}(\mathbf r,z)
\right].
\label{eq_uv_required_per_hydrogen}
\end{equation}
Here, $x_{e,R}^{\rm X}(\mathbf r,z)$ denotes the X-ray contribution to the free-electron fraction averaged over a spherical region of comoving radius $R$ centered at $\mathbf r$,
\begin{equation}
x_{e,R}^{\rm X}(\mathbf r,z)
=
\int d^3\mathbf y\,
W_R(|\mathbf r-\mathbf y|)
x_e^{\rm X}(\mathbf y,z).
\label{eq_xeX_smoothed}
\end{equation}
The quantity $n_{{\rm rec},R}(\mathbf r,z)$ is the cumulative number of recombinations per hydrogen nucleus averaged over the same region and is defined in Appendix~\ref{app:inhomogeneous_recombination}.
In Eq.~\eqref{eq_uv_required_per_hydrogen}, the first factor,
$[1-x_{e,R}^{\rm X}(\mathbf r,z)]$, gives the fraction of
hydrogen that remains to be ionized after accounting for X-rays. The second factor, $[1+n_{{\rm rec},R}(\mathbf r,z)]$, accounts for the additional ionizations required to compensate for subsequent recombinations.

The ionization condition is evaluated on progressively smaller comoving length scales. The calculation begins at
$R_{\max}=15\,{\rm Mpc}$, corresponding to the adopted maximum mean free
path of ionizing UV photons, and proceeds down to the cell scale
$R_{\rm cell}\simeq\Delta x_{\rm lo}=2\,{\rm Mpc}$. For each cell centered
at $\mathbf r$, the code determines whether there exists a scale
$R\in[R_{\rm cell},R_{\max}]$ for which
\begin{equation}
n_{\rm ion}(\mathbf r,z,R)
\geq
\left[
1-x_{e,R}^{\rm X}(\mathbf r,z)
\right]
\left[
1+n_{{\rm rec},R}(\mathbf r,z)
\right].
\label{eq_ionization_condition}
\end{equation}
If Eq.~\eqref{eq_ionization_condition} is satisfied at any of the
smoothing scales considered, the central cell is flagged as fully
ionized. Since the X-ray contribution has already ionized a fraction
$x_e^{\rm X}(\mathbf r,z)$ of the hydrogen in the cell, the additional
contribution assigned to UV photons is
$x_{e}^{\rm UV}(\mathbf r,z)=1-x_e^{\rm X}(\mathbf r,z)$.

If Eq.~\eqref{eq_ionization_condition} is not satisfied at any smoothing
scale, the partial UV contribution to the ionized fraction is evaluated
at the cell scale as
\begin{equation}
x_{e}^{\rm UV}(\mathbf r,z)
=
n_{\rm ion}(\mathbf r,z,R_{\rm cell}).
\label{eq_xHII_UV_partial}
\end{equation}
No additional correction for inhomogeneous recombinations is applied in
this cell-scale prescription. Combining the two cases, the UV contribution
is given by
\begin{equation}
x_{e}^{\rm UV}(\mathbf r,z)
=
\begin{cases}
1-x_e^{\rm X}(\mathbf r,z),
&
\text{if Eq.~\eqref{eq_ionization_condition} is satisfied for some }R,
\\[6pt]
\displaystyle
n_{\rm ion}(\mathbf r,z,R_{\rm cell}),
&
\text{otherwise}.
\end{cases}
\label{eq_xHII_UV}
\end{equation}

\begin{figure}[t!]
    \centering
    \includegraphics[width=1.0\linewidth]{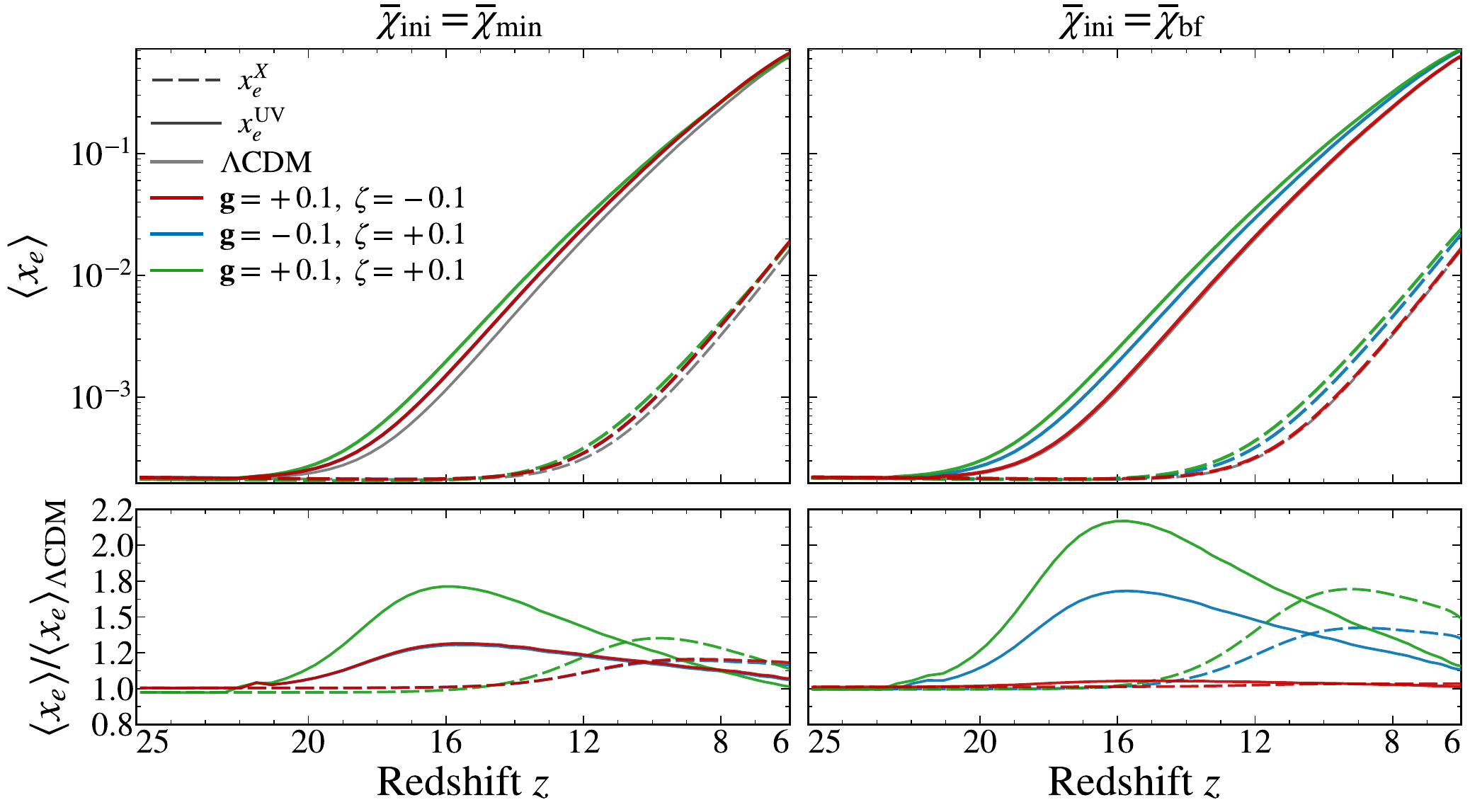}
    \caption{Evolution of the volume-averaged X-ray free-electron fraction and UV free-electron fraction for
    $\bar\chi_{\rm ini} = \bar\chi_{\rm min} = 74.010$ (left) and $\bar\chi_{\rm ini} = \bar\chi_{\rm bf}=74.213$ (right). The dashed curves represent $\langle x_e^{\rm X}\rangle$, while the solid curves show
    $\langle x_{e}^{\rm UV}\rangle$, with their ratios to the corresponding \LCDM\ values. 
    The gray curves represent \LCDM, while the red, blue, and green curves denote $(\bfg,\zeta)=(+0.1,-0.1)$, $(-0.1,+0.1)$, and
    $(+0.1,+0.1)$, respectively.}
    \label{fig:xeX_xeUV}
\end{figure}

Fig.~\ref{fig:xeX_xeUV} shows the redshift evolution of the
volume-averaged UV and X-ray contributions to the free-electron
fraction, $\langle x_e^{\rm UV}\rangle$ and
$\langle x_e^{\rm X}\rangle$, respectively. The UV contribution is
determined by $x_e^{\rm UV}$ in
Eq.~\eqref{eq_xHII_UV}, while the X-ray contribution follows the
evolution equation in Eq.~\eqref{eq_xe_evolution}. Toward lower redshift, source-hosting halos become more abundant, increasing the cumulative number of ionizing UV photons. The resulting photoionization of
neutral hydrogen drives the rapid increase in
$\langle x_e^{\rm UV}\rangle$.

After the first sources form, the UV contribution rapidly becomes much
larger than the X-ray contribution. For the fiducial EOS2021
astrophysical parameters~\cite{Munoz:2021psm} adopted in
Eq.~\eqref{eq:astro_params_schematic}, $N_\gamma=5000$
UV photons are produced per stellar baryon. The
adopted value of $L_X^{\rm II}/{\rm SFR}$, together with the assumed
X-ray spectrum, corresponds to approximately $0.57$ X-ray photons per
stellar baryon. The number of
ionizing UV photons produced per stellar baryon is therefore nearly
four orders of magnitude larger than the corresponding number of
X-ray photons.

The ordering of the ionized fractions is primarily inherited from the
model-dependent growth of matter perturbations and the resulting
abundance of source-hosting halos. At a fixed redshift, a larger halo
abundance increases the cosmic star-formation-rate density, raising both
the cumulative production of ionizing UV photons and the X-ray
emissivity. For $\bar\chi_{\rm ini}=\chimin$, the green case consequently
shows the earliest rise and the largest enhancement in both
$\langle x_e^{\rm UV}\rangle$ and $\langle x_e^{\rm X}\rangle$. The red
and blue cases are nearly degenerate and show smaller enhancements
relative to \LCDM. The differences are more pronounced for
$\bar\chi_{\rm ini}=\chibf$. In this case, the ionized fractions follow
the ordering green, blue, and red, with the red case remaining close to
\LCDM.

The preceding discussion has focused on the volume-averaged UV and
X-ray contributions to the free-electron fraction. The corresponding
ionization fields also differ in their spatial distributions because UV
and X-ray photons have very different mean free paths in the IGM.
X-ray photons have much higher
energies than UV photons. 
As shown in Eq.~\eqref{eq_sigma_hydrogenic}, the
photoionization cross section decreases rapidly with increasing photon
energy. UV photons typically have energies near the hydrogen
ionization threshold, ranging from $13.6\,{\rm eV}$ to several tens of
electronvolts. Their relatively large photoionization cross sections
cause them to be absorbed close to their sources, resulting in short
mean free paths. X-ray photons in the energy range
$E\simeq0.5$--$2\,{\rm keV}$ have much smaller photoionization cross
sections. They consequently have a lower probability of interacting
with neutral hydrogen and can travel much greater distances before
being absorbed. X-ray photons
have more than sufficient energy to ionize hydrogen, while their high
energies make interactions with neutral hydrogen less frequent.

The energy deposition following absorption also differs between UV and
X-ray photons. An absorbed UV photon generally produces one primary
photoionization, and the resulting photoelectron carries relatively
little kinetic energy. An absorbed X-ray photon produces a highly
energetic photoelectron that can generate multiple secondary
ionizations. The photoelectron energy is divided among ionization,
atomic excitation, and heating according to Eq.~\eqref{eq_Fi_definition}.

\subsubsection{Small Scale Structure of the IGM for Inhomogeneous Recombinations}
\label{app:inhomogeneous_recombination}

In this appendix, we summarize the sub-grid prescription implemented
in \FirstClass\ to compute the cumulative recombination correction
$n_{{\rm rec},R}(\mathbf r,z)$ entering
Eq.~\eqref{eq_ionization_condition}. This prescription follows
Refs.~\cite{Miralda-Escude:1998adl,Sobacchi:2014rua,Park:2018ljd}.
Gas within an ionized simulation cell is not uniform below the grid scale. Dense
unresolved gas recombines more rapidly than gas at the mean density,
while sufficiently dense regions can absorb the incident ionizing UV
radiation and remain partially neutral. The prescription models these
effects using the unresolved gas-density distribution and the local
photoionization rate.

We first define the dimensionless sub-grid hydrogen overdensity as
\begin{equation}
\Delta
\equiv
\frac{n_{H}^{\rm sg}}
{\bar n_{H}(z)},
\label{eq:subgrid_hydrogen_density}
\end{equation}
where $n_{H}^{\rm sg}$ is the hydrogen-nucleus number density of
an unresolved gas element and $\bar n_{H}(z)$ is the background
mean hydrogen-nucleus number density in Eq.~\eqref{eq_mean_nH_nHe}. The corresponding resolved
cell-averaged overdensity is
\begin{equation}
\Delta_{\rm cell}(\mathbf r,z)
\equiv
\frac{n_{H}(\mathbf r,z)}
{\bar n_{H}(z)}
=
1+\delta_\ssB^{\rm NL}(\mathbf r,z),
\label{eq:resolved_cell_overdensity}
\end{equation}
where $n_{H}(\mathbf r,z)$ is given in
Eq.~\eqref{eq_local_nH}. The quantity $\Delta_{\rm cell}$ specifies
the mean hydrogen density of a resolved cell, whereas $\Delta$
describes the distribution of unresolved gas densities within that
cell.

The unresolved gas density is modeled using the volume-weighted
Miralda-Escudé--Haehnelt--Rees (MHR) distribution
\cite{Miralda-Escude:1998adl,Sobacchi:2014rua,Park:2018ljd},
\begin{equation}
P_V(\mathbf r,z,\Delta)
=
A_{\rm MHR}(\mathbf r,z)
\exp\left[
-\frac{
\left[
\Delta^{-2/3}-C_0(\mathbf r,z)
\right]^2
}{
2\left[
2\delta_0(\mathbf r,z)/3
\right]^2
}
\right]
\Delta^{-2.5}.
\label{eq:MHR_pdf}
\end{equation}
The quantity
$P_V(\mathbf r,z,\Delta)d\Delta$ is the fraction of the cell volume
occupied by unresolved gas with overdensity between $\Delta$ and
$\Delta+d\Delta$. The differential forms of
the MHR distribution are shown in Fig.~1 of
Ref.~\cite{Miralda-Escude:1998adl}. The width parameter is evaluated as
\begin{equation}
\delta_0(\mathbf r,z)
=
\frac{7.61}
{(1+z)\Delta_{\rm cell}^{1/3}(\mathbf r,z)}.
\label{eq:MHR_delta0}
\end{equation}
The exponent $2.5$ follows the tabulated MHR redshift
dependence at $z\geq6$. The coefficients
$A_{\rm MHR}(\mathbf r,z)$ and $C_0(\mathbf r,z)$ are determined from
\begin{equation}
\int_0^\infty d\Delta\,
P_V(\mathbf r,z,\Delta)
=
1,
\qquad
\int_0^\infty d\Delta\,
\Delta P_V(\mathbf r,z,\Delta)
=
\Delta_{\rm cell}(\mathbf r,z).
\label{eq:MHR_normalization}
\end{equation}
These conditions normalize the total cell volume and reproduce the
resolved cell-averaged hydrogen density.

The ionization state of gas at a given $\Delta$ depends on the
ionizing UV background incident on the cell. We approximate the UV
spectrum and the hydrogen photoionization cross section of Eq.~\eqref{eq:sigma_rescale_nu} as
\begin{equation}
J_\nu^{\rm UV}
\propto
\left[
\frac{\nu}{\nu_{\rm HI}(z)}
\right]^{-\alpha_{\rm UV}},
\qquad
\sigma_{\rm HI}(\nu,z)
\simeq
\sigma_{\rm HI,0}(z)
\left[
\frac{\nu}{\nu_{\rm HI}(z)}
\right]^{-\beta_{\rm HI}},
\qquad
\nu\geq\nu_{\rm HI}(z),
\label{eq:UV_spectrum_cross_section}
\end{equation}
where
\begin{equation}
\nu_{\rm HI}(z)
=
\frac{E_{\rm HI}^{\rm th}(z)}{h},
\qquad
\sigma_{\rm HI,0}(z)
\equiv
\sigma_{\rm HI}\!\left[\nu_{\rm HI}(z),z\right].
\label{eq:UV_threshold_quantities}
\end{equation}
We adopt $\alpha_{\rm UV}=5.0$ and $\beta_{\rm HI}=2.75$. To leading
order in the electron-mass variation, the threshold cross section is
rescaled according to
\begin{equation}
\sigma_{\rm HI,0}(z)
=
\sigma_{\rm HI,0}^{(0)}
\left[
\frac{m_e(z)}{m_{e,0}}
\right]^{-2},
\qquad
\sigma_{\rm HI,0}^{(0)}
=
6.3\times10^{-18}\,{\rm cm^2}.
\label{eq:UV_threshold_cross_section}
\end{equation}

For a candidate UV-ionized region of comoving radius $R$, the mean
photoionization rate before self-shielding is approximated by
\begin{equation}
\Gamma_{\rm HII}(\mathbf r,z,R)
\simeq
(1+z)^2
\lambda_{\rm mfp}
\sigma_{\rm HI,0}(z)
\frac{\alpha_{\rm UV}}
{\alpha_{\rm UV}+\beta_{\rm HI}}\left[\bar n_H(0)+\bar n_{\rm He}(0)\right]\,\frac{H(z)}{t_\ast}\,n_{\rm ion}(\mathbf r,z,R),
\label{eq:Gamma_HII}
\end{equation}
where $\Gamma_{\rm HII}$ has units of ${\rm s}^{-1}$. The comoving
mean free path is approximated by the size of the candidate HII
region,
\begin{equation}
\lambda_{\rm mfp}=R.
\label{eq:UV_mean_free_path}
\end{equation}
The quantity $n_{\rm ion}(\mathbf r,z,R)$ defined in
Eq.~\eqref{eq_nion} is the cumulative number of ionizing photons
produced per hydrogen nucleus within the region. The ionizing photon production rate per hydrogen nucleus is $H(z)\,n_{\rm ion}(\mathbf r,z,R)/t_\ast$. Here $t_\ast$ is the star-formation timescale as a fraction of the Hubble time, Eq.~\eqref{eq:astro_params_schematic}.
The factor
$\sigma_{\rm HI,0}(z)\alpha_{\rm UV}/
(\alpha_{\rm UV}+\beta_{\rm HI})$
is the photoionization cross section averaged over the adopted UV
spectrum.

Dense unresolved gas can become self-shielded and experience a local
photoionization rate below $\Gamma_{\rm HII}$. We use the fitting
function
\begin{equation}
\begin{aligned}
\frac{
\Gamma_{\rm local}(\mathbf r,z,\Delta,R)
}{
\Gamma_{\rm HII}(\mathbf r,z,R)
}
={}&
0.98
\left[
1+
\left(
\frac{\Delta}
{\Delta_{\rm ss}(\mathbf r,z,R)}
\right)^{1.64}
\right]^{-2.28}
\\
&+
0.02
\left[
1+
\frac{\Delta}
{\Delta_{\rm ss}(\mathbf r,z,R)}
\right]^{-0.84},
\end{aligned}
\label{eq:Gamma_local}
\end{equation}
where the characteristic self-shielding overdensity is
\begin{equation}
\Delta_{\rm ss}(\mathbf r,z,R)
=
26.7
\left(
\frac{T_{\rm ion}}{10^4\,{\rm K}}
\right)^{0.17}
\left(
\frac{1+z}{10}
\right)^{-3}
\left[
\frac{\Gamma_{\rm HII}(\mathbf r,z,R)}
{10^{-12}\,{\rm s}^{-1}}
\right]^{2/3}.
\label{eq:Delta_ss}
\end{equation}
We adopt $T_{\rm ion}=10^4\,{\rm K}$ for gas inside UV-ionized
regions. Gas with $\Delta\ll\Delta_{\rm ss}$ remains nearly
optically thin and experiences
$\Gamma_{\rm local}\simeq\Gamma_{\rm HII}$. Self-shielding
progressively suppresses $\Gamma_{\rm local}$ at
$\Delta\gtrsim\Delta_{\rm ss}$.

The neutral fraction of an unresolved gas element is defined as
\begin{equation}
x_{\rm HI}^{\rm sg}(\mathbf r,z,\Delta,R)
\equiv
\frac{
n_{\rm HI}^{\rm sg}(\mathbf r,z,\Delta,R)
}{
\bar n_{H}(z)\Delta
}.
\label{eq:subgrid_neutral_fraction}
\end{equation}
Assuming local photoionization equilibrium, it satisfies
\begin{equation}
x_{\rm HI}^{\rm sg}(\mathbf r,z,\Delta,R)
\Gamma_{\rm local}(\mathbf r,z,\Delta,R)
=
\chi_{\rm HeII}\,
\Delta\bar n_{H}(z)\,
\alpha_B(T_{\rm ion},z)
\left[
1-x_{\rm HI}^{\rm sg}(\mathbf r,z,\Delta,R)
\right]^2.
\label{eq:subgrid_photoionization_equilibrium}
\end{equation}
Here, $\chi_{\rm HeII}\simeq1.08$ accounts for free electrons
contributed by singly ionized helium. The case-B recombination
coefficient is evaluated as
\begin{equation}
\alpha_B(T_{\rm ion},z)
=
F\,\alpha_B^{(0)}(T_{\rm ion})
\left[
\frac{m_e(z)}{m_{e,0}}
\right]^{-3/2},
\label{eq:subgrid_alphaB_rescaling}
\end{equation}
with $F=1.125$ as in Eq.~\eqref{eq:alphaB_fit_me} and
$\alpha_B^{(0)}(10^4\,{\rm K})
\simeq2.6\times10^{-13}\,{\rm cm^3\,s^{-1}}$.
The origin of the leading electron-mass dependence in
Eq.~\eqref{eq:subgrid_alphaB_rescaling} is discussed in
Appendix~\ref{sec:recombination}. Solving
Eq.~\eqref{eq:subgrid_photoionization_equilibrium} determines
$x_{\rm HI}^{\rm sg}$ at each sub-grid density.

Since $P_V$ is volume weighted, the volume-averaged recombination
rate is proportional to $\Delta^2$. We define the recombination rate
per background mean hydrogen nucleus as
\begin{equation}
\frac{d\mathcal N_{\rm rec}}{dt}(\mathbf r,z,R)
=
\bar n_H(z)
\int_0^\infty d\Delta\,
P_V(\mathbf r,z,\Delta)\,
\Delta^2
\alpha_B(T_{\rm ion},z)
\left[
1-x_{\rm HI}^{\rm sg}(\mathbf r,z,\Delta,R)
\right]^2.
\label{eq:subgrid_recombination_rate}
\end{equation}
The integral combines the recombination rates of the unresolved gas
elements weighted by their volume fractions. Increasing $\Delta$
raises the recombination rate through the larger ionized hydrogen and
free-electron densities. Self-shielding raises
$x_{\rm HI}^{\rm sg}$ in dense gas and reduces the ionized fraction
available for recombination.

For a cell first UV ionized at $z_{\rm ion}(\mathbf r)$, the
cumulative number of recombinations per background mean hydrogen
nucleus is
\begin{equation}
\mathcal N_{\rm rec}(\mathbf r,z)
=
\int_z^{z_{\rm ion}(\mathbf r)}
dz'\,
\frac{
d\mathcal N_{\rm rec}/dt
\left[
\mathbf r,z',R_{\rm HII}(\mathbf r,z')
\right]
}{
(1+z')H(z')
}.
\label{eq:cumulative_recombinations}
\end{equation}
Here, $R_{\rm HII}(\mathbf r,z')$ is the largest comoving length
scale at which the cell satisfies
Eq.~\eqref{eq_ionization_condition} at redshift $z'$. It lies within
$R_{\rm cell}\leq R_{\rm HII}\leq R_{\max}$. The factor
$[(1+z')H(z')]^{-1}$ converts the redshift interval into a time
interval. We set $\mathcal N_{\rm rec}(\mathbf r,z)=0$ before the
cell becomes UV ionized.

To obtain the cumulative number of recombinations per hydrogen
nucleus within a spherical region of radius $R$, we first define its
mean hydrogen overdensity,
\begin{equation}
\Delta_R(\mathbf r,z)
=
\int d^3\mathbf y\,
W_R(|\mathbf r-\mathbf y|)
\Delta_{\rm cell}(\mathbf y,z).
\label{eq:smoothed_hydrogen_overdensity}
\end{equation}
The recombination correction entering
Eq.~\eqref{eq_ionization_condition} is then
\begin{equation}
n_{{\rm rec},R}(\mathbf r,z)
=
\frac{
\displaystyle
\int d^3\mathbf y\,
W_R(|\mathbf r-\mathbf y|)
\mathcal N_{\rm rec}(\mathbf y,z)
}{
\Delta_R(\mathbf r,z)
}.
\label{eq:smoothed_recombinations}
\end{equation}
The denominator converts the recombination count per background mean
hydrogen nucleus into the count per hydrogen nucleus contained in
the region.

In the axiodilaton cosmology, the background mean number density
$\bar n_{H}(z)$ retains its standard $(1+z)^3$ evolution because
the dilaton dependence of the baryon mass density cancels that of the
proton mass. The recombination correction nevertheless acquires
several axiodilaton dependences. Changes in the resolved baryon
density field modify $\Delta_{\rm cell}$ and the local MHR
distribution. Changes in the matter perturbations alter
$f_{\rm coll}^{\rm eff}$ and the ionizing photon production rate,
which propagate to $\Gamma_{\rm HII}$, $\Delta_{\rm ss}$, and
$x_{\rm HI}^{\rm sg}$. The varying electron mass changes
$\sigma_{\rm HI,0}(z)$ and $\alpha_B(T_{\rm ion},z)$. The modified
expansion rate $H(z)$ changes the time accumulated per redshift
interval. The ionization history also changes
$z_{\rm ion}(\mathbf r)$ and $R_{\rm HII}(\mathbf r,z)$.

\begin{figure}
    \centering
    \includegraphics[width=1\linewidth]{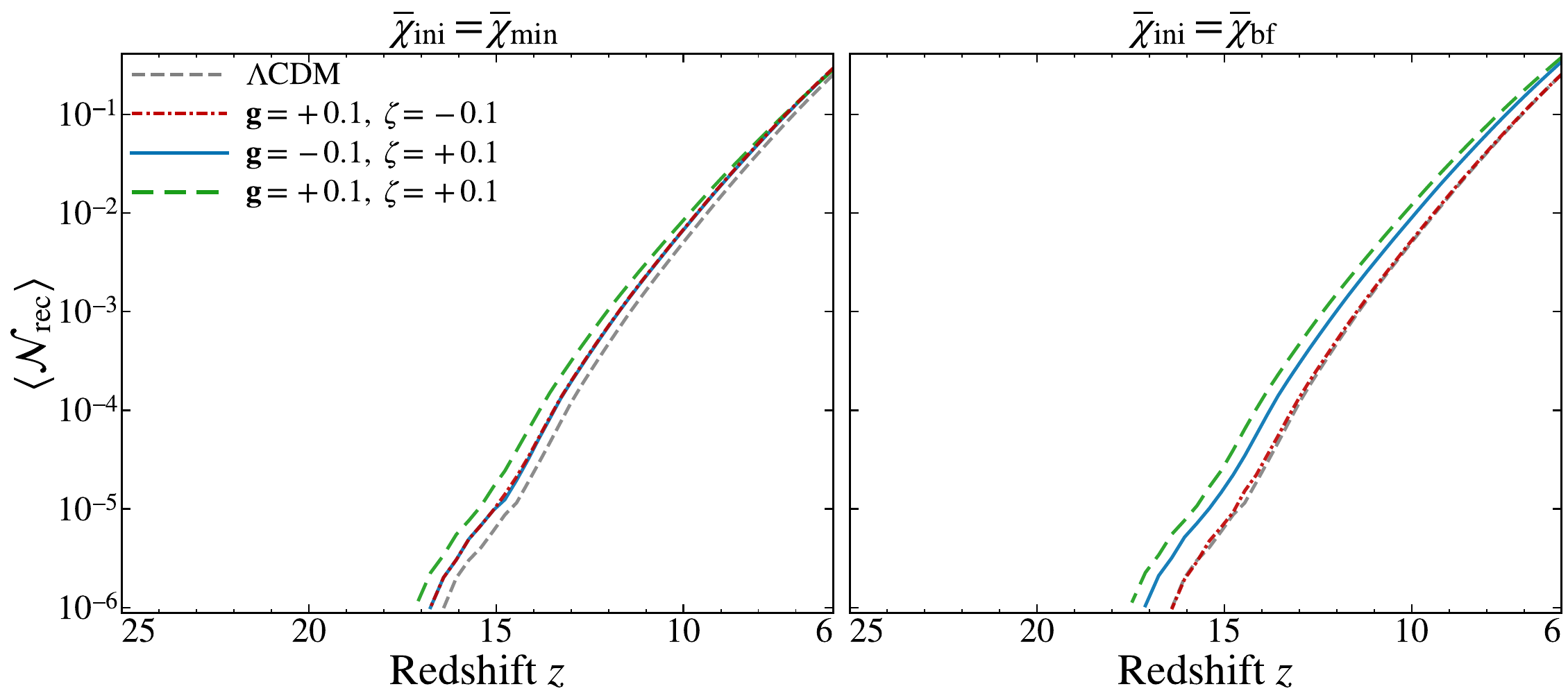}
    \caption{Evolution of the volume-averaged cumulative number of recombinations $\langle\mathcal N_{\rm rec}\rangle$, Eq.~\eqref{eq:cumulative_recombinations}, for $\bar\chi_{\rm ini}=\chimin$ (left) and $\bar\chi_{\rm ini}=\chibf$ (right). The gray dashed curves represent \LCDM. The blue solid, green dashed, and red dash-dotted curves denote $(\bfg,\zeta)=(-0.1,+0.1)$, $(+0.1,+0.1)$, and $(+0.1,-0.1)$, respectively.}
    \label{fig:Nrec}
\end{figure}

We retain the MHR density distribution and the numerical
self-shielding fit in their fiducial \FirstClass\ forms. These
prescriptions were calibrated using simulations with standard atomic
properties and a \LCDM\ matter distribution. A dedicated calibration
could modify the unresolved density distribution and the
self-shielding threshold beyond the explicit axiodilaton
dependences included above. Such a calibration would require
high-resolution hydrodynamical and radiative-transfer simulations and
is left for future work.

Fig.~\ref{fig:Nrec} shows $\langle\mathcal N_{\rm rec}\rangle$. The cumulative count $\mathcal N_{\rm rec}$ is zero in a cell until the cell becomes UV ionized, Eq.~\eqref{eq:cumulative_recombinations}. The volume average $\langle\mathcal N_{\rm rec}\rangle$ then grows with the UV-ionized volume. The green model has the largest $\langle\mathcal N_{\rm rec}\rangle$ for both initial values, as $\langle x_e^{\rm UV}\rangle$ does in Fig.~\ref{fig:xeX_xeUV}.

\subsection{Ly$\alpha$ Scattering}
\label{app:lya_scattering}

\subsubsection{Scattering Rate}
\label{app:lya_rate}

Let $I_\nu(\mathbf{r},\hat{\mathbf{n}},z)$ be the specific intensity of the UV background at position $\mathbf{r}$ and redshift $z$ along direction $\hat{\mathbf{n}}$. Dividing by the photon energy $h\nu$, we obtain the differential photon number flux, $I_\nu(\mathbf{r},\hat{\mathbf{n}},z)/(h\nu)$, which has units ${\rm s}^{-1}\,{\rm cm}^{-2}\,{\rm Hz}^{-1}\,{\rm sr}^{-1}$.
Consider photons in the frequency interval $[\nu,\nu+d\nu]$ arriving from solid angle element $d\Omega$ around $\hat{\mathbf{n}}$.
The number of photons crossing a unit area per unit time in this phase-space element, which has units of ${\rm s}^{-1}\,{\rm cm}^{-2}$ is
\begin{equation}
\left(\frac{I_\nu(\mathbf{r},\hat{\mathbf{n}},z)}{h\nu}\right)\,d\nu\,d\Omega.
\end{equation}
For a single atom with Ly$\alpha$ absorption cross section $\sigma_\alpha(\nu,\mathbf{r},z)$, the probability per unit time to scatter (absorb and re-emit) one of these photons is given by 
\begin{equation}
dP_\alpha
=
\left(\frac{I_\nu(\mathbf{r},\hat{\mathbf{n}},z)}{h\nu}\right)\,
\sigma_\alpha(\nu,\mathbf{r},z)\,
d\nu\,d\Omega.
\label{eq:dPalpha_def}
\end{equation}
Integrating over all arrival directions and photon frequencies yields the total scattering rate per atom, with units of ${\rm s}^{-1}$,
\begin{equation}
P_\alpha(\mathbf{r},z)
=
\int d\Omega \int d\nu\,
\left(\frac{I_\nu(\mathbf{r},\hat{\mathbf{n}},z)}{h\nu}\right)\,
\sigma_\alpha(\nu,\mathbf{r},z).
\label{eq:Palpha_from_Inu}
\end{equation}
It is convenient to express this in terms of the angle-averaged specific intensity, with units of ${\rm erg}\,{\rm s}^{-1}\,{\rm cm}^{-2}\,{\rm Hz}^{-1}$$\,{\rm sr}^{-1}$,
\begin{equation}
J_\nu(\mathbf{r},z)\equiv \frac{1}{4\pi}\int d\Omega\, I_\nu(\mathbf{r},\hat{\mathbf{n}},z) \;.
\label{eq:Jnu_def}
\end{equation}
We also introduce the Ly$\alpha$ background intensity evaluated at $\nu=\nu_\alpha(z)$,
\begin{equation}
J_\alpha(\mathbf{r},z)\equiv J_\nu(\mathbf{r},z)\big|_{\nu=\nu_\alpha(z)} .
\label{eq:Jalpha_def}
\end{equation}
In general, this quantity receives contributions from two physically distinct channels,
\begin{equation}
J_\alpha(\mathbf{r},z)=J_{\alpha,*}(\mathbf{r},z)+J_{\alpha,X}(\mathbf{r},z),
\label{eq:Jalpha_split}
\end{equation}
where $J_{\alpha,*}$ denotes the contribution sourced by stellar UV radiation, while $J_{\alpha,X}$ denotes the contribution generated indirectly by X-ray photoelectrons through collisional excitations of neutral hydrogen. A more detailed discussion of $J_{\alpha,*}$ and $J_{\alpha,X}$ is provided in
Appendix~\ref{app:Jalpha}.
Substituting Eq.~(\ref{eq:Jnu_def}) into Eq.~(\ref{eq:Palpha_from_Inu}) gives the total scattering rate per atom,
\begin{equation}
P_\alpha(\mathbf{r},z)
=
4\pi \int d\nu\,
\frac{J_\nu(\mathbf{r},z)}{h\nu}\,
\sigma_\alpha(\nu,\mathbf{r},z).
\label{eq:Palpha_def}
\end{equation}

It is customary to factor the frequency dependence of the Ly$\alpha$ cross section into an overall strength and a normalized line profile
\begin{equation}
\sigma_\alpha(\nu,\mathbf{r},z)=\sigma_{\alpha,0}(z)\,\phi_\alpha(\tilde\nu,\mathbf{r},z),
\qquad
\Big({\rm with}~\int d\tilde \nu\,\phi_\alpha(\tilde \nu ,\mathbf{r},z)=1 \Big)
\label{eq:sigma_alpha_profile}
\end{equation}
where $\phi_\alpha(\tilde \nu,\mathbf{r},z)$ with $\tilde \nu \equiv \nu - \nu_\alpha(z)$ describes how the absorption probability is distributed around the Ly$\alpha$ line-center frequency $\nu_\alpha(z)$ of neutral hydrogen, namely the resonance frequency of the $1s\leftrightarrow 2p$ transition in the local rest frame of the gas.
The line-center frequency $\nu_\alpha$ is given by
\begin{equation}
    \nu_\alpha(z)
    =
    \nu_{\alpha,0}
    \left[
        \frac{m_e(z)}{m_{e,0}}
    \right]
    =
    \nu_{\alpha,0}e^{\bfg\Delta\bar\chi(z)},    
    \label{eq:nualpha_mass_scaling}
\end{equation}
where $\nu_{\alpha,0} \simeq 2.47\times10^{15}\,{\rm Hz}$.
In Eq.~\eqref{eq:sigma_alpha_profile}, $\sigma_\alpha(\nu,\mathbf{r},z)$ is the frequency-dependent absorption (scattering) cross section with units of area, $[\sigma_\alpha] = {\rm cm}^2$.
The line profile carries units of an inverse frequency, $[\phi_\alpha] = {\rm Hz}^{-1}$. Accordingly, the prefactor $\sigma_{\alpha,0}$ has units
$[\sigma_{\alpha,0}] = {\rm cm}^2\,{\rm Hz}$, and it is fixed by the oscillator strength $g_\alpha$,
\begin{equation}
    \sigma_{\alpha,0}(z)
    =
    \frac{\pi e^2}{m_e(z)c}\,g_\alpha
    =
    \sigma_{\alpha,0}^{(0)}
    \left[
        \frac{m_e(z)}{m_{e,0}}
    \right]^{-1},
    \qquad
    \sigma_{\alpha,0}^{(0)}
    \equiv
    \frac{\pi e^2}{m_{e,0}c}\,g_\alpha .
    \label{eq:sigma_alpha0}
\end{equation}
For the hydrogen Ly$\alpha$ transition, $g_\alpha\simeq 0.4162$.
Thus, $\sigma_{\alpha,0}(z)$ sets the overall strength of the transition, 
while all line-shape information is contained in $\phi_\alpha$.  
The detailed form of the Ly$\alpha$ line profile, including Doppler broadening, natural broadening, the resulting Voigt profile, and their leading mass scalings in axiodilaton cosmology, is presented in Appendix~\ref{app:lya_profile}.

We now return to the total scattering rate in Eq.~(\ref{eq:Palpha_def}). Substituting Eq.~(\ref{eq:sigma_alpha_profile}) into Eq.~(\ref{eq:Palpha_def}), we obtain
\begin{equation}
P_\alpha(\mathbf{r},z)
=
4\pi\,\sigma_{\alpha,0}(z)
\int d\nu\,
\frac{J_\nu(\mathbf{r},z)}{h\nu}\,
\phi_\alpha\!\bigl(\nu-\nu_\alpha(z),\mathbf{r},z\bigr).
\label{eq:Palpha_profile_form}
\end{equation}
Because the Ly$\alpha$ line profile is sharply peaked around the resonance frequency $\nu_\alpha(z)$, the integral is dominated by a narrow interval near line center, over which $J_\nu(\mathbf{r},z)/(h\nu)$ may be treated as approximately constant. Using the normalization of the line profile, one finds the narrow-line estimate
\begin{equation}
P_\alpha(\mathbf{r},z)
\simeq
4\pi\,
\frac{J_{\alpha}(\mathbf{r},z)}{h\nu_\alpha(z)}
\,\sigma_{\alpha,0}(z)\,
S_\alpha(\mathbf{r},z),
\label{eq:Palpha_narrowline}
\end{equation}
where $J_{\alpha}(\mathbf{r},z)$ is given in Eq.~\eqref{eq:Jalpha_def} and $S_\alpha(\mathbf{r},z)$ is a dimensionless factor of order unity that accounts for the fact that, after many Ly$\alpha$ scatterings, the radiation spectrum near the line center is no longer perfectly smooth. Recoil slightly shifts photons to lower frequencies, while thermal motions of atoms cause photons to diffuse in frequency space. Together, these effects modify the local spectral shape around $\nu_\alpha$ and $S_\alpha(\mathbf{r},z)$ summarizes their impact.
The detailed expression is given in Appendix~\ref{sec:Salpha}.

\subsubsection{Line Profile}
\label{app:lya_profile}
In this appendix, we summarize the detailed form of the Ly$\alpha$ line profile used in the main text. We describe the Doppler broadening induced by the thermal motion of hydrogen atoms, the natural broadening associated with the finite lifetime of the upper level, and the resulting Voigt profile. We also record the leading mass scalings relevant in axiodilaton cosmology.

If the atoms in the gas undergo thermal motions due to the kinetic temperature $T_k$, then in the local rest frame of the gas (i.e.\ the frame in which the bulk velocity of the IGM patch vanishes) different atoms have different line-of-sight thermal velocities. Because of the Doppler effect, a photon with frequency $\nu$ in this gas rest frame is seen in the rest frame of an individual atom at a slightly shifted frequency $\nu'$, so that the resonance condition $\nu'=\nu_\alpha$ is satisfied for slightly different values of $\nu$. Equivalently, some atoms scatter photons with frequencies slightly higher than $\nu_\alpha$, while others scatter photons with slightly lower than $\nu_\alpha$. The resulting line profile in the gas rest frame is broadened into a Gaussian profile
\begin{equation}
\phi_D(\tilde\nu,\mathbf{r},z)=\frac{1}{\Delta\nu_D(\mathbf{r},z)\sqrt{\pi}}
\exp\!\left[-\frac{\tilde \nu^2}{\Delta\nu_D(\mathbf{r},z)^2}\right],
\end{equation}
where the quantity $\Delta\nu_D(\mathbf{r},z)$ denotes the Doppler width of the Ly$\alpha$ line
\begin{equation}
\Delta\nu_D(\mathbf{r},z)
=\frac{\nu_\alpha(z)}{c}\,
\sqrt{\frac{2k_B T_k(\mathbf{r},z)}{m_H(z)}}\;,
\label{eq:Doppler_width_def}
\end{equation}
where $m_H(z)$ is the hydrogen-atom mass, which we approximate as $m_H(z)\simeq m_p(z)$.
In an axiodilaton cosmology, the Doppler width inherits the mass dependence through $\nu_\alpha(z)$ and $m_H(z)$, giving the leading scaling
\begin{equation}
\Delta\nu_D(\mathbf{r},z)
\ \propto\ 
m_e(z)\,\sqrt{\frac{T_k(\mathbf{r},z)}{m_p(z)}}\;,
\label{eq:Doppler_width_scaling_axiodilaton}
\end{equation}
with additional implicit dependence through the evolved gas temperature $T_k(\mathbf{r},z)$.

Even for an atom at rest, an absorption or emission line is not perfectly sharp.
The upper level of the transition has a finite lifetime $t_\alpha$, which implies an intrinsic energy uncertainty, $\Delta E \sim \hbar/t_\alpha$. It translates into a finite linewidth in frequency space.
This intrinsic broadening is called natural (or lifetime) broadening and produces a Lorentzian line profile.
We parametrize the natural line profile by
\begin{equation}
\phi_L(\tilde \nu,z)=\frac{1}{\pi}\,
\frac{\Delta\nu_L(z)}{\tilde \nu^2+\Delta\nu_L^2(z)},
\label{eq:phiL_def}
\end{equation}
where $\Delta\nu_L(z)$ is related to the total decay rate of the upper level, $\Gamma_\alpha(z)\equiv t_\alpha^{-1}$, by
\begin{equation}
\Delta\nu_L(z)=\frac{\Gamma_\alpha(z)}{4\pi}.
\label{eq:DnuL_Gamma_relation}
\end{equation}
The natural width is governed by the total decay rate of the upper level.
For the hydrogen Ly$\alpha$ transition, the $2p\to1s$ channel dominates, so we take
\begin{equation}
\Gamma_\alpha(z)\ \simeq\ A_{2p\to 1s}(z),
\label{eq:Gamma_total}
\end{equation}
where $A_{2p\to 1s}$ is the Einstein coefficient for spontaneous emission in the Ly$\alpha$ transition. It is defined as the probability per unit time for an atom in the upper state $|2p\rangle$ to decay to $|1s\rangle$ while emitting a photon. 
For an electric-dipole allowed transition, it can be written as
\begin{equation}
A_{2p\to 1s}(z)
=
\frac{4\,\omega_{\alpha}^3(z)}{3\hbar c^3}\,
\bigl|\mathbf{d}_{\alpha}(z)\bigr|^2,
\label{eq:A21_def_specific}
\end{equation}
where $\omega_{\alpha}(z)\equiv (E_{2p}(z)-E_{1s}(z))/\hbar = 2\pi\nu_\alpha(z)$ is the transition angular frequency.
Here $\mathbf{d}_{\alpha}(z)$ denotes the transition electric-dipole matrix element for the Ly$\alpha$ line,
\begin{equation}
\mathbf{d}_{\alpha}(z)\equiv \langle 1s|\,\mathbf{d}\,|2p\rangle,
\qquad
\Big({\rm with}~\mathbf{d}\equiv -e\,\mathbf{r}_{ep}\Big)
\end{equation}
where $\mathbf{d}$ is the electric dipole moment operator of the hydrogen atom, and $\mathbf{r}_{ep}$ is the electron position operator in the proton rest frame.
The appearance of the dipole matrix element follows from the fact that, at wavelength much larger than the atomic size, the leading coupling of an atom to the radiation field is given by the electric--dipole interaction,
\begin{equation}
H_{\rm int}\simeq -\,\mathbf{d}\cdot \mathbf{E},
\end{equation}
where $\mathbf{E}$ is the electric-field operator of the ambient electromagnetic radiation evaluated at the atomic position.
For the Ly$\alpha$ transition, one finds the standard result
\begin{equation}
|\mathbf{d}_{\alpha}(z)|^2
=
\frac{2^{15}}{3^{10}}\,
\bigl(e\,a_0(z)\bigr)^2.
\label{eq:dalpha_exact}
\end{equation}
The Bohr radius scales as $a_0\propto m_e^{-1}$, so in an axiodilaton cosmology, to leading order,
\begin{equation}
|\mathbf{d}_{\alpha}(z)|\ \propto m_e^{-1}(z).
\label{eq:dalpha_me_scaling}
\end{equation}
Combining this with the scaling $\omega_\alpha=2\pi\nu_\alpha\propto m_e(z)$, we obtain
\begin{equation}
A_{2p\to 1s}(z)\ \propto\ \omega_\alpha^3(z)\,|\mathbf{d}_{\alpha}(z)|^2 \propto\  m_e(z),
\label{eq:A21_me_scaling}
\end{equation}
and hence 
\begin{equation}
\Delta\nu_L(z)  \simeq\ \frac{A_{2p\to 1s}(z)}{4\pi} \propto m_e(z).
\label{eq:DnuL_me_scaling}
\end{equation}

When both Doppler and natural broadening operate, the resulting line profile is not the sum $\phi_D+\phi_L$.
Instead, it is the convolution of the two independent broadening kernels, which yields the Voigt profile,
\begin{equation}
\phi_\alpha(\tilde\nu,\mathbf{r},z)
=
\int_{-\infty}^{\infty} d\tilde \nu'\,
\phi_D\!\bigl(\tilde \nu - \tilde \nu',\mathbf{r},z\bigr)\,
\phi_L(\tilde \nu',z),
\qquad
\Big({\rm with}~\int d\tilde \nu\,\phi_\alpha(\tilde \nu,\mathbf{r},z)=1 \Big)
\label{eq:voigt_convolution_tildenu}
\end{equation}
We introduce the dimensionless frequency variable and damping parameter,
\begin{equation}
x\equiv \frac{\tilde\nu}{\Delta\nu_D(\mathbf{r},z)},
\qquad
a_V\equiv \frac{\Delta\nu_L(z)}{\Delta\nu_D(\mathbf{r},z)}.
\label{eq:voigt_xa_def_tildenu}
\end{equation}
We then change integration variables in Eq.~\eqref{eq:voigt_convolution_tildenu} to
\begin{equation}
y \equiv \frac{\tilde\nu'}{\Delta\nu_D(\mathbf{r},z)},
\qquad
\tilde\nu'=\Delta\nu_D(\mathbf{r},z)\,y,
\qquad
d\tilde\nu'=\Delta\nu_D(\mathbf{r},z)\,dy \;.
\label{eq:voigt_changevar_t}
\end{equation}
Using $x-y =(\tilde\nu-\tilde\nu')/\Delta\nu_D(\mathbf{r},z)$ and $\Delta\nu_L(z)=a_V\,\Delta\nu_D(\mathbf{r},z)$, we obtain
\begin{equation}
\phi_\alpha(\tilde\nu,\mathbf{r},z)
=
\frac{1}{\Delta\nu_D(\mathbf{r},z)\sqrt{\pi}}
\left[
\frac{a_V}{\pi}\int_{-\infty}^{\infty} dy\,
\frac{e^{-(x-y)^2}}{y^2+a_V^2}
\right].
\label{eq:voigt_intermediate}
\end{equation}
This motivates the standard definition of the (dimensionless) Voigt (Hjerting) function,
\begin{equation}
H(a_V,x)\equiv
\frac{a_V}{\pi}\int_{-\infty}^{\infty} dy\,
\frac{e^{-(x-y)^2}}{y^2+a_V^2}.
\label{eq:Hax_def1}
\end{equation}
By the change of variables $u\equiv x-y$ (so that $y=x-u$ and $dy=-du$), one obtains the equivalent representation
\begin{equation}
H(a_V,x)=
\frac{a_V}{\pi}\int_{-\infty}^{\infty} du\,
\frac{e^{-u^2}}{(x-u)^2+a_V^2} \;.
\label{eq:Hax_def2}
\end{equation}
The Voigt profile in Eq.~\eqref{eq:voigt_convolution_tildenu} takes the compact form
\begin{equation}
\phi_\alpha(\tilde\nu,\mathbf{r},z)
=
\frac{1}{\Delta\nu_D(\mathbf{r},z)\sqrt{\pi}}\,
H(a_V,x)\;.
\label{eq:voigt_final}
\end{equation}
Note that $a_V$ is typically small in the IGM, so the line core is close to Gaussian, $H(a_V,x)\to e^{-x^2}$ for $a_V\to 0$. In an axiodilaton cosmology, the leading explicit scaling of $a_V=\Delta\nu_L/\Delta\nu_D$ becomes
\begin{equation}
a_V\ \propto\ \sqrt{\frac{m_p(z)}{T_k(\mathbf{r},z)}}.
\label{eq:a_scaling}
\end{equation}
%

\subsubsection{Spectral Suppression $S_\alpha$ and Color Temperature $T_\alpha$}
\label{sec:Salpha}
The scattering rate $P_\alpha(\mathbf{r},z)$ in Eq.~\eqref{eq:Palpha_narrowline} is proportional to the radiation spectrum at the Ly$\alpha$ line center, and repeated resonant scattering suppresses the spectrum exactly there. The correction factor $S_\alpha(\mathbf{r},z)\leq 1$ is this suppression factor~\cite{Chen:2003gc,Hirata:2005mz}. \FirstClass\ evaluates $S_\alpha(\mathbf{r},z)$ with the fitting formula of Hirata~\cite{Hirata:2005mz}, whose coefficients were calibrated at the present-day masses. We therefore replace the fitting formula with the wing approximation of Chuzhoy and Shapiro~\cite{Chuzhoy:2005wv}, in the series form of Ref.~\cite{Mittal:2020kjs}, with the mass dependence kept explicit. The wing approximation expresses $S_\alpha(\mathbf{r},z)$ through three quantities, the number of scatterings, the recoil shift per scattering, and the strength of spin exchange.
 
A Ly$\alpha$ photon emitted above $\nu_\alpha(z)$ redshifts into the narrow line profile in Eq.~\eqref{eq:sigma_alpha_profile}, and the mean number of scatterings during the passage through the line equals the Gunn--Peterson optical depth~\cite{Gunn1965,Furlanetto:2006jb},
\begin{equation}
\tau_{\rm GP}(\mathbf{r},z)
=
\frac{3\,c^3\,n_{\rm HI}({\bf r},z)\,A_{2p\to 1s}(z)}
{8\pi\,\nu_\alpha^3(z)\,H(z)},
\label{eq:tauGP}
\end{equation}
where $n_{\rm HI}(\mathbf{r},z)=x_{\rm HI}(\mathbf{r},z)\,n_{H}(\mathbf{r},z)$ is the neutral hydrogen number density. The scaling follows from $\nu_\alpha(z)\propto m_e(z)$ in Eq.~\eqref{eq:nualpha_mass_scaling} and from $A_{2p\to 1s}(z)\propto m_e(z)$ in Eq.~\eqref{eq:A21_me_scaling}.
 
Each scattering shifts the photon to a slightly lower frequency through the recoil of the atom. Over the $\tau_{\rm GP}(\mathbf{r},z)$ scatterings these shifts add up and push photons through the line center faster than the redshift alone, so photons spend less time there and the spectrum at the line center is suppressed. The recoil shift per scattering, in units of the Doppler width of Eq.~\eqref{eq:Doppler_width_def}, defines the recoil parameter
\begin{equation}
\eta(\mathbf{r},z)
=
\frac{h\,\nu_\alpha^2(z)}{m_H(z)\,c^2\,\Delta\nu_D(\mathbf{r},z)}.
\label{eq:recoil_eta}
\end{equation}
The scaling follows from $\nu_\alpha^2(z)\propto m_e^2(z)$, from $m_H(z)\simeq m_p(z)$, and from $\Delta\nu_D(\mathbf{r},z)$ in Eq.~\eqref{eq:Doppler_width_scaling_axiodilaton}.
 
A scattering can also flip the hyperfine state, and the photon then gains or loses the energy $h\nu_{21}(z)$. This spin exchange adds a frequency shift on top of the recoil and modifies the suppression, with the size set by the spin-exchange temperature,
\begin{equation}
T_{\rm se}(z)
=
\frac{2}{9}\,T_k(\mathbf{r},z)
\left[\frac{\nu_{21}(z)}{\Delta\nu_D(\mathbf{r},z)}\right]^{2}
=
\frac{m_H(z)\,c^2}{9\,k_B}
\left[\frac{\nu_{21}(z)}{\nu_\alpha(z)}\right]^{2}.
\label{eq:Tse}
\end{equation}
The factor $T_k(\mathbf{r},z)$ cancels in the second equality, so $T_{\rm se}(z)$ is an atomic constant with the value $0.4055\,{\rm K}$ at the present-day masses. The scaling follows from $\nu_{21}(z)\propto m_e^2(z)\,m_p^{-1}(z)$ in Eq.~\eqref{eq:nu21_mass_scaling} and from $\nu_\alpha(z)\propto m_e(z)$.
 
The suppressed spectrum follows from the radiative transfer equation near the line center~\cite{Chuzhoy:2005wv}. In the wing region $|x|\gg 1$, where $x=[\nu-\nu_\alpha(z)]/\Delta\nu_D(\mathbf{r},z)$ is the frequency offset from the line center in Doppler widths, the Voigt function in Eq.~\eqref{eq:Hax_def2} reduces to $H(a_V,x)\simeq a_V/(\sqrt{\pi}\,x^2)$, with $a_V=\Delta\nu_L(z)/\Delta\nu_D(\mathbf{r},z)$ the ratio of the natural to the Doppler width in Eq.~\eqref{eq:voigt_xa_def_tildenu}. The transfer equation with this wing profile has an analytic solution, and the three quantities enter the solution through
\begin{equation}
\gamma'(\mathbf{r},z)
=
\frac{\tau_{\rm GP}^{-1}(\mathbf{r},z)}{1+T_{\rm se}(z)/T_k(\mathbf{r},z)},
\qquad
\eta'(\mathbf{r},z)
=
\eta(\mathbf{r},z)\,
\frac{1+T_{\rm se}(z)/T_s(\mathbf{r},z)}{1+T_{\rm se}(z)/T_k(\mathbf{r},z)}
-\frac{\Delta\nu_D(\mathbf{r},z)}{\nu_\alpha(z)},
\label{eq:Salpha_wing_params}
\end{equation}
where $T_s(\mathbf{r},z)$ is the spin temperature in Eq.~\eqref{eq:Ts}. The closed form of Ref.~\cite{Mittal:2020kjs} depends on the single variable
\begin{equation}
\beta(\mathbf{r},z)
=
\eta'(\mathbf{r},z)
\left[\frac{3\,a_V}{2\pi\,\gamma'(\mathbf{r},z)}\right]^{1/3},
\label{eq:Salpha_beta}
\end{equation}
and gives
\begin{equation}
S_\alpha(\mathbf{r},z)
\simeq
1
-\frac{4\pi}{3\sqrt{3}\,\Gamma(2/3)}\,\beta(\mathbf{r},z)
+\frac{8\pi}{3\sqrt{3}\,\Gamma(1/3)}\,\beta^2(\mathbf{r},z)
-\frac{4}{3}\,\beta^3(\mathbf{r},z).
\label{eq:Salpha_wing}
\end{equation}
 
The mass dependence of $S_\alpha(\mathbf{r},z)$ enters through the scalings of $\tau_{\rm GP}(\mathbf{r},z)$, $\eta(\mathbf{r},z)$, $T_{\rm se}(z)$, and $a_V$ in Eq.~\eqref{eq:a_scaling}, with additional implicit dependence through the evolved $T_k(\mathbf{r},z)$ and $T_s(\mathbf{r},z)$. The simulation evaluates Eqs.~\eqref{eq:tauGP}--\eqref{eq:Salpha_wing} with the full redshift-dependent masses.

The quantity $T_\alpha(\mathbf{r},z)$ appearing in Eq.~(\ref{eq:Ts}) is the color temperature of the radiation field in the vicinity of the Ly$\alpha$ resonance. It is defined from the local spectral slope of the radiation field as
\begin{equation}
\frac{1}{T_\alpha(\mathbf{r},z)}
\equiv
-\,\frac{k_B}{h}\,
\left.
\frac{d\ln J_\nu(\mathbf{r},z)}{d\nu}
\right|_{\nu=\nu_\alpha(z)}.
\label{eq:Talpha_def}
\end{equation}
Repeated resonant scatterings drive the photon spectrum toward a quasi-thermal shape around line center. As a result, the color temperature tends to approach the gas kinetic temperature
\begin{equation}
T_\alpha(\mathbf{r},z)\ \to\ T_k(\mathbf{r},z),
\label{eq:Talpha_to_Tk}
\end{equation}
up to a correction from spin exchange.
\begin{equation}
\frac{1}{T_\alpha(\mathbf{r},z)}
=
\frac{1}{T_k(\mathbf{r},z)}
+
\frac{T_{\rm se}(z)}{T_k(\mathbf{r},z)}
\left[
\frac{1}{T_s(\mathbf{r},z)}
-\frac{1}{T_k(\mathbf{r},z)}
\right].
\label{eq:Talpha_Tse}
\end{equation}
\FirstClass\ evaluates this relation in the spin-temperature iteration with the rescaled $T_{\rm se}(z)$. Both $S_\alpha(\mathbf{r},z)$ and $T_\alpha(\mathbf{r},z)$ enter the spin temperature in Eq.~\eqref{eq:Ts}, with $S_\alpha(\mathbf{r},z)$ acting through $x_\alpha(\mathbf{r},z)$ in Eq.~\eqref{eq:xalpha_explicit} and $T_\alpha(\mathbf{r},z)$ entering directly.

\subsection{Ly$\alpha$ Background Intensity $J_\alpha$}
\label{app:Jalpha}

In this section, we derive detailed expressions for the two contributions
to the Ly$\alpha$ intensity introduced in Eq.~\eqref{eq:Jalpha_split}:
the stellar UV contribution $J_{\alpha,*}$ and the X-ray-induced
contribution $J_{\alpha,X}$. 
For the calculations below, it is convenient to introduce the corresponding
photon-number specific intensity,
\begin{equation}
\widehat J_\alpha(\mathbf r,z)
\equiv
\frac{J_\alpha(\mathbf r,z)}{h\nu_\alpha(z)} .
\label{eq:Jalpha_hat_def}
\end{equation}
The quantity $\widehat J_\alpha$ has units
${\rm s}^{-1}\,{\rm cm}^{-2}\,{\rm Hz}^{-1}\,{\rm sr}^{-1}$.
Applying the same definition to each contribution, the total photon-number
specific intensity can be written as
\begin{equation}
\widehat J_\alpha(\mathbf r,z)
=
\widehat J_{\alpha,*}(\mathbf r,z)
+
\widehat J_{\alpha,X}(\mathbf r,z).
\label{eq:Jalpha_split_app}
\end{equation}
The quantity $\widehat J_{\alpha,*}$ represents the intergalactic
Ly$\alpha$ background produced as stellar UV photons emitted between
Ly$\alpha$ and the Lyman limit.
Here, the Lyman limit corresponds to the ionization threshold of
ground-state hydrogen.
This contribution arises through two channels. First, photons emitted between the Ly$\alpha$ ($2p\leftrightarrow 1s$) and Ly$\beta$ ($3p\leftrightarrow 1s$) frequencies simply redshift into the local Ly$\alpha$ resonance without encountering another Lyman-series resonance. 
Second, photons emitted between Ly$\beta$ and the Lyman limit are typically absorbed in a higher Lyman-series transition, Ly$n$ ($1s\leftrightarrow np$) with $n\ge 3$, and subsequently cascade through lower levels. Only a fraction of these cascades eventually passes through the Ly$\alpha$ transition and therefore contributes to the Wouthuysen--Field coupling $x_\alpha$.

\subsubsection{Stellar Contribution}
\label{sec:lyman_alpha_flux}

The proper Ly$\alpha$ intensity from stellar sources is given by~\cite{Furlanetto:2006jb}:
\begin{equation}
\bar{\widehat J}_{\alpha,*}(z) = \frac{(1+z)^2}{4\pi} \sum_{n=2}^{n_{\rm max}} f_{\rm recycle}(n) \int_z^{z_n^{\rm max}} dz' \frac{c}{H(z')} \epsilon_\alpha(\nu'_n, z'),
\label{eq:J_alpha_stellar}
\end{equation}

The comoving Ly$\alpha$ emissivity can be written as:
\begin{equation}
\epsilon_\alpha(\nu, z) = \epsilon_*(\nu) \cdot \rho_*(z)
\label{eq:epsilon_alpha}
\end{equation}
where $\epsilon_*(\nu)$ is the spectral distribution function and $\rho_*(z)$ is the comoving star formation rate density, counted in baryons per unit comoving volume per unit time.

The spectral distribution follows a power law between Lyman lines:
\begin{equation}
\epsilon_*(\nu) = N_n \frac{\beta_n + 1}{\nu_{n+1}^{\beta_n+1} - \nu_n^{\beta_n+1}} \nu^{\beta_n},
\label{eq:epsilon_spectrum}
\end{equation}
for $\nu_n \leq \nu \leq \nu_{n+1}$, where $\beta_n$ is the spectral index.

At fixed $N_n,\beta_n, z_n^{\rm max}, H(z'),$ and $\rho_*$, Eqs.~\eqref{eq:J_alpha_stellar}--\eqref{eq:epsilon_spectrum} give
\begin{equation}
\bar{\widehat J}_{\alpha,*}(z) \propto m_e^{-1}.
\label{eq:J_alpha_approx}
\end{equation}

Eq.~\eqref{eq:J_alpha_stellar} is the homogeneous limit of the shell form below, with $\delta_{\ssB,R_{\rm sh}}^{\rm NL}\to0$ and $f_{\rm coll,*}^{\rm eff}$ replaced by its volume average. \FirstClass\ evaluates the shell form.

It is therefore useful to decompose the stellar contribution into the sum over higher Lyman resonances,
\begin{equation}
\widehat J_{\alpha,*}(\mathbf r,z)
=
\sum_{n=2}^{n_{\rm max}}
\widehat J_{\alpha}^{(n)}(\mathbf r,z),
\label{eq:Jalpha_star_sum}
\end{equation}
with
\begin{equation}
\widehat J_{\alpha}^{(n)}(\mathbf r,z)
=
f_{\rm recycle}(n)
\int_{z}^{z_n^{\rm max}}
dz'\,
\frac{1}{4\pi}
\frac{1}{4\pi r_p^2(z,z')}
\frac{d\phi_{*,e}(\nu'_n,\mathbf r,z,z')}{dz'} .
\label{eq:Jalpha_star_n}
\end{equation}
Here $f_{\rm recycle}(n)$ is the probability that a photon absorbed in the
Ly$n$ resonance eventually produces a Ly$\alpha$ photon through radiative
cascade. For the direct Ly$\alpha$ channel, $f_{\rm recycle}(2)=1$, while
higher Lyman resonances have smaller recycling probabilities determined by
atomic branching ratios.
The geometric dilution factor $[4\pi r_p^2(z,z')]^{-1}$ is expressed in
terms of the proper radial separation $r_p(z,z')$, defined in
Eq.~\eqref{eq_rp_def}, between a source at redshift $z'$ and the target
location at redshift $z$.

The quantity
$d\phi_{*,e}(\nu'_n,\mathbf r,z,z')/dz'$ denotes the contribution to the stellar UV photon emission rate, with units
${\rm s}^{-1}\,{\rm Hz}^{-1}$, from all sources in the shell
$[z',z'+dz']$. 
These sources emit photons at frequency $\nu'_n=\nu'_n(z')$, which
subsequently redshift into the local Ly$n$ resonance at the target point
$(\mathbf r,z)$. To determine this emission frequency, we first identify
the photon frequency $\nu_n(z)$ corresponding to the local Ly$n$
resonance at redshift $z$. Tracing this frequency back to the source
redshift $z'$ using cosmological redshifting gives
\begin{equation}
\nu'_n(z')
=
\nu_n(z)
\left(\frac{1+z'}{1+z}\right).
\label{eq:nuprime_n_app}
\end{equation}
The differential
photon flux contributed by sources at redshift $z'$ is given by
\begin{equation}
\frac{d\phi_{*,e}(\nu_n',\mathbf r,z,z')}{dz'}
=
f_\ast^{\rm II}\,
\epsilon_*(\nu_n')
\Bigg(
\frac{\Omega_{\ssB,0}\bar\rho_{\rm crit,0}}{m_{p,0}}
\left[
1+\delta_{\ssB,R_{\rm sh}}^{\rm NL}(\mathbf r,z')
\right]
\frac{dV(z,z')}{dz'}
\Bigg)
\dfrac{H(z')}{t_\ast}\,f_{\rm coll,*}^{\rm eff}(\mathbf r,z',R_{\rm sh})
\label{eq_dphistar_dz}
\end{equation}
Here $f_\ast^{\rm II}$ denotes the Pop II star-formation efficiency
defined in Eq.~\eqref{eq:astro_params_schematic}. The function
$\epsilon_*(\nu_n')$ gives the number of stellar UV photons emitted
per unit frequency per baryon converted into stars, evaluated at the
emission frequency $\nu_n'$, and has units ${\rm Hz}^{-1}$. For
$\epsilon_*(\nu_n')$, we adopt the Pop II spectral template
implemented in \FirstClass. This template is the piecewise power law of Eq.~\eqref{eq:epsilon_spectrum}. The code takes $\beta_n$ from a table and sets $N_n$ to $N_\gamma$ of Eq.~\eqref{eq:astro_params_schematic} times a tabulated number for each band.
The factor $\left[\Omega_{\ssB,0}\bar\rho_{\rm crit,0}/m_{p,0}\right] \left[1+\delta_{\ssB,R_{\rm sh}}^{\rm NL}(\mathbf r,z')\right]dV/dz'$ 
gives the number of baryons contained in the source shell per unit
source redshift. \FirstClass\ counts baryons as $\bar n_H(0)+\bar n_{\rm He}(0)$ in this factor. The shell scale $R_{\rm sh}$ and the
volume element $dV(z,z')/dz'$ are defined in
Eq.~\eqref{eq:source_shell_geometry}.
The effective collapsed fraction
$f_{\rm coll,*}^{\rm eff}$ is defined in
Eq.~\eqref{eq_fcoll_star}. Here $t_\ast$ is the star-formation timescale as a fraction of the Hubble time, Eq.~\eqref{eq:astro_params_schematic}.

In axiodilaton cosmology, the proper propagation distance $r_p(z,z')$ and the shell volume element $dV(z,z')/dz'$ depend on the modified Hubble rate $H(z)$, while the source modulation and abundance are altered through the baryon overdensity $\delta_{\ssB,R_{\rm sh}}^{\rm NL}$ and the collapsed fraction $f_{\rm coll,*}^{\rm eff}$. In the present implementation, however, we do not attempt to recompute the stellar population model entering $\epsilon_*(\nu'_n)$ under varying particle masses, and instead keep $N_n$ and $\beta_n$ in Eq.~\eqref{eq:epsilon_spectrum} fixed. The band edges $\nu_n(z)$ in Eq.~\eqref{eq:nu_n_rydberg} scale as $m_e(z)$, so $\epsilon_*(\nu_n')$ scales as $m_e^{-1}(z)$.

The emission frequency $\nu'_n(z')$ at which $\epsilon_*(\nu'_n)$ is evaluated
is determined by the local Ly$n$ resonance frequency $\nu_n(z)$ through
Eq.~\eqref{eq:nuprime_n_app}.
For the Lyman series, whose lines correspond to transitions
between the ground state and the excited $np$ levels,
$1s\leftrightarrow np$, the resonance frequencies obey the Rydberg relation
\begin{equation}
\nu_n(z)
=
\nu_{\rm HI}(z)
\left(1-\frac{1}{n^2}\right),
\label{eq:nu_n_rydberg}
\end{equation}
For $n=2$,
Eq.~\eqref{eq:nu_n_rydberg} gives the Ly$\alpha$ frequency,
\begin{equation}
\nu_{n=2}(z)
\equiv
\nu_\alpha(z)
=
\frac{3}{4}\,\nu_{\rm HI}(z).
\end{equation}
In the axiodilaton scenario, all Lyman-series resonance
frequencies vary with redshift according to
\begin{equation}
\nu_n(z)
=
\nu_{n,0}
\left[
    \frac{m_e(z)}{m_{e,0}}
\right]
=
\nu_{n,0}e^{\bfg\Delta\bar\chi(z)},
\end{equation}
where $\nu_{n,0}$ denotes the present-day rest-frame frequency of the
corresponding transition.

As indicated by the upper integration limit $z_n^{\rm max}$ in Eq.~\eqref{eq:Jalpha_star_n}, not all photons emitted at frequency $\nu'_n(z')$ contribute to the Ly$n$ channel at redshift $z$.
For such a photon to contribute to the Ly$n$ channel, it must not encounter the next
higher Lyman resonance, Ly$(n+1)$, before reaching Ly$n$. This requires the emitted
frequency at redshift $z'$ to lie below the local Ly$(n+1)$ resonance frequency at the same redshift $z'$,
\begin{equation}
\nu'_n(z')<\nu_{n+1}(z').
\label{eq:znmax_ineq1}
\end{equation}
Using Eq.~\eqref{eq:nuprime_n_app}, this condition can be rewritten as
\begin{equation}
\frac{\nu_n(z)}{1+z}
<
\frac{\nu_{n+1}(z')}{1+z'}.
\label{eq:znmax_ineq2}
\end{equation}
The maximum source redshift $z_n^{\rm max}$ is determined by
\begin{equation}
\frac{\nu_n(z)}{1+z}
=
\frac{\nu_{n+1}\!\left(z_n^{\rm max}\right)}{1+z_n^{\rm max}} .
\label{eq:zmax_general}
\end{equation}
In the standard $\Lambda$CDM limit, substituting
Eq.~\eqref{eq:nu_n_rydberg} into Eq.~\eqref{eq:zmax_general} gives
\begin{equation}
1+z_n^{\rm max}(z)
=
(1+z)\,
\frac{1-(n+1)^{-2}}{1-n^{-2}} .
\label{eq:zmax_LCDM}
\end{equation}
In axiodilaton cosmology, the Lyman-series resonance frequencies vary
with redshift according to
$\nu_n(z)=\nu_{n,0}e^{\bfg\Delta\bar\chi(z)}$. Substituting this relation into
Eq.~\eqref{eq:zmax_general} gives
\begin{equation}
1+z_n^{\rm max}(z)
=
(1+z)\,
\frac{1-(n+1)^{-2}}{1-n^{-2}}\,
e^{\bfg\left[
\Delta\bar\chi\!\left(z_n^{\rm max}\right)
-\Delta\bar\chi(z)
\right]}.
\label{eq:zmax_axiodilaton}
\end{equation}
The maximum source redshift $z_n^{\rm max}(z)$ can then be obtained by
solving this implicit equation numerically for each $z$ and $n$.

The recycling fraction $f_{\rm recycle}(n)$ in Eq.~\eqref{eq:Jalpha_star_n} denotes the probability that a photon absorbed in the Ly$n$ resonance eventually produces a Ly$\alpha$ photon through the subsequent radiative cascade. It is therefore determined by the atomic selection rules and by the branching ratios among the allowed decay channels of the excited hydrogen levels. For Ly$\alpha$ itself, one has $f_{\rm recycle}(2)=1$, since excitation to the $2p$ state is followed by the allowed decay $2p\rightarrow 1s$, which is precisely the Ly$\alpha$ transition. By contrast, Ly$\beta$ absorption excites the $3p$ state, whose allowed radiative decays do not pass through the $2p\rightarrow 1s$ Ly$\alpha$ channel. Instead, the decay proceeds either directly back to $1s$ or through the $2s$ state, so that no Ly$\alpha$ photon is produced, implying $f_{\rm recycle}(3)=0$. For higher Lyman-series transitions with $n\ge 4$, some cascade paths pass through $2p$ while others do not, so that $0<f_{\rm recycle}(n)<1$. For large $n$, the recycling fractions approach an asymptotic value $f_{\rm recycle}(n)\simeq 0.36$. In practice, the sum over $n$ converges rapidly, and one may truncate Eq.~(\ref{eq:Jalpha_star_sum}) at a finite $n_{\rm max}$.

In the present work, we do not recompute the recycling fractions in the axiodilaton theory. Since $f_{\rm recycle}(n)$ is determined by dimensionless branching ratios and atomic selection rules, we treat it as unchanged at leading order under the universal rescaling of the hydrogenic energy levels considered here.

\subsubsection{X-ray-Induced Contribution}
\label{app:Jalpha_X}

The X-ray-induced contribution to the Ly$\alpha$ background is obtained in close analogy with the X-ray heating rate discussed in Appendix~\ref{subsec:xray}. The same X-ray photon field that heats and ionizes the  neutral IGM also produces energetic primary electrons through photoionization of HI, HeI, and HeII. A fraction of the kinetic energy of these primary electrons is subsequently deposited into atomic excitations, and part of those excitations ultimately yields HI Ly$\alpha$ photons. 
The resulting X-ray-induced contribution to the Ly$\alpha$ photon number
intensity in Eq.~\eqref{eq:Jalpha_split_app} can be written as
\begin{align}
\widehat J_{\alpha,X}(\mathbf r,z)
=
&
\frac{c\,n_H(\mathbf r,z)(1+z)}
{4\pi \nu_\alpha(z)}
\int_{\nu_X}^{\infty} d\nu
\sum_{i\in\{{\rm HI},{\rm HeI},{\rm HeII}\}}
x_i^X(\mathbf r,z)\,
\sigma_i(\nu,z)\,
\widetilde{\mathcal J}_X(\mathbf r,\nu,z)
\nonumber\\
&
\times
\left(h\nu-E_i^{\rm th}(z)\right)
\frac{
f_{\rm Ly\alpha}
\left(
h\nu-E_i^{\rm th}(z),
x_e^{\rm X}(\mathbf r,z)
\right)
}
{h\nu_\alpha(z)},
\label{eq_Jalpha_X_hat}
\end{align}
where the X-ray photon field $\widetilde{\mathcal J}_X(\mathbf r,\nu,z)$ is defined in Eq.~\eqref{eq_JX_def} and $f_{\rm Ly\alpha}(E,x_e)$ is the dimensionless fraction of the primary
photoelectron energy deposited into Ly$\alpha$ photons, evaluated at the
local free-electron fraction. For $f_{\rm Ly\alpha}$, we follow the standard
tabulated prescription adopted in \FirstClass~\cite{Furlanetto:2009uf}. As for $f_{\rm heat}$ in Appendix~\ref{subsec:xray}, we keep this table at the present-day masses. The code evaluates the frequency integral in Eq.~\eqref{eq_Jalpha_X_hat} with the same limits as the integral in Eq.~\eqref{eq_eps_X_shell}.

\begin{figure}[t!]
    \centering
    \includegraphics[width=1.0\linewidth]{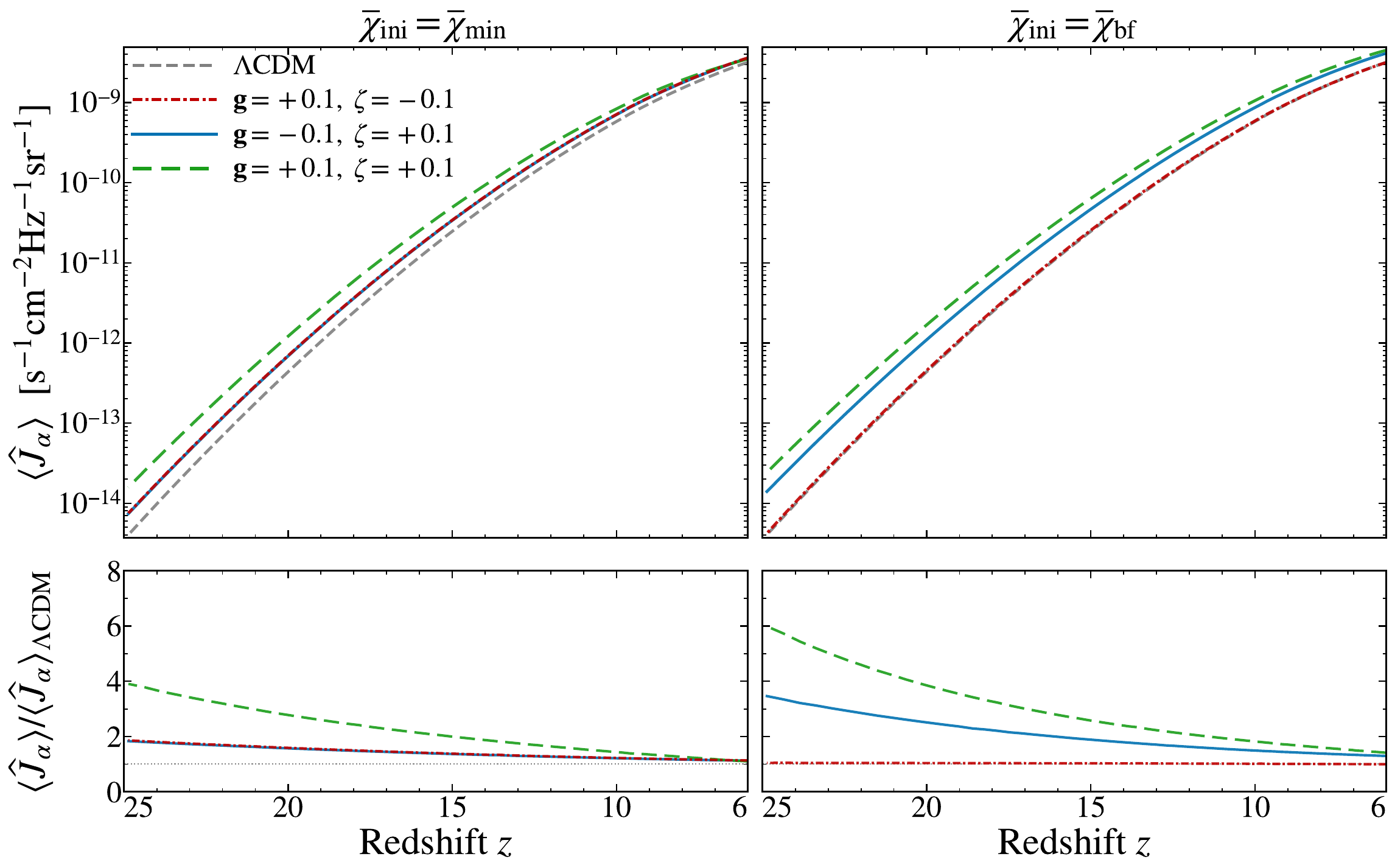}
    \caption{Redshift evolution of the volume-averaged Ly$\alpha$ photon number
    intensity, $\langle\widehat J_\alpha\rangle$, defined in
    Eq.~\eqref{eq:Jalpha_hat_def}, for
    $\bar\chi_{\rm ini} = \bar\chi_{\rm min} = 74.010$ (left) and $\bar\chi_{\rm ini} = \bar\chi_{\rm bf}=74.213$ (right).
    The upper panels show the total Ly$\alpha$ background in units of
    ${\rm s}^{-1}\,{\rm cm}^{-2}\,{\rm Hz}^{-1}\,{\rm sr}^{-1}$, 
    while the lower panels show its ratio to the \LCDM\ result. The gray dashed
    curves represent \LCDM. The red dash-dotted, blue solid, and green dashed
    curves correspond to $(\bfg,\zeta)=(+0.1,-0.1)$, $(-0.1,+0.1)$, and
    $(+0.1,+0.1)$, respectively.}
    \label{fig:Jalpha_hat}
\end{figure}

Fig.~\ref{fig:Jalpha_hat} shows that the volume-averaged Ly$\alpha$
background increases rapidly toward lower redshift in all models as the
population of star-forming halos grows and produces larger stellar UV and
X-ray emissivities. The axiodilaton models generally predict a larger
Ly$\alpha$ background than \LCDM. For
$\bar\chi_{\rm ini}=\chimin$, the red and blue curves are nearly degenerate,
reaching approximately twice the \LCDM\ value at $z\simeq25$, while the
green curve reaches a ratio of about four. The similarity between the red
and blue cases follows their equal coupling magnitudes and common negative
value of $\bfg\zeta$. In the green case, $\bfg\zeta>0$ enhances the mixed
axion--baryon gravitational coupling, $G_{aB}>G_N$, in Eq.~\eqref{eq:effective_gravitational_couplings} leading to a larger
collapsed fraction and greater source emissivity. 

The separation becomes more pronounced for
$\bar\chi_{\rm ini}=\chibf$. The red curve remains close to the \LCDM\
result, whereas the blue and green curves reach ratios of approximately
$3.5$ and $6$, respectively, at $z\simeq25$. The overall ordering is
governed primarily by the altered growth of matter perturbations and the
resulting source formation history. Toward lower redshifts, source formation
becomes less sensitive to these differences in growth, and the ratios
gradually approach unity.

\subsection{Ly$\alpha$ Coupling and Heating}
\label{app:lya_coupling_heating}

\subsubsection{Wouthuysen--Field Coupling Coefficient}
\label{subsec:xalpha}

We now turn to the Wouthuysen--Field coupling coefficient $x_\alpha$ that enters the spin-temperature relation in Eq.~(\ref{eq:Ts}). The essential point is that Ly$\alpha$ scattering does not produce a 21\,cm photon directly. Instead, a hydrogen atom initially in one of the two $1s$ hyperfine ground states can absorb a Ly$\alpha$ photon, be excited to the $2p$ states, and then decay back to either of two $1s$ hyperfine states while emitting another UV photon in the Ly$\alpha$ line. As a result, a single Ly$\alpha$ absorption--re-emission cycle can convert a triplet atom into a singlet atom, or vice versa. The role of this process is therefore to redistribute the populations of the ground-state triplet and singlet levels. Since the ratio of these populations, $n_1/n_0$ determines the spin temperature $T_s$, Ly$\alpha$ scattering provides an efficient mechanism for driving $T_s$ away from the CMB temperature and toward the color temperature of the Ly$\alpha$ radiation field.

Since $P_\alpha(\mathbf{r},z)$ in Eq.~\eqref{eq:Palpha_narrowline} denotes the total Ly$\alpha$ scattering rate per atom, with units of ${\rm s}^{-1}$, only a fraction of these scatterings leads to the indirect transition from the triplet to the singlet state. In the standard IGM limit, this gives
\begin{equation}
P_{10}(\mathbf{r},z)
=
\frac{4}{27}\,P_\alpha(\mathbf{r},z).
\label{eq:P10_Palpha_relation}
\end{equation}
The Wouthuysen--Field coupling coefficient $x_\alpha$ measures the strength with which Ly$\alpha$ scattering mixes the hyperfine populations, relative to the intrinsic 21\,cm radiative transition. 
Since $P_{10}(\mathbf{r},z)$ denotes the Ly$\alpha$-induced indirect transition rate from the triplet to the singlet state, while $A_{10}(z)$ in Eq.~\eqref{eq:A10_mass_scaling} sets the spontaneous 21\,cm decay rate, the natural dimensionless measure of this coupling is
\begin{equation}
x_\alpha(\mathbf{r},z)
\equiv
\frac{T_\star(z)}{\bar T_\gamma(z)\,A_{10}(z)}\,P_{10}(\mathbf{r},z)
=
\frac{4\,T_\star(z)}{27\,\bar T_\gamma(z)\,A_{10}(z)}\,
P_\alpha(\mathbf{r},z).
\label{eq:xalpha_def}
\end{equation}
This has the same structure as the collisional coupling coefficient in Eq.~\eqref{eq:x_coll}, with the Ly$\alpha$-induced transition rate $P_{10}(\mathbf{r},z)$ replacing the collisional rate.
Substituting Eq.~(\ref{eq:Palpha_narrowline}) yields
\begin{equation}
x_\alpha(\mathbf{r},z)
\simeq
\frac{16\pi}{27}\,
\frac{T_\star(z)}{A_{10}(z)\,\bar T_\gamma(z)}\,
\frac{J_\alpha(\mathbf{r},z)}{h\nu_\alpha(z)}\,
\sigma_{\alpha,0}(z)\,
S_\alpha(\mathbf{r},z).
\label{eq:xalpha_expanded}
\end{equation}
Using Eq.~(\ref{eq:sigma_alpha0}), this may also be written as
\begin{equation}
x_\alpha(\mathbf{r},z)
\simeq
\frac{16\pi^2 e^2 g_\alpha}{27\,m_e(z)c}\,
\frac{T_\star(z)}{A_{10}(z)\,\bar T_\gamma(z)}\,
\frac{J_\alpha(\mathbf{r},z)}{h\nu_\alpha(z)}\,
S_\alpha(\mathbf{r},z).
\label{eq:xalpha_explicit}
\end{equation}

As summarized in the middle panel of Fig.~\ref{fig:x_Ts_Tb}, at high
redshifts, $z\gtrsim20$, $\langle x_\alpha\rangle\ll1$, and coupling
to the CMB keeps $\langle T_s\rangle$ close to $\bar T_\gamma$. As the
first radiative sources establish a Ly$\alpha$ background,
$\langle x_\alpha\rangle$ rises and exceeds unity around
$z\simeq13$--$15$. Repeated Ly$\alpha$ scattering then couples
$\langle T_s\rangle$ to the Ly$\alpha$ color temperature, which closely
approaches $\langle T_k\rangle$. Since the gas remains colder than the
CMB during this period, $\langle T_s\rangle$ falls below
$\bar T_\gamma$, deepening the 21\,cm absorption signal.

\subsubsection{Ly$\alpha$ heating}
\label{subsubsec:lya_heating}

Photons that enter the Ly$\alpha$ resonance undergo an enormous number of resonant scatterings with neutral hydrogen in the IGM. 
This induces a net energy exchange between the photons and the gas, producing a Ly$\alpha$ heating term $\epsilon_{\mathrm{Ly}\alpha}$
in Eq.~(\ref{eq_Tk_evolution}).

In each resonant scattering, the finite mass of the hydrogen atom implies a small recoil of the scatterer, which induces a small energy exchange between the radiation field and the gas. 
A Ly$\alpha$ photon carries momentum $p_\gamma(z)=h\nu_\alpha(z)/c$. The resonant scattering induces a small recoil of the atom. The characteristic energy exchange per scattering is therefore set by the kinetic energy associated with a momentum transfer of order $p_\gamma$,
\begin{equation}
\Delta E_{\rm recoil}(z)\sim \frac{p_\gamma^2(z)}{2m_H(z)}
=
\frac{1}{2m_H(z)}
\left(\frac{h\nu_\alpha(z)}{c}\right)^2 \propto
\frac{m_e^2(z)}{m_p(z)}.
\label{eq:Erecoil_from_p}
\end{equation}
The net Ly$\alpha$ heating rate per hydrogen nucleus, with units of ${\rm erg}\,{\rm s}^{-1}$, can be written as
\begin{equation}
\epsilon_{\mathrm{Ly}\alpha}(\mathbf{r},t)
\simeq
x_{\rm HI}(\mathbf{r},z)\,
P_\alpha(\mathbf{r},z)\,
\Delta E_{\rm recoil}(z)\,
\Bigg( 1-\frac{T_k(\mathbf{r},z)}{T_\alpha(\mathbf{r},z)} \Bigg),
\label{eq:eps_Lya_def}
\end{equation}
where the last factor in parentheses is an order-unity factor that accounts for the fact that the actual direction and efficiency of the energy flow depend on how close the radiation field is to kinetic equilibrium with the gas. When $T_\alpha(\mathbf{r},z)>T_k(\mathbf{r},z)$, the radiation field near the Ly$\alpha$ line center is effectively hotter than the gas, and the net energy flow is from photons to atoms, so the gas is heated. In contrast, when $T_\alpha(\mathbf{r},z)<T_k(\mathbf{r},z)$, the net energy flow is reversed and the gas is cooled. Finally, when $T_\alpha(\mathbf{r},z)=T_k(\mathbf{r},z)$, the radiation field is locally in kinetic equilibrium with the gas near the resonance, and the net Ly$\alpha$ heating vanishes even though resonant scattering continues to occur. The simulation does not evaluate Eq.~\eqref{eq:eps_Lya_def}. \FirstClass\ takes the energy that one Ly$\alpha$ photon transfers to the gas from the tables of Ref.~\cite{Reis:2021nqf}. The tables are functions of $T_k$, $T_s$, and $\tau_{\rm GP}$ and were computed at the present-day masses. The code applies the tables only to the stellar Ly$\alpha$ photons of Appendix~\ref{sec:lyman_alpha_flux}. We write $\mathcal{R}(z)=[m_e(z)/m_{e,0}]^2\,[m_{p,0}/m_p(z)]$ for the ratio of $\Delta E_{\rm recoil}(z)$ in Eq.~\eqref{eq:Erecoil_from_p} to $\Delta E_{\rm recoil}$ at the present-day masses. The code multiplies the table value at $T_k/\mathcal{R}(z)$, $T_s/\mathcal{R}(z)$, and $\tau_{\rm GP}\,m_{e,0}\,m_p(z)/[m_e(z)\,m_{p,0}]$ by $\mathcal{R}(z)$.

The corresponding energy deposited per unit decrease in redshift is
\begin{equation}
    \widetilde{\epsilon}_{\mathrm{Ly}\alpha}(\mathbf r,z)
    =
    -\epsilon_{\mathrm{Ly}\alpha}\!\left(\mathbf r,t\right)
    \frac{dt}{dz}.
    \label{eq:Lya_heating_per_redshift}
\end{equation}

As shown by the dashed curves in Fig.~\ref{fig:heat_terms},
$\langle\widetilde{\epsilon}_{\mathrm{Ly}\alpha}\rangle$ is extremely
small at $z\gtrsim20$, when the Ly$\alpha$ background is still weak.
It rises rapidly as the first radiative sources develop and the increasing
Ly$\alpha$ scattering rate enhances the energy transfer to the gas.
Despite this growth, Ly$\alpha$ heating remains several orders of
magnitude smaller than X-ray heating throughout the redshift range shown.
For $\bar\chi_{\rm ini}=\chimin$, the green branch shows the earliest
increase, while the other branches remain close to \LCDM. For
$\bar\chi_{\rm ini}=\chibf$, the enhancement occurs earliest and is
largest for the green branch, followed by the blue branch, whereas the
red branch stays closest to \LCDM.

\section{SKA1-Low Sensitivity Calculation}
\label{app_ska_sensitivity}

In this appendix, we describe the calculation of the SKA1-Low 
uncertainty $\sigma_{\rm SKA}(k,z)$ used in Figs.~\ref{fig:P21_chi}, \ref{fig:chi2_contour}, and \ref{fig:chi2_theoretical_error}. The calculation
is performed with \textsc{21cmSense}~\cite{Murray:2024} using the SKA1-Low
AA4 station layout~\cite{SKAO:2023}.

Let $\mathbf r_i$ and $\mathbf r_j$ denote the positions of two
SKA1-Low stations, each of which is treated as a single interferometric
element.
The baseline associated with the station pair is
\begin{equation}
    \mathbf b_{ij}
    =
    \mathbf r_i-\mathbf r_j .
    \label{eq_baseline_vector}
\end{equation}
We include only station pairs with baseline lengths shorter than 1 km. For a line-of-sight direction $\hat{\mathbf n}$, we choose two orthonormal
basis vectors $\hat{\mathbf e}_u$ and $\hat{\mathbf e}_v$ on the plane
perpendicular to $\hat{\mathbf n}$. The projected baseline coordinates,
measured in units of the observed wavelength, are
\begin{equation}
    u_{ij}
    =
    \frac{\mathbf b_{ij}\cdot\hat{\mathbf e}_u}
    {\lambda_{\rm obs}},
    \qquad
    v_{ij}
    =
    \frac{\mathbf b_{ij}\cdot\hat{\mathbf e}_v}
    {\lambda_{\rm obs}},
    \label{eq_uv_coordinates}
\end{equation}
where $ \lambda_{\rm obs}(z)= (1+z)\lambda_{21}(z)$.
We suppress the station-pair indices below and write
$\mathbf u=(u,v)$. The coordinates $u$ and $v$ span the
two-dimensional instrumental Fourier plane transverse to the line of
sight, conventionally called the $uv$ plane. At a given observing time and frequency, each station pair $(i,j)$ is
associated with a point on this plane.

The distance of this point from the origin of the $uv$ plane is related
to the transverse comoving wavenumber by
\begin{equation}
    k_\perp
    =
    \frac{2\pi|\mathbf u|}{X(z)},
    \label{eq_kperp_u}
\end{equation}
where $X(z)$ is the transverse comoving distance,
\begin{equation}
    X(z)
    =
    \int_0^z
    \frac{c\,dz'}{H(z')}.
    \label{eq_X_def}
\end{equation}

The line-of-sight coordinate is obtained from the observed frequency. The observed frequency is $\nu_{\rm obs}(z) = \nu_{21}(z)/(1+z)$, where in axiodilaton cosmology $\nu_{21}(z)$ varies with redshift, Eq.~\eqref{eq:nu21_mass_scaling}.
The frequency-to-distance conversion factor is defined by
\begin{equation}
    Y(z)
    \equiv
    \left|
    \frac{dX}{d\nu_{\rm obs}}
    \right|
    =
    \frac{c}{H(z)}
    \left|
    \frac{d\nu_{\rm obs}}{dz}
    \right|^{-1}.
    \label{eq_Y_def_general}
\end{equation}
Substituting $\nu_{\rm obs}(z)$ into Eq.~\eqref{eq_Y_def_general} gives
\begin{equation}
    Y(z)
    =
    \frac{c(1+z)^2}
    {H(z)
    \left|
    \nu_{21}(z)
    -
    (1+z)\dfrac{d\nu_{21}(z)}{dz}
    \right|}.
    \label{eq_Y_conversion_axdil}
\end{equation}
We calculate the noise only for the $\Lambda$CDM reference model, for which $\nu_{21}(z)=\nu_{21,0}$ and $\lambda_{\rm obs}(z)=(1+z)\lambda_{21,0}$. Eq.~\eqref{eq_Y_conversion_axdil} then reduces to
\begin{equation}
    Y(z)
    =
    \frac{c(1+z)^2}
    {\nu_{21,0}H(z)}.
    \label{eq_Y_conversion_standard}
\end{equation}
We denote by $\eta$ the Fourier variable conjugate to the observing frequency $\nu_{\rm obs}$. Thus, $\eta$ has units of ${\rm Hz}^{-1}$ and characterizes the inverse frequency scale over which the signal varies.
For a Fourier mode with frequency-space scale
$\Delta\nu_{\rm obs}\sim|\eta|^{-1}$, the corresponding line-of-sight
comoving wavenumber is
\begin{equation}
    k_\parallel
    =
    \frac{2\pi\eta}{Y(z)}.
    \label{eq_kparallel_eta}
\end{equation}
Therefore, the instrumental coordinates $(u,v,\eta)$ are mapped to the
cosmological wavenumbers $(k_\perp,k_\parallel)$, with
\begin{equation}
    k
    =
    \sqrt{k_\perp^2+k_\parallel^2}.
    \label{eq_k_spherical_ska}
\end{equation}

We label an individual $uv\eta$ cell by $\alpha$ and denote it by $(\mathbf u_\alpha,\eta_\alpha)$. This cell is associated with the cosmological wavenumber magnitude $k_\alpha=\sqrt{k_{\perp,\alpha}^2+k_{\parallel,\alpha}^2}$.
For this cell, the thermal-noise contribution to the dimensional
21\,cm power spectrum is
~\cite{Parsons:2011ew, Pober:2013jna}
\begin{equation}
    P_N(\mathbf u_\alpha,\eta_\alpha,z)
    =
    X^2(z)Y(z)\,
    \Omega_{\rm eff}
    \frac{T_{\rm sys}^2(\nu_{\rm obs})}
    {2t_{\mathbf u_\alpha}^{\rm tot}}.
    \label{eq_21cmsense_noise_cell}
\end{equation}
Here $P_N$ has the same units as $P_{21}$,
${\rm mK}^2\,{\rm Mpc}^3$. The quantity $t_{\mathbf u_\alpha}^{\rm tot}$ is the total integration
time of all measurements assigned to the $\mathbf u_\alpha$ cell. The effective beam solid angle
$\Omega_{\rm eff}$ characterizes the angular area of the sky sampled
after the directional response of the stations is taken into account.
Its thermal-noise contribution to the dimensionless power spectrum is
then
\begin{equation}
    \Delta_N^2(k_\alpha,z)
    =
    \frac{k_\alpha^3}{2\pi^2}
    P_N(\mathbf u_\alpha,\eta_\alpha,z).
    \label{eq_DeltaN_cell_def}
\end{equation}

To combine the thermal-noise contributions from the individual
$uv\eta$ cells, we group them into a collection of analysis bins labeled
by $(k,z)$, which we denote by ${\cal B}$. 
We first estimate the characteristic transverse and line-of-sight
wavenumber spacings, $\Delta k_{\perp,{\rm cell}}$ and
$\Delta k_\parallel$, together with the redshift-bin width $\Delta z$.

The effective diameter $D_{\rm eff}=35\,{\rm m}$ characterizes the size
of each SKA1-Low station and sets the $uv$-cell spacings adopted in
\textsc{21cmSense},
$\Delta u=\Delta v=D_{\rm eff}/\lambda_{\rm obs}(z)$. 
The corresponding
transverse wavenumber spacings are
\begin{equation}
    \Delta k_x
    =
    \Delta k_y
    \equiv
    \Delta k_{\perp,{\rm cell}}
    =
    \frac{2\pi D_{\rm eff}}
    {X(z)\lambda_{\rm obs}(z)}.
    \label{eq_kperp_spacing}
\end{equation}

We divide each $8\,{\rm MHz}$ analysis band into $N_{\rm ch}=300$ adjacent frequency channels. The resulting
channel width is
\begin{equation}
    \delta\nu_{\rm ch}
    =
    \frac{8\,{\rm MHz}}{300}
    \simeq
    26.7\,{\rm kHz}.
    \label{eq_frequency_channel_width}
\end{equation}
For each power-spectrum calculation, these $N_{\rm ch}$ adjacent frequency channels form a single analysis band with bandwidth
\begin{equation}
    B
    =
    N_{\rm ch}\delta\nu_{\rm ch}{=}
    8\,{\rm MHz}.
    \label{eq_analysis_bandwidth}
\end{equation}
This frequency interval corresponds approximately to a redshift width
\begin{equation}
    \Delta z
    \simeq
    \frac{B}
    {\left|d\nu_{\rm obs}/dz\right|}.
    \label{eq_redshift_bin_width}
\end{equation}
For reference, using the standard rest-frame 21\,cm frequency, the
redshift width is approximately 
\begin{equation}
\Delta z\simeq(2.48,\ 1.44,\ 0.68)
\quad\text{at}\quad
z=(20,\ 15,\ 10),
\label{eq:zbin}
\end{equation}
respectively.
Using Eq.~\eqref{eq_kparallel_eta} with $\Delta\eta=1/B$, the
corresponding line-of-sight wavenumber spacing is
\begin{equation}
    \Delta k_\parallel
    =
    \frac{2\pi\Delta\eta}{Y(z)}
    =
    \frac{2\pi}{Y(z)B}.
    \label{eq_kparallel_spacing}
\end{equation}
For reference, using the standard rest-frame 21\,cm frequency, this spacing is approximately 
\begin{center}
$\Delta k_\parallel\simeq(3.1,\ 3.5,\ 4.2)\times10^{-2}\,
{\rm Mpc}^{-1}$ at $z=(20,\ 15,\ 10)$, respectively.
\end{center}

In addition to the mode spacing, we next consider the lower limit on
the allowed values of $|k_\parallel|$ imposed by foreground
contamination.
Let $\hat{\mathbf n}$ denote the pointing direction. We denote the
magnitude of the projection of the station-separation vector
$\mathbf b_{ij}$ onto the plane perpendicular to $\hat{\mathbf n}$ by
$b_\perp$, with the station-pair indices suppressed. From
Eq.~\eqref{eq_uv_coordinates}, it follows that
$b_\perp=\lambda_{\rm obs}|\mathbf u|$.
Foreground emission arriving from a direction $\hat{\mathbf s}$ forms
an angle $\theta$ with the pointing direction, where
$\cos\theta=\hat{\mathbf s}\cdot\hat{\mathbf n}$. When foreground
emission is considered over the entire visible sky out to the horizon,
the maximum angular separation is $\theta_{\max}=\pi/2$. The largest
geometric arrival-time difference within this angular region is
\begin{equation}
    \tau^{\rm hor}
    =
    \frac{b_\perp}{c}.
    \label{eq_horizon_delay}
\end{equation}
Since $\eta$ is Fourier conjugate to the observing frequency, a
geometric delay $\tau$ corresponds to a Fourier mode with
$|\eta|=\tau$. Substituting the horizon delay
$\tau^{\rm hor}$ into Eq.~\eqref{eq_kparallel_eta}, and using
$b_\perp=\lambda_{\rm obs}|\mathbf u|$, we obtain
\begin{equation}
    k_\parallel^{\rm hor}(k_\perp,z)
    =
    \frac{2\pi\tau^{\rm hor}}{Y(z)}
    =
    \frac{2\pi b_\perp}{cY(z)}
    =
    \frac{X(z)\lambda_{\rm obs}(z)}
    {cY(z)}k_\perp
    \equiv
    {\cal S}_{\rm hor}(z)k_\perp,
    \label{eq_horizon_boundary}
\end{equation}
where
\begin{equation}
    {\cal S}_{\rm hor}(z)
    =
    \frac{X(z)\lambda_{\rm obs}(z)}
    {cY(z)}
    \label{eq_horizon_slope}
\end{equation}
is the dimensionless slope of the horizon-delay boundary in the
$(k_\perp,k_\parallel)$ plane. In the moderate foreground treatment,
only modes satisfying 
\begin{equation}
    |k_\parallel|
    \geq
    k_\parallel^{\rm hor}(k_\perp,z)
    +
    0.1\,h\,{\rm Mpc}^{-1}
    \label{eq_moderate_foreground_cut}
\end{equation}
are retained. The additional term extends the excluded region by
$0.1\,h\,{\rm Mpc}^{-1}\simeq0.068\,{\rm Mpc}^{-1}$, providing an additional buffer.

In the optimistic foreground treatment, foreground emission is assumed
to affect the measurement only within the narrower angular region where
the station has appreciable sensitivity. For this treatment,
$\theta_{\max}$ is determined by the angular extent of the station
response and satisfies $\theta_{\max}<\pi/2$. The retained modes satisfy
\begin{equation}
    |k_\parallel|
    \geq
    {\cal S}_{\rm hor}(z)
    \sin\theta_{\max}\,k_\perp.
    \label{eq_optimistic_foreground_cut}
\end{equation}
Since $\sin\theta_{\max}<1$, this lower boundary lies below the one used
in the moderate treatment, allowing more modes to contribute to the
sensitivity calculation. The value of $\theta_{\max}$, $0.61\,\lambda_{\rm obs}(z)/D_{\rm eff}\simeq2.3^\circ$--$4.4^\circ$ for $z=10$--$20$, and the resulting
boundary are calculated by \textsc{21cmSense} from the adopted station
response.

The finite channel width $\delta\nu_{\rm ch}$ in Eq.~\eqref{eq_frequency_channel_width} sets an upper limit on the
allowed values of $|k_\parallel|$,
\begin{equation}
    |k_\parallel|
    \leq
    k_{\parallel,{\rm max}}(z)
    =
    \frac{\pi}
    {Y(z)\delta\nu_{\rm ch}}.
    \label{eq_kparallel_nyquist}
\end{equation}
For $\delta\nu_{\rm ch}\simeq26.7\,{\rm kHz}$, this limit is approximately
$6.4$, $5.3$, and $4.6\,{\rm Mpc}^{-1}$ at $z=10$, $15$, and $20$,
respectively.

In \textsc{21cmSense}, the one-dimensional $k$ bins are constructed
using the $|k_\parallel|$ grid. The lower limit of the allowed
$|k_\parallel|$ range is set by
Eq.~\eqref{eq_moderate_foreground_cut} or
\eqref{eq_optimistic_foreground_cut}, for the moderate and optimistic treatments respectively, and the upper limit by Eq.~\eqref{eq_kparallel_nyquist}, while the bin spacing is
given by $\Delta k_\parallel$ in
Eq.~\eqref{eq_kparallel_spacing}.

The redshift-bin width associated with the analysis bandwidth
$B=8\,{\rm MHz}$ is given by Eq.~\eqref{eq_redshift_bin_width}.
Requiring the full frequency interval to lie within the SKA1-Low
observing range restricts its central frequency to $54\,{\rm MHz} \leq \nu_{\rm obs} \leq 346\,{\rm MHz}$.
The corresponding redshift range is
\begin{equation}
    z_{\rm min}
    =
    \frac{\nu_{21,0}}{346\,{\rm MHz}}-1
    \simeq 3.1,
    \qquad
    z_{\rm max}
    =
    \frac{\nu_{21,0}}{54\,{\rm MHz}}-1
    \simeq 25.3.
\end{equation}

Together with the $k$ bins defined above, these redshift bins define the
complete set of analysis bins,
\begin{equation}
    {\cal B}
    \equiv
    \left\{(k_p,z_q)\right\},
\end{equation}
where $p$ and $q$ denote the $k$- and redshift-bin indices,
respectively.

For each analysis bin $(k_p,z_q)\in{\cal B}$, we denote the set of
$uv\eta$-cell indices assigned to that bin by
${\cal C}_{pq}$. The corresponding cell-level thermal uncertainties in Eq.~\eqref{eq_DeltaN_cell_def} are
combined in inverse quadrature, giving
\begin{equation}
    {\sigma^{\rm th}_{\rm SKA}}(k_p,z_q)
    =
    \left[
    \sum_{\alpha\in{\cal C}_{pq}}
    \frac{1}
    {\left[\Delta_N^2(k_\alpha,z_q)\right]^2}
    \right]^{-1/2}.
    \label{eq_sigma_SKA_bin}
\end{equation}
Eq.~\eqref{eq_sigma_SKA_bin} is the thermal-noise part of $\sigma_{\rm SKA}$. The $\sigma_{\rm SKA}$ used in this work also includes the sample variance of the $\Lambda$CDM 21\,cm power spectrum. \textsc{21cmSense} adds the two parts for each wavenumber pair $(k_\perp,k_\parallel)$ before the sum over the bin. Let ${\cal A}(k_\perp,k_\parallel)$ be the set of cells in ${\cal C}_{pq}$ with transverse wavenumber $k_\perp$ and line-of-sight wavenumber $\pm k_\parallel$. For each pair, the thermal part is
\begin{equation}
    \sigma^{\rm th}(k_\perp,k_\parallel,z)
    =
    \left[
    \sum_{\alpha\in{\cal A}(k_\perp,k_\parallel)}
    \frac{1}
    {\left[\Delta_N^2(k_\alpha,z)\right]^2}
    \right]^{-1/2},
    \label{eq_sigma_th_pair}
\end{equation}
and the sample-variance part is
\begin{equation}
    \sigma^{\rm sv}(k_\perp,k_\parallel,z)
    =
    \frac{\Delta^2_{21,\Lambda{\rm CDM}}(k,z)}
    {\sqrt{N_s(k_\perp,k_\parallel)}},
    \label{eq_sigma_sv_pair}
\end{equation}
where $k=\sqrt{k_\perp^2+k_\parallel^2}$. The number of independent samples $N_s(k_\perp,k_\parallel)$ is the number of cells in ${\cal A}(k_\perp,k_\parallel)$ times the number of independent sky fields observed per day. \textsc{21cmSense} adds the two parts,
\begin{equation}
    \sigma(k_\perp,k_\parallel,z)
    =
    \sigma^{\rm th}(k_\perp,k_\parallel,z)
    +
    \sigma^{\rm sv}(k_\perp,k_\parallel,z).
    \label{eq_sigma_pair}
\end{equation}
Let ${\cal P}_{pq}$ be the set of pairs $(k_\perp,k_\parallel)$ of the cells in ${\cal C}_{pq}$. The pairs are combined in inverse quadrature,
\begin{equation}
    \sigma_{\rm SKA}(k_p,z_q)
    =
    \left[
    \sum_{(k_\perp,k_\parallel)\in{\cal P}_{pq}}
    \frac{1}
    {\sigma^2(k_\perp,k_\parallel,z_q)}
    \right]^{-1/2}.
    \label{eq_sigma_SKA_total}
\end{equation}

The thermal uncertainty in Eq.~\eqref{eq_21cmsense_noise_cell} is
controlled primarily by the total integration time
$t_{\mathbf u_\alpha}^{\rm tot}$ and the system temperature $T_{\rm sys}$.
We consider total observing times of
\begin{equation}
    t_{\rm obs}
    =
    833\times6\,{\rm hr}
    \simeq
    5000\,{\rm hr}
    \quad\text{and}\quad
    1667\times6\,{\rm hr}
    \simeq
    10000\,{\rm hr}.
    \label{eq_tobs_SKA}
\end{equation}
Both observing times are adopted for each of the moderate and optimistic forecast scenarios described below. As the projected station-pair coordinates evolve during each observing session,
\textsc{21cmSense} assigns the AA4 station-pair tracks to the $uv$ grid and
accumulates the observing time associated with each cell. This procedure
determines the cell-dependent integration time
$t_{\mathbf u_\alpha}^{\rm tot}$ in Eq.~\eqref{eq_21cmsense_noise_cell}. The detailed gridding of the station-pair tracks
and the calculation of $t_{\mathbf u_\alpha}^{\rm tot}$ follow the
\textsc{21cmSense} sensitivity algorithm; see
Refs.~\cite{Pober:2013jna, Murray:2024}.

The thermal noise in Eq.~\eqref{eq_21cmsense_noise_cell} depends
quadratically on the system temperature, $T_{\rm sys}$. We calculate
$T_{\rm sys}$ using the SKA1-Low receiver model,\footnote{See the
SKA1 Imaging Science Performance document, available at
\url{https://indico.skatelescope.org/event/270/attachments/1919/2406/SKA1_Science_PerformanceRevA_draft2.pdf}.}
\begin{equation}
    T_{\rm sys}(\nu_{\rm obs})
    =
    40\,{\rm K}
    +
    1.1T_{\rm sky}(\nu_{\rm obs}).
    \label{eq_Tsys_SKA}
\end{equation}
The constant term represents the receiver noise of the instrument, while the second term describes the effective sky-noise contribution received through the station beam. At the low observing frequencies, the sky noise is dominated by diffuse Galactic synchrotron emission.

We consider two models for this sky contribution.
For the moderate forecast, we use the sky-temperature model
\cite{Thompson:2017isra,Mesinger:2016ddl,Greig:2015qca}
\begin{equation}
    T_{\rm sky}(\nu_{\rm obs})
    =
    60\,{\rm K}
    \left(
    \frac{\nu_{\rm obs}}{300\,{\rm MHz}}
    \right)^{-2.55},
    \label{eq_Tsky_moderate}
\end{equation}
whereas for the optimistic forecast, we use
\cite{DeBoer:2016tnn,Mason:2022obt}
\begin{equation}
    T_{\rm sky}(\nu_{\rm obs})
    =
    120\,{\rm K}
    \left(
    \frac{\nu_{\rm obs}}{150\,{\rm MHz}}
    \right)^{-2.55}.
    \label{eq_Tsky_optimistic}
\end{equation}
For the optimistic forecast, the receiver temperature is a constant $100\,{\rm K}$, so that $T_{\rm sys}=100\,{\rm K}+T_{\rm sky}(\nu_{\rm obs})$.

\bibliographystyle{JHEP}
\bibliography{draft}

\end{document}